
\input harvmac
\def\figflag{I}
\global\newcount\figno \global\figno=1
\newwrite\ffile
\def\pfig#1#2{Fig.~\the\figno\pnfig#1{#2}}
\def\pnfig#1#2{\xdef#1{Fig. \the\figno}%
\ifnum\figno=1\immediate\openout\ffile=figs.tmp\fi%
\immediate\write\ffile{\noexpand\item{\noexpand#1\ }#2}%
\global\advance\figno by1}

\def\tfig#1{Fig.~\the\figno\xdef#1{Fig.~\the\figno}\global\advance\figno by1}

\def\figI{I}
\newdimen\tempszb \newdimen\tempszc \newdimen\tempszd
\newdimen\tempsze
\ifx\figflag\figI
\input epsf
\def\epsfsize#1#2{\expandafter\epsfxsize{
 \tempszb=#1 \tempszd=#2 \tempsze=\epsfxsize
     \tempszc=\tempszb \divide\tempszc\tempszd
     \tempsze=\epsfysize \multiply\tempsze\tempszc
     \multiply\tempszc\tempszd \advance\tempszb-\tempszc
     \tempszc=\epsfysize
     \loop \advance\tempszb\tempszb \divide\tempszc 2
     \ifnum\tempszc>0
        \ifnum\tempszb<\tempszd\else
           \advance\tempszb-\tempszd \advance\tempsze\tempszc \fi
     \repeat
\ifnum\tempsze>\hsize\global\epsfxsize=\hsize\global\epsfysize=0pt\else\fi}}
\epsfverbosetrue
\fi
\def\ifigure#1#2#3#4{
\midinsert
\vbox to #4truein{\ifx\figflag\figI
\vfil\centerline{\epsfysize=#4truein\epsfbox{#3.eps}}\fi}
\narrower\narrower\noindent{\bf #1:} #2
\endinsert
}

\Title{\vbox{\hbox{HUTP--93/A025,}}
       \vbox{\hbox{RIMS--946,}} \vbox{\hbox{SISSA--142/93/EP}}      }
{\vbox{\centerline{Kodaira--Spencer
Theory of Gravity} \vskip .1in \centerline{and} \vskip .1in
\centerline{Exact
Results for Quantum String Amplitudes}}}
\vskip .2in

\centerline{M. Bershadsky, S. Cecotti,\foot{On leave from
SISSA--ISAS and INFN sez. di Trieste, Trieste, Italy.\hfill\break
Present Address: Direzione Regionale per
 le Autonomie Locali, via A. Caccia 17, 33100 Udine, Italy.}
H. Ooguri\foot{Permanent Address:
RIMS, Kyoto University, Kyoto 606-01,
Japan.} and C.
Vafa}
\vskip .2in \centerline{Lyman Laboratory of Physics, Harvard University}
\centerline{Cambridge, MA 02138, USA}
\vskip .3in
We develop techniques to compute higher loop string amplitudes for
twisted $N=2$ theories with $\hat c=3$ (i.e. the critical case).
An important ingredient is the discovery of an anomaly at every genus
in decoupling of BRST trivial states, captured to all
orders by a master anomaly equation.  In a particular realization
of the $N=2$ theories, the resulting string field theory is equivalent
to a topological theory in six dimensions,  the
Kodaira--Spencer theory, which
may be viewed as the closed string analog of the Chern--Simon theory.
Using the mirror map this leads to computation of the `number'
of holomorphic
curves of higher genus curves in Calabi--Yau manifolds.
It is shown that topological
amplitudes can also be reinterpreted as computing
corrections to superpotential terms appearing in the
effective 4d theory resulting from compactification
of standard 10d superstrings on the corresponding $N=2$ theory.
Relations with $c=1$ strings are also pointed out.

\Date{9/93}

\centerline{\bf{Table of Contents}}
\vskip .1in

\line{1.~~Introduction\hfil}

\line{2.~~Review of twisted $N=2$ theories\hfil}
\itemitem{\it 2.1~}{\it Vacuum geometry and twisting of $N=2$ theories}
\itemitem{\it 2.2~}{\it Examples}
\itemitem{\it 2.3~}{\it Special geometry and Calabi--Yau $3$--folds}
\itemitem{\it 2.4~}{\it Coupling twisted $N=2$ theory to gravity}
\itemitem{\it 2.5~}{\it Properties of $n$--point functions and the
holomorphicity paradox}
\itemitem{\it 2.6~}{\it Canonical coordinates and special coordinates}

\line{3.~~Holomorphic anomaly\hfil}
\itemitem{\it 3.1~}{\it Holomorphic anomalies of partition functions}
\itemitem{\it 3.2~}{\it Holomorphic anomalies of correlation functions}

\line{4.~~Comments on the open string version\hfil}
\itemitem{\it 4.1~}{\it $tt^*$ in the open string case}
\itemitem{\it 4.2~}{\it Holomorphic anomaly at one--loop}
\itemitem{\it 4.3~}{\it Holomorphic anomaly at higher loops}

\line{5.~~What are the topological amplitudes computing? \hfil}
\itemitem{\it 5.1~}{\bf Kodaira--Spencer theory as a string field theory of
B--model}
\itemitem{\it 5.2~}{\it Deformations of complex structure}
\itemitem{\it 5.3~}{\it Kodaira--Spencer theory as the string field theory}
\itemitem{\it 5.4~}{\it BV formalism and closed string field theory}
\itemitem{\it 5.5~}{\it Open string field theory}
\itemitem{\it 5.6~}{\bf Kodaira--Spencer theory at one--loop}
\itemitem{\it 5.7~}{\it Holomorphic Ray--Singer torsion}
\itemitem{\it 5.8~}{\it KS theory at one--loop and RS torsion}
\itemitem{\it 5.9~}{\it One--loop topological open string amplitudes}
\itemitem{\it 5.10~}{\bf Geometrical information encoded in $F_g$ for
the A--model}
\itemitem{\it 5.11~}{\it Contribution to $F_g$ from an isolated genus g curve}
\itemitem{\it 5.12~}{\it Contribution to $F_g$ from a continuous family of
curves}
\itemitem{\it 5.13~}{\it Contribution to $F_g$ from constant maps}

\vfill
\eject

\line{6.~~Solutions to the anomaly equation and the Feynman rules for $F_g$
\hfil}
\itemitem{\it 6.1~}{\it Feynman rules at $g=2, 3$}
\itemitem{\it 6.2~}{\it Feynman rules for arbitrary $g$}
\itemitem{\it 6.3~}{\it Construction of propagators}

\line{7.~~Examples --- the experimental evidence \hfil}
\itemitem{\it 7.1~}{\it Orbifold}
\itemitem{\it 7.2~}{\it Quintic}
\itemitem{\it 7.3~}{\it Other models}

\line{8.~~Physical implications of topological amplitudes\hfil}
\itemitem{\it 8.1~}{\it Type II superstring interpretation}
\itemitem{\it 8.2~}{\it Open superstring interpretation}
\itemitem{\it 8.2~}{\it Threshold corrections for heterotic strings}

\line{9.~~Open Problems\hfil}

\line{10.~~Appendix A. The bubbled torus \hfil}

\line{11.~~Appendix B. Further analysis on the master anomaly equation\hfil}
\vfill
\eject

\newsec{Introduction}

Despite the fact that string theory has been investigated
very intensively in particular in the past decade,
many of its fundamental principles and symmetries
remain as elusive as ever.
This lack of understanding of the fundamental principles
renders questions of selection of vacua and
 non--perturbative aspects of string theory
out of reach.  Actually
the problem runs deeper:
Not only the problem has to do with understanding
the underpinnings of string theory, but also not even
many {\it perturbative} computations are practical,
even though in principle many should be computable.

There are some exceptions to the above:  First of all,
thanks to the matrix models and topological theories, for
non-critical strings with dimension $d\leq 2$ one can compute
the partition function of the string theory to all order in
perturbation theory summarized as solutions to interesting
equations belonging to integrable hierarchies.  Nevertheless such
computations are usually viewed as toy models, not necessarily of relevance
to more realistic critical string theories.
One of the more useful results coming from these theories was
the realization that there is an alternative
topological string reformulation of bosonic strings.
The word topological signifies the fact that
in these theories, before coupling to gravity,
the correlation functions are independent of the worldsheet metric.
Actually the topological framework is more general
than the conventional view of bosonic strings
as there are some topological string theories which
do not correspond to bosonic strings formulated
as matter coupled to gravity.  To see this, one
has to note that the most interesting subclass of topological
theories can be obtained by {\it twisting} an $N=2$ superconformal
theory, and in such case, the string BRST operator will correspond
to the supercharge $Q=G^+$ and the $b$ operator will correspond
to the supercharge $G^-$.  That this is more
general than the usual bosonic string set up is easy to see from the fact
that for the standard
formulation of bosonic string the $b$-cohomology is trivial
but for the $N=2$ theory the $G^-$ cohomology is generally non-trivial.
The question naturally arises as to whether the non-triviality
of the $b$--cohomology introduces new phenomena for bosonic strings.
We will see in this paper that the non-triviality of the
$b$--cohomology has dramatic consequences in string theory.
The $b$--cohomology elements
can be used to form $Q$--trivial perturbations of the theory,
that nevertheless do not decouple.  In other words, we shall find
an {\it anomaly in decoupling of BRST trivial states from the
physical amplitudes}.

Anomalies of various kinds have played a key role
in the development of quantum field theories and string
theory.  The existence of anomalies
means that a computation that on formal grounds would be expected
to be zero turns out to be non--zero due to subtleties of the
quantum field theory (QFT) in question.
For example,
the famous $U(1)$ chiral anomaly, explains why the
mass of the meson singlet in massless QCD is
non-vanishing, and the existence of conformal anomalies in 2d QFT's leads
to the fact that the critical dimension of string theory
is 26 (or 10) rather than 0.
The anomalies are in one way or another related to topological
aspects of the theory in question and have been one of the
most fruitful areas of interaction between physics and mathematics.
All these anomalies can be related to index computations
in mathematics which can in turn be effectively understood
in the physical set up in terms of
1d supersymmetric sigma--models. The
topological strings obtained from twisting supersymmetric sigma--model
may be viewed as a more fancy 2d version of these index theorems
which combine the geometry of moduli of Riemann surfaces with
the geometry of target space.  Viewed in this way, it is perhaps
ironic that the
very object usually employed to compute anomalies has itself anomalies!

Topological string theories that are obtained from twisting
an $N=2$ superconformal theory have a central charge
${\hat c}$ which can be viewed as the complex dimension of these
theories.  It turns out that topological string partition functions
vanish for all genus
(except $g=1$) unless the critical dimension ${\hat c=3}$
is achieved.  As far as topological theories obtained from
twisting {\it unitary} $N=2$ theories are concerned there
are very few other cases of interest.  In particular
for unitary twisted theories with ${\hat c}>3$
(with integral charges) not only the
partition function, but all the correlation functions vanish as well.
For ${\hat c <3}$ one
must make special choices of operator insertions
to have non-vanishing amplitudes.
It is thus clear that the most
interesting case is the case of ${\hat c} =3$.

There are two other reasons to be interested in this particular value
of ${\hat c}$.  One reason is that perhaps the most interesting
case of non-critical bosonic string corresponds to strings
propagating in 2 dimensions, which is the
 lower critical dimension of bosonic strings, and this turns out to be related
to a topological theory with ${\hat c}=3$.  The other reason
to be interested in this particular value of $\hat c$ is that
in constructing more or less realistic superstring models
compactifying from 10 dimensions
down to 4, one has to introduce a 6 dimensional internal theory
with ${\hat c}=3$ such as is the case for a Calabi--Yau $3$-fold.  It is thus
exciting that the critical
topological theory may be related to more or less realistic
string compactifications and indeed we will see that the {\it topological
amplitudes of} ${\hat c}=3$ {\it topological string theories effectively
compute superpotential terms in the effective low energy field theory of
$4$--dimensional theories obtained by compactifying superstring on the
corresponding internal theory}.  This is an exciting link which
thus makes the computation of topological amplitudes more than just an
academic exercise.

As if these are not enough reasons to consider critical topological
string theories there are many more:
In a particular realization of the critical topological strings,
the classical limit of the string field theory turns out to describe
the classical deformation of the complex
structure of Calabi--Yau
manifolds (and the related variation of Hodge structure),
i.e. the Kodaira--Spencer theory.  This relation can be
summarized by writing an action whose classical solution
correspond to all possible deformation of the complex structure
of the Calabi--Yau manifold.  This field theory we call the
{\it Kodaira--Spencer (KS) theory of gravity}.  It is a
gravitational theory in $6$ real dimensions with vacua
being Calabi--Yau $3$--folds and
which gauges the complex structure of the manifold.
The {\it Kodaira--Spencer
 theory can be viewed as the closed string field theory for the
critical topological string on a Calabi--Yau}.  This is
thus a rather simple realization of a closed string field theory
which may be helpful for further understanding of closed string
field theory in more general cases.
One can also consider the quantum Kodaira--Spencer theory, i.e.
the higher loops on it which are the same as the partition
function at higher genus of topological strings.
In particular at one--loop the partition function can be related
to an appropriate combination of determinants of various operators
which turns out to be related to the Ray--Singer holomorphic torsion.
In this context the anomaly in decoupling of BRST--trivial states
at one--loop becomes identical to the Quillen anomaly.
Thus the higher genus anomaly that we have found in the
string set up may be viewed as generalization of the
holomorphic Quillen anomaly
for the Kodaira--Spencer theory to higher loops.  As far as we
know no analog of Quillen type anomaly was previously known
for higher loops, and our derivation of the anomaly relies heavily
on string theory techniques.

The partition function of the critical topological strings in
 another realization, which
is the mirror transformed version of the KS theory, at the classical level
`counts' the number of holomorphic maps from
sphere to the Calabi--Yau manifold.
The counting of holomorphic
maps from Riemann surfaces of genus $g$ gets  `mirror mapped' to the
$g$--th
loop computation in the quantum Kodaira-Spencer theory.

The open string version of the critical topological string theories
are also rather interesting.  In particular, in
one version of these theories (the `A' version) the string field theory
one obtains is the
ordinary Chern-Simon theory in $3$ real dimensions.  By
mirror map these should be related (in certain cases) to computation
of open strings on $3$--complex--dimensional Calabi--Yau manifolds.

In this paper we develop techniques for computations of correlation
functions of
twisted $N=2$ theories coupled to gravity with
${\hat c}=3$.    The fact that one can actually compute
the integral of certain realistic string amplitudes over the moduli
space and write the answer in a closed form is a
pleasant surprise.  In particular
in more or less realistic string theories there are
no known computations beyond tree and one--loop level
that can be done in the string theory.  For example
the bosonic string theories do not make sense beyond
tree level
(the amplitudes diverge due to tachyons). Superstring theories
have vanishing partition
function (due to supersymmetry) to all orders in perturbation
theory, and there is a general method to formulate the computations
of scattering amplitudes
(modulo subtleties with the question of integration over
supermoduli space). But no method to explicitly carry them out
exists in these theories.
In fact there is no reason why they should be simple
at all given the presumption that essentially all infinitely
many stringy modes
should be relevant for such computations, and thus probably
it is too much to expect an exact solution.

The idea which leads to the computation of these amplitudes is as follows:
Formally the partition function at genus $g$ of these theories
would depend holomorphically on parameters $t^i$ which characterize
moduli of the theory.  However we will find that, even though the
conjugate fields, which are in the $b$-cohomology, are $Q$
trivial, they do not decouple, and so we end up with
${\overline \partial}_i F_g \not=0$.  The
fact that the integrand of this dependence is a total derivative on
moduli space allows us to go to the boundary to pick up the contributions
which will thus involve lower genus computations. In this way
we get a recursive equation which we solve for the ${\overline t}^i$
dependence of $F_g$.  The holomorphic dependence cannot be fixed
from the anomaly equation alone, however from the fact that they
are modular forms of appropriate weight, and
making heavy use of the properties
of $F_g$ at the boundary of the moduli space
(making use of the KS theory)
they can be determined up to a few constants.
In explicit examples, in particular the quintic
$3$--fold, for low genus we fix these constants
using the mirror map by using the interpretation of leading
terms as counting holomorphic maps.

The organization of this paper is as follows:  In section 2 we
review aspects of $N=2$ theories and their twisting.
This is a basic section for review of old material but some
of it from a new light.  This includes
a review of the geometry of vacua captured
by $tt^*$ equations (which in a special case corresponds
to special geometry).  We also discuss examples of $N=2$
theories obtained from sigma--models
on Calabi--Yau manifolds.   These theories
admit two different ways to twist, depending on whether
the physical fields correspond
to deformation of K\"ahler classes ($A$--model) or
complex structure ($B$--model).  We also discuss coupling of
these theories to gravity, i.e. how to get string theories from them,
and the notion of the critical dimension.
We also point out why the topological partition functions are sections
of line bundles and why inserting physical fields corresponds to taking
covariant derivatives.  Also we show how one can choose {\it topological
coordinates} for moduli (as well as a trivialization for the line
bundle) so that as far as topological observables
are concerned we can replace covariant derivatives with ordinary ones.
This turns out to explain the ansatz used in construction of mirror map.

In section 3 we derive the basic anomaly equation.  This includes
the anomaly both for partition function as well as correlation functions.
It is shown how this equation can be rewritten as a master equation for
the full partition function (i.e. summed over all $g$) of the theory.
We will see that the integrability of the master equation is
true but non--trivial and is
a consequence of $tt^*$ equations.

In section 4 some aspects of open string theory are discussed.
This includes discussion of $tt^*$ equation in this context, as well
as derivation of anomaly equation at one--loop and aspects
of the anomaly equation at higher loop.

In section 5 we discuss what topological strings compute for the
$A$-- and $B$--models.  In particular for the $B$--model we derive the
Kodaira--Spencer theory as the string field theory for topological
strings.  This includes discussion of the
symmetries of KS theory as well as the background
(in)dependence of it with respect to a
 choice of base point for complex structure of the manifold.
We use the Batalin-Vilkovisky (BV)
formalism to quantize the KS theory.  We also
discuss the one--loop computations of the KS theory and show how
the computation is equivalent to computation of
a particular combination of the Ray--Singer
torsion for the Calabi--Yau manifold.  Also
discussed there is the behaviour of the partition function
of the KS theory near the boundary of moduli space of complex structures
--- a result which will imply that there are only a finite
number of coefficients needed to determine the purely holomorphic dependence
of $F_g$ on moduli parameters.
In the context of $A$--model we show that the topological partition
function in a particular limit
($\overline t \rightarrow \infty$) computes the number of holomorphic
maps (or more generally the Euler class of a
particular bundle on the moduli space of holomorphic maps).  A particularly
important case
discussed there is the contribution of constant maps
to the topological string amplitude.

In section 6 we discuss how to solve the recursive anomaly equation
by introducing an auxiliary space consisting of the massless
modes of the conformal theory.  We show that certain
Feynman graph rules involving fields corresponding to this
auxiliary space can be used to solve the anomaly equation recursively.
The vertices of this theory involve $n$--point functions of lower
genus topological theory, and the propagator involves a covariantly
defined prepotential and its first two derivatives.
These rules can be summarized as a path--integral
(which in our case is just a finite dimensional integral) over the
auxiliary space.  The emergence of this way to solve the anomaly
equation is somewhat mysterious, but we try to understand it in the
context of the Kodaira--Spencer theory.

Section 7 is the experimental verification of the paper.  In that
section we give examples of computation of topological amplitudes,
including orbifolds and the quintic $3$--fold.
In particular for the quintic we compute the genus $2$
partition function explicitly and use the mirror map to translate
it to the `counting' of holomorphic maps from genus $2$ to the quintic.

Section 8 is where topological strings meet realistic
string models.
We show how the partition function of topological
strings can be reinterpreted as particular computations in conventional
type II and open
superstrings compactified on the corresponding internal theory.
In particular we show that the closed and open string versions
of the topological theory compute the dependence of the coefficients
of particular terms in the superpotential on the moduli of the
internal theory.  In the context of open strings in particular this
term will have a bearing on the question of gaugino condensates
and is worth further investigation for its phenomenological implications
for supersymmetry breaking.  Also in this section we show that the
one--loop computation of threshold corrections
for heterotic strings
in the context of $(2,2)$ compactifications
can be directly related to the one--loop amplitude of topological strings.
Using properties established for this amplitude we show that quite
independently of which Calabi--Yau manifold one chooses the
effective unification scale is rather sensitive to the change
of volume of the manifold and the dependence is such as to push
the unification scale up as we increase the volume of Calabi--Yau
from Planck scale.  The sign is fixed by the fact that $c_2>0$
for any Calabi--Yau manifold.

Finally in section 9 we discuss open problems and prospects
for future work.

In appendix A we discuss computation of contribution of bubblings
of sphere to topological amplitudes.  In appendix B we
present some preliminary analysis on the master anomaly equation.

Perhaps it is useful to summarize the organization of this paper
with the following flow chart.

\pageinsert
\vfil
\centerline{\epsfbox{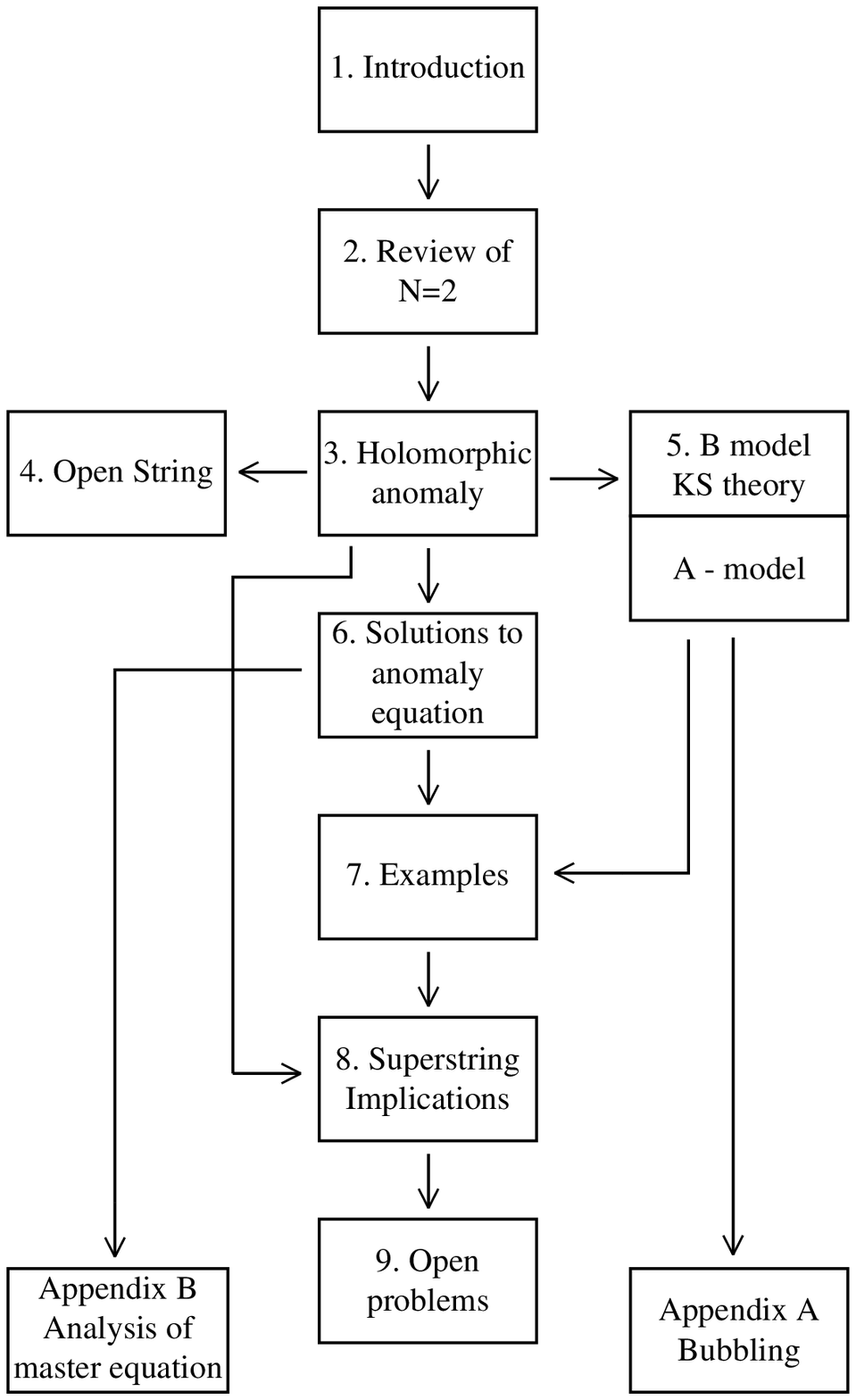}}
\vfil
\endinsert

\vfill
\eject

\newsec{Review of twisted N=2 theories}

In this section we review aspects of twisted $N=2$ theories.  This section
will also serve to set some of the notation we will later use, as well
as motivate some of the issues which are discussed later in the paper.
In subsection 2.1 we discuss the topological nature of $N=2$ theories by
reviewing the structure of its vacua and chiral rings.  We then
discuss the geometry of vacuum bundle as a function of moduli
of $N=2$ theories ($tt^*$ equations).  We then consider examples of
$N=2$ theories in the context of
sigma--models.  Next we specialize some of the discussions
to the Calabi--Yau $3$-folds and review special geometry.
In subsection 2.4 we discuss how to make a string theory out of
twisted $N=2$ theories, which is known as `coupling to topological
gravity'.  Also discussed there is why Calabi--Yau $3$-folds
enjoy a special status among such string models
(i.e. why dimension 3 is critical).  We will also
discuss how the partition function of these theories are
not numbers but rather sections
of bundles, and how inserting chiral fields is equivalent to
taking {\it covariant\/} derivatives of the partition
function; a fact which will be heavily used
in the rest of the paper.  Also discussed there
is the fundamentally important notion of `canonical coordinates'
which turns out to explain the ansatz used in the construction of
mirror maps.
 We will then give a formal argument
for the decoupling of anti-chiral fields from correlation functions, but
argue why this formal argument cannot be correct by showing
its inconsistency
with special geometry, which leads us to the notion of anomalies
discussed in section 3.

\vskip .15in

\subsec{Vacuum geometry and twisting of $N=2$ theories}
$N=2$ supersymmetric theories in 2 dimensions
have a very rich structure.  We will restrict our attention
below to superconformal ones
even though most of what we say can be
easily generalized to massive $N=2$ theories
(we will use the notations and the results of
\ref\lvw{W. Lerche, C. Vafa and N. P.
Warner, Nucl.Phys. B324 (1989) 427}, \ref\cv{S. Cecotti and C. Vafa,
Nucl. Phys. B367 (1991) 359-461} which the reader may consult for
more detail).

Superconformal $N=2$ theories have four supercharges:  Two
conjugate left--movers ($G^\pm$) and two conjugate
right--movers ($\overline G^\pm$), and
two $U(1)$ currents, one left--moving ($J$) and one right--moving
($\overline J$). The $\pm$ sign over $G$'s signify their $U(1)$
charge with respect to the corresponding current.
 Among the important commutation relations of the $N=2$ algebra
is the zero-mode commutators, which we denote by the same
label as the fields:
$$(G^{\pm})^2=0$$
$$\{ G^+,G^-\} =2H_L$$
$$[G^{\pm},H_L]=0$$
where $H_L$ denotes the
left--moving hamiltonian, and similarly for the right--movers.
Also all left--moving operators (anti-)commute with the right--moving ones.
{}From the nilpotency of the $G$'s it follows that we can define
the notion of $G$ cohomologies both for the fields and for the states.
  If we wish to get a finite
dimensional space for cohomology group we need to consider
suitable addition of left-- and right--moving $G's$.  There
are two inequivalent choices, up to conjugation, and they are given
by
$$Q_1=G^++\overline G^+$$
$$Q_2=G^++\overline G^-$$
As far as the cohomology states are concerned $Q_1$ and $Q_2$
and their conjugates all
give rise to the {\it same} space, spanned by the supersymmetric
ground states of the theory ($H_L=H_R=0$).  However as far as the cohomology
of the fields are concerned, i.e., fields which satisfy
$$[Q,\phi ]=0 \qquad \phi \sim \phi+ [Q,\Lambda]$$
even though they can be set into 1-1 correspondence
with ground states of the theory, they are not equivalent
with each other as operators.  The cohomology operators for $Q_1$
are called $(c,c)$, i.e., (chiral, chiral) fields and those of
$Q_2$ are called $(c,a)$, i.e., (chiral, anti-chiral) fields
(where the two entries correspond to the cohomology condition
for left-- and right--moving charges respectively).
Those of $Q_1^\dagger$ and $Q_2^\dagger$ are the conjugate
fields and are called $(a,a)$ and $(a,c)$ fields respectively.
Since the discussion
for the two choices of $Q$'s is essentially identical, as they differ only
by a convention dependent choice
of sign $(\overline J\rightarrow -\overline J)$,
we will restrict ourselves to $Q_1$ and its conjugate.
We will also drop the subscript from $Q_1$ and denote it
simply by $Q$.
Also, to simplify terminology we will call the $(c,c)$ fields
simply as chiral fields and the $(a,a)$ fields as the anti-chiral fields.

Let us choose a basis for chiral fields representing
the $Q$--cohomology by $\phi_i$,
and denote the conjugate anti-chiral fields by $\overline \phi_i$ .  The
$N=2$ algebra implies that the (left,right) dimension of $ \phi_i$,
$(h_i,\overline h_i)$
is half its charge $(q_i,\overline q_i)$
$$(h_i,\overline h_i)={1\over 2}(q_i, \overline q_i)$$
and that the range for the $q_i$ are bounded by the central charge
$\hat c$ of the $N=2$ algebra:
$$0\leq q_i,\overline q_i \leq \hat c$$
The dimensions of the anti-chiral fields $\overline \phi_i$
are the same as $\phi_i$ but their $U(1)$ charge is minus that of $\phi_i$.

The chiral fields form a ring, the {\it chiral ring}, defined by
$$\phi_i \phi_j =C_{ij}^k \phi_k +[Q,\,\cdot\, ]$$
Using the $N=2$ algebra it is easy to see that this definition
of the ring is independent of the points of insertion of the fields
on the worldsheet.
Sometimes we will view $C_{ij}^k$ as a matrix $C_i$ with
component $(C_i)_j^k$.  The corresponding ring for anti-chiral
fields differs only by complex conjugation of $\overline{C_{ij}^k}
=(C_{ij}^k)^*$.

 As is well known,
viewing chiral fields as the first component of a superfield,
we can modify the action
by perturbing with them:
$$t^i\int d^2z\, d^2\theta^+
\phi_i +{\overline t^i}\int d^2z\, d^2\theta^- {\overline
\phi_i}=t^i\int d^2z\, \phi_i^{(2)}+\overline t^i\int d^2z\, \overline
\phi_i^{(2)}$$
where $\phi_i^{(2)}=\{ G^-,[\overline G^-, \phi_i] \}$ and
$t^i$ are complex parameters. If we wish to have
a unitary theory we need $\overline t^i=(t^i)^*$ (later in the
paper we will relax this condition).  It is known
\ref\dix{L. Dixon in Proc. of the 1987 {ICTP Summer Workshop in
High Energy Physics and Cosmology}, Trieste}\ that the only
criterion needed for preserving the conformal invariance is that
$(h_i,\overline h_i)=(1/2,1/2)$, i.e. that the charge of $\phi_i $
be $(1,1)$.

As mentioned before there is a 1-1 correspondence between the chiral
fields and the supersymmetric ground states of the theory, which
follows by general considerations of QFT relating operators to states.
However it is more useful to do this rather explicitly, which along
the way leads to the notion of defining a topological theory.  If
we consider a hemisphere (see \tfig\FigureTwoOne) with the field $\phi_i$
inserted
on it then one is tempted to identify the state obtained
at the boundary of this region by the
path integral, as the cohomologically non-trivial state representing
the supersymmetric ground state corresponding to $\phi_i$.

\ifigure\FigureTwoOne{Inserting the chiral field $\phi_i$ on the
hemisphere and doing the twisted path integral on it will
result in a state $|i\rangle$ at the boundary.  This state
is BRST equivalent to a ground state and can be made
an exact ground state by pulling the neck infinitely long.}{Fig21}{.75}

However this
is not correct:  One reason for this is that the state
we get at the boundary is in the wrong Hilbert space, i.e. the NS
sector.  For the supersymmetric vacuum we need to be in the Ramond sector.
Another difficulty is that we need to argue that the state we get
is annihilated by $Q$ and to do this we have to make sure
that $Q$ is a scalar charge (especially when we have more non-trivial
Riemann surfaces). To solve both these problems one {\it twists} the
supersymmetric theory to obtain a {\it topological} theory,
 by introducing a background gauge field $A$
which couples to the $U(1)$ current \ref\wittop{
E. Witten, Comm. Math. Phys. 117 (1988) 353\semi
E. Witten, Comm. Math. Phys. 118 (1988) 411\semi
T. Eguchi and S. -K. Yang, Mod. Phys. Lett. A5 (1990) 1693}
$$S\rightarrow S+\int J \bar A +\bar J A$$
and one sets $A=\omega /2 $, $\bar A =\overline \omega /2$
where $\omega$ is the spin connection. Introducing this gauge
field has the effect of shifting the spin of charged fields
by half their charge.  Thus $Q$ becomes a scalar charge.  At the
same time $G^-(z)$ and $\overline G^-(\overline z)$ become spin (2,0)
and (0,2) currents respectively.
Introducing this twisting,  on the hemisphere of Fig.1,
has the effect of converting the state obtained at the boundary
 to a state in the Ramond
sector, which is annihilated by $Q$ (where we use the fact that
$\phi_i $ commutes with $Q$).  Also the dimensions of all
the fields will shift by $h\rightarrow h-{q\over 2}$, thus $\phi_i$
become dimension zero and $\overline \phi_i$ become dimension $(1,1)$ for
marginal directions.
We could obtain the exact
ground state, not just a state cohomologically
equivalent to a ground state, simply by doing the path integral on the
hemisphere with the neck pulled infinitely long.  We denote
the corresponding ground state by $|i\rangle$.  There is a canonical
vacuum which is obtained by not inserting any field at all. This
we will denote by $|0\rangle$.
Note that we can write
$$|i\rangle =\phi_i |0\rangle +Q|\,\cdot\,\rangle$$
by using the fact that moving $\phi_i$ to the boundary
is a $Q$--trivial operation.   Also note that
$$\phi_i |j\rangle =(C_i)_j^k|k\rangle +Q|\,\cdot\,\rangle$$
We can consider also the conjugate twist
(the {\it anti-topological} theory) in which $Q^\dagger$
becomes a scalar.  In this case we can parametrize {\it the same
vacua} using the anti-chiral fields, which we denote by $|
\overline i\rangle $.  We must thus have a change of basis
transformation relating the two:
$$\langle \overline i |=\langle j| M_{\overline i}^j$$
which by CPT satisfies $MM^*=1$.
Note that we have thus two natural
inner products, the topological one $\eta$ and the hermitian one
$g$ defined respectively by
$$\eta_{ij}=\langle j|i\rangle$$
$$g_{i\overline j}=\langle \overline  j| i\rangle $$
which satisfy
$$g_{i\overline k}=\eta_{ij}M_{\overline k}^j $$
Also note that in the topological theory
$$\langle \phi_i \phi_j \phi_k \rangle=
\langle 0|\phi_i \phi_j \phi_k |0\rangle
=\langle i|\phi_j |k\rangle =C_{jk}^l\langle i|l\rangle =
C_{jk}^l \eta_{il}=C_{jki}$$
which implies that $C_{ijk}$ is totally symmetric
in indices (for indices of even fermion number).

We are interested in seeing how the structure
of vacua and chiral fields deform as we perturb
the theory by marginal chiral fields.  As discussed
the parameter space is locally given
by $(t^i,\overline t^i)$.  We would like to study the geometry of
the {\it vacuum bundle}, i.e. how the collection of vacuum
states $\{ |i(t,\overline t)\rangle \}$ varies as a function of the
parameters and in particular find the
dependence of $C_{ijk}$,$g$ and $\eta$ on $(t^i,\overline t^i)$.
These are studied in \cv\ with the following results:
First, using
the fact that insertion of anti-chiral fields in the action
modifies the theory by $Q$-trivial terms, it follows that
\eqn\holo{\overline \partial_l C_{ijk}=0}
i.e., $C_{ijk}$ is a symmetric holomorphic function of moduli.
One introduces
a connection on the vacuum bundle so that
$D_i|j\rangle \equiv
(\partial /\partial t^i -A_i)|j\rangle$ and $D_{\overline i}
|j\rangle$ are orthogonal to all the
vacua.  Then the following equations, the $tt^*$ equations, hold
$$[D_i,D_j]=[\overline D_i,\overline D_j]=[D_i,\overline C_j]=
[\overline D_i, C_j]=0$$
$$[D_i,C_j]=[D_j,C_i]\quad [\overline D_i,\overline C_j]=
[\overline D_j,\overline C_i]$$
\eqn\ttstar{[D_i,\overline D_j]=-[C_i,\overline C_j]}
One can also arrange, by a judicious choice of coordinates,
for $\eta$ to be constant \ref\ver{R. Dijkgraaf, H. Verlinde and
E. Verlinde, Nucl. Phys. B352 (1991) 59}.
It is also possible to choose the holomorphic (or topological)
gauge
\foot{This choice of gauge is implicit in eq.\holo.},
in which
$A_{\overline i}=0$ and in which
\eqn\nameit{A_i=-g\partial_i g^{-1}.}
  In this gauge, which
is the natural gauge chosen by the twisted path--integral,
the chiral vacuum states $|i(t)\rangle$ depend holomorphically on the
moduli.  In this gauge the third line of
eq.\ttstar\ can be written as
\eqn\holgauge{
\overline \partial_j (g\partial_i g^{-1})=[C_i,g (C_j)^\dagger g^{-1}].}

We would like to digress slightly to discuss some ambiguities
in the twisted path--integral.  Ambiguities arise in the normalization
of path--integral when there are zero modes to be absorbed.  When
we twist a topological theory by coupling the $U(1)$ current
to a background gauge field, the axial $U(1)$ current ($J+\overline J$)
becomes anomalous, and so there are fermion zero modes to absorb.
In fact to be precise, using the fact that the OPE of $J$'s have
a central term
$$J(z) J(0)\sim {\hat c\over z^2},$$
in genus $g$ the twisting will give rise to a charge violation of
\eqn\chvi{\Delta(q,\overline q)=\hat c (g-1,g-1).}
This ambiguity in how to absorb the zero mode translates to the
ambiguity in defining the normalization of the chiral states $|i\rangle$.
Since they are all related by applying the operators $\phi_i$ on
$|0\rangle$ it suffices to discuss ambiguities for $|0\rangle$.  We can
choose the absorption of fermion zero modes to be consistent with
the holomorphic dependence of $|0\rangle$ on moduli.  But we cannot
completely fix the ambiguity.  Consider the line bundle $\cal L$
over the moduli space, generated by the vacuum state $|0\rangle $.  Then
what we are saying is that a holomorphic choice of normalization
of the twisted path--integral is equivalent to a choice of holomorphic
section of $\cal L$.  Two different normalizations of the path-integral
give differently normalized vacua as:
$$|0\rangle \rightarrow f(t^i)|0\rangle $$
Note that this freedom in redefining the normalization
of $|0\rangle$ holomorphically translates to a change
in the connection $A_{i0}^0\rightarrow A_{i0}^0+\partial_i f$.
Since we have chosen a holomorphic gauge
$$A_{i0}^0 ={\partial_i \langle \overline 0 |0\rangle \over
\langle \overline 0 |0\rangle}=-\partial_i K,$$
where
$${\rm exp}(-K)=\langle \overline 0 |0\rangle .$$
The fact that $|0\rangle $ is a section of $\cal L$ translates
to the statement that the genus zero partition function $Z_0$
(with operators inserted to avoid vanishing by charge conservation)
will be a section of $\CL^2$ (in addition to having
properties induced from insertion of operators).  Similarly
by sewing axioms of topological theory it follows that $Z_g$  is a section of
$\CL^{2-2g}$.
Needless to say, all physically interesting quantities should
be independent of how we choose to fix this normalization
ambiguity.

\vskip .15in

\subsec{Examples}
Even though our discussions in the paper will be for the general
case we will occasionally specialize the results to some interesting
classes of $N=2$
SCFT.  The one which we will use most
in this paper are the supersymmetric sigma--models.  It is known
that the sigma--model on a K\"ahler manifold $M$ give rise to an $N=2$
QFT \ref\alfre{L. Alvarez-Gaume and D. Z. Freedman,
Phys.Lett. 94B (1980) 171}.
The action is given by
$$S=\int d^2z \Big[\omega_{i\overline j}
\partial X^i \overline \partial\overline X^j+\omega^*_{i\overline j}
\overline \partial X^i {\partial}\overline X^j\Big]
 +{\rm Fermionic}~~{\rm terms.}$$
where $\omega_{i\overline j}$ denotes a complexified K\"ahler class.
If we denote an integral basis for $H^{(1,1)}(M,Z)$ by $\omega_i$,
we have
$$\omega =t^i\omega_i$$
and thus $t^i$ parametrize the moduli of this theory.  The fermionic terms
in the action are there to make the above supersymmetric.  Apart from
the kinetic term, the fermionic terms include the four fermion interaction
term
$$\int {R_{i\overline j}}^{k\overline l}\chi^i\overline \chi^j \psi_k
\overline
\psi_l$$
which will prove crucial for us later on.

One
can twist the fermion number, as discussed in the previous section, to
obtain  a topological theory \wittop . The effect on the action is
only to modify the spin of the fermions, making the $\chi$'s scalar,
the $\psi_i$ is a (1,0) form and $\overline \psi_j$ is a (0,1) form.
 To obtain the observables of this
theory it is convenient to go to the large volume limit first
$t^i,\overline t^i\rightarrow \infty$.  In this limit the Hilbert
space of the
theory can be represented by differential forms on $M$
where the (left, right) $U(1)$ charge of the state
can be identified with (holomorphic, anti-holomorphic)
degree of the form.  Moreover
on this Hilbert space we
get the following dictionary
$$G^+\leftrightarrow \partial$$
$$\overline G^+ \leftrightarrow \overline \partial$$
and so $Q_1=G^+ +\overline G^+=d$
and thus the observables $\phi_i$
in this theory are in 1-1 correspondence with the cohomology elements
of $M$ represented by
$$\phi^{(a)} =A^{(a)}_{i_1...i_r{\overline j_1}...{\overline j_s}}
\chi^{i_1} \cdots \chi^{i_r}
\overline {\chi^{j_1}} \cdots \overline {\chi^{j_s}}$$
  The chiral ring for $t,\overline t\rightarrow \infty$ is the
same as the cohomology ring, but for finite $t$'s it is in general
a deformation of the cohomology
ring of $M$, taking into account the holomorphic
instantons from sphere to $M$, i.e.,
rational curves on $M$.  This deformed ring is called the quantum
cohomology ring of $M$.  The precise form of how
the instantons contribute to this ring will very much depend
on the first  Chern class of $M$.  The most interesting case
is when $c_1(M)=0$, i.e., the Calabi--Yau case.  In this case,
which is also the case needed to obtain a conformal theory out
of the sigma--model, the instanton of arbitrary large
degree affect the chiral ring.  In particular if one is
interested in the quantum cohomology ring for a $3$-fold Calabi--Yau,
if we let $i,j,k$ to denote three (1,1) classes, then the ring
structure coefficients are given by
$$C_{ijk}(t)=\sum_r r_ir_j r_k d_{r_1...r_n}{q_1^{r_1}...q_n^{r_n}
\over 1-q_1^{r_1}...q_n^{r_n}}$$
where $n={\rm dim}\, H^{(1,1)}(M)$,  $q_r={\rm exp}(-t^r)$, and
$d_{r_1...r_n}$ are the number of primitive holomorphic
instantons of degree $(r_1,...,r_n)$.  The denominator
is the contribution of multi-coverings of primitive
instantons \ref\cand{P. Candelas, Xenia C. de la Ossa,
Paul S. Green and L. Parkes, Nucl. Phys. B359 (1991) 21}, \ref\aspmor{P. S.
Aspinwall and D. R. Morrison Commun. Math. Phys. 151 (1993) 245-262}.
Note that the structure constants of this ring depend on the
choice of the K\"ahler class, but are independent of the complex
structure of $M$.

Now, as discussed in the previous section superconformal theories,
have two natural rings \lvw , the $(c,c)$ and $(a,c)$ and thus also there
exists two
ways to twist the theory.  In particular in the case of Calabi--Yau
case for which both the fermion number and the axial fermion
number are conserved, we can twist in two different ways,
 depending on which of these rings we
wish to be the physical ring.  The choice discussed above corresponds
to twisting the fermion number current and gives rise to (say) the
$(c,c)$ ring as being the topological one.  The other choice
of twisting, corresponding to axial fermion number twisting has also been
studied \ref\witten{E. Witten, in {\it Essays on Mirror Manifolds},
ed. by S. T. Yau,  International Press, 1992},
\ref\vafa{C. Vafa, in {\it Essays on Mirror Manifolds},
ed. by S. T. Yau,  International Press, 1992}.   Again it turns out to be
easier to study
the model in the large volume limit.  In this limit again the
Hilbert space
can be identified with anti-holomorphic forms wedged with holomorphic
vectors, i.e.,
\eqn\hilb{{\cal H}=\bigoplus_{p,q}\wedge^p \overline T_M^*
\otimes \wedge^q T_M}
where in here and in the following $T_M, {\overline T}_M$ denote the
holomorphic
and anti-holomorphic tangent bundles respectively
and $T^*_M, \overline T^*_M$ denote
the holomorphic and anti-holomorphic cotangent bundles.
We can obtain from this Hilbert space the Hilbert space of forms
simply by contracting the vector indices with holomorphic $n$-form
which always exists for a Calabi--Yau $n$-fold.  This converts
the $(q,p)$ sector above to a differential form of degree $(n-q,p)$.
On this Hilbert space ${\cal H}'$ the dictionary for the supercharges
turn out to be
$$G^+={1 \over 2}({\partial}^\dagger + {\bar \partial})$$
\eqn\dicb{{\bar G}^-={1 \over 2}({\bar \partial} - {\partial}^\dagger)}
Note that $Q_2=G^++\overline G^-\leftrightarrow \overline \partial $
and so again the observables can be identified with the
cohomology elements of $M$.  For later use in the paper
we need also expression for left-- and right--moving fermion numbers.
Unlike the previous case, the fact that \dicb\ mixes the
holomorphic and anti--holomorphic degrees in this non--trivial way,
implies that the fermion numbers in this case are not simply
identified as the left and right degrees of form, as that
would lead to a wrong commutation relation with $G^+$.
To fix this, we should recall \ref\gh{P. Griffiths and J. Harris
{\it Principles of Algebraic Geometry} New York, Wiley, 1978.}\
that for K\"ahler manifolds there is an $sl(2)$ action
on the forms, generated by wedging with the K\"ahler class $k$,
contracting with $k$ which we represent by $k^\dagger $ and
the shifted total degree of the form $(p+q-n)/2$.
Also we have
$$\eqalign{&[k,\partial^\dagger ]=i\overline \partial \qquad [k^\dagger
,\overline
\partial ]=-i\partial^\dagger\cr
&[k^\dagger ,
\partial ]=i\overline\partial^\dagger \qquad [k,\overline\partial^\dagger
]=-i\partial. \cr}$$
Using this we can write $F_{L,R}$  as
\eqn\flrb{F_{L,R}={1\over 2}(ik-ik^\dagger \pm(p-q))}
the reader can check that the right--hand side (r.h.s.)
is CPT odd and has the correct
commutation properties with the $G$'s.

Even though the chiral fields are again in 1-1 correspondence
with the cohomology elements,
the chiral ring for this twisting is very different
from the previous one.   Let $\Omega$ denote the holomorphic $n-$form.
For example
if the Calabi--Yau is a $3$--fold, then
the structure constants for the marginal directions,
which are parametrized by elements of $H^{(2,1)}(M)$ can be written as
$$C_{ijk}=-\int_M \Omega\wedge \partial_i\partial_j\partial_k \Omega$$
Note that these structure constants are independent
of the K\"ahler class so they continue to hold for
finite volume as well. But they do depend on the complex structure
of $M$, which is parametrized by elements of $H^{(2,1)}(M)$ as
will be discussed in great detail later in this paper.

So to summarize,
we see that in the first case of twisting, i.e. the $(c,c)$ twisting,
the topological correlation functions are sensitive to the
K\"ahler class of the manifold and compute the rational curves
in the Calabi--Yau.  This twisting is called the {\it A--twisting}
or the {\it K\"ahler twisting}.  On the other hand, in the case of
$(a,c)$ twisting we see that the topological correlation
functions are only sensitive to the complex structure of the manifold
which is encoded in how the holomorphic three form varies (or the
variation of Hodge structure).  This twisting is called the
{\it B--twisting\/}
or the {\it complex twisting}.

\vskip .15in

{\bf NOTE}: For convenience of keeping the same notation in the rest
of the paper when we deal with the $B$-- or $A$--model we denote
the supercharge always by $Q=G^+ +\overline G^+$ by a trivial
change of conventions on the right--moving $U(1)$ charge
if necessary.  This
will not cause confusion as we rarely talk about both models
at the same time.

\vskip .15in

\subsec{Special geometry and Calabi--Yau $3$--folds}

In a unitary superconformal theory there is only
one chiral primary with $q=0$, namely the identity operator $1$. We consider
the normalized metric\foot{Here $\beta$ is the perimeter of the circle
used to define the states $|i\rangle$ (cf. Fig. 1).}
\eqn\zamolodchikov{G_{i\bar j}= \left.{g_{i\bar j}\over \langle \bar
0|0\rangle}\right|_{\beta=1},}
(here and in the following the indices $i$, $\bar j$ are restricted to the
marginal directions in coupling constant space whereas indices $a$,
$\bar b$ go through all chiral primaries).
It is easy to see that $G_{i\bar j}$
is equal to the usual Zamolodchikov metric
\ref\zamoM{A.B. Zamolodchikov, JETP Lett. 43 (1986) 730.}
(restricted to the marginal directions).
Indeed the definition of the $tt^\ast$ metric can be
written as
$$\langle \bar b|a\rangle_\beta=\langle R^\dagger_{\bar b}(\beta)
R_a(0)\rangle_{\rm sphere},$$
where $R_a$ is the operator creating the Ramond vacuum $|a\rangle$ out of the
$SL(2,{\bf C})$--invariant vacuum. Now, the topological map
$|a\rangle\rightarrow \phi_a$ differs only for the overall normalization from
the {\it unitary} spectral flow operator $U$ which maps the $R$ sector into the
$NS$ one. Since the unitary operator
$U$ preserves inner products, we have
\eqn\unspfl{\langle \bar\phi_{\bar b}(\beta)\phi_a(0)\rangle_{\rm
sphere}={\langle\bar b|a\rangle_\beta\over \langle\bar 0|0\rangle_\beta},}
where we used that the correct (unitary) normalization of $U$ is just the one
for which $\langle 1\rangle_{\rm sphere}=1$. Eq.\unspfl\ is true for any chiral
primary fields. In the particular case in which $\phi_i$ has charge $1$ (and
dimensions $(1/2,1/2)$) from \zamolodchikov\ we get
\eqn\spdep{\langle \bar\phi_j(z)\phi_i(0)\rangle_{\rm sphere}=
{G_{i\bar j}\over z\bar z}.}
Let $\Phi_i(z,\theta)$ be the $N=2$ chiral superfield whose first component is
$\phi_i$. From\foot{We use the shorthand notation
$\theta_{12}^\pm=\theta^\pm_1-\theta^\pm_2$ and $\tilde z_{12}=
z_1-z_2-\theta^+_1\theta^-_2-\theta^-_1\theta^+_2$.}
\spdep\ and $N=2$ supersymmetry we get
$$\langle\Phi_i(z_1,\theta_1)\ \bar\Phi_j(z_2,\theta_2)\rangle_{\rm
sphere}={G_{i\bar j}\over \tilde z_{12} \tilde
z_{12}^\ast}\left[1+{\theta^-_{12}\theta^+_{12}\over \tilde z_{12} }\right]
\left[1+{\bar\theta^-_{12}\bar\theta^+_{12}\over \tilde z_{12}^\ast}\right].$$
Then, if $\phi^{(2)}_i\equiv \int d^2\theta\, \Phi_i$ is the marginal
operator multiplying the coupling $t^i$ in the action, one has
$$G_{i\bar j}=\langle \phi_i^{(2)}(1)\,
\bar\phi^{(2)}_{\bar j}(0)\rangle_{\rm
sphere},$$
which is the original definition of the Zamolodchikov metric \zamoM.

The Zamolodchikov metric $G_{i\bar j}$ has remarkable geometric properties.
The
most interesting situation is when $\hat c=3$. In this case the
metric $G_{i\bar j}$ satisfies a set of constraints which define the
so--called
{\it special (K\"ahler) geometry}.
A hermitian metric $G_{i\bar j}$ is said to be {\it special K\"ahler} if:
\item{\it i)}{It is a restricted K\"ahler metric, i.e.
a K\"ahler metric such that the corresponding K\"ahler form is $2\pi$ times the
Chern class of a line bundle $\CL$. Locally this means
\eqn\hodgemet{\eqalign{&G_{i\bar j}=\bar\partial_{\bar j}\partial_i K,\cr
&{\rm with}\quad \| 1\|_{\CL}^{\ 2}= e^{-K}\cr} }
where $1$ is a local holomorphic section trivializing $\CL$.}
\item{\it ii)}{There is a holomorphic symmetric tensor $C_{ijk}$ with
coefficients in $\CL^2$ satisfying\foot{Here $D_i$ is covariant both with
respect the Christoffel connection of $G_{i\bar j}$ and the canonical
connection
on the bundle $\CL$,  i.e.
$A_i=-\partial_i K$.}
\eqn\integra{\bar \partial_i C_{jkl}=0 \qquad D_i C_{jkl}=D_j C_{ikl},}
such that the Riemann curvature of $G_{i\bar j}$ reads
\eqn\specialgeo{{R_{i\bar jk}}^l\equiv -\bar\partial_{\bar j}\Gamma_{ki}^l =
G_{k\bar j}\delta_i^{\ l}+ G_{i\bar j}\delta_k^{\ l}-e^{2K} C_{ikn} G^{n\bar n}
C^\ast_{\bar j\bar m\bar n} G^{\bar m l}.}

It follows from the $tt^\ast$ equations that the condition {\it i)} is
satisfied by the Zamolodchikov metric $G_{i\bar j}$ of any critical $N=2$
theory
,
with the bundle $\CL$ identified with the vacuum line bundle defined by the
identity operator. Moreover,
if $\hat c=3$ the condition {\it ii)} holds as well with
\eqn\deftensor{C_{ijk}=\langle 0|\phi_i(\infty)\, \phi_j(1)\,
\phi_k(0)|0\rangle_{\rm top.}.}
$\hat c=3$ is crucial here because only in this case $C_{ijk}$
is non-vanishing for marginal fields (1+1+1=3).

In view of eq.\hodgemet, the first assertion is equivalent to saying that
$G_{i\bar j}$ is K\"ahler with potential
\eqn\Kpot{K=-\log\,\langle \bar 0|0\rangle.}
Let us show this. The index $0$ will denote the identity operator, while
$i,j,...=1,\dots, m$ denote the marginal directions (i.e. chiral primary
fields with charge $1$). Then $U(1)$ charge conservation gives
$$g_{0\bar k}=g_{k\bar 0}=0\hskip 1.2cm \big(g C^\dagger_i g^{-1}\big)_k^{\
0}=0.$$
Let us project the $tt^\ast$ equation in the identity sector
$$-\bar\partial_{\bar j}\partial_i \log\,\langle\bar
1|1\rangle=\big[\bar\partial_{\bar j}(g\partial_i g^{-1})\big]_0^{\ 0}=
(C_i)_0^{\ k}g_{k\bar l}{C^\ast_{\bar j\bar 0}}^{\bar l}g^{\bar 0 0}=g_{i\bar
j}/g_{0\bar 0}\equiv G_{i\bar j},$$
where we used the definition of the identity operator i.e.
\eqn\iddef{C_{i0}^{\ k}=C_{0i}^{\ k}=\delta_i^{\ k}.}
This shows {\it i)}. To show {\it ii)} let us notice that, if $\hat c=3$, the
tensor defined by eq.\deftensor\ has all the required
properties: From \deftensor\ we see that it is a section\foot{Here and below
$T$ denotes the $(1,0)$ tangent bundle of the coupling constant (moduli)
space.} of ${\rm Sym}^3 T\otimes \CL^2$: Indeed $\phi_i$ is an operator valued
section of $T$ while $|0\rangle$ and $\langle 0|$ are Hilbert space--valued
sections of $\CL$.
$C_{ijk}$ is holomorphic because, as we saw in section 2.1, the topological
$3$--point function has this property. Finally, it satisfies the first
condition in \integra\ as a
consequence of  equation \holo.
It also satisfies the second condition
in \integra\ because it differs from the  corresponding equation in
\ttstar\ only by the fact that in $D_iC_j$ the derivative should also act on
the
 $j$
index by the Christoffel connection but that is clearly symmetric in its
indices.
Here it is crucial that the $tt^\ast$
connection
is equal to the Zamolodchikov connection plus the canonical connection for the
bundle $\CL$ as it follows from the equation (compare eqs.\zamolodchikov,
\Kpot)
\eqn\relmets{g_{i\bar j}= e^{-K} G_{i\bar j},}
the definition of the $tt^\ast$ connection \nameit , and the definitions of the
Zamolodchikov and line bundle connections which are given by
$\Gamma_{ki}^l\equiv G_{k\bar m}\partial_i G^{\bar m l}$ and $-\partial_i K$,
respectively.

Now we are ready to show the main identity, eq.\specialgeo.
Using the well--known
formula for the Riemann curvature in K\"ahler geometry, we have
\eqn\longcomp{\eqalign{-{R_{i\bar j k}}^l&=\bar\partial_{\bar j}\big(G_{k\bar
m}\partial_i G^{\bar m l}\big)=\cr
&=\bar\partial_{\bar j}\big(g_{i\bar m}\partial_i g^{\bar m
l}\big)-\big(\bar\partial_{\bar j}\partial_i K\big) \delta_k^{\ l}=\cr
&=[C_i,\bar C_{\bar j}]_k^{\ l}- G_{i\bar j}\, \delta_k^{\ l}=\cr
&=e^{2K} C_{ikn} G^{n\bar n} C^\ast_{\bar j\bar m\bar n} G^{\bar m l}
-G_{k\bar j}\delta_i^{\ l}- G_{i\bar j}\delta_k^{\ l},\cr}  }
where we used \relmets, \iddef\ and the $tt^\ast$ equations together with the
CPT constraint $\eta^{-1}g=(g^{-1})^t \eta^\ast$.

Special geometry originally was discovered in two seemingly unrelated contexts:
The geometry of periods on a Calabi--Yau $3$--fold
\ref\bryant{R. Bryant and P. Griffiths, in {\it Arithmetic and Geometry},
papers dedicated to I.R. Shafarevitch, eds. M. Artin and J. Tate,
(Boston, Birkh\"auser, 1983), vol.2 p. 77.}
and $N=2$ supergravity in four
dimensions
\ref\dewit{B. de Wit and A. van Proeyen, Nucl. Phys. B245 (1984) 89\semi
B. de Wit, P.G. Lauwers and A. van Proyen, Nucl. Phys. B255 (1985) 569\semi
E. Cremmer, C. Kounnas, A. van Proeyen, J.--P. Derendinger, S. Ferrara,
B. de Wit and L. Girardello, Nucl. Phys. B250 (1985) 385\semi
S. Cecotti, Comm. Math. Phys. 131 (1990) 517.}.
 The ground state geometry of $\hat c=3$ superconformal theories
combines these two topics together in a natural way.
In the present paper we shall use quite heavily the relationship of special
geometry with the complex geometry of Calabi--Yau $3$--folds. In order to be
self--contained, we recall the basic facts about this connection.

Before doing this, it is convenient to formulate the above special geometry in
a slightly more abstract way. Since special geometry is equivalent to the
$tt^\ast$ geometry for a family of $\hat c=3$ superconformal theories, we shall
use the terminology arising in this last context\foot{Without losing any real
generality, we can also assume that all the chiral primary fields have integral
$U(1)$ charges. This assumption will be implicit throughout the paper. For a
discussion of `special geometry' in presence of fractional charges, see
Ref.\ref
\cecot{S. Cecotti, Int. J. Mod.
Phys. A6 (1991) 1749 \semi
S. Cecotti, Nucl. Phys. B355 (1991) 755}}.

Consider the `improved' connection
\eqn\improved{\nabla_i=D_i-C_i,\qquad\bar\nabla_{\bar j}=
\bar D_{\bar j}-\bar
C_{\bar j},}
acting on the  vector bundle $\cal V$ of ground states of equal left--right
charge.
The $tt^\ast$ equations are equivalent to the statement that the `improved'
connection is flat. Hence  we can identify all fibers of $\cal V$ with the one
at a given base point by parallel transport with respect to this improved
connection. In this way (apart for aspects related to global monodromies) we
can see $\cal V$ as a product bundle with fiber the fixed ground state vector
space $V$. In this gauge, the `improved' derivatives $\nabla_i$ and $\bar
\nabla_{\bar j}$ reduce to the ordinary ones $\partial_i$ and
$\bar\partial_{\bar j}$. To the fibers of $\cal V$ we can give a real structure
by declaring real the ground states which are mapped into themselves by CPT.
Since
$$(C_i)_{\bar k}^{\ \bar l}M_{\bar l}^{\ m}= M_{\bar k}^{\ k}(C_i)_k^{\ l},$$
the real  structure is invariant under parallel transport, i.e. the fixed
vector space $V$ has a natural
real structure. We fix once and for all a basis of $V$ whose elements
$|\alpha\rangle$ ($\alpha=1,\dots,2m+2$) are real vectors. In this basis CPT
acts by the usual complex conjugation.
The $tt^\ast$ metric is not invariant under parallel transport by the
$\nabla$--connection;
however, if $q$ is the $U(1)$ charge operator, the following real
skew--symmetric (symplectic) metric
$$Q_{\alpha\beta}=\langle \alpha|(-1)^{q+3/2}|\beta\rangle$$
{\it is} invariant because the matrix $C_i Q$ is skew--symmetric. In the gauge
in which the `improved' connection vanishes, this symplectic form is just a
constant matrix; we can choose our real basis $\{|\alpha\rangle\}$ so that it
is the standard symplectic unit $E$.

At a given point in coupling space, the ground state space $V$ admits a
decomposition into subspaces corresponding to states having definite $U(1)$
charges. However, as we change the couplings $t^i$, this charge decomposition
changes, since parallel transport by the
$\nabla$--connection does not preserve
charge. This is obvious from \improved\ since the matrix $C_i$, representing
multiplication by the field $\phi_i$, increases the charge by $1$. Special
geometry just describes how the states of given charge rotate in the fixed
space $V$ as we vary the couplings.

At a given point in coupling constant space $V$ decomposes into a
one--dimensional subspace corresponding to $|0\rangle$ having degree\foot{For
later convenience, we define the `degree' $l$ of a ground state to be the
$U(1)$ charge of the corresponding NS state, i.e. $l=q+3/2$.} $0$, an
$m$--dimensional subspace spanned by the vectors $\phi_i|0\rangle$ having
degree $1$, and their dual subspaces (with respect to the symplectic form $Q$)
having degrees 3 and 2, respectively. As we vary $t$, the states of degree $0$
form a line subbundle of the trivial vector bundle with fiber the fixed space
$V$. This line subbundle is just our vacuum bundle $\CL$.
In the same way, the
 states of degree 1, $\{\phi_i|0\rangle\}$, span the fibers of the vector
bundle $(T\otimes \CL)$, those of degree 2 the dual space
$(T\otimes\CL)^\ast$, and finally those of degree 3 the dual line bundle
$\CL^\ast$. Thus we have
the charge decomposition
\eqn\decomposition{V= \CL\ \oplus\  (T\otimes\CL)\ \oplus\
(T\otimes\CL)^\ast\ \oplus\ \CL^\ast.}
This decomposition satisfies four main properties. Let $\xi$ and $\zeta$ be two
sections of $V$ with definite degrees; then
\item{1.}{$\xi^t E \zeta=0$ unless $l(\xi)+l(\zeta)=3$.}
\item{2.}{$(-1)^{l(\xi)} \xi^\dagger E\xi>0$. Indeed, comparing with the
definition of $g$, we see that this is just the squared norm of the vacuum
state corresponding  to $\xi$.}
\item{3.}{$\partial_i \xi$ is a sum of two pieces, one with $l=l(\xi)$ and one
with $l=l(\xi)+1$. This property is evident from \improved\ which also gives
\eqn\torsion{\partial_i \xi\Big|_{l(\xi)+1}= -C_i\xi.}
\item{4.}{$\CL$ is a {\it holomorphic} line subbundle of $V$. Indeed,
$\bar\nabla_{\bar j}$ acting on a degree $0$ state produces a pure degree $0$
state; hence the flat connection $\bar\nabla_{\bar j}$ induces a holomorphic
structure on $\CL$. Since in the present gauge the
$\nabla$--connection is trivial,
this is just the canonical holomorphic structure for a subbundle of $V$.}

Working in the symplectic basis, the only non--trivial datum is how the
original ground states $|\phi_a\rangle$ are written in terms of the symplectic
ones $|\alpha\rangle$, i.e. we must know the coefficients
of the expansion
\eqn\chasymp{|\phi_a\rangle= V_{\ a}^\alpha|\alpha\rangle.}
{}From these coefficients we can easily recover the $tt^\ast$ metric
\eqn\othermet{g_{a\bar b}= (-1)^{l_a} V_b^\dagger E V_a,}
while the matrices $C_i$ can be extracted from \torsion\ which can be rewritten
as
$$ \partial_i V_a^\alpha = -(C_i)_a^{\ b}V_b^\alpha +{\rm terms\ with\ lower\
charge}.$$
Giving $V_{\ a}^\alpha$ is equivalent to giving the decomposition in
eq.\decomposition. The matrix $V_a^\alpha$ is restricted by the above four
conditions. Conversely, given any decomposition \decomposition\ satisfying
these conditions we can construct a metric $g$ satisfying the $tt^\ast$
equations with respect to the $C_i$ defined by eq.\torsion\ and having all the
properties discussed above. Indeed the $tt^\ast$ equations are equivalent to
the flatness of the connection $\nabla$, which is automatic in such a
construction.

In fact, it is enough to know $V^\alpha\equiv V_{\ 0}^\alpha$, i.e. how the
vacuum line bundle $\CL$ sits in $V$. Indeed from \iddef\ one has
$$\partial_i V^\alpha= -C_{i0}^k V_k^\alpha+\dots =- V_i^\alpha\ {\rm mod.}\,
V^\alpha,$$
so we can read $V_i^\alpha$ (i.e. the degree 1 subbundle) from the derivatives
of $V^\alpha$. The degree 2 and 3 subbundles then can be recovered by duality.
Acting $\bar\nabla$ on both sides of eq.\chasymp, we see that $V^\alpha$ should
be a holomorphic function of the $t$'s; this is property 4. above.
Then the above four conditions are automatically satisfied if and only if
$V^\alpha$ is holomorphic and satisfies\foot{In particular, any special
manifold is a Legendre submanifold of a complex contact manifold.}
\eqn\contact{V^tEV= V^t E\, \partial_i V=0,\qquad V^t E V>0.}
So given any holomorphic function $V^\alpha(t)$ satisfying \contact\ we
construct a special geometry. In particular, from \othermet\ we see that the
K\"ahler potential is
\eqn\otherK{e^{-K}=g_{0\bar 0}= V^t E V.}
Consider $\partial_i\partial_j\partial_k V^\alpha$. The component of top degree
is given\foot{The
index $\rho$ labels the unique chiral primary field of charge
$3$, normalized so that $\langle\rho\rangle=1$.}
by $-(C_iC_jC_k)_0^{\rho}V_\rho^\alpha$. Hence
\eqn\otherC{V^t\, E\partial_i\partial_j\partial_k V= -(C_iC_jC_k)_{00}\equiv
-C_{ijk},}
which is the most convenient way to define $C_{ijk}$.

The above discussion applies to any $N=2$ conformal model with $\hat c=3$. Now
we
specialize to the $B$--model based on a Calabi--Yau $3$--fold $M$ (which we
assume to be simply--connected). In this case the chiral primary fields of
$U(1)
$
charge $q$ are in one--to--one correspondence with the elements of
$H^{3-q,q}(M)$. This follows from eq.\flrb\ and the fact that for a simply
connected Calabi--Yau
$3$--fold the relevant vacua correspond to primitive cohomology
classes in degree $3$, which are annihilated both by $k$ and $k^\dagger$.
All these spaces are subspaces of the ordinary de Rham group
$H^3(M,{\bf C})$; since the de Rham cohomology depends only on the topology of
$M$, this group is independent of the couplings $t^i$ (which control the
complex structure of $M$).
$H^3(M,{\bf C})$ can be seen as a fixed space while the definite
charge subspaces $H^{3-q,q}(M)$ do move as we move the $t^i$'s. Then the
constant space $H^3(M,{\bf C})$ is easily identified with the space $V$ of the
abstract $N=2$ theory. Notice that this space has a natural real structure,
namely a class in $H^3(M,{\bf C})$ is {\it real} iff it belongs to $H^3(M,{\bf
R})$; this structure coincides with the one defined by CPT in the $B$--model.
Then the charge decomposition is identified with the Hodge decomposition
$H^3(M)=\oplus_q H^{3-q,q}(M)$. On $H^3(M)$ there is a natural symplectic
form given by
$$Q(\alpha,\beta)=\int\limits_M \alpha\wedge\beta,$$
which is also independent of $t^i$ and real since it is topologically defined.
With respect to this pairing and real structure, the Hodge decomposition
satisfies 1. and 2. (the Riemann bilinear relations). That the Hodge
decomposition also satisfies conditions 3. and 4. is a consequence of the
Kodaira--Spencer theory of complex deformations. Let $\mu_i$ be
the element\foot{$T_M$ denotes the holomorphic tangent bundle (sheaf) of the
Calabi--Yau
manifold $M$.} of $H^1(M,T_M)$ associated with an infinitesimal variation
$\delta t^i$  of the complex structure, and let $\omega$ be any harmonic
$(3-q,q)$ form. Then
\eqn\kstan{\partial_i \omega=\mu_i\wedge\omega+ \beta_i,}
where $\beta_i$ is a closed $(3-q,q)$ form, and $\mu_i$ acts on forms as
contraction for the vector index and exterior multiplication for the form
index. Thus $\mu_i\wedge \omega$ is a $(2-q,q+1)$ form. Eq.\kstan\ is nothing
else than condition 3. The same argument applied to $\bar\partial_{\bar
j}\omega$ implies condition 4.

According to our previous discussion, we can recover the ground state geometry
for the $B$--model provided we know how the space $H^{3,0}(M)$ (which we
identified with the line subbundle $\CL$) sits in $H^3(M,{\bf C})$, i.e. if we
know for each point in moduli space which de Rham class corresponds to the
$(3,0)$ form $\Omega$. Choosing $\Omega$ to depend holomorphically on $t^i$, we
can rewrite eq.\otherK\ and \otherC\ in the form
\eqn\intermsom{\eqalign{& e^{-K}=\int_M \bar\Omega\wedge\Omega\cr
&  C_{ijk}=-\int_M\, \Omega\wedge \partial_i\partial_j\partial_k\Omega.\cr} }

As a symplectic basis of vacua we can take the states associated to the (real)
3--forms which are dual to a canonical set of
$3$--cycles, i.e. a set of cycles $\gamma_\alpha$ ($\alpha=1,\dots,2m+2$) such
that their intersection pairing has the canonical form
$$\#(\gamma_\alpha,\gamma_\beta)=
\delta_{\beta,\alpha+m+1}-\delta_{\beta+m+1,\alpha}.$$
Then our basic vector $V^\alpha$ becomes
\eqn\periods{V^\alpha=\int\limits_{\gamma_\alpha}\Omega,}
so that $V^\alpha$ is just given by the periods of the holomorphic $(3,0)$
form. Eqs.\otherK\ and \otherC\ allow to write all the relevant geometric
quantities in terms of these periods. One can also show that the metric
$G_{i\bar j}=\partial_i\bar\partial_{\bar j}K$ is equal to the Weil--Petersson
metric on the Calai--Yau
moduli space. To see this, consider eq.\kstan\ with $\omega$
replaced by $\Omega$. Since $H^{3,0}(M)$ is one--dimensional, $\beta_i$ is
cohomologous to $f_i\Omega$ where $f_i$ is some holomorphic function of the
moduli $t^i$. Using the first of eqs.\intermsom, we have
\eqn\ttan{\eqalign{G_{i\bar j} &=\partial_i\bar\partial_{\bar
j}K=-\partial_i\bar\partial_{\bar j}\log\, \int_M \bar\Omega\wedge\Omega\cr
&=-{\int \bar \partial_{\bar j}\bar\Omega \wedge \partial_i\Omega\over \int
\bar\Omega\wedge \Omega}+{\int \bar\partial_{\bar j}\bar\Omega\wedge \Omega \
\int\bar\Omega\wedge \partial_i\Omega\over \left(\int
\bar\Omega\wedge\Omega\right)^2}.\cr} }

Now, eq.\kstan\ together with type considerations give
$$\eqalign{\int_M\bar\Omega\wedge \partial_i\Omega &= f_i\int_M\bar\Omega\wedge
\Omega,\cr
\int_M \bar\partial_{\bar j}\bar\Omega \wedge \partial_i\Omega &= \int_M
\big(\bar\mu_{\bar j}\wedge \bar\Omega\big)\wedge
\big(\mu_i\wedge\Omega)+ f_i\bar f_{\bar j} \int_M
\bar\Omega\wedge\Omega.\cr}$$
Inserting these back into \ttan\ we get
\eqn\weilp{G_{i\bar j}= -{1\over \int_M \bar\Omega\wedge\Omega}\, \int\limits_M
\big(\bar\mu_{\bar j}\wedge \bar\Omega\big)\wedge
\big(\mu_i\wedge\Omega) \equiv \big(\bar\mu_{\bar j}\, ,\, \mu_i),}
where $(\, \cdot\, ,\, \cdot\, )$ is the inner product on the bundle
$\Omega^{0,1}\otimes T_M$ induced by (any) Calabi--Yau metric on $M$.
[The last equality in \weilp\ is most easily seen by writing down the explicit
index structure of the various tensors involved]. By definition, the r.h.s. of
\weilp\ is the Weil--Petersson metric on the Calabi--Yau
moduli space. Comparing this
result with eq.\zamolodchikov, we also see that
$$g_{i\bar j}=-\int\limits_M \big(\bar\mu_{\bar j}\wedge \bar\Omega\big)\wedge
\big(\mu_i\wedge\Omega).$$

We stress that $\Omega$ is well defined by the above conditions only up to
multiplication by a holomorphic function of the $t^i$'s. Clearly, two
$\Omega$'s differing for such a factor define the same Hodge decomposition;
therefore this ambiguity is immaterial in all the above discussion and has no
physical consequence
(in particular the Zamolodchkikov metric is independent of these
choices). Instead the tensor $C_{ijk}$ gets multiplied by $f(t)^2$ when
$\Omega\rightarrow f(t)\Omega$. This behaviour reflects the basic fact that
$C_{ijk}$ is a tensor with coefficients in the line bundle $\CL^2$.

For future reference, we give an alternative expression for the $B$--model
$3$--point functions $C_{ijk}$.
As before, we start from the basic identity
\eqn\idone{\partial_i\Omega=\mu_i\wedge \Omega+\dots,}
where the dots stand for a closed form of type $(3,0)$.
Taking a second derivative
\eqn\idtwo{\partial_i\partial_j\Omega= \mu_i\wedge\mu_j\wedge\Omega+ \dots}
where now the dots denote closed forms of type $(2,1)$ and $(3,0)$.
{}From \idtwo\ and considerations of type we see that
$$\int\limits_M \Omega\wedge\partial_i\partial_j \Omega=0.$$
Taking the derivative of this identity, we get
\eqn\idthree{\int_M\partial_i\Omega\wedge
\partial_j\partial_k\Omega=      -\int_M
\Omega\wedge\partial_i\partial_j\partial_k\Omega=C_{ijk}.}
Note that replacing the derivatives in the l.h.s.
of \idthree\  by their explicit
expressions \idone\ and \idtwo\ the $\dots$ do not contribute because of type
considerations. So for our present purposes we can ignore them.

We introduce the following notation: A prime$\phantom{a}^\prime$ means
contraction of the vector indices with the holomorphic $3$--form $\Omega$,
while a superscript
$\phantom{a}^{\vee}$ means contraction of the form indices with
the
unique holomorphic $3$--vector dual to $\Omega$. Obviously these two operations
are each others inverse  $(A^\prime)^{\vee}=A$, and $(B^{\vee})^\prime=B$.
In this notation \idone\ and \idtwo\ read
$$(\partial_i\Omega)^\vee=\mu_i+\dots$$
$$\partial_i\partial_j\Omega=(\mu_i\wedge\mu_j)^\prime+\dots=
\big[(\partial_i\Omega)^\vee\wedge(\partial_j\Omega)^\vee\big]^\prime+\dots$$
Replacing the second equation in \idthree\ we get the formula we are looking
for\foot{Recall that $\wedge$ means exterior product with respect to the form
indices and {\it contraction with respect to the vector indices}.}
\eqn\TheFormula{C_{ijk}=\int\limits_M \partial_i\Omega\wedge
\big[(\partial_j\Omega)^\vee\wedge(\partial_k\Omega)^\vee\big]^\prime
=\int\limits_M\partial_i\Omega\wedge (\partial_j\Omega)^\vee \wedge
\partial_k\Omega.}

The case of a Calabi--Yau $n$--fold ($n>3$) is rather similar. As we saw, the
Zamolodchikov metric $G_{i\bar j}$ is K\"ahler also in this case, again with
potential $-\log\langle\bar 0|0\rangle$. On the other hand the basic identity
for its curvature, eq.\specialgeo\ is going to change. To get the
corresponding identity for an $n$--fold, we have to go through the
same steps as
in \longcomp. All the step are unchanged, except for the last one.
Then the following formula is valid for all $n$ (we use capital latin letters
to denote charge $2$ chiral primaries)
$$-{R_{i\bar j k}}^l= e^K C_{ik}^I\, g_{I\bar J}\, C^{\ast\, \bar J}_{\bar
j\bar m}
G^{\bar m l}- G_{k\bar j}\delta_i^l-G_{i\bar j}\delta_k^l.$$

\vskip .15in

\subsec{Coupling twisted $N=2$ theory to gravity}

Bosonic string theory is in many ways like a twisted $N=2$ theory
\ref\gat{B. Gato-Rivera and A.M. Semikhatov, Phys. Lett. B293 (1992) 72-80},
\ref\bw{M. Bershadsky, W. Lerche, D. Nemeschansky, N.P. Warner,
{\it Extended N=2 Superconformal Structure of Gravity and W Gravity
Coupled to Matter} CERN-TH-6694-92, hep-th/9211040}.
It has a {\it scalar}
supercharge $Q_{BRST}=Q+\overline Q$, which is the usual BRST operator.  It has
anti-ghosts,  $b,\overline b$ of spin (2,0) and (0,2), with the property
$$Q^2=b_0^2=0$$
$$\{ Q,b_0\} =H_L $$
and it has two $U(1)$'s, $G,\bar G$  corresponding to the left and right ghost
numbers.  Identifying
$$2j_{BRST}\leftrightarrow G^+$$
$$ b\leftrightarrow G^-$$
$$bc\leftrightarrow J$$
and similarly for right-movers.
Thus the notion of a physical state in the bosonic string becomes
exactly the same as that of a chiral state in the twisted theory.
Thus we can define coupling of twisted $N=2$ theory to gravity
by integrating correlation functions of chiral fields over moduli
space of Riemann surface, with the insertion of $G^-$'s folded
with $3g-3$ Beltrami differentials. In particular the partition
function of the twisted $N=2$ theory coupled to gravity at genus
$g>1$, $F_g$, can be defined by\foot{For the case of $g=2$ one
has to put a factor of 1/2 in front because all the $g=2$ curves
have a ${\bf Z}_2$ symmetry.}
$$F_g=\int_{{\cal M}_g} \Big\langle \prod_{k=1}^{3g-3}(\int G^-\mu_k)(\int\bar
G^-\bar\mu_k)\Big\rangle $$
where $\mu_i$ denote the Beltrami differentials, and ${\cal M}_g$
denotes the moduli space of genus $g$ Riemann surfaces.  For $F_1$
the answer can be written using the corresponding analysis
of the bosonic string case \ref\pol{J. Polchinski,
Commun. Math. Phys. 104 (1986) 37}. To do this
note that in bosonic string one inserts $bc\overline b\overline c$ to
absorb the ghost zero modes.  This is translated in the twisted
theory to the insertion of left and right fermion number currents.
Also, to fix the normalization it is best to write the answer
in the operator formulation which is particularly convenient
for torus:
\eqn\fone{F_1={1\over 2}\int {d^2\tau \over \tau_2}
\Tr\left[(-1)^FF_L F_R q^{H_L}
{\overline q}^{H_R}\right] }
where the factor of 1/2 in front takes care of the fact that there
is a ${\bf Z}_2$ reflection symmetry for all tori (this
normalization is different from the one used in \ref\bcov{M. Bershadsky, S.
Cecotti,
H. Ooguri and C. Vafa, {\it Holomorphic Anomaly in Topological Field Theories},
HUTP-93/A008, RIMS-915, hep-th/9302103}).
For genus 0, the 0,1, and 2 point functions are zero, as is
the case in bosonic string,
 and the three
point functions can be written as
$$\langle \phi_i \phi_j \phi_k \rangle =C_{ijk}.$$
In other words the chiral fields $\phi_i$
which after twisting have dimension zero play the same
role as $c\overline c V_i$ in the usual bosonic strings, and
$\phi_i^{(2)}$ plays the same role as $V_i$.

It is a rather nice property of twisted {\it unitary} $N=2$
theories that $F_g$ thus defined is finite
and thus well defined.  The only potential
divergence would have come from the regions near the boundary
of moduli space of Riemann surfaces.  But in such cases, the fact
that the propagator on a long tube
is given by $G_0^-{1\over L_0+\bar L_0 } \bar G_0^-$
and  that this annihilates the massless modes, imply that
only the massive modes propagate and thus the integrand
in $F_g$ is exponentially small in these regions (the coefficient
of exponent being fixed by the first non-vanishing
eigenvalue of $L_0=\bar L_0$).

Despite an almost complete parallel between bosonic string
and twisted $N=2$ theories coupled to gravity,
there are two notable differences.  The first one
is that the ghost number violation in bosonic string at genus $g$
is universal and is given by $3g-3$, whereas for twisted $N=2$
theories it is given by \chvi\ as $\hat c (g-1)$.  In
particular we see that $\hat c=3$ is a critical case in that it
gives the same degree of charge violation as bosonic string.  So in
particular this suggests that Calabi--Yau $3$-folds are a specially
interesting class to consider \ref\ewittop{E. Witten, Nucl. Phys. B340
(1990) 281\semi R. Dijkgraaf and E. Witten, Nucl. Phys. B342 (1990)
486.}.
Note that only
for $\hat c=3$ the $F_g$ has a chance to be non-zero for $g>1$,
by $U(1)$ charge conservation.  For all the other values of
$\hat c$, the only way to get non-zero result is by introducing
other correlators.  The correlators involving chiral fields
may be used to prevent vanishing of correlation
functions only for $1<\hat c $;   For $\hat c\leq 1$ the charges
of all $\phi_i^{(2)}$ are negative (the maximum being $\hat c -1$)
and so cannot be used to balance charges.  In these cases, which
happen to be intensively studied in connection with matrix models,
one needs to include the
full topological gravity multiplet and construct gravitational
descendants which give rise to non-vanishing correlation functions
\ewittop \ref\Vers{E. Verlinde and H. Verlinde, Nucl. Phys. B348 (1991) 457}
\ref\Li{K. Li, Nucl. Phys. B354 (1991) 725-739}.
Also for $\hat c>3$ one needs fractional chiral fields $\phi_i$
with charges between
$0<q<1$ in order to have a chance of balancing the charges
(gravitational descendants do not help in this case as they
contribute $+N$ to charge violation condition).  In particular
for Calabi-Yau manifolds, which have no fractional chiral states,
 with dimension bigger than $3$ all the
correlations vanish, and thus the theory is not very interesting
(except possibly for three point functions on the sphere with
$\sum q_i=n$ and
$F_1$ which is computable and non-zero in general).  For a Calabi-Yau
$2$-fold, which is either $K3$ or $T^4$, the situation is hardly more
interesting:  all the correlation functions vanish in the case of $T^4$
due to too many fermion zero modes and on $K3$ because
of charge conservation\foot{In the case of $K3$ in principle there
 was a chance that genus $g$ correlation function for $g>1$
with $g-1$ insertions
of the highest charge chiral field which balances
the charge lead to non-vanishing of the amplitude.  But in fact
using the techniques in this paper, i.e. the holomorphic
anomaly equation in this case, one can show the amplitude
still vanishes even with this insertion.}.
Finally for the case of one--dimensional Calabi--Yau manifolds
only $F_1$ is non--zero (as studied in \bcov ), due to charge conservation
unless one introduces gravitational descendants.

So clearly as far as the twisted sigma--models coupled
to gravity are concerned, the most interesting case is the $3$--fold
Calabi--Yau case which will be the focus of our examples.  Actually
we will consider the more general possibility of a unitary
SCFT with $\hat
c=3$ with integral $U(1)$ charges for chiral fields.  Many of the results
we will discuss can be easily generalized to other similar cases and
would be interesting to study.

The other major difference between the bosonic string and critical
topological strings, even if we choose ${\hat c}=3$, is that in
the case of topological strings obtained by twisting unitary $N=2$
theories, the $G^-$ cohomology is generically non--trivial whereas
absolute $b$--cohomology is always trivial in the bosonic strings.  This,
as we will see in the rest of the paper is a crucial difference,
and it leads to the anomaly discussed in detail in the next section.

The case of Calabi--Yau $3$--fold
as a string theory has already been studied in \ref\witcs{E. Witten,
{\it Chern--Simons Gauge Theory as a String Theory}, IASSNS-HEP-92/45,
hep-th/9207094}\ for both open and closed strings.
In particular it was discovered there that in the case of the open
string theory, the target space physics is equivalent
to three dimensional Chern--Simon theory.  In the case of closed
strings there were some puzzles raised which we resolve in connection
with our discussion on the Kodaira--Spencer theory in section 5.

\vskip .15in

\subsec{Properties of $n$--point functions and the holomorphicity paradox}
\vglue 10pt

Consider the $n$--point functions of the $N=2$ twisted
theory coupled to gravity at genus $g$
\eqn\npointg{\eqalign{&C^g_{i_1i_2\dots i_n}=\cr
&\quad = \int_{{\cal M}_g} \Big\langle \int\phi_{i_1}^{(2)}\cdots
\int\phi_{i_n}^{(2)} \prod_{k=1}^{3g-3}(\int G^-\mu_k)(\int\bar
G^-\bar\mu_k)\Big\rangle.\cr}}
We would like to relate this to $F_g$.  At first sight one may
think that $C^g_{i_1i_2\dots i_n}$ can be written simply as
$$C^g_{i_1i_2\dots i_n}=\partial^n_{i_1i_2\dots i_n}F_g$$
However this formula cannot possibly be true since, as it is evident from its
definition \npointg\ (and discussed in section 2.1), $F_g$ is a section of
$\CL^{2-2g}$ and
$C^g_{i_1i_2\dots i_n}$ is a section of the {\it non--trivial} vector bundle
${\rm Sym}^nT\otimes \CL^{2-2g}$ sitting over coupling space. Therefore acting
with $\partial_{i_k}$'s on $F_g$ makes no sense at all. Geometrically it is
clear that the correct relation should have the form
\eqn\covariantization{C^g_{i_1i_2\dots i_n}=\CD_{i_n}\dots
\CD_{i_1}F_g}
where $\CD_i$ is some suitable connection compatible with the transition
functions for the appropriate bundles. On  ${\rm Sym}^nT\otimes \CL^{2-2g}$
there is a natural connection  $D_i$ i.e. the one induced by the Zamolodckikov
connection on $T$ plus the canonical connection on $\CL$, see section
2.3. It is
natural to guess that \covariantization\ holds with $\CD_i$ replaced by this
natural connection $D_i$.  This is what we will presently argue.
This will imply that the following recursion relation
(for $2g+n-3>0$) holds
\eqn\expcon{\eqalign{e^{2(1-g)K} G^{\bar j_i i_1} G^{\bar j_2 i_2}&\cdots
G^{\bar j_{n-1} i_{n-1}}\, C^g_{i_1i_2\dots i_{n-1}i_n}=\cr
&=\partial_{i_n}\Big(e^{2(1-g)K} G^{\bar j_i i_1} G^{\bar j_2 i_2}\cdots
G^{\bar j_{n-1} i_{n-1}}\, C^g_{i_1i_2\dots i_{n-1}}\Big).\cr}}

To show this, there are two things that we will have to argue:
One is covariantization with respect to the Zamolodchikov metric
and the other is covariantization with respect to the natural
connection on $\CL$.  Both will arise from contact terms.
 First let us discuss how the covariantization
with respect to the Zamolodchikov metric arises. This actually has
been done in full generality for marginal operators of
conformal theory in \ref\kut{D. Kutasov, Phys. Lett. B220 (1989) 153}\
leading to the
following
contact term in our case
$$\phi^{(2)}_i(z)\, \phi^{(2)}_j(0)\sim \delta^2 (z)
\Gamma_{ij}^k\phi^{(2)}_k(0)$$
where $\Gamma_{ij}^k$ is defined in section 2.3
and is the connection for the Zamolodchikov metric.  This amounts
to the first thing we wished to show.  However we would also like
to rederive this result using the $tt^*$ machinery:  In defining the
amplitude in \npointg\ we have to be careful to
regularize the computation by making sure that two operators
do not get closer to each other than
distance $\epsilon$.  However, when we take derivative
of $F_g$ with respect to $t^i$ this region is not excluded. Therefore
the difference between the explicit meaning of the correlation and
the derivative with respect to $t^i$ will include the regularization
of the integration of $\phi^{(2)}_i$ in a small neighborhood of
$\phi^{(2)}_j$ for all $j$'s.  Since this is a local computation
we may as well do it on a hemisphere, where we can apply $tt^*$
equations.  In such a case the integral of $\phi^{(2)}_i$ over
the hemisphere including the field $\phi^{(2)}_j$ in it minus the
one with $\phi^{(2)}_j$  outside of the hemisphere,
is equal to, as far as the
topological states are concerned \cv\
$$G_{-1}^-{\overline G}_{-1}^-\partial_i |j\rangle-\phi^{(2)}_j \partial_i |0
\rangle \rightarrow
\Big[(A_i)_j^k \phi^{(2)}_k-(A_i)_0^0\cdot \phi^{(2)}_j\Big]|0\rangle$$
$$=[(g^{-1})^{k\overline j}\partial_i g_{j\overline j} +\partial_i K
\delta_j^k]\phi^{(2)}_k|0\rangle=
\Gamma_{ij}^k\phi^{(2)}_k|0\rangle.$$
Therefore we see that the insertion of $\phi^{(2)}_i$ in the correlation
is equivalent to
$$\partial_i -\Gamma_i$$
as was to be shown.

Now we turn to the second covariantization, i.e., with respect
to the line bundle connection on $\CL$.  This arises from the
hidden contact term between the term we added to the action
to twist the supersymmetric theory
$${1\over 2}\int J \bar \omega+c.c.={1\over 2}\int R \varphi
+{\rm total}\quad {\rm derivatives}$$
 and the operator $\phi^{(2)}_i$, as $J$ has a contact term with it.
Here $\omega$ is the spin connection and $\varphi$ is the scalar which
bosonizes the
$U(1)$ current.
To study this term, again we use the fact that it is local, and that we can
thus first study the case of the hemisphere to which we can apply
$tt^*$ considerations.  The same argument as above leads
to the contact term
$$(A_i)_0^0\cdot 1=-\partial_i K\cdot 1$$
More generally, since the contribution due to this term
is proportional to $\int R$,
and the above computation was done on hemisphere with net $\int R=2\pi$,
we can write the above term more generally as
$$-\partial_i K {R\over 2\pi }$$
On genus $g$ surface this leads to an integrated contact term
$$-\partial_i K (2-2g)$$
and so insertion of $\int \phi^{(2)}_i$ in the correlation
amount to covariantization also with respect to the line bundle
$\CL^{2-2g}$ which leads us to our final answer for the operator
insertion
\eqn\ficov{\int \phi^{(2)}_i\rightarrow
\partial_i -\Gamma_i -(2-2g)\partial_i K }
Note that this is consistent with the
fact that the insertion of $\int \phi^{(2)}_i$
should not lead to any further ambiguities in fixing
the normalization of the path integral.  In particular
$\exp[(2g-2)K ]\cdot F_g$ is independent of such ambiguity
as far as holomorphic derivatives are concerned as
the $K$ varies precisely to compensate the ambiguity
of $F_g$ as discussed in section 2.1.
Note in particular that the symmetry
in exchange of $\phi^{(2)}_i$ is consistent
with the above relation with covariantization since
$$D_{i_1}D_{i_2}\dots D_{i_{n-3}} C^{(g)}_{i_{n-2} i_{n-1} i_n}$$
is symmetric in all its $n$ indices. This is a consequence of  $[D_i,D_j]=0$
(i.e. the curvature of the natural connection has type $(1,1)$)
(for the case of genus 0 we also need the fact that $D_iC_{jkl}$ is totally
symmetric in its four indices, see eq.\integra ).

Now that we have understood how covariantizations arise
we are going to present a formal argument for decoupling of
anti-chiral operators ${\bar \phi}^{(2)}_{\bar i}$ from
the correlation functions.  Let us consider the correlations
on the sphere:
\eqn\npointzero{C_{i_1 i_2\dots i_n}=\Big\langle \phi_{i_1}(0)\,
\phi_{i_2}(1)\, \phi_{i_3}(\infty)\, \int \phi^{(2)}_{i_4}\, \cdots\,
\int \phi^{(2)}_{i_n}\Big\rangle.}
If the BRST--trivial states do decouple from the physical amplitudes, then the
$n$--point functions $C_{i_1\dots i_n}$ should depend holomorphically on the
couplings $t^i$. Indeed,one has
\eqn\npointf{\eqalign{&\bar\partial_{\bar j} C_{i_1i_2\dots  i_n}=\cr
& =\int \sqrt{g}\, d^2 z\Big\langle \phi_{i_1}(0)\, \phi_{i_2}(1)\,
\phi_{i_3}(\infty)\, \int \phi^{(2)}_{i_4}\, \cdots\,
\int \phi^{(2)}_{i_n} \oint_{C^\prime_z} G^+ \oint_{C_z} \bar G^+
\bar\phi_{\bar j}(z)\Big\rangle,\cr}  }
where $C_z$ and $C^\prime_z$ are small contours enclosing the point $z$. Now
we can deform the $C^\prime_z$ countour around the other operator insertions.
Since
$$\eqalign{& \oint_{C_w} G^+ \phi_i(w) =0,\cr
& \oint_{C_w} G^+ \phi_i^{(2)}= \partial \phi^{(0,1)}_i= d\phi^{(0,1)}_i,\cr}$$
we get
\eqn\naiveargum{\eqalign{\bar\partial_{\bar j} C_{i_1i_2\dots i_n}&=
-\sum_{k=4}^n\int \sqrt{g} d^2 z\Big\langle
\phi_{i_1}(0)\phi_{i_2}(1)\phi_{i_3}(\infty)\, \int \phi^{(2)}_{i_4} \cdots\
\times\cr
&\hskip 2.6cm\times\ \int d \phi^{(0,1)}_{i_k} \cdots
\int \phi^{(2)}_{i_n} \oint_{C_z} \bar G^+ \bar\phi_{\bar j}(z)\Big\rangle\cr
&\buildrel {\rm formally} \over =\ 0,\cr}  }
since $\int d\phi^{(0,1)}_i =0$. This formal manipulation can be extended to
the $n$--point functions at genus $g$ \npointg .
Since the BRST variation of $(\int G^-\mu_k)$ produces the energy momentum
tensor folded in the Beltrami differential, the additional terms arising from
the deformation of the countour have the form of the derivative of some
correlation function with respect to the moduli of the complex surface just as
it happens in the bosonic string theory; hence they are also expected to vanish
upon integration over the (compactified) moduli space of genus $g$ Riemann
surfaces ${\cal M}_g$.

However, this consequence of the decoupling of BRST--trivial states is in
contradiction with what we know from $tt^\ast$ geometry as we
will now see. This contradiction leads to a paradox that will be resolved by
the discovery of a new `holomorphic' anomaly, which will be discussed in the
next section.
 The point is that holomorphicity of the $n$--point functions
(for $n>3$) is not consistent
with the recursion relation \covariantization. Indeed, $\bar\partial_{\bar j}$
does not commute with $D_i$: Rather $[\bar\partial_{\bar j},D_i]$ is the
non--vanishing curvature of the natural connection.
For instance, consider the $4$--point function $C_{ijkl}=D_lC_{ijk}$. Since
$\bar\partial C_{ijk}=0$, we have
$$\eqalign{\bar\partial_{\bar m} C_{ijkl}&=\bar\partial_{\bar
m}D_lC_{ijk}=[\bar\partial_{\bar m},D_l]C_{ijk}=\cr
&=
2G_{l\bar m} C_{ijk} - ({R_{l\bar m i}}^n C_{njk}+{\rm 2\ permutations}),\cr}$$
where ${R_{l\bar m i}}^n$ is the curvature of the Zamolodchikov metric. To see
that the r.h.s. is indeed not zero, replace this curvature by its explicit
expression given by special geometry.

As we shall see in the next section, in higher genus the situation is even
worse, since there the partition function $F_g$ is also not holomorphic.

What is the way out of this paradox, i.e. where is the loop--hole in the naive
argument around eq.\naiveargum? The point is that although it
 is true that we
remain with a sum of terms each with an operator $\int d\phi^{(0,1)}_i$
inserted (cf.\naiveargum), these terms do not vanish upon integration over the
Riemann surface, because the corresponding integral gets a non--trivial
boundary term when the field $\phi^{(0,1)}_i$ approaches a point where some
other operator is inserted. Indeed the $n$--point function on the sphere should
be written more invariantly as an integral over the moduli space ${\cal
M}_{0,n}$ of a sphere with $n$ punctures. The configurations where two points
get close together make the boundary of this space.
Then, taking $\bar\partial_{\bar j}$ of the $n$--point function and deforming
contours as in eq.\naiveargum\ we get the integral over ${\cal M}_{0,n}$ of an
{\it exact} top form. But, since the boundary of ${\cal M}_{0,n}$ is not empty
(for $n>3$),
this does not mean that the integral itself vanishes. However,
$\bar\partial_{\bar j}C_{i_1\dots i_n}$ may get a contribution only from the
boundary of ${\cal M}_{0,n}$. Since the boundary corresponds to two operators
colliding, we see that the $n$--point function may fail
 to be holomorphic only because
of contact terms. This is precisely what we found by the explicit computation
above.

In the next section we shall see how this holomorphic anomaly appears in higher
genus. There again we shall find that $\bar\partial_{\bar j}C^g_{i_1\dots i_n}$
gets contribution only from the boundary of the moduli space of genus $g$
surfaces with $n$ punctures ${\cal M}_{g,n}$. However, since in this case the
boundary has more components, new interesting phenomena will appear.

\vglue 14pt
2.6. {\it Canonical coordinates and special coordinates}
\vglue 10pt

Before turning to the next section we would like to make one comment
about covariantization which will be both useful for us later as well as
clarifying the
relation to some work already in the literature:  In topological theories
it is well known that the insertion of chiral fields can be represented
by ordinary derivatives \ref\verd{R. Dijkgraaf, E. Verlinde and H. Verlinde,
Nucl. Phys. B352 (1991) 59} . This is also
implicitly used in the discussion of counting of holomorphic curves
using the topological sigma--models \wittop .  What we have said
so far seems to be at odds with these work.  In fact not only there
is no contradiction but actually clarifying the relation of our work
to these will explain some of the ansatz made in the
study of mirror map \cand.  The point is that
if we start from a unitary $N=2$ theory which we denote by
$S_0$, twist it and then perturb {\it only} by topological
fields
$$S=S_0(t_0,\overline t_0)+\int d^2z d^2\theta\, \delta t^i \phi _i$$
there are no infinities and one could be naive in taking
field insertions on the world sheet.  In other words
we do not have to prevent points approaching each other by
cutting out small discs around each field insertion.
In fact as we mentioned
it was rather important in derivation of various properties
of the topological correlations to take ordinary derivatives
of partition function to obtain topological correlations.  It turns
out that this is possible as long as we are only interested
in topological correlations {\it and for a fixed} ${\overline t^i_0}$.
This is possible because of \ttstar , in other words, the fact that
$$[D_i,D_j]=0$$
means that for a fixed base point $(t_0,\overline t_0)$, shifting $t$
alone can be accomplished by ordinary derivative, i.e.,
we can choose a gauge in which $D_i\rightarrow \partial_i$.
This being true means that we are taking a choice of coordinates
on the moduli space as well as a gauge for the line bundle on the moduli
space so that
\eqn\gchoi{
\partial_{i_1}...\partial_{i_n}\Gamma_{ij}^k\Big|_{(t_0,\overline t_0)}=0
=\partial_{j_1}...\partial_{j_r}K\Big|_{(t_0,\overline t_0)}}
It turns out that these conditions {\it fix} the choice of coordinates
and line bundle section $|0\rangle $ up to linear transformation,
and can be done for {\it arbitrary} K\"ahler manifolds with
arbitrary line bundles on it having real analytic metrics.

Let us first talk about the line bundle.  Consider a local
section near $t_0$ and let the norm of this section be $e^{-{K}}$.
In arbitrary coordinate system $K$ has the
following expansion $K(z, \bar z)=K_0(z)+ \bar z^m F_{\bar m}(z)+o(\bar z^2)$
where we take the $t_0$ to correspond to $z=0$.  By redefining
the choice of the local holomorphic section we can get rid of $K_0$
which is purely holomorphic.  With this choice of local section all holomorphic
derivatives of $K$ are equal to zero at the origin.  So such
a section exists.   Moreover it is unique up to
multiplication by a constant because any $z$ dependence will
give rise to a non-constant $K_0$ which will thus violate
the condition that holomorphic derivatives of $\partial_i K$
vanish at the origin.

Now we show that the same can be done for the Christoffel connection.
On a K\"ahler manifold there is locally a K\"ahler potential,
which we again denote by $K$, and which can be expanded as above, and
with a choice of gauge $K_0$ can be chosen to be zero.
The expansion for the metric follows from the expansion for
K\"ahler potential
$G_{k \bar m}=\partial_k F_{\bar m}(z)+ o(\bar z)$.
Making the {\it holomorphic} change of variables $z^i \rightarrow t^i$ such
that
\eqn\holfl{{\partial z^k \over \partial t^i}
\partial_k F_{\bar m}(z) =C_{i \bar m}=const,}
or, explicitly, $t^i= C^{i\bar m}F_{\bar m}(z)$, one reduces metric to the form
$G_{k \bar m}=C_{i \bar m}+o(\bar z)$. The ambiguity in $t$-coordinates
is in the choice of constant matrix $C_{i \bar m}$ and it is parametrized by
$GL(n)$.
In $t$-coordinates holomorphic Christoffel symbols
vanish at the origin ($\bar z=0$) together with all holomorphic derivatives
as was to be shown. We shall call the coordinates $t^i$
and the choice of local trivialization of the line bundle in which \gchoi\
holds {\it canonical coordinates\/} with respect
to the base--point $t_0$.

To relate to some comments made in the literature, we would like
to draw attention to a natural base point in the case of $A$--model,
and that is infinite volume $\overline t_0=\infty$.  In this case
the gauge choice \gchoi\ implies that we must have
$\partial_iK\Big|_{\overline t \rightarrow \infty}$ vanish which explains the
gauge choice
made in finding the mirror map in \cand\ (in particular the
normalization of the holomorphic three form one needs
is in the gauge where $\partial_i K
=\partial_i {\rm log \langle \overline 0|0\rangle }\Big|_{
\overline t \rightarrow \infty}=0$). It is in this gauge that
the path integral for
the $A$--model is given by a sum over holomorphic maps and thus this is the
right gauge in order to count these maps.  We will now discuss this
in more detail.

The crucial property of the canonical coordinates with base point at
infinity\foot{By `point at infinity' we mean the following.
For the $A$--model a
point where the volume of the Calai--Yau
 manifold is infinite, i.e. the weak coupling
limit. For the $B$--model we mean a degeneration point in the complex moduli
space around which the nilpotent part of the monodromy is maximal. As it is
well known, this is `infinite volume' from the mirror viewpoint.} is that, for
an appropriate choice of the matrix $C_{i \bar m}$, they coincide with the
special coordinates (in the sense of special geometry). Since eq.\gchoi\
completely characterize the canonical coordinates, it  is enough to show that
the special coordinates satisfy  this equation with $\bar t_0=\infty$.
For convenience, we show this in the context of the $B$--model, using the
periods of the holomorphic $3$--form. The argument can be easily extended to
the general case using the more abstract methods of section 2.3.
The key formula is \otherK\
\eqn\defK{e^{-K}= \varpi^\dagger E \varpi,}
where $\varpi^\alpha$ are the periods in a symplectic basis. $\varpi$ depends
holomorphically on the moduli.
Let $s^i$ ($i=1,2,\dots, m$) be the `special coordinates' and put $X_i= X_0
s_i$,  where $X_0$ is a holomorphic coordinate along the fiber of $\CL$.
Then the periods  take the form
$$\varpi^t=(X_0,X_i,\partial_{X_0} \CF,\partial_{X_i} \CF),$$
 where $\CF$ is a holomorphic function of $X_0$, $X_i$ homogeneous of degree
$2$.
We have introduced the homogeneous coordinates $X_I$ ($I=0,1,\dots, m$) because
$\varpi$ takes value in the line bundle $\CL$;
then $X_0$ corresponds to the freedom in the
choice of trivialization of $\CL$. The condition that $\varpi$ is a section of
$\CL$ also explains the homogeneity condition on $\CF$.
We want to take the limit $\bar s^j\rightarrow\infty$ in eq.\defK\ while
keeping $s^i$ generic.  We need
 the behaviour of the periods $\varpi^*(\bar s)$ as $\bar s\rightarrow \infty$.
 This behaviour is described by the Schimd orbit theorems \ref\schmid{W.
Schmid, Invent. Math. 22 (1973) 211.}: As $s\rightarrow\infty$ one has
$$\CF(X_0,X_i)={d_{ijk}X_iX_jX_k\over X_0}+c X_0^2+O(e^{2\pi is}),$$
for some non--degenerate numbers  $d_{ijk}$.
Given the `factorized' form of eq.\defK,  taking the limit
$\bar s^j\rightarrow\infty$ while keeping $s^i$ fixed is a well defined
procedure. More precisely, we make $\bar s^i\rightarrow \bar \lambda \bar s^i$
and send $\bar \lambda$ to infinity. In this limit one has (up to exponentially
small terms)
$$\varpi^\dagger=\bar X_0 (1, \bar \lambda\bar s^i, -\bar \lambda^3 d_{ijk}\bar
s^i \bar s^j \bar s^kk+2c^*,3\bar d_{ijk}\bar \lambda^2 \bar d_{ijk}\bar
s^j\bar s^k).$$
Therefore
\eqn\asym{e^{-K}=\sum_{r=0}^3\bar \lambda^r A_r,}
with
$$\eqalign{& A_3=|X_0|^2 \bar d_{ijk}\bar s^i\bar s^j\bar s^k\cr
& A_2=-3|X_0|^2 \bar d_{ijk} s^i \bar s^j \bar s^k\cr
& A_1=\bar X_0 \bar s^i \partial_i\CF\cr
& A_0=\bar X_0 \CF-2c^*|X_0|^2.\cr}$$
Notice that in the special coordinates $A_3$ and $A_2$ take a universal form
(that is, they are independent of $\CF$). From \defK, \asym\ one has
$$K=-\log X_0-\log \bar X_0 -\log[\bar \lambda^3 d_{\bar i\bar j\bar k}\bar
s^i\bar s^j\bar s^k - 3 \bar \lambda^2 d_{\bar i\bar j\bar k}s^i\bar s^j\bar
s^k]+O(\bar \lambda^{-1}),$$
and then the Zamolodchikov metric reads
\eqn\useful{G_{i\bar j}=\partial_i\overline\partial_{\bar j}K={1\over
\lambda^2}
L_{i\bar j}(\bar s) +O(\bar \lambda^{-3}),}
where $L_{i\bar j}(\bar s)$ is a non--degenerate anti--holomorphic matrix. On
the other hand,
\eqn\asyK{K=-\log X_0 + ({\rm anti-holomorphic})+ O(\bar \lambda^{-1}).}
Thus,  taking the base point $\bar t_0$ be at infinity, the canonical gauge for
$\CL$ defined by the second of \gchoi\ is
just $X_0=1$, which is the standard gauge in special geometry.
Moreover as $\bar \lambda\rightarrow\infty$ the $(1,0)$ part of the Christoffel
connection $D_i$ becomes the trivial one. Indeed
using \useful\
\eqn\coTwo{D_i=G\partial_i G^{-1}= \partial_i +O(\bar \lambda^{-1}),}
which shows that the special coordinates $s^i$ satisfy \gchoi\ at infinity
and hence can be identified with the canonical coordinates $t^i$ with respect
to this base point.

\vfill
\eject

\newsec{Holomorphic anomaly}

In the topological theory, the BRST invariance would imply
that partition functions and correlation functions
are holomorphic on the moduli space of the theory
since variation with respect to the anti-holomorphic moduli
$\bar{t}^i$ inserts the BRST trivial operator
$\bar{\phi}^{(2)}_{\bar i}=\{ G^+ ,[ \overline{G}^+ ,
 \overline{\phi}_{\bar i}] \}$. This indeed is the
case for the Yukawa coupling. However we saw in the previous
section that the holomorphicity is, in general, not consistent
with the covariance on the moduli space. This means that there
is something wrong with the assumption on the BRST invariance.
What we saw there is reminiscent of the chiral anomaly in the Yang-Mills
theory where one finds that it is not possible to preserve both the vector
and the chiral gauge invariances of the theory. Thus we call
this phenomenon the holomorphic anomaly.
In this section, we will uncover a subtle breakdown of the BRST invariance
in the twisted $N=2$ model coupled to the gravity,
and rederive the non-holomorphicity we found in the previous
section as a special case.

\vskip .15in

\subsec{Homolorphic anomalies of partition functions}

Let us first examine the partition function $F_g$ for $g \geq 2$.
The naive BRST invariance would imply  $\overline{\partial}_{\bar i} F_g =0$.
We are going to show that this is not the case. The derivative
with respect to $\overline{t}^i$ is generated by an insertion of
the anti-chiral field $\overline{\phi}_{\bar i}$ as
\eqn\derivative{
\eqalign{ {\partial \over \partial \overline{t}^i} F_g &=
 \int_{{\cal M}_g} [d m] \int d^2 z \langle
\oint_{C_z} G^+ \oint_{C'_z} \overline{G}^+ \overline{\phi}_{\bar i}(z)
\prod_{a=1}^{3g-3}
        \int \mu_a G^- \int \overline{\mu}_a
\overline{G}^- \rangle_{\Sigma_g} \cr
&=\int_{{\cal M}_g} [d m] \sum_{b,\bar{b}=1}^{3g-3}
 \langle \int \overline{\phi}_{\bar i} \int 2 \mu_b T
   \int 2 \overline{\mu}_{\bar b} \overline{T} \prod_{a \neq b}
        \int \mu_a G^-
 \prod_{\bar{a} \neq \bar{b}} \int \overline{\mu}_{\bar a}
\overline{G}^-
 \rangle_{\Sigma_g} \cr
 &=
\int_{{\cal M}_g} [d m]  \sum_{b,\bar{b}=1}^{3g-3}4 {\partial^2 \over
 \partial m_b \partial \overline{m}_{\bar b}}
 \langle \int
\overline{\phi}_{\bar i} \prod_{a \neq b}
        \int \mu_a G^-
 \prod_{\bar{a} \neq \bar{b}} \int \overline{\mu}_{\bar{a}}\overline{G}^-
  \rangle_{\Sigma_g}. \cr} }
In the first line of this equation, the contours $C_z$ and $C'_z$
are around the point $z$ where the anti-chiral $\overline{\phi}_{\bar i}$
is inserted. We then
moved these contours around the Riemann surface $\Sigma_g$,
and picked up the commutators, $\oint_{C_w} G^+ \cdot G^-(w)  = 2 T(w)$
and $\oint_{C_w} \overline{G}^+ \cdot
\overline{G}^-(\bar{w})
= 2 \overline{T}(\bar{w})$. The insertions
of $T$ and $\overline{T}$ are then converted into the derivatives
with respect to the moduli $m, \bar{m}$ of $\Sigma_g$.
Using the Cauchy theorem, we can reduce the
r.h.s. to an integral on the boundary
of the moduli space ${\cal M}_g$.

The boundary of ${\cal M}_g$ consists of $([ {1 \over 2} g ]+1)$
irreducible components ${\cal D}_g^r$ ($r=0,1,...,[ {1 \over 2} g
]$) each of which consists of surfaces with nodes. Surfaces
belonging to ${\cal D}_g^0$ are such that they become connected
surfaces of genus $(g-1)$ with two punctures upon removal of the nodes.
On the other hand,
${\cal D}_{g}^{r}$ ($r \geq 1$) consists of surfaces which
become, upon removal of the nodes, two disconnected surfaces, one of
genus $r$ and one of genus $(g-r)$, each with one puncture.

A surface which sits in the neighbourhood of ${\cal D}_g^{0}$ has
a long tube which becomes a node as the surface approaches ${\cal D}_g^{0}$.
Thus we can choose coordinates near ${\cal D}_{g}^{0}$ as $4$--tuple
$(\tau, m', z,w)$
where $\tau$ is the length and the
twist of the tube and it serves as a transverse coordinate to
${\cal D}_{g}^{0}$ (the surface approaches ${\cal D}_g^{0}$ as
$\tau \rightarrow \infty$), while $(m',z,w)$ are moduli
of a genus--$(g-1)$ surface with two punctures (where
$z$ and $w$ denote the moduli corresponding to the two
punctures) which is obtained
by removing the node from the surface.

The contribution of the boundary component ${\cal D}_{g}^{0}$
to $\overline{\partial}_{\bar i} F_g$ is given as follows.
Because of the second--order derivative in r.h.s. with respect to
$m_b$ and $\overline{m}_{\bar b}$, at the boundary we will be left with
a derivative in the direction normal to ${\cal D}_{g}^{0}$.
In the coordinates $(\tau, m', z, w)$, the normal derivative
is expressed as ${\partial \over \partial {\rm Im}\, \tau}$.
In the limit $\tau \rightarrow \infty$,
the Beltrami-differentials $\mu^{(z)}$ and $\mu^{(w)}$ associated
to the moduli $z$ and $w$
become localized near the punctures, i.e.
$$ \int \mu^{(z)} G^- \rightarrow \oint_{C_z} G^-  ,$$
while those associated to the moduli $m'$ reduces to the
Beltrami-differentials $\mu'$ on the genus--$(g-1)$ surface $\Sigma_{g-1}$.
Thus the contribution of ${\cal D}_{g}^{0}$ is given by
\eqn\zerocont{ \eqalign{\int_{{\cal D}_{g}^{0}} [dm',dz,dw]
{\partial \over \partial {\rm Im}\, \tau}
\langle \int_{\Sigma_g} \overline{\phi}_{\bar i}& \oint_{C_z} G^-
\oint_{C'_z}  \overline{G}^-  \oint_{C_w} G^-
\oint_{C'_w} \overline{G}^-  \times \cr &
\times \prod_{a=1}^{3g-6} \int_{\Sigma_{g-1}} \mu_a' G^-
\int_{\Sigma_{g-1}} \overline{\mu}_a' \overline{G}^- \rangle_{\Sigma_g} \cr}}

Let us examine the integrand of \zerocont.
Since the operator
$\overline{\phi}_{\bar i}$ is integrated over the entire surface $\Sigma_g$
it either sits on the tube which will be stretched out
in the limit $\tau \rightarrow \infty$ or lies outside of the tube
which becomes the genus--$(g-1)$ surface $\Sigma_{g-1}$ in this limit.
When $\overline{\phi}_{\bar i}$ sits outside of the tube (see
\tfig\FigureThreeOneA), states which
propagate on the tube are projected onto the ground states in the
limit $\tau \rightarrow \infty$.
Since the ground states are generated by the
chiral fields, the effect of a node on the degenerate surface
can be represented by insertions of $\phi_j(z)$ and $\phi_k(w)$
on the points $z$ and $w$
where the node is attached, and the node itself is replaced by
the ground state metric $\eta^{jk}$.
In the coordinates $(\tau, m , z, w)$,  the integrand becomes
$$
\eqalign{   {\partial \over \partial {\rm Im}\, \tau} \eta^{jk}
 \Big\langle   &
 \oint_{C_z} G^- \oint_{C'_z} \overline{G}^- \phi_j(z)
    \oint_{C_w} G^- \oint_{C'_w} \overline{G}^- \phi_k(w)  \times \cr &\times
 \int_{\Sigma_{g-1}} \overline{\phi}_{\bar i}
  \prod_{a=1}^{3g-6} \int_{\Sigma_{g-1}}
 \mu_a' G^- \int_{\Sigma_{g-1}} \mu_a' \overline{G}^-
 \Big\rangle_{\Sigma_{g-1}}. \cr}
$$
This turns out to be zero
since the correlation function in the above is defined on $\Sigma_{g-1}$
and does not depend on the coordinate $\tau$. Thus, when
$\overline{\phi}_{\bar i}$ lies outside of the tube, there is no
contribution from the component ${\cal D}_{g}^{0}$ to
 $\overline{\partial}_{\bar i} F_g$.

\ifigure\FigureThreeOneA{Contributions from the boundary
of moduli space where $\overline{\phi}_{\bar i}$ is outside the long
tube vanishes.}{Fig31a}{1.8}

Let us turn to the case when
$\overline{\phi}_{\bar i}$
sits on the tube (see \tfig\FigureThreeOneB). Suppose
$\overline{\phi}_{\bar i}$ is away from both ends of the tube.
In this case, states on both sides of $\overline{\phi}_{\bar i}$
on the tube are projected onto the ground states. Thus
the effect of the node is represented by an insertion of
$$  \phi_{j}(z) \eta^{jj'}
    \langle j' | \int \overline{\phi}_{\bar i} | k' \rangle
                \eta^{k' k} \phi_{k}(w) $$
on $\Sigma_{g-1}$. Here the integral $\int \overline{\phi}_{\bar i}$
is over the tube away from both ends. Since
$$ \langle j | \overline{\phi}_{\bar i} | k \rangle
   =    \langle \bar{j} | \bar{\phi}_{\bar i}
       | \bar{k} \rangle M_j^{\bar j} M_k^{\bar k}
   = \overline{C}_{\bar{i}\bar{j}\bar{k}}
  e^{2K} G^{\bar j j'} G^{\bar k k'}\eta_{j'j} \eta_{k'k}
 $$
is independent of the position of $\overline{\phi}_{\bar i}$, we can
replace the integral by the multiplication of the volume
of the domain of the integral which can be approximated by
the volume ${\rm Im}\, \tau$ of the tube when $\tau \rightarrow \infty$.
When $\overline{\phi}_{\bar i}$ is close to one of the ends
of the tube,
the amplitude does not scale like the volume ${\rm Im}\, \tau$, and
such a configuration
can be neglected in this approximation. The integrand of \zerocont\
then becomes
\eqn\zeroint{
     \eqalign{ \overline{C}_{\bar{i}\bar{j}\bar{k}} e^{2K} G^{j \bar j}
G^{k \bar k}
     \Big\langle \oint_{C_z}& G^- \oint_{C'_z} \overline{G}^- \phi_j(z)
             \oint_{C_w} G^- \oint_{C'_w} \overline{G}^- \phi_k(w) \times\cr
&\times \prod_{a=1}^{3g-6}
       \int \mu_a' G^- \int \overline{\mu}_a' \overline{G}^-
       \Big\rangle_{\Sigma_{g-1}} \cr}}
where the volume factor ${\rm Im}\, \tau$ is cancelled by the normal derivative
${\partial \over \partial {\rm Im}\, \tau}$.
This remains finite in the limit $\tau \rightarrow \infty$.

\ifigure\FigureThreeOneB{The contribution from the
boundary of moduli space comes from the configuration
where $\overline{\phi}_{\bar i}$ is on the long tube
of length ${\rm Im} \tau$ as $\tau \rightarrow \infty$.}{Fig31b}{1.7}

We need to integrate \zeroint\ over the boundary component
${\cal D}_{g}^{0}$ which is parametrized by $m' \in {\cal M}_{g-1}$
and $z,w \in \Sigma_{g-1}$. Since the interchange of the two
points $z$ and $w$ does not change the complex structure
of the punctured surface, we should include a factor of $(1/2)$
if we are to integrate $z$ and $w$ over the entire surface $\Sigma_{g-1}$
without a constraint. The contribution of the
boundary component ${\cal D}_{g}^{0}$ to $\overline{\partial}_{\bar i} F_g$
is then expressed as
\eqn\zerointegral{
 {1 \over 2} \overline{C}_{\bar{i}\bar{j}\bar{k}} e^{2K} G^{j \bar j}
             G^{k \bar k} \int_{{\cal M}_{g-1}} [dm']
          \Big\langle \int \phi^{(2)}_j
               \int \phi^{(2)}_k
                 \prod_{a=1}^{3g-6} \int \mu_a' G^-
            \int \overline{\mu}_a' \overline{G}^- \Big\rangle_{\Sigma_{g-1}}}

The expression \zerointegral\ can be further simplified by the condition
$\hat{c}=3$. In general, when $\overline{C}_{\bar{i}\bar{j}\bar{k}} \neq 0$,
the left and the right $U(1)$ charges of the three chiral fields
$\phi_i$, $\phi_j$ and $\phi_k$
should sum up to be $\hat{c}$.
$$ q_j + q_k + q_i =
  \overline{q_j} + \overline{q_k} + \overline{q_i}
 = \hat{c} $$
In the present situation, $\hat{c}=3$
and $q_i=\overline{q_i} = 1$.
Therefore we must have $q_{j} + q_{k} =\overline{q}_j
+\overline{q}_k = 2$.
Furthermore, if $q_{j}=0$ or $\overline{q}_j = 0$,
$\phi^{(2)}_j=\{ G^- , [ \overline{G}^- , \phi_j] \}=0$
since a chiral state with $q=0$
is annihilated by both $G^+$ and $G^-$. Therefore we can restrict $j$ and
$k$ in \zeroint\ to those with $(q_j,\overline{q}_j)=(q_k,
\overline{q}_k)=(1,1)$.
These are the ones which correspond to the marginal deformations
of the twisted $N=2$ model, and we can replace the insertions of
$\int \phi_j^{(2)}$ in \zerointegral\ by
a derivative $D_j$.
The contribution \zerointegral\ of the boundary component
${\cal D}_{g}^{0}$ to $\overline{\partial}_{\bar i} F_g$ is then
expressed as
\eqn\zerofinal{{1 \over 2} \overline{C}_{\bar{i}\bar{j}\bar{k}}
e^{2K} G^{j \bar j}
G^{k \bar k} D_j D_k F_{g-1}.}

Let us turn to the other boundary components ${\cal D}_{g}^{r}$
($r=1,...,[{1 \over 2} g]$).
A surface in the neighborhood of ${\cal D}_{g}^{r}$,
has a long tube which connects two disconnected surfaces
$\Sigma_r$ and $\Sigma_{g-r}$ of genus $r$ and genus $(g-r)$.
Thus we can choose coordinates near ${\cal D}_{g}^{r}$
as $5$--tuple
$(\tau, m', z , m'',w)$
where $\tau$ characterizes the tube connecting the two surfaces,
and $(m',z) \in {\cal M}_{r,1}$ and $(m'',w) \in {\cal M}_{g-r,1}$.
As in the case of ${\cal D}_{g}^{0}$ discussed in the above,
a non-vanishing contribution to $\overline{\partial}_{\bar i} F_g$
comes from the region where the amplitude scales like
${\rm Im}\, \tau$. This is the case when the operator
$\overline{\phi}_{\bar i}$ is on the tube  (see \tfig\FigureThreeTwo).
The factor ${\rm Im}\, \tau$ is cancelled by the derivative
operator ${\partial \over \partial {\rm Im}\, \tau}$
 and the effect of the tube
is represented by the operator
$$ \phi_j(z) \eta^{jj'} \langle j'
| \overline{\phi}_{\bar i} | k' \rangle \eta^{k'k} \phi_k(w)
 = \overline{C}_{\bar{i}\bar{j}\bar{k}} e^{2K} G^{j\bar j} G^{k \bar k}
      \phi_j(z)\phi_k(w)$$
where $\phi_j(z)$ is inserted on $\Sigma_{r}$ and
$\phi_k(z)$ is on $\Sigma_{g-r}$. The contribution of
${\cal D}_{g}^{r}$ to $\overline{\partial}_{\bar i} F_g$ is then given by
$$ \eqalign{&
 \overline{C}_{\bar{i}\bar{j}\bar{k}} e^{2K} G^{j \bar j}
             G^{k \bar k} \int_{{\cal M}_{r}} [dm']
          \langle \int \phi_j^{(2)}
              \prod_{a=1}^{3r-3} \int \mu_a' G^-
            \int \overline{\mu}_a' \overline{G}^- \rangle_{\Sigma_{r}}
\times \cr &~~~~~~
    \times \int_{{\cal M}_{g-r}} [dm'']
           \langle \int \phi_k^{(2)}
              \prod_{a=1}^{3(g-r)-3} \int \mu_a'' G^-
            \int \overline{\mu}_a''
\overline{G}^- \rangle_{\Sigma_{g-r}} =\cr
&= \overline{C}_{\bar{i}\bar{j}\bar{k}} e^{2K} G^{j\bar j} G^{k \bar k}
    D_j F_r D_k F_{g-r} .\cr } $$

\ifigure\FigureThreeTwo{Another component of the boundary
of moduli space where the Riemann surface splits to
two Riemann surfaces connected by a long tube; to get
a nonvanishing contribution
$\overline{\phi}_{\bar i}$ is inserted on the tube.}{Fig32}{1.3}

Extra care is required when $g$ is even and $r={1 \over 2}g$. In this case
there is a ${\bf Z}_2$ symmetry between the two surfaces $\Sigma_r$
and $\Sigma_{g-r}$, and we must include a factor $(1/2)$ to take
into account this symmetry. The contributions of the boundary
components ${\cal D}_{g,k}$ ($k=1,...,[{1 \over 2} g]$) are
then
$$
 \sum_{r=1}^{[{1 \over 2} g]}
     \overline{C}_{\bar{i}\bar{j}\bar{k}} e^{2K} G^{j \bar j}
                G^{k \bar k} D_j F_r D_k F_{g-r} $$
if $g$ is odd and
$$  \sum_{r=1}^{{1 \over 2}g-1}
   \overline{C}_{\bar{i}\bar{j}\bar{k}} e^{2K} G^{j\bar j}
                   G^{k \bar k} D_j F_r D_k F_{g-r}
+{1 \over 2} \overline{C}_{\bar{i}\bar{j}\bar{k}}
   e^{2K} G^{j\bar j}  G^{k \bar k} D_j F_{{1 \over 2}g} D_k
   F_{{1 \over 2}g} $$
if $g$ is even. They can be summarized in a single equation as
$$ {1 \over 2} \sum_{r=1}^{g-1}
    \overline{C}_{\bar{i}\bar{j}\bar{k}} e^{2K} G^{j\bar j}
                   G^{k \bar k} D_j F_r D_k F_{g-r} $$

By combining this with \zerofinal , we obtain the holomorphic
anomaly of the genus--$g$ partition function as
\eqn\anomaly{
  \overline{\partial}_{\bar i} F_g =
 {1 \over 2} \overline{C}_{\bar{i}\bar{j}\bar{k}}
 e^{2K} G^{j \bar j} G^{k \bar k}
 \left( D_j D_k F_{g-1} + \sum_{r=1}^{g-1}
          D_j F_r D_k F_{g-r} \right). }
This gives a recursion relation for $F_g$ with respect to the genus $g$.
In fact, it is possible to solve this equation iteratively, and we
will present a systematic method to do so in Section 6.

The holomorphic anomaly equations of $F_g$'s for all $g \geq 2$ can be
combined into a single equation by introducing a formal sum of $F_g$'s as
\eqn\partitionsum{ {\cal F}(\lambda; t,\overline{t}) = \sum_{g=1}^\infty
        \lambda^{2g-2} F_g .}
Since each $F_g$ is a section of a line-bundle
${\cal L}^{2-2g}$ over the moduli space of the topological theory,
${\cal F}(\lambda; t, \overline{t})$ should be regarded as a function on
the total space of ${\cal L}$, with $\lambda$ being a coordinate
on the fiber of ${\cal L}$. We then consider the following equation.
\eqn\master{
   \left( \overline{\partial}_{\bar i} - \overline{\partial}_{\bar i} F_1
          \right) \exp({\cal F})
   = {\lambda^2 \over 2} \overline{C}_{\bar{i}\bar{j}\bar{k}}
     e^{2K} G^{j\bar{j}} G^{k\bar{k}}
     \hat{D}_j \hat{D}_k \exp({\cal F}) }
where
$$ \eqalign{
   \hat{D}_j {\cal F}(\lambda ; t, \overline{t}) & \equiv
     \sum_g \lambda^{2g-2} D_j F_g \cr
   &= \sum_g \lambda^{2g-2} (\partial_j - (2g-2) \partial_j K) F_g \cr
   &= (\partial_j - \partial_j K \lambda \partial_\lambda)
      {\cal F}(\lambda ; t, \overline{t}) .\cr} $$
By expanding both-hand sides of \master\ in power series of $\lambda$
and by comparing each term in the expansion, we recover the holomorphic
anomaly equation \anomaly .
We call this the {\it master anomaly equation} of the topological string
theory.  It is satisfying to see that the holomorphic anomalies for all
$g \geq 2$ are summarized in a single equation. Later we will further
improve this equation to incorporate the genus--$1$ anomaly equation.

As we will solve the holomorphic anomaly equation \anomaly\
later, it is instructive to check the integrability of the equation
here. Since the holomorphic anomaly is summarized in the master
equation \master , it is sufficient to prove
$$ \eqalign{ & [ d_{\bar i} , d_{\bar j} ] = 0 \cr
 & d_{\bar i}= \overline{\partial}_{\bar i} - \overline{\partial}_{\bar i}
F_1 -{\lambda^2 \over 2} \overline{C}_{\bar{i}\bar{j}\bar{k}}
e^{2K} G^{j \bar j}
     G^{k \bar k}    \hat{D}_{j} \hat{D}_k . \cr} $$
By using the special geometry relation
\eqn\speccial{ [ \overline{\partial}_{\bar i} , D_j ]_k^l
    = - G_{\bar{i} j} \delta_k^l - G_{\bar{i} k} \delta_j^l
      + C_{jkm} \overline{C}_{\bar{i}\bar{l}\bar{m}} e^{2K} G^{m \bar m}
        G^{\bar{l} l} }
and the properties of the Yukawa coupling
$$\eqalign{ \overline{C}_{\bar{i}\bar{j}\bar{k}} = &
\overline{C}_{\bar{j}\bar{i}\bar{k}},~~~
   D_{\bar{i}} \overline{C}_{\bar{j}\bar{k}\bar{l}} =
     D_{\bar{j}} \overline{C}_{\bar{i}\bar{k}\bar{l}} \cr
      & \partial_i \overline{C}_{\bar{j}\bar{k}\bar{l}} = 0,\cr} $$
we find the commutator to be
$$\eqalign{ [ d_{\bar i} , d_{\bar j } ]
& =  \lambda^2
  \overline{C}_{\bar{i}\bar{k}\bar{l}} e^{2K} G^{k \bar{k}} G^{l \bar{l}}
  (\partial_k \overline{\partial}_{\bar j} F_1 -
   {1 \over 2} {\rm Tr} C_k \overline{C}_{\bar j})  \partial_l
  - (\bar{i} \leftrightarrow \bar{j}) + \cr
&~~~~+{\lambda^2 \over 2}
 \overline{C}_{\bar{i}\bar{k}\bar{l}} e^{2K} G^{k \bar k}G^{l \bar l}
   D_k \partial_l \overline{\partial}_{\bar j} F_1
 - (\bar{i} \leftrightarrow \bar{j}) .\cr } $$
That the r.h.s. of this equation is zero is the
consequence of the holomorphic anomaly equation\foot{
The holomorphic anomaly equation for $F_1$ here differs by
a factor $(1/2)$ to the one presented in \bcov\ due to the
different normalization of $F_1$.} of $F_1$
\eqn\genusoneanomaly{
  \partial_i \overline{\partial}_{\bar j} F_1
  ={1 \over 2} {\rm Tr}\, C_i \overline{C}_{\bar j}
 - {\chi \over 24} G_{i \bar j}}
where $\chi = {\rm Tr}(-1)^F$.
Substituting this into the above, we obtain
$$  [ d_{\bar i}, d_{\bar j} ]  =
   {\lambda^2 \over 4} \left[
  \overline{C}_{\bar{i}\bar{k}\bar{l}} e^{2K} G^{k\bar k}G^{l \bar l}
   {\rm Tr}( D_k C_l ) \overline{C}_{\bar j}  -
  (\bar{i} \leftrightarrow \bar{j}) \right] = 0 .$$
Here we also used $D_i C_{jkl} = D_j C_{ikl}$.
It is curious to see that both the special geometry relation
and the genus--$1$ anomaly equation play important roles in proving
the consistency of the holomorphic anomaly at $g \geq 2$.
This in fact is not without a reason. We will see later that
the special geometry relation can be regarded as a holomorphic
anomaly equation at genus--$0$, and the anomalies at all genera
including $g=0$ and $1$ can be described in a single framework.

Now that we found the BRST--trivial operator $\{ G^+ ,[
\overline{G}^+, \overline{\phi}_{\bar i}] \}$
does not decouple from $F_g$, one
might wonder whether $F_g$ is sensitive to still other types
of BRST-trivial deformations of
the topological theory. The
$(c,c)$ field $\phi_i$ which
generate the truly marginal deformation of the topological theory satisfies
$$[ G^+ , \phi_i ] = [ \overline{G}^+ , \phi_i ] = 0,$$ and the $(a,a)$
field $\overline{\phi}_{\bar i}$ which is complex conjugate to $\phi_i$
obeys
$$[ G^-, \overline{\phi}_{\bar i} ] = [ \overline{G}^- ,
\overline{\phi}_{\bar i}]=0.$$
However the topological theory realized by the twisted $N=2$ model
may also contain a $(a,c)$ field $\widetilde{\phi}$
subject to
$$[ G^- , \widetilde{\phi} ]
= [ \overline{G}^+ , \widetilde{\phi} ] =0$$
and its conjugate $(c,a)$ field.
Thus we would like to know if $F_g$ is sensitive to a deformation
generated by these operators.
We show here that, in fact,  the operator $\{ G^+ , [
\overline{G}^- , \widetilde{\phi} ] \}$ and its conjugate
decouple from $F_g$.

If we insert such an operator on $\Sigma_g$, we can deform the contour
of $G^+$ surrounding
the operator $\widetilde{\phi}$ and pick up the
commutator of $G^+$ with
$\int \mu_a G^-$ ($a=1,...,3g-3$) inserted on
$\Sigma_g$. The commutator produces the energy-momentum tensor $T$ which is
then converted into a derivative with respect to the moduli $m$.
$$ \eqalign{ &
 \int_{{\cal M}_g} [dm] \int d^2 z
   \langle \oint_{C_z} G^+ \oint_{C'_z} \overline{G}^-
  \widetilde{\phi}(z) \prod_{a=1}^{3g-3} \int \mu_a G^-
      \int \overline{\mu}_a \overline{G}^- \rangle = \cr
  &= \int_{{\cal M}_g} [dm]  \sum_{b=1}^{3g-3} 2{\partial \over
   \partial m_b} \int d^2 z \langle \int_{C'_z}
\overline{G}^-\widetilde{\phi}(z)
      \prod_{a \neq b} \int \mu_a G^- \prod_{\bar{a}=1}^{3g-3}
        \int \overline{\mu}_a \overline{G}^- \rangle .\cr}  $$
Due to the derivative with respect to $m_b$, this becomes
an integral on the boundary of ${\cal M}_g$. It turns out that
the boundary term vanishes for the following reason. So far
we have not touched $\int \overline{\mu}_a \overline{G}^-$
in the right-moving
sector, and there are still $(3g-3)$ of them on $\Sigma_g$.
In the neighbourhood of ${\cal D}_g^r$ ($r=0,1,...,[{1 \over 2}g]$),
one of them becomes a contour integral of $\overline{G}^-$ around the tube
which becomes a node on ${\cal D}_g^r$. Namely we have the $\overline{G}^-$
charge inserted on the tube. As the surface approaches the boundary,
states which propagate on the tube are projected onto the ground
states, all of which are annihilated by $\overline{G}^-$. It does not
matter whether the operator $[ \overline{G}^- , \widetilde{\phi} ]$
is on or off the tube since it anti-commutes with
$\overline{G}^-$. In this way, the boundary term vanishes due to
the $\overline{G}^-$ charge which comes from one of $\int \overline{\mu}_a
\overline{G}^-$. Since there is no boundary term, the operator
$\{ G^+,[ \overline{G}^-, \widetilde{\phi} ] \}$ decouples
from $F_g$. Similarly $F_g$ is invariant under the deformation
generated by $(c,a)$ fields.

In the case of the topological
sigma--model of $A$--type described in Section 2,
the $(c,c)$ and $(a,a)$ fields generate
deformations of the K\"ahler class on the Calabi-Yau manifold $M$
while the $(a,c)$ and $(c,a)$ fields correspond to deformations
of the complex structure. The result here suggests that $F_g$ in this
case is independent of the complex structure of $M$, but depends on
the K\"ahler class on $M$. The anti-holomorphic dependence of $F_g$
on the K\"ahler moduli
is determined by the holomorphic anomaly equation \anomaly . The situation
is opposite in the case of the $B$--model. In this case, $F_g$ does not
depend on the K\"ahler moduli of $M$. Especially $F_g$ is independent
of the volume of $M$. This fact becomes important in section 5.

\vskip .15in

\subsec{Holomophic anomalies of correlation functions}

So far, we have studied the holomorphic anomaly of partition functions.
Let us now turn to correlation functions $C_{i_1 \cdots i_n}^{(g)}$ of
the chiral fields given by
$$\eqalign{ C_{i_1 \cdots i_n}^{(g)}& =
  \int_{{\cal M}_g} \langle \prod_{r=1}^n \int
    \phi^{(2)}_{i_r} \prod_{a=1}^{3g-3}
    \int  \mu_a G^- \int \overline{\mu}_a \overline{G}^- \rangle \cr
 &= D_{i_1} \cdots D_{i_n} F_g. \cr}$$
As in the case of the partition function $F_g$,
the derivative $\overline{\partial}_{\bar i}$ brings down the BRST trivial
operator $\{ G^+ , [ \overline{G}^+ , \overline{\phi}_{\bar i} ] \}$,
and the commutators of $G^+$ and $\overline{G}^+$ with $G^-$ and
$\overline{G}^-$ in $C^{(g)}_{i_1,...,i_n}$
generate second--order derivatives
with respect to $(m,\overline{m}) \in {\cal M}_g$ and $(z_r,\overline{z}_r)
\in \Sigma_g$ where $z_r$ ($r=1,...,n$) are the positions of the
chiral  fields $\phi_{i_r}$.
We can then apply the Cauchy theorem to reduce the computation
to a boundary integral. The boundary in this case consists of
two types; one is the boundary of the moduli space
${\cal M}_{g,n}$ of a genus--$g$ surface with $n$--punctures.
Another contribution arises in a limit when one of the chiral fields
$\phi_{i_r}$ approaches $\overline{\phi}_{\bar i}$.

The computation on
the boundary of the first--type is a straightforward generalization of
the one for $\overline{\partial}_{\bar i}F_g$ we did in the above.
The boundary of the moduli space ${\cal M}_{g,n}$
consists of irreducible components ${\cal D}_{(g,n)}^{(0)}$
and ${\cal D}_{(g,n)}^{(r,s)}$
each of which
consists of surfaces with punctures and nodes.
Here, for ${\cal D}_{(g,n)}^{(r,s)}$,
$r$ and $s$ run from $0$ to $g$ and from
$0$ to $n$ respectively,  ${\cal D}_{(g,n)}^{(0,0)}$
and ${\cal D}_{(g,n)}^{(0,1)}$
are empty, and ${\cal D}_{(g,n)}^{(r,s)}$ is identified with
${\cal D}_{(g,n)}^{(g-r,n-s)}$. Surfaces belonging
to ${\cal D}_{(g,n)}^{(0)}$ become connected surfaces of genus $(g-1)$
with $(n+2)$ punctures upon removal of the nodes
(see \tfig\FigureThreeThreeA). On the other
hand, ${\cal D}_{(g,n)}^{(r,s)}$ consists
of surfaces which become, upon removal of the nodes, two disconnected
surfaces, one of genus $r$ with $(s+1)$ punctures and another of
genus $(g-r)$ with $(n-s+1)$ punctures
(see \tfig\FigureThreeThreeB).
As in the case of the partition
function $F_g$, contributions of these boundary
components to $\overline{\partial}_{\bar i} C^{(g)}_{i_1 \cdots i_n}$ come
from the region where the operator $\overline{\phi}_{\bar i}$ sits
on the tube which becomes the node at the boundary
of ${\cal M}_{g,n}$, and are expressed as
\eqn\firsttype{ \eqalign{&
 {1 \over 2}\overline{C}_{\bar{i}\bar{j}\bar{k}} e^{2K} G^{j \bar j}
   G^{k \bar k}   C^{(g)}_{j k i_1 \cdots i_n}+ \cr
  + {1 \over 2}\overline{C}_{\bar{i}\bar{j}\bar{k}} & e^{2K} G^{j \bar j}
   G^{k \bar k} \sum_{r=0}^{g} \sum_{s=0}^n
  {1 \over s! (n-s)!} \sum_{ \sigma \in S_n}
       C^{(r)}_{j i_{\sigma(1)} \cdots i_{\sigma(s)}}
    C^{(g-r)}_{k i_{\sigma(s+1)} \cdots i_{\sigma(n)}}
    \cr}   }
where
$$ \eqalign{ & C_{i_1 \cdots i_n}^{(0)} =
     D_{i_1} \cdots D_{i_{n-3}} C_{i_{n-2}i_{n-1}i_n} ~~~(n \geq 3) \cr
     &C^{(0)}=0,~~C^{(0)}_i=0,~~C^{(0)}_{ij}=0 . \cr} $$

\ifigure\FigureThreeThreeA{One of the boundary components
of moduli space with fields inserted.  The contribution
again comes from the insertion of the $\overline{\phi}_{\bar i}$
on the tube.}{Fig33a}{2.1}

\ifigure\FigureThreeThreeB{The contribution from another component
of moduli space where again the operator $\overline{\phi}_{\bar i}$
is inserted on the tube.}{Fig33b}{1.7}

The boundary of the second type arises since
there is a singularity in the operator product of
$\phi_{j}$ ($j=i_1,...,i_n$) and $\overline{\phi}_{\bar i}$.
\eqn\divergence{ \phi_{j}(z) \overline{\phi}_{\bar i}(w) \sim
   {G_{j \bar i} \over |z-w|^2 }~~~~
     (z \rightarrow w) }
How we regularize this divergence is a part of the
definition of the theory. In the
perturbed $N=2$ theory given by the action
$S = S_0(t_0,\overline{t}_0) + \delta t^i \int \phi^{(2)}_i +
\delta \overline{t}^i \int
\overline{\phi}_{\bar i}^{(2)}$, we assume that the original theory
with the action $S_0$ has the $N=2$ superconformal
invariance, which in particular means that the theory is finite.
In order to perturb the theory while maintaining
the superconformal symmetry, we must specify how to deal with
the short distance singularity between $\phi^{(2)}_j$
and $\overline{\phi}_{\bar i}^{(2)}$
\eqn\divergencetwo{ \phi^{(2)}_j(z) \overline{\phi}_{\bar i}^{(2)}(w) \simeq
4 \partial_z \overline{\partial}_{\bar z}
\phi_j(z) \overline{\phi}_{\bar i}(w) \simeq
{4G_{j \bar i} \over |z-w|^4 }~~~~
       (z \rightarrow w) .}
This divergence which arises from this short distance
singularity is power in $|z-w|$ and is not universal.
Thus we can simply subtract it away (one can renormalize
the divergence into the cosmological constant if one wishes).
Once we subtract the singularity in the operator product
between $\phi_j$ and $\overline{\phi}_{\bar i}$,
the boundary of the second type
does not contribute to $\overline{\partial}_{\bar i} C_{i_1 \cdots
i_n}^{(g)}$.

This is the case when the world-sheet is a flat infinite plane.
When the world-sheet is compact, there are subleading
divergences in \divergence\ and \divergencetwo\ which
generate non-vanishing contributions for the boundary
of the second type. The subleading divergences depend
linearly on the curvature of $\Sigma$, and they can
be derived from the short distance expansion of
the Green's function on $\Sigma$. We can also understand
this effect from the topological field theoretical
point of view as follows. Let us choose a metric on
$\Sigma$ as $|\nu(z)|^4$ where $\nu$ is a meromorphic
${1 \over 2}$--differential on $\Sigma$ with
a pole and $g$--zeros at the Riemann divisor.
Since the theory is conformally invariant,
we are free to use any metric we like.
In this metric, the curvature has delta-function like
singularities each of which
carries $\int R = \pm 4 \pi$ ($+4\pi$ at the pole
of $\nu$ and $-4\pi$ at the zeros of $\nu(z)$).
When the operators $\phi_j(z)$ and $\overline{\phi}_{\bar i}(w)$
are away from the support of the curvature, there
is no contribution from the boundary of the second type
since the computation is the same as in the case of the flat
infinite plane. On the other hand, near the curvature singularity,
we must take into account the fact that, due to the twisting,
there is an operator $e^{\pm \varphi}$ inserted
there where $\varphi$ is the bosonized $U(1)$ current.
The operator $e^{\varphi}$ is the chiral field of the
maximum charge $(3,3)$ (corresponding to the holomorphic
$3$--form on the Calabi-Yau $3$--fold), and $e^{-\varphi}$
is its conjugate anti-chiral field.
Thus we can evaluate the boundary term as in the case
of the boundary of the first type discussed in the above.
We then obtain $\pm 2 \sum_{s=1}^n G_{\bar{i} i_s}C_{i_1 \cdots i_{s-1}
i_{s+1} \cdots i_n}^{(g)}$ from each of the curvature singularities.
We should also take into account the effect of the punctures
on the surface. This can be done most easily by noting that
the final result should be linear in the integral of the
curvature $\int R = - 2 \pi (2g-2 +n-1)$ on the genus--$g$
surface with $(n-1)$--punctures. The contribution from the
boundary of the second type is then
\eqn\secondtype{ - (2g-2+n-1) \sum_{s=1}^n G_{\bar{i} i_s}
         C_{i_1 \cdots i_{s-1} i_{s+1} \cdots i_n}^{(g)} .}

By combining \firsttype\ and \secondtype , we obtain
\eqn\corranomaly{\eqalign{
&\overline{\partial}_{\bar i} C_{i_1 \cdots i_n}^{(g)}
= {1 \over 2}\overline{C}_{\bar{i}\bar{j}\bar{k}} e^{2K} G^{j \bar j}
   G^{k \bar k}   C^{(g-1)}_{j k i_1 \cdots i_n} +\cr &
{}~~+
 {1 \over 2}\overline{C}_{\bar{i}\bar{j}\bar{k}} e^{2K} G^{j \bar j}
   G^{k \bar k}  \sum_{r=0}^{g} \sum_{s=0}^n
 {1 \over s! (n-s)!} \sum_{ \sigma \in S_n}
   C^{(r)}_{j i_{\sigma(1)} \cdots i_{\sigma(s)}}
  C^{(g-r)}_{k i_{\sigma(s+1)} \cdots i_{\sigma(n)}}
        - \cr
 &~~- (2g-2+n-1) \sum_{s=1}^n G_{\bar{i} i_s}
         C_{i_1 \cdots i_{s-1} i_{s+1} \cdots i_n}^{(g)} . \cr} }
Especially when $n=0$, this equation reduces to the anomaly equation
\anomaly\ of $F_g$.
The derivation of this equation is valid also
for $g=0$ ($n \geq 4$) and $g=1$ ($n \geq 2$).
The anomaly equation in the case of $g=1, n=1$
is given by \genusoneanomaly\ and is slightly
different from the above\foot{The genus-$1$ one-point function
may be included in the above
equation if we allow the substitution
$(2g-2) C^{(g)} \rightarrow (\chi /24 -1)$ for $g \rightarrow 1$.}.

To understand
the structure of this equation better, let us take a look at
the simplest case of $g=0, n =4$. In this case, the equation
becomes
$$ \eqalign{\overline{\partial}_{\bar i} C_{i_1 i_2 i_3 i_4}
 = & \overline{C}_{\bar{j}\bar{i}\bar{k}}
     e^{2K} G^{j \bar j} G^{k \bar k}
  \left( C_{j i_1 i_2}C_{k i_3 i_4} + C_{j i_1 i_3} C_{k i_4 i_2}
          + C_{j i_1 i_4}C_{k i_2 i_3} \right)
  - \cr
  & - G_{\bar{i} i_1} C_{i_2i_3i_4}
  - G_{\bar{i} i_2} C_{i_3i_4i_1} - G_{\bar{i}i_3} C_{i_4i_1i_2}
   - G_{\bar{i} i_4} C_{i_1i_2i_3}.\cr} $$
We can rederive this equation by computing $\overline{t}^i$-derivative
of $C_{i_1i_2i_3i_4} = D_{i_1} C_{i_2i_3i_4}$ directly by using
the holomorphicity of the Yukawa coupling $\overline{\partial}_{\bar i}
C_{ijk} = 0$ and the special geometry relation \speccial\ for
the commutator
$[ \overline{\partial}_{\bar i} , D_j ]$. In general, at $g=0$,
one can deduce  the anomaly equation \corranomaly\
$n \geq 4$ from the special geometry relation
and the holomorphicity of $C_{ijk}$ by mathematical induction in $n$.
Similarly the anomaly equation
\corranomaly\ for $g \geq 1$ is a consequence of
the special geometry and the holomorphic
anomaly \anomaly\ of $F_g$.
Thus we come to view that that {\it the special geometry is also
one of the aspects
of the holomorphic anomaly in the topological string theory}.

Previously we found that the holomorphic anomalies of
the partition functions $F_g$
($g \geq 2$) can be summarized
in the form of  the master anomaly equation \master.
It is also possible to combine them with
the anomalies of the correlation
functions \corranomaly\ into a single set of equations.
It turns out that the equations also contain the genus-$1$ anomaly
equation \genusoneanomaly .
For this purpose, we introduce the following object.
\eqn\generating{
  W(\lambda, x; t, \overline{t}) =
    \sum_{g=0}^\infty \sum_{n=0}^\infty
          {1 \over n!} \lambda^{2g-2} C^{(g)}_{i_1 \cdots i_n}
                x^{i_1} \cdots x^{i_n} +
 \left( {\chi \over 24} - 1 \right)\log \lambda  }
where $C^{(g)}_{i_1 \cdots i_n} = 0$ for $(2g-2+n) \leq 0$.
This may be regarded as
a generating function for the correlation functions.
Because of the $\log \lambda$ term in r.h.s.,
$\exp(W)$ transforms like a section of ${\cal L}^{({\chi
\over 24} -1)}$.
Let us consider the following equation
\eqn\delbarequation{ \eqalign{&
  {\partial \over \partial \overline{t}^i} \exp ( W ) = \cr & =
  \left[ {\lambda^2  \over 2} \overline{C}_{\bar{i} \bar{j} \bar{k}}
            e^{2K} G^{j \bar j} G^{k \bar k}
      {\partial^2 \over \partial x^j \partial x^k}
       - G_{\bar{i} j} x^j
   \big( \lambda {\partial \over \partial \lambda}+
       x^k {\partial \over \partial x^k}\big) \right]
   \exp ( W ).\cr} }
Substituting \generating\ into the above and expanding it
in powers of $\lambda$ and $x^i$'s, one recovers
the anomaly equation \corranomaly\ for the correlation functions.
One also finds that the genus-$1$ equation \genusoneanomaly, which
in the case of $\hat{c}=3$ can be written as
$$ \partial_i \overline{\partial}_{\bar j} F_1 =
 {1 \over 2} C_{ikl} \overline{C}_{\bar{j}\bar{k}\bar{l}} e^{2K}
G^{k \bar k} G^{l \bar l} - \left({\chi \over 24} -1 \right) G_{i\bar j} $$
is also contained in this equation. Here the sums over $k$ and $l$
are over those with $(q_k,\overline{q}_k)=(q_l, \overline{q}_l) =
(1,1)$.

The equation \delbarequation\ will prove to be crucial
in section 6 when we solve the anomaly equation \anomaly\
and derive explicit expressions for $F_g$.

Since the anomaly equation is summarized in
\delbarequation , one may try to solve it directly.
However we must also remember that $W$ has the structure
of \generating\ with
$$ C^{(g)}_{i_1 \cdots i_n} =\cases{ D_{i_1} \cdots
D_{i_n} F_g & for $g \geq 1$ \cr
D_{i_1} \cdots D_{i_{n-3}} C_{i_{n-2}i_{n-1}i_n} & for $g=0$ \cr} $$
and
$$ C^{(g)}_{i_1 \cdots i_n} = 0 ~~~~{\rm for}~~ 2g-2+n \leq 0 . $$
This property of $W$ can also be summarized in a single equation
as
\eqn\delequation{
\eqalign{&
 \left[ {\partial \over \partial t^i} + \Gamma_{ij}^k x^j
 {\partial \over \partial x^k} + \partial_i K
 ({\chi \over 24} -1 -\lambda
 {\partial \over \partial \lambda} )
 \right] \exp (W) = \cr
& = \left[ \sum_{g=0}^\infty \sum_{n=0}^\infty
     {1 \over n!} \lambda^{2g-2} D_i C_{i_1 \cdots i_n}^{(g)}
         x^{i_1} \cdots x^{i_n} \right] \exp(W) \cr
&=    \left[ {\partial \over \partial x^i} - \partial_i F_1
             - {1 \over 2 \lambda^2} C_{ijk} x^j x^k \right]
        \exp(W) . \cr}}
The two equations, \delbarequation\ and \delequation\ combined,
are equivalent to all the holomorphic anomaly equations.
In Appendix B, we analyse
the two equation directly to all order in $g$. The order-by-order
solution of the anomaly equation is presented in section 6.

Recently Witten
\ref\witt{E. Witten, {\it Quantum Background
Independence in
String Theory},
IAS preprint, IASSNS-HEP-93/29, hep-th/9306122} discussed the implication
of the holomorphic anomaly, which we had previously announced in \bcov, to the
background
(in)dependence of the string theory. There he
also derived two equations, one involving $\partial_{\bar{t}^{\bar i}}$
and the other involving $\partial_{t^i}$, for some finite
dimensional quantum system associated to the Calabi-Yau manifold
which resemble the two equations, \delbarequation\ and
\delequation , derived in the above. It would be
interesting to understand the precise connection between
them.

\vfill
\eject

\newsec{Comments on the open string version}
\def\CO{{\cal O}}

The topological field theories obtained by twisting $N=2$ supersymmetry can
also be defined on Riemann surfaces $\Sigma$
having boundaries. In order to preserve topological invariance, one has to
impose appropriate $Q$--invariant boundary conditions.
Generally speaking, in the open string
case the methods of sects.2, and 3 are much
less powerful than in the closed one: The reason being that all our arguments
rest on manipulations
involving the {\it two\/} scalar supercharges of the twisted theory. In the
open case the boundary condition is chosen so that the combination
$Q=G^++\overline G^+$ is preserved;  however the other combination
$G^+-\overline G^+$ does
not leave the boundary condition invariant and hence is not conserved
any longer. Then some of the manoveurs do not extend to the open case. In
particular in the open case the curvature of the Zamolodchikov metric does not
satisfy the special geometry relation. This is related to the fact, that while
closed strings compactified on a Calabi--Yau $3$--fold
lead to $N=2$ space--time
supergravity, in the open case they lead to $N=1$. In the first case special
geometry is implied by space--time supersymmetry \dewit, while the only
requirement
from $N=1$ supersymmetry is that the Zamolodchikov metric should be K\"ahler.

A particular realization of open strings satisfying
the appropriate boundary conditions can be described in the
context of the sigma--models.  The corresponding boundary conditions,
with either $A$-- or $B$--twisting, were described
in ref.\witcs. In the $A$--model one
picks a Lagrangian submanifold $L_i\subset M$ for each
component $C_i$ of $\partial\Sigma$.
Let $TL_i$ and $NL_i$ be the tangent and normal bundles of $L_i$ in $M$. Then
one requires that the boundary $C_i$ is mapped
into the submanifold $L_i$; at the boundary the
normal derivative of the bosonic field $X$ takes values in
$X^*(NL_i)$; instead $\chi$ and the pullback of $\psi$ to $L_i$ take value in
$X^*(TL_i)$. For the $B$--model one requires that the normal derivative of $X$
to vanish on $\partial\Sigma$, and that $\theta$ vanishes on the boundary as
well as the pullback of $\star\rho$ to $\partial\Sigma$.

Just as the topological field theory on surfaces without boundaries defines a
closed string theory, if we allow boundaries the topological model will
define
an open string theory. We can also couple to this string theory
target--space gauge fields by introducing Chan--Paton factors as usual, i.e.
coupling the
gauge fields to charges which propagate along the boundary. In the $B$--model
this results in a coupling of the open string to a rank $N$ holomorphic bundle
$E$ over the Calabi--Yau $3$--fold $M$ having structure group $U(N)$.

Given the deep analogy between the open and closed cases, it may be appropriate
to pause a while to discuss how
the results of sections 2 and 3 get modified in the open case.

\vskip .15in

\subsec{$tt^*$ in the open string case}

Let us start with the open analog of the $tt^\ast$ equation. We consider the
following geometry (see \tfig\FigureFourOne): a flat strip of width $\pi$ and
length $L$
with a half--disk attached at one end. On the boundary, except for the segment
$l$ at the opposite end, we impose the appropriate open string boundary
condition, as discussed above. On the circle arc we insert the open string
topological observable $\CO_\alpha$.

\ifigure\FigureFourOne{By inserting the topological observable
$\CO_\alpha$ on the circle arc of the open string world sheet and
doing the twisted path integral on the half--disk we get a state
$|\alpha \rangle $ at the boundary.  If we take
$L\rightarrow \infty$ the state thus obtained is a ground
state in the open string Ramond sector.}{Fig41}{.9}

The topological path integral in this
geometry --- viewed as a functional of the boundary values of the fields on the
segment $l$ --- defines a state in the open string Hilbert space which we call
$|\alpha\rangle$. Notice that this state is automatically in the Ramond
sector. To see this, observe that the twisting introduces an extra holonomy
factor for the fermions equal to
\eqn\extraholo{\exp\left[\pm {i\over 2}\oint \omega\right],}
where $\omega$ is the spin--connection. Given that the boundary in Fig. 7 has
a geodesic deviation of $\pi$, \extraholo\ gives an additional factor $(-1)$
which transforms the $NS$ sector into the $R$ one. Just as in the closed case,
for
each topological state $|\alpha\rangle$ we can find a representative which is
an actual vacuum for the untwisted theory defined in the strip. This vacuum is
obtained simply by letting $L\rightarrow\infty$ in the definition of
$|\alpha\rangle$. If $\theta$ is the CPT operation for the untwisted theory,
$|\bar\alpha\rangle\equiv \theta|\alpha\rangle$ is also a vacuum. This allows
us to introduce in the open case a real structure matrix
${M_{\bar\beta}}^\alpha$ analogous to that for the closed case,
and then a
hermitian $tt^\ast$ metric
$$g_{\alpha\bar\beta}= \eta_{\alpha\gamma}\, {M_{\bar\beta}}^\gamma,$$
where $\eta_{\alpha\gamma}$ is the open case topological metric, defined by the
topological path integral performed
in the geometry of \tfig\FigureFourTwo

\ifigure\FigureFourTwo{Open string topological metric $\eta_{\alpha \beta}$
can be defined by gluing the two topological path-integrals
on the two half--discs with the chiral operators inserted
at the two end boundaries.}{Fig42}{.5}

Going through the same argument used in the closed case, we introduce the
natural {\it metric} connection
$$A_{i\ \alpha\beta}=\langle\beta|\partial_i|\alpha\rangle,\hskip 1.8cm
\overline A_{\bar j\ \alpha\beta}=\langle\beta|\overline\partial_{\bar
j}|\alpha\rangle.$$
Again topological invariance implies that $\overline A_{\bar j\alpha\beta}=0$
and hence\foot{Notice that this equation is consistent with the fact that the
combined Zamolodchikov metric for open {\it and} close string operators should
be K\"ahler. As mentioned in the text, this is the only condition on the
Zamolodchikov metric which is expected for the open case.}
$$(A_i)_\alpha^{\ \beta}=- g_{\alpha\bar\gamma}\partial_i
g^{\bar\gamma\beta}.$$

We wish to compute the curvature of this connection. Repeating word--for--word
the closed case analysis \cv, we see that the curvature
can be represented by the
$L\rightarrow\infty$ limit of the
difference of the two contributions represented in \tfig\FigureFourThree .

\ifigure\FigureFourThree{The computation of the curvature
of the $tt^*$ metric in the path integral formulation
involves the difference of the two path--integrals shown here. }{Fig43}{2.1}

 In both cases we
perform the path--integral with the twisted action on a long strip with half
disks attached to the two ends on whose boundaries we insert the topological
observables $\CO_\alpha$ and $\CO_\beta$, respectively. In the first term
the integral of $\phi_i^{(2)}$ over the right half of the
`rounded strip' $D$ is also inserted
while the insertion\foot{Here and below
$\overline\phi_{\bar j}^{[1]}$ is defined to be equal to $\half [(G^+-\overline
G^+),\overline\phi_{\bar j}]$.} of
$\{Q,\overline \phi_{\bar j}^{[1]}\}$  is integrated over the left half. In the
second term the two halves interchange their role. Let us consider the first
contribution.
By topological invariance, we can deform the contours such that $Q$ acts on
$\phi_i^{(2)}$, giving $d\phi^{(1)}$. Then the integral in the right half of
$D$ gives just the line integral $\int_l \phi_i^{(1)}$ where $l$ is the segment
separating the two halves of the figure 9\foot{The integral
$\int \phi_i^{(1)}$ along the other components of the boundary
vanishes because of the boundary conditions.}. The second term can be handled
in
 the
same way. But this time we get $-\int_l \phi_i^{(1)}$ because the orientation
is the opposite one. Then the difference is just
$$ -R_{i\bar j \alpha\beta}=\Big\langle \CO_\beta(+\infty)\, \int\limits_l
\phi_i^{(1)}\, \int\limits_D d^2z\, \overline\phi_{\bar j}^{[1]}(z)\,
\CO_\alpha(-\infty)\Big\rangle_{\rm strip},$$
where $\overline\phi_{\bar j}^{[1]}$ is integrated over
the full `rounded
strip' $D$, and the limit $L\rightarrow \infty$ is
implied. Then the open version of the $tt^*$ equations read
\eqn\opentts{\overline\partial_{\bar j} \big(g_{\alpha\bar\gamma}\partial_i
g^{\bar\gamma\beta}\big)=
\Big\langle \CO^\beta(+\infty)\, \int\limits_l \phi_i^{(1)}\, \int\limits_D
d^2z\, \overline\phi_{\bar j}^{[1]}(z)\,
\CO_\alpha(-\infty)\Big\rangle_{\rm strip}\Bigg|_{L\rightarrow\infty}.}

Eq.\opentts\ is much less useful than its closed counterpart 
because it is not in the form of a closed differential equation for the metric
$g_{\alpha\bar\beta}$.
However, eq.\opentts\  can be used to relate the holomorphic anomaly for the
open case to the
$tt^*$ metric $g_{\alpha\bar\beta}$ much in the same spirit as we did in
section 3 for the closed case.

\vskip .15in

\subsec{Holomorphic anomaly at one--loop}

In the open case the one--loop partition function for the topological theory
coupled to gravity is given by
the following quantity
\eqn\openFone{F_1=\int_0^\infty {dL\over L} \Tr\big[ (-1)^F F e^{-L H}\big],}
which is represented by a path integral (see \tfig\FigureFourFour) over a flat
cylinder of
length $\pi$ and perimeter $L$ with the Fermi current integrated along a
generator $l$.

\ifigure\FigureFourFour{Open one-loop partition function,
is represented by a cylinder with perimeter $L$.  The integration
over moduli involves integration over
$L$ with the Fermion number current
inserted on the line $l$}{Fig44}{.75}

Taking the derivative of $F_1$ with respect to the complex modulus $t^i$ and
going through the standard manipulations, we get
\eqn\openoneP{\partial_i F_1=\int_0^\infty dL\, \Big\langle \int\limits_l
(G^-+\overline G^-)
\int\limits_{l^\prime} \phi_i^{(1)}\Big\rangle_{{\rm cylinder}}^L.}
By definition the r.h.s. of \openoneP\ is the one `point' function
$$\left\langle \int_l \phi_i^{(1)}\right\rangle_{1-\rm loop}$$
for the (open) topological theory coupled to gravity. This quantity, being
topological, is a holomorphic function of the $t^i$'s except possibly for
anomalies associated to failure in the decoupling of $Q$--exact states. Thus
$\overline\partial_{\bar j}\partial_i F_1$ measures the holomorphic anomaly at
one--loop for the open case.

Let us compute $\overline\partial_{\bar j}\partial_i F_1$. We have
$$\eqalign{&\overline\partial_{\bar j}\partial_i F_1=\cr
&\quad=\int\limits_0^\infty dL\,
\Big\langle \int\limits_l (G^-+\overline G^-)
\int\limits_{l^\prime} \phi_i^{(1)}\,\int d^2z\,  \{Q,
\overline\phi^{[1]}_{\bar j}(z)\}\Big\rangle_{{\rm cylinder}}^L+\dots\cr
&\quad=-\int\limits_0^\infty dL \,
\Big\langle \int\limits_l (T+\bar T)
\int\limits_{l^\prime} \phi_i^{(1)}\,\int d^2z\, \overline\phi^{[1]}_{\bar
j}(z)\Big\rangle_{{\rm cylinder}}^L+\dots\cr}$$
where $\dots$ stands for the contribution from the contact term between
$\phi_i^{(1)}$ and $\overline\phi_{\bar j}^{[1]}$.
In the next section we shall introduce much more powerful techniques to deal
with such contact terms in one--loop
stringy computations. For this reason we defer the discussion of such terms
until we have developed the right tools.

The insertion of the operator $\int_l(T+\bar T)\equiv H$ is equivalent to
taking the derivative with respect $L$. Then we have
\eqn\twoDer{\eqalign{&\overline\partial_{\bar j}\partial_i F_1=\cr
&\quad=-\int\limits_0^\infty dL\,
{d\phantom{a}\over dL}\Big\langle
\int\limits_{l^\prime} \phi_i^{(1)}\,\int d^2z\, \overline\phi^{[1]}_{\bar
j}(z)\Big\rangle_{{\rm cylinder}}^L+\dots\cr
&\quad=-\Big\langle
\int\limits_{l^\prime} \phi_i^{(1)}\,\int d^2z\, \overline\phi^{[1]}_{\bar
j}(z)\Big\rangle_{{\rm cylinder}}\Bigg|_{\ L\rightarrow \infty}+\dots\cr
&\quad=-\int d^2z \lim_{L\rightarrow\infty} \Tr\Bigg[(-1)^F
\int\limits_{l^\prime} \phi_i^{(1)}\,  \overline\phi^{[1]}_{\bar j}(z)
e^{-LH}\Bigg]+\dots,\cr}}
where the contact--like contribution from the boundary at $L=0$ is absorbed in
the dots to be
discussed in the next section.

As $L\rightarrow \infty$ only the vacuum contributions survive in \twoDer, and
the trace in the Hilbert space can be replaced by a trace over the open string
vacua.
Then
\eqn\twoDDa{\eqalign{&\overline\partial_{\bar j}\partial_i F_1=\cr
&= - \sum_{\alpha\beta} (-1)^{F_\alpha} \eta^{\alpha\beta} \Bigg\langle
\CO_\alpha(+\infty) \int_l \phi^{(1)}_i \int d^2z\, \overline\phi^{[1]}_{\bar
j}(z)\, \CO_\beta(-\infty)\Bigg\rangle_{\rm
strip}^L\Bigg|_{L\rightarrow\infty}+\dots\cr
&= \tr\big[(-1)^F R_{i\bar j}\big]+\dots,\cr}}
where in the last step we used the open $tt^*$, eq.\opentts.
This is the anomaly equation we are looking for. It can be rewritten as
\eqn\openanfin{\partial_i\overline\partial_{\bar j}\big(F_1- \tr[(-1)^F\log
g]\big)=\dots,}
where $\dots$ again denotes the short distance contributions that will be
described from a more geometrical perspective in the next section.

\vskip .15in

\subsec{The holomorphic anomaly at higher loops}

At higher loops the situation with anomaly is similar to the one--loop case,
although more complicated. Consider, for instance, the case of
a surface $\Sigma$ with $h+1$ boundaries and genus $0$ (see
\tfig\FigureFourFive).

\ifigure\FigureFourFive{An open string diagram with no handles $g=0$
and $h=5$ boundaries.}{Fig45}{.75}

In this
case we have $3h-6$ real moduli, and $F_h^0$ is given by
\eqn\Fho{F_h^0=\int\limits_{{\cal M}_h^0}\Big\langle \prod_{k=1}^{3h-6}
\int \mu_k (G^-+\overline G^-)\Big\rangle_h^0,}
where ${\cal M}_h^0$ is the moduli space of genus zero surfaces with $h+1$
boundaries and $\mu_k$ are the corresponding Beltrami differentials. Taking the
derivative
$\overline\partial_{\bar j}F_h^0$ inserts in the r.h.s.
of eq.\Fho\ the operator
$\int d^2z\, \{Q,\overline\phi_{\bar j}^{[1]}\}$.
Integrating $Q$ by parts we get a sum of terms in which $Q$ acts on  $\int
\mu_k (G^-+\overline G^-)$
resulting in an insertion of $\int \mu_k (T+\bar T)$, which is then replaced by
a derivative with respect to the corresponding modulus $m^k$.

Then  the r.h.s. of eq.\Fho\ is reduced to a sum of contributions from the
boundary of the moduli space ${\cal M}_h^0$. This boundary has many components.
There are components like those in \tfig\FigureFourSix\
which corresponds to open surfaces
with a smaller number of boundaries and involving a sum over intermediate open
string vacua $|\alpha\rangle$, but also components as the one in
\tfig\FigureFourSeven\ in
which the degeneration of the surface involves a sum over the closed string
vacua $|i\rangle$. Collecting all contributions we get the anomaly formula for
the open case which will involve the derivative of lower $h,g$
partition function with respect to open or closed string couplings
with the operator $\int \overline\phi_{\bar j}^{[1]}$ inserted
in the lower genus amplitude (cf. eq.\twoDDa ).
Because of this operator insertion, in the open string case
 the holomorphic anomaly is a much less powerful tool
 than in
the closed string case, and the anomaly equation has not
the form of a recursion relation for the $F_h^g$'s.

\ifigure\FigureFourSix{The boundary of open string worldsheet
may involve degeneration of the surface connected
by long strips, represented here conformally by a black dot.
The intermediate state on the long strip is an open string state.
These contributions lead to insertion of open string vertices
on the lower $h$ Riemann surfaces.}{Fig46}{2.25}

\ifigure\FigureFourSeven{The degeneration may also include contributions
where the intermediate state is a closed string state.}{Fig47}{1.2}

This additional operator insertion
in the anomaly equation is problematic in the sense that it is not topological,
and hence it seems that the result of its insertion for a general model
cannot be computed by TFT methods.
The $tt^*$ methods are somehow more powerful: e.g. on the strip they allow to
compute such a correlation in terms of the derivatives of the metric
$g_{\alpha\bar\beta}$, see eq.\opentts. It is plausible that all the lower $h$
and $g$ correlations arising in the computation of $\overline\partial_{\bar j}
F_h^g$ can be computed in a similar fashion by an extension  of the $tt^*$ idea
to geometries other than the strip.

\vfill
\eject

\newsec{What are the topological amplitudes computing?}

In section 2 we discussed two classes of examples of twisted
$N=2$ theories coupled to gravity, the $A$-- and the $B$--model
and discussed what they compute at the tree level.  In this
section we give an interpretation of what
the topological amplitudes are computing in these two cases
after coupling to 2d gravity, i.e. the higher genus
interpretation of the amplitudes for these theories.
We will first discuss the case of the $B$--model, where we will see
that the target space field theory interpretation of the model
is related to a theory of gravity on Calabi--Yau $3$--folds
which for reasons to be explained we will call the Kodaira--Spencer
theory of gravity.  The tree level amplitudes in this case
are related to the classical theory of variation of Hodge
structures, i.e. the special geometry that we discussed in section
2.  The one--loop amplitude of this theory is related to the
holomorphic Ray--Singer torsion.  The geometric meaning of higher--loop
amplitudes is less clear, though can be formally defined
and may be viewed as quantum corrections to special geometry.
In the case of the $A$--model, the target space field theory
interpretation is far more difficult.  It should be again
a theory of gravity on Calabi--Yau manifolds, but a very non-standard
one, which requires the loop space of Calabi--Yau even
for the formulation of the theory.  However the interpretation
of what the $A$--model is computing is rather simple for any
genus $g$.  In fact the $A$--model, in the limit $\overline
t\rightarrow \infty$, computes the number
(or the appropriate Euler character) of holomorphic maps
from a genus $g$ surface to the Calabi--Yau.  In this
sense the $A$-- and $B$-- models have complementary
virtues.  The meaning of the computations are more
clear in the $A$--model but the formulation of the target
space theory is very clear for the $B$--model.  We will use
both models, in conjunction with mirror symmetry, later
in the paper to solve explicit examples at higher loops.

In subsections 5.1-5.9 we discuss the case of $B$--model
and KS gravity, and in subsections 5.10-5.13 we discuss
the case of the $A$--model.

\vskip .2in

\subsec{\bf Kodaira--Spencer Theory as a String Field Theory
of the B--Model}

The computations of topological $B$--model before coupling to gravity,
can be related to classical questions in variation
of Hodge structure, i.e. the complex structure of
Calabi-Yau and how it varies.  In the language
of sigma models this is related to the fact that the $B$--model
topological theory is independent of the volume of the manifold.
Rescaling the volume to infinity implies that in the
topological $B$--theory, not coupled to gravity, the path--integral
configurations are dominated by constant maps, thus leading
to classical geometry questions, and in particular the questions
of variation of Hodge structure of Calabi--Yau $3$-folds.   As we will discuss
later in the section this is essentially true
(modulo a crucial subtlety) even after we couple
to 2$d$-gravity, where we discuss the closed string field theory of the
$B$--model.  Before doing this we wish to discuss some mathematical
aspects of the Kodaira--Spencer
theory of deformations of complex structure which turn out to
correspond to target space physics of the $B$--model.
In other words we will argue why the $g$--th loop correction for the
Kodaira--Spencer theory is the same as $F_g(t^i, \bar t^i)$ defined
in section 2.   We will explicitly check this correspondence at
genus zero and one, and also show that in the case of genus
one our anomaly coincides with the Quillen anomaly for the Ray--Singer
torsion.

\vskip .15in

\subsec{Deformations of complex structure}

As it was discussed in Section (2.1) the observables in the $B$--model are in
one to one correspondence with cohomology elements $H^p(\wedge^q T_M)$,
where $T_M$ is the holomorphic tangent bundle.
The two forms  ${\phi}_A ^{(2)}$ are possible perturbations of the Lagrangian.
In case $p=1,~q=1$ operators ${\phi}_A^{(2)}$ for $A \in H^{(0,1)}(T_M)$
correspond to
marginal deformations of the $B$--model and are in one to correspondence with
deformations of complex structure of Calabi--Yau $3$--fold $M$.
In the spirit of string theory one expects that
$A \in \Omega^{(0,1)}(T_M)$ should be the basic field in the
field theory in question. This field theory is closely related to the
mathematical theory
of deformations of complex structures. Before proceeding further we first
review
some elements of this theory.

The complex structure on manifold $M$ is determined by the
$\bar \partial$ operator. To the first order the change of complex
structure is described
by deformation of $\bar \partial$ operator $\bar \partial
\rightarrow  \bar \partial  +  A^i {\partial}_i$
\ref\kodsp{K. Kodaira and D.C. Spencer, Annals of Math. 67
(1958) 328\semi
K. Kodaira, L. Niremberg and D.C. Spencer,
Annals of Math. 68 (1958) 450\semi
K. Kodaira and D.C. Spencer, Acta Math. 100 (1958) 281\semi
K. Kodaira and D.C. Spencer, Annals of Math. 71 (1960) 43.}.
This is a deformation of $\bar \partial$ operator acting on functions.
One can describe not only the infinitesimal deformations of complex structure
but
a finite one.
The new complex structure is described by
requiring that functions satisfying the condition
\eqn\perturbed{(\bar \partial + A^i {\partial}_i) f=0~,}
%
are holomorphic in the new complex structure.
In other words the kernel of the deformed $\bar \partial$ coincides with
with kernel of \perturbed.
The integrability condition
$$\bar \partial (\bar \partial f + A^i {\partial}_i f)=
(\bar \partial A^j+A^i \partial_i A^j )\partial_j f=0$$
is equivalent to the Kodaira--Spencer (KS) equation \kodsp
\eqn\ks{\bar \partial A + {1 \over 2} [A,A]=0~~.}
Once again $A$ is $(1,0)$ vector field with coefficients in $(0,1)$ forms
and the brackets $[\, ,\,]$ mean the commutator of two vector fields and
wedging.
Two solutions of \ks\ correspond to the same complex structure if they differ
by a diffeomorphism. In the linear approximation Kodaira Spencer equation
reduces to $\bar \partial A=0$.
The solution is defined modulo diffeomorphisms generated by
vector fields $A \rightarrow A+\bar \partial \epsilon$, and thus
$A$ has to be a cohomology element. The ambiguity in the choice of
cohomology representative is
promoted to the ambiguity in the solution of Kodaira Spencer equation.

Before fixing the ambiguity in question let us mention that for any
Calabi--Yau manifold there is an isomorphism
\eqn\pr{':~~\Omega^{(0,p)}(\wedge^q T_M) \rightarrow \Omega^{(3-q,p)}(M)}
given by the product with the holomorphic $(3,0)$ form.
Sometimes we use the notation $(A \cdot \Omega)=A'$.
Without lack of generality we impose the constraint
\eqn\constr{\partial A'=0~.}
To fix the ambiguity, $A \rightarrow A+\bar \partial \epsilon$,
we impose the gauge condition
\eqn\gauge{\bar \partial ^{\dagger} A'=0~.}
This gauge condition requires the choice of metric on the
Calabi--Yau manifold.
It will be clear later that these conditions fix the solution uniquely.

Let $A,~B$ be vector fields with the coefficients in $(0,1)$ forms
which satisfy the gauge condition $\partial A'=\partial B'=0$.
It was proven by Tian
\ref\tian{G. Tian, in {\it Essays on Mirror manifolds, ed. by S. T. Yau},
International Press, 1992\semi
G. Tian, in {\it Mathematical aspects of String theory, ed. by S.
T. Yau}, World Scientific, Singapore, 1987.}
that
\eqn\tn{[A,B]'=\partial (A \wedge B)'~,}
Later we will need the generalization
 of Tian's lemma where
$A,B$ belong to $ {\Omega}^p(\wedge^q T_M)$ \ref\gtian{G. Tian, private
communication}.
Using this lemma we can rewrite the KS equation in Tian form
$$\bar \partial A'+{1 \over 2} \partial (A \wedge A)'=0.$$

The tangent space to the moduli space of complex structures is
given by $H^{(0,1)}(T_M)$.
Let $A_1$ be an infinitesimal deformation of complex structure
satisfying conditions \gauge\ \constr.
Then for any $A_1$ one can ``exponentiate'' the deformation of complex
structure by constructing the solution to the KS equation
$$A=\sum_{n=1}^\infty  {\epsilon}^n A_n, $$
where $\epsilon$ is a
formal expansion parameter (we put $\epsilon=1$ later).
We will show that
it is possible to get a unique solution of the KS equation satisfying the
gauge condition $\bar \partial^\dagger  A'=0$ such that $A_n'$ is
$\partial$--exact for $n>1$.  Note that
this latter condition automatically implies
that we can use Tian's form of the KS equation.  This choice means that $A_1'$
is a harmonic form, which we will call {\it massless},
and $A_n'$ for $n>1$ can be written as a linear combination
of eigenstates of Laplacian with positive eigenvalue.  We will call
these states the {\it massive} states.

Let us see how we can construct the solution recursively (following
the work of \tian, \ref\Tod{A.N. Todorov, Comm. Math. Phys. 126 (1989)
325.})
making sure that at each stage $\bar \partial A_n'=0$ and that $A_n'$ is
$\partial$--exact for $n>1$.
Let $A_1$ satisfy the gauge  condition \gauge\ together with constraint
\constr.
Thanks to Tian's Lemma the equation for $A_2'$ becomes
$$\bar \partial A_2'+{1 \over 2} \partial (A_1 \wedge A_1)'=0.$$
Note that the solution to this equation for $A_2'$ is unique
up to addition of $ \bar \partial \nu$.  In order to get rid of this
ambiguity we will consider the gauge condition $\bar
{\partial}^{\dagger} A_2'=0$.  Then the solution can be written as
\eqn\pertwo{A_2'=-{\bar \partial}^{\dagger} {1\over \Delta}\partial (A_1 \wedge
A_1)'~.}
where
$$\Delta =2[\bar \partial \bar \partial^\dagger +\bar \partial^\dagger \bar
\partial ]$$
is the Laplacian.  To see that the above is a solution, first note that it
is well defined, because $\partial$ annihilates the kernel of $\Delta$.  Then
acting by $\bar \partial$ and using the fact that $\bar \partial (A_1\wedge
A_1)'=0$  (because $A_1$ is $\bar \partial$ closed and $\bar \partial$ commutes
with the operation of $'$ since $\Omega$ is holomorphic) one checks
that it is a solution to the equation.  It also satisfies the conditions
of being $\partial$--exact (because $\partial$ and $\bar \partial^\dagger$
anticommute for a K\"ahler manifold) and $\bar \partial^\dagger$ closed.
The fact that there is always a solution to the above
equation is also known as $\partial \bar \partial$-Lemma
\gh\foot{The
$\partial \bar \partial$--Lemma reads:
if $\omega$ is any
$\bar \partial$ closed form and
$\omega$ is also $\partial$ exact, then
$\omega=\partial \bar \partial \phi$.}.
This in particular means that with the gauge condition we have
chosen
\eqn\prop{{-1\over 2\bar \partial}\partial \equiv -{\bar \partial}^{\dagger}
{1\over \Delta}\partial}
and it can be viewed as a propagator for massive modes.
The equation \pertwo\ describes interaction between two massless
modes and a massive one and then further propagation of the massive state.
It can be represented as the diagram of \tfig\FigureFiveOne.

\ifigure\FigureFiveOne{The first order perturbation
computation for solving the KS equation in Tian's
gauge.  Two massless modes represented by wavy lines
join to give a massive mode whose
propagator is represented by a solid line.}{Fig51}{1.5}

The equation for the
next iteration becomes $\bar \partial A_3'+\partial (A_2
\wedge A_1)'=0$.
The second term in this equation is $\bar \partial$ closed
$\bar \partial \partial (A_2 \wedge A_1)' \sim \partial([A_1,A_1] \wedge A_1)'
\sim [[A_1,A_1],A_1]=0$
and therefore one may use the above propagator again.
$$A_3'=2{\bar \partial}^{\dagger} {1\over \Delta}\partial (A_1 \wedge
({\bar \partial}^{\dagger} {1\over \Delta}\partial (A_1 \wedge
A_1)')^{\vee})'$$
where $(A')^{\vee}=A$.  Note that this solution satisfies the
required conditions.
Again this contribution has a clear interpretation.
Two massless states go to a massive one (as before), but now the
propagator receives corrections due to the coupling with the massless state
in the background. This contribution corresponds to the diagram of
\tfig\FigureFiveTwo.

\ifigure\FigureFiveTwo{Second order perturbation
computation for solving the KS equation.  Here two massless
modes join to give a massive mode, emit a massless mode
which finally gives rise to the massive mode $A_3$.  Note
that the only propagators involve massive modes, and the
massless modes are like the background fields.}{Fig52}{1.5}

\vskip .15in

 It is already clear that
$\partial{\bar \partial}^{\dagger}/\Delta$ is a propagator for the massive
states for the field theory in question. The massless modes play the role of
the background. It is quite remarkable that the KS equation
reproduces the perturbation series of a $\phi^3$ theory.

At $n-$th iteration step  all $A_1,...A_{n-1}$ satisfy the conditions
$\partial A_1'= \cdots = \partial A_{n-1}'=0$ and the KS equation
becomes
\eqn\rec{\bar \partial A'_n+{1 \over 2}
\sum_{i=1}^n \partial (A_{n-i} \wedge  A_i)'=0}
The second term of this equation is $\bar \partial$ closed.
This follows from the equations satisfied for $\bar \partial A_i'$
dictated by induction and the Jacobi identity
for the vector fields with coefficients in $(0,1)$ forms
and Tian's lemma
\eqn\jac{ \partial ([A,B]\wedge C)'+\partial ([C,A]\wedge B)'+\partial
([B,C]\wedge A)'=0~.}
It follows from the above arguments therefore that equation
\rec\ has a solution and it is $\partial$--exact.
The perturbation theory described above is convergent in some open
neighborhood of the origin \tian .

We just proved that for any $x \in H^{(0,1)}(T_M)$ there is a map
$x \rightarrow A[x]$ given by the solution of the KS equation,
 with
$A_1=x$. This choice of terminology is consistent with the
definition of $x^i$ given in section 3, and can
basically be viewed as shifting the complex structure of the
Calabi-Yau labeled by $(t,\bar t)\rightarrow (t+x,\bar t)$.
For later convenience we will write $A[x]\rightarrow
x + A(x)$.
Decomposition into $x$ and $A(x)$ is quite natural.
A cohomology element $x$ represent a {\it massless} mode while $A(x)
= \sum_{n=2}^\infty A_n$ contains the
{\it massive} modes of the field.

Under the deformation of complex structure the holomorphic $(3,0)$ form get
changed. For infinitesimal deformation the deformed holomorphic form
is equal to $\Omega_0 + x'$. For the finite deformations the holomorphic
$(3,0)$ form mixes with $(2,1)$, $(1,2)$ and $(0,3)$ and  it satisfies the
equation
$$\bar \partial \Omega + {1 \over 2}\partial (\Omega^{\vee} \wedge
A)'=0$$
where prime and check are defined with respect to the fixed holomorphic
three form $\Omega_0$.
It follows from Tian's lemma that the deformed holomorphic $(3,0)$
form is given as follows \Tod
\eqn\meg{\Omega = \Omega_0 + A' + (A \wedge A)' + (A \wedge A \wedge A)'~.}

Coordinates in  $H^{(0,1)}(T_M)$, denoted by $x$, may serve as affine
coordinates on some open neighborhood
of the moduli space of complex structures (see also
\ref\kur{M. Kuranishi, Annals of Math 75 (1962) 536.}) thanks
to the Tian's mapping.
These coordinates are in fact very special (not to be confused with special
coordinates
except for the particular case of base point at infinity, as discussed
in section 2.6)
and corresponds to the canonical coordinates discussed in
full generality in section 2.6.
In this coordinate the K\"ahler potential is
given as follows
\eqn\kahl{\eqalign{e^{-K(x,\bar x)}&=
\int_M \Omega_0 \wedge \Omega_0 +
\int_M A' \wedge \bar A'+
\int_M (A \wedge A)'
\wedge (\bar A \wedge \bar A)'+\cr
&\int_M  (A \wedge A \wedge A)' \wedge
(\bar A \wedge \bar A \wedge \bar A)' ~,\cr}}
where $A[x]=x+A(x)$. Taking into account that $A(x)=O(x^2)$ and $x$ and
$A(x)$ are orthogonal to each other we get
the expansion
$$e^{-(K(x, \bar x)-K_0)}=1+x^i \bar x^j G^{(0)}_{ij} +
 O(x^2 \bar{x}^2)$$
It immediately follows from this expansion that in this coordinates $\partial_i
K=0=\Gamma_{ij}^k$ vanishes at the origin together
with all holomorphic derivatives. Therefore in this coordinate the covariant
holomorphic derivatives at the origin coincides with the ordinary derivatives
$$D_iD_j...D_k F=\partial_i \partial_j...\partial_k F ~.$$
This property is very important and was the defining property
of canonical coordinates discussed in full generality
in section 2.6. Let us clearly state that the
canonical coordinates
are uniquely determined by the point in the moduli space (the origin of the
coordinate system) and the choice of the basis in $H^{(0,1)}(T_M)$.

It is instructive to consider the example of canonical
 coordinates in the case $T^2 \times T^2 \times T^2$, where $T^2$ is a
two
dimensional torus.
The complex structure of each torus is described by one complex parameter
$\tau_i$.
One can carry out the construction of canonical coordinates
for each torus separately.
Let us parametrize each torus
using coordinates $(\sigma_1,\sigma_2)$, where $(\sigma_1,\sigma_2)$
runs over unit square. In this parametrization $\partial$, $\bar \partial$ are
given as follows
$$\partial={1 \over (\bar \tau - \tau)}
(\bar \tau \partial_1 -\partial_2) ~~,~~~~
\bar \partial={1 \over (\bar \tau - \tau)}
(-\tau \partial_1 + \partial_2)$$
Now, let us choose the base point $(a, \bar a)$. The holomorphic flat
coordinate $x$ around
$(a, \bar a)$ is defined as follows $\bar \partial(\tau)=\bar \partial(a) + x
\partial(a)$.
It implies the relation $x(\tau)$
$$x={ \tau -a \over \tau -\bar a}~,$$
i.e. the upper--half plane gets mapped into the open unit disk.
The K\"ahler potential in this coordinate
is equal to
\eqn\ex{\eqalign{e^{-K(x,\bar x)}=&\int
\prod_i (dz_i-x_i d \bar z_i) \wedge (d \bar z_i - \bar x_i dz_i)=\cr
=\prod_i (1-x_i \bar x_i) &(a_i-\bar a_i)=
\prod_i (\tau_i-\bar \tau_i) \left|i{a_i-
\bar a_i \over \tau_i - \bar a_i}\right|^2 \cr}}
The $x$ dependence is quite remarkable. It is clear that all derivatives with
respect
to $x$ are proportional to $\bar x$ and therefore identically equal to zero at
the origin.
The factor inside the absolute value is the gauge factor $f(a_i)=-i\prod_i
(a_i- \bar a_i)/(\tau_i - \bar a_i)$.

\vskip .15in

\subsec{Kodaira--Spencer theory as the string field theory}

So far we have discussed what seems to be a perturbative field theory
which describes the perturbation of complex structure of Calabi--Yau
manifolds starting from a base--point.  Since the $B$--model describes
the deformation of the complex structure,
the effective string field theory of the $B$--models must be
this underlying field theory, which we shall call the
Kodaira--Spencer theory of gravity.
We have two options in writing this field theory:  We can either
use the Kodaira-Spencer equation in the Tian gauge to write
the action giving rise to these equations, or directly use
the rules for constructing closed string field theory along
general lines discussed in the literature
(see \ref\zw{B. Zwiebach, {\it Closed String
Field Theory: Quantum Action and the B-V Master equation}, IASSNS-HEP-92/41,
MIT-CTP-2102, hep-th/9206084}
for a thorough review of the literature).  We will follow
the first line and see why it is the same as the second.

To write an action we first need to fix some data: the point $P$
(which we sometimes denote also by $(t^i_0,\bar t^i_0)$)
in the moduli space of complex structures (background)
and a cohomology element $x \in H^{(0,1)}(T_M)$.
The physical field  $A$ in the KS theory is a vector field  with coefficients
in $(0,1)$ forms which is also constrained to satisfy condition $\partial
A'=0$. For
reasons that will be clear in a moment we assume that $A$ includes only
massive modes.
This means that $A$ lies in the subspace
${\cal H} \subset \Omega^{(0,1)}(T_M)$ orthogonal to $H_{ \bar
\partial}^{(0,1)}(T_M)$,
or in other words
$$\int_M A' \wedge \bar z' =0$$
for any $\bar z \in H_{\partial}^{(1,0)}(T^{*})$.
Thanks to constraint \constr, this definition is independent of the choice
of
representative in cohomologies.

The Kodaira--Spencer action is given as follows
\eqn\KSact{ {\lambda}^2 S(A,x|P)={1 \over 2 }
\int_M A' {1 \over \partial} \bar \partial A' +
{1 \over 6 } \int_M ((x+A) \wedge (x+A))' (x+A)'~,}
where $\lambda^2$ is the coupling constant.
In spite of the non-local kinetic term this action is well defined.
Indeed, it follows from the $\partial \bar \partial$-Lemma
that $\bar \partial A'=\partial \bar \partial v$ and therefore
${\partial}^{-1} \bar \partial A'=\bar \partial v + \partial \rho +\bar z$,
where $\rho$ and $z$ summarize the ambiguities and $z \in
H_{\partial}^{(1,0)}(T^{*})$.
The condition that $A'$ is massive together
with the constraint it satisfies implies that
$\rho$ and $z$ do not contribute to the action
which therefore is well defined.  Note that to define
the action we did not use the metric on Calabi--Yau manifold.
We just used its complex structure\foot{
To see that the action is well defined and independent
of the choice of metric on $M$, we can also use the $\partial$ constraint
to write $A'=\partial \phi$ and substitute it in the action to get
a local action for $\phi$.}.
This is just like the Chern--Simon theory.
Thus the KS theory is a
topological theory (or more properly it could be called
a holomorphic topological theory in the sense that it does depend on
the complex structure of the Calabi--Yau).
Varying the
KS action with respect to $A$ we recover the Kodaira--Spencer equation
in Tian's form
\eqn\eqm{\bar \partial A' + {1 \over 2}\partial ((x+A) \wedge (x+A))'=0}
The existence of this action explains the fact that in the
perturbation expansion for $A(x)$
discussed before one naturally gets Feynman rules of some field theory.
In fact
they are nothing but the tree
level diagrams of KS theory.  Note that the propagator for KS
action $\overline{\partial}^{-1} \partial$ is given by
\prop\ in the appropriate gauge.

We now wish to see why the action \KSact\ is the same as what
we would have gotten from the target space theory of the
$B$--model.  For this, we employ the arguments of Witten \witcs .
He used the fact that volume perturbation for the Calabi--Yau
is BRST trivial in the $B$--model set up, to take the infinite
volume limit.  In this case the worldsheet configurations
{\it for a fixed worldsheet modulus} is dominated by
constant maps.   But as noted in \witcs\ this is not the full
story.  The reason is that we are discussing a theory of 2d gravity
which means we are integrating over the moduli of Riemann surfaces.
No matter how large a volume of Calabi--Yau we choose if we go
close enough to the boundary of the moduli space we can get
finite action.  In other words the worldsheets which will
have finite action are the ones concentrated in long thin
tubes, which means that we are going to end up with an ordinary field
theory as an {\it exact} field theory of string model (i.e. all
the stringy massive modes are irrelevant because of topological
triviality of these modes).  Indeed this argument applies even
taking into account potential anomalies, because as discussed
in section 3 there is no anomaly for the
decoupling of the K\"ahler-moduli in the $B$--model.

So to fix the string field we have to recall that the field in question
should have charge $(1,1)$ which in our case translates to the
fact that $A'$ should belong to $\bar T^*_M\wedge T_M$.  Let us also recall the
dictionary developed in section 2:
In the large volume limit operator $\bar \partial$ is identified  with BRST
operator
$\bar \partial=Q=G^+ _0 + {\bar G}^+ _0$, while $\partial=G^- _0 -{\bar G}^-
_0=b^- _0$.
The string field $A'$ should satisfy two constraints
\eqn\sc{\partial A'=b^- _0 A'=0 ~~{\rm and}~~~
(L_0 - \bar L_0)A'=(\Delta - \bar \Delta)A'=0~.}
 In case of the KS theory the second constraint is
trivial consequence of K\"ahlerian geometry and amazingly the
first condition is precisely Tian's condition which led
to the simplification and proof of integrability of the KS
equation in the case of Calabi--Yau $3$-fold.
In order to borrow the machinery of closed
string field theory we need to find an expression for
$c_0^-=c_0-\bar c_0$.  However there is no such object
just because the $b$-cohomology is not trivial.   What
is true instead is that on the {\it massive} states of the theory, we
can in fact define a
$$c_0^-={1\over \del}= {\partial^{\dagger} \over \Delta} $$
which satisfies
$$\{ c_0^-,b_0^-\}=1$$
and we are thus forced to write down the action {\it only for the massive
modes}.  Therefore, the kinetic piece of the KS action
coincides with the free part of the standard string field theory action
$${1 \over 2} \int A' { 1 \over \partial} \bar \partial A'={1 \over 2}(A', c^-
_0 Q A')$$
 The gauge $\bar \partial^{\dagger} A'=0$ is nothing else but the
Siegel gauge
in which both $b_0^-=\partial$ and $ b_0^+=\bar \partial^{\dagger}$
annihilates the physical fields.
In this gauge the propagator takes the familiar form
$${b^+ _0 b^- _0 \over (L_0 +\bar L_0)}={ \bar \partial^{\dagger}\partial
 \over \Delta}$$
Magically enough this is identical with the Kodaira--Spencer
kinetic term and the propagator.
  The cubic interaction term is quite standard and gives
rise to the interaction term of the Kodaira--Spencer action.

Thus the KS action is nothing else
but the closed string field theory action at least up to cubic
order.
One of the main difficulty of the closed string theory is the absence of a
decomposition
of the moduli space  of Riemann surfaces compatible with Feynman rules. To
avoid this problem one should introduce higher string vertices and as a
result the
closed string field theory becomes non--polynomial (see \zw\ and references
there). The contribution to these
higher string vertices comes entirely from the internal domains of the moduli
space of Riemann surfaces.
Quantized KS theory is defined as the large volume limit of topological
sigma--model
and
as a topological theory it gets contribution entirely
from the boundary of moduli space of Riemann
surfaces. Therefore, the higher vertices should be absent in quantized KS
theory.  It is quite satisfactory that we thus end up with precisely
the KS theory as the string field theory of the $B$--model\foot{It
is amusing to note that the closed string field theory of $N=2$ strings
\ref\ov{H. Ooguri and C. Vafa, Nucl. Phys. B361 (1991) 469,
Mod. Phys. Lett. A5 (1990) 1389}\  also has a cubic action,
the $4$-real-dimensional
action for the Plebanski equation
describing Ricci--flat K\"ahler metric in $2$ complex dimensions,
which is very similar to the KS action given above.}.  This is further
confirmed in section 5.4 where we will find that the KS theory,
with the ghost fields added, already satisfies the BV master equation
and needs no further corrections.

Let us now discuss the gauge symmetries of Kodaira--Spencer theory.
As a string field theory we certainly expect it to have such
symmetries.
Being a theory of gravity the Kodaira--Spencer
theory should be invariant under
diffeomorphisms (we will make this statement precise in a moment).
Put differently, the fact that the variation of $\bar \partial$
can also be affected by diffeomorphisms, and we do not wish to
take this as a physical variation, we need to consider
the theory as a gauge theory with respect to diffeomorphism
group.
The kinetic part of the action is clearly invariant under the shift of
$A$ by $\bar \partial$-exact term which means
$\delta A=\bar \partial \epsilon=Q \epsilon$. This linearized gauge
transformation can be extended to a full non linear gauge transformation
which turns out to be nothing else but
an $\Omega$-{\it preserving} diffeomorphism
$$z^i \longrightarrow z^i + \epsilon^i (z, \bar z)$$
The condition that $\epsilon$ is $\Omega$ preserving diffeomorphism means
that it satisfies the constraint $\partial \epsilon'=0$. The full gauge
transformation of the Kodaira-Spencer field $A$,
which can be deduced from the variation of $\bar \partial$ under
the diffeomorphism, is given as follows
$$\delta A=\bar \partial \epsilon - [\epsilon, (x+A)]~$$
and using the Tian's lemma it can be rewritten in a more familiar form
$\delta A'=\bar \partial \epsilon' - \partial (\epsilon \wedge (x+A))'$.
One can verify that this transformation is a symmetry of the action.
Indeed the variation of the action is equal to
\eqn\var{\eqalign{\lambda^2 \delta S=&-\int_M A' \bar \partial ((x+A) \wedge
\epsilon)'+
{1 \over 2} \int_M ((x+A) \wedge (x+A))' \bar \partial \epsilon' \cr
     & - {1 \over 2} \int_M ((x+A) \wedge (x+A))'
\partial ((x+A) \wedge \epsilon)'}}
The first two terms cancel each other
as can be seen by integrating by parts. The vanishing of the third term
follows from the Jacobi identity.
Indeed, the last term can be rewritten as follows
\eqn\ident{\eqalign{\int_M ([(x+A) ,\epsilon]\wedge & (x+A))' (x+A)'=
{1 \over 2}\int_M  ([(x+A) ,(x+A)] \wedge \epsilon)' (x+A)'=\cr
&{1 \over 2}\int_M  ([(x+A) ,(x+A)] \wedge (x+A))' \epsilon'=0 \cr}}

To formulate the KS theory we fixed some data: point in the moduli
space
$P$ and the cohomology element $x$.  Note that the
fact that $x$ cannot be written as part of the kinetic term
is because of the $\partial^{-1}$ in the kinetic term, which
renders the appearance of $x$ meaningless.  So the KS theory
{\it does not have the degree of freedom to shift the complex
structure} as a dynamical field in the theory.  Instead the existence
of the coupling with $x$ {\it as a background field}
in the interaction term is there to take
care of this.
One may ask how the theory changes if we choose a different base point $P$.
We parametrize the position of the base point $P$ in canonical
coordinates $P=P(t,\bar t)$. Ignoring the
holomorphic anomaly the KS action depends
only on $t$ and is independent of $\bar t$. The shift in $t$ coordinate can be
achieved by shifting the field $A$ by the solution of the
KS equation (let $A_0(x)$
be the solution of KS equation). Then, consider the following identity
\eqn\backgr{\eqalign{&{\lambda}^2 S(A+A_0 (x),x|t,\bar t)=
  \int_M A ' {1 \over \partial}
         \left(  \bar \partial  A_0' + {1 \over 2} \partial
                    (x+A_0) \wedge (x+A_0))' \right) + \cr
& ~~~~{1 \over 2 } \int_M A_0' {1 \over \partial} \bar \partial A_0' +
{1 \over 6 } \int_M ((x+A_0) \wedge (x+A_0))' (x+A_0)' +\cr
&~~~~~~~{1 \over 2 } \int_M A' {1 \over \partial}
\Big[ \bar \partial A' + \partial ((x+A_0) \wedge A)'  \Big] +
{1 \over 6 } \int_M (A \wedge A)' A'  \cr}}

The first term vanishes due to the equation of motion.
The second and the third terms are naturally combined into
the classical KS action evaluated on the solution of KS equation.
The two remaining terms have an interpretation
as the KS action around the {\it new background}. Indeed
the combination in the square brackets coincides with the deformed
$\bar \partial$ operator around the new background. There is still one
subtlety, the {\it prime} operation is defined with respect to old background.
In the new background the prime operation should be defined by contraction with
 the deformed holomorphic $3$--form given by \meg. Noticing, that only
projection on $(3,0)$--forms contributes to the action one can replace the
prime operation around the old background by the prime operation around the new
background. As a result of this formal manipulation we obtain the relation
\eqn\naivback{S(A+A_0 (x),x|t,\bar t)=S(A_0(x),x|t,\bar t) + S(A,0|t+x,\bar
t)~ .}
In the original definition of the KS theory $t$ and $\bar t$ are complex
conjugate to each other. Without the
holomorphic anomaly, the KS action is independent
of $\bar t$  and one can replace $S(~|t+x,\bar t)$ by $S(~|t+x,\bar t+\bar x)$.
If such arguments were true they imply the background independence of
the KS theory
or background independence of
the corresponding closed string field theory.
The dependence of the KS action on $\bar t$ destroys background independence.
In other words the holomorphic anomaly governs the background dependence of
the KS
action (see also discussion in \witt).
In the presence of holomorphic anomaly relation \backgr\ may serve
as the definition of the KS action where the condition $t=\bar t^*$ is relaxed.

We now come to a puzzle raised by Witten in his study of this theory
\witcs .  It was pointed out in \witcs\ that the
fact that three point function
$C_{ijk}$ is not zero seems to be at odds with the fact that
there is no obstruction to deforming by the marginal operators.
The resolution of this puzzle in the context of the KS theory is simply
that the massless fields, i.e. the string modes, {\it are not dynamical
fields} and so there is no reason that the classical
value of action is independent of their expectation value (as we will
discuss in more detail below).  Thus the fact that the kinetic
term cannot be defined unless we delete the massless modes means
in particular that $C_{ijk}$ may be non-zero even
if the massless modes can be
given arbitrary expectation value.

Being a quantum theory in six dimension it
is not easy to explicitly compute
higher loop amplitudes in the KS theory.
In particular this $6$-dimensional field theory  looks
highly non-renormalizable from the simple power counting argument.
It is quite remarkable that topological string theory of the $B$--model
provides a prescription to quantize the Kodaira-Spencer theory.
The properly regularized Kodaira-Spencer theory
should satisfy
$$e^{W (\lambda ,x|t, \bar t)}=\int DA e^{S(x ,A|t, \bar t)}~,$$
where the effective action $W (\lambda ,x|t, \bar t)$ was
defined in section 3.
We also introduce the notation $x=x^i \mu_i$, where ${\mu_i}$ is some
basis in $H^{(0,1)}(T)$.
Even though the r.h.s. of this equation is to
be properly defined at higher loop, it is well defined
as it stands for the tree level. Let us prove this
relation at least at the tree
level. Later in this section we will see that it also continues
to hold at one--loop.

At the tree level, the contribution of the path-integral
simply gives rise to the classical action evaluated
for solutions to the field equations.  Let us denote
this action by $S_0(x ,A_0|t, \bar t)$ where
$A_0(x)$ is such that $A_0(x)+x$ satisfies the KS equation
(expanded about the base point $(t,\bar t)$).  Thus we need to show
\eqn\rel{W_0 (x|t, \bar t) = \lambda ^2 S_0(x ,A_0(x)|t, \bar t)~,}
where $W_0$ is the tree level contribution to $W$ (i.e. the coefficient
of $\lambda^{-2}$).
Note that in the $x$-coordinate which is a canonical one,
$W_0$ of section 3 is defined by the condition
$$\partial_i\partial_j \partial_k W_0=C_{ijk}(x)=\sum_{n=0}^{\infty} {1\over
n!}
  C^0 _{ijks_1...s_n} x^{s_1}...x^{s_n}$$
and also $W_0$ has no linear or quadratic dependence on $x$.
We see simply from the definition of $S_0$ that up to $O(x^3)$
they are thus equal.  We need to show that it holds to all orders.
Let us compare the third derivatives of both sides of \rel.
The third derivative of the classical action is given as follows
\eqn\der{\eqalign{{d^3 S_0 \over dx^i dx^j dx^k}&=
\big[(\delta_A S  )\partial_i \partial_j \partial_k A \big]+
\big[(\delta_A ^2 S  ) \partial_i A +
(\delta_A \partial_i S  ) \big] \partial_j  \partial_k A +\cr
\big[(\delta_A ^3 S  )\partial_i A \partial_j A \partial_k A+&
3 (\delta_A ^2 \partial_{i} S ) \partial_j A \partial A_{k} +
3  (\delta_A \partial_{i} \partial_k S )\partial A_{k} +
    \partial_i \partial_j \partial_k S \big] \cr}}
where $\delta_A$ is variational derivative with respect to $A$ and
$\partial_i=\partial/{\partial x^i}$
and symmetrization with respect to $ijk$ is implicit. The first two
terms vanish:  the first one vanishes because $\delta_A S=0$ by
the equations of motion which is the definition of $A_0(x)$. The
second term vanishes by taking derivative of $\delta_A S$,
along the classical solution, with
respect to $x_i$ and expanding to the third order term.
Finally the last term can be rewritten as
$${d^3 S_0 \over dx^i dx^j dx^k}=\int \big( (\mu_i +\partial_i A_0)\wedge
(\mu_j +\partial_j A_0)\big )'\wedge (\mu_k +\partial_k A_0)'=C_{ijk}(x)$$
where the last equality follows from the alternative definition of Yukawa
coupling discussed in section 2 (see \TheFormula).
This proves the equation \rel .

$W_0(x|t,\bar t)$ may be viewed as the effective action for the
massless modes $x$ having integrated out the massive modes.
It is quite amazing that integrating the massive modes has only
the effect of taking derivatives of the Yukawa coupling.  For
example (see \tfig\FigureFiveThree) the four point function gives
rise to $\partial_lC_{ijk}$, the five point function to $\partial_s
\partial_l C_{ijk}$ and the six point function to $\partial_r
\partial_s \partial_l C_{ijk}$.

\ifigure\FigureFiveThree{Tree level computations in the KS theory
as a function of the background fields (the wavy lines which
represent the massless modes).  As argued in the text the $n$--point
functions at the tree level can also be computed by taking
appropriate number of derivatives of the Yukawa coupling.  Here
the four point function (a), five point function (b),
and six point function (c),
are represented and can be most easily computed by taking
the first, second and third derivative of Yukawa
couplings respectively.}{Fig53}{4.0}

 In fact we will
now use this fact to estimate the behavior of the partition
function at genus $g$ of the KS theory to all loops,
as we approach the boundary of moduli space.
This will be needed in conjunction with the anomaly equation to constrain the
global properties of the partition function of the topological string theory
and will be heavily used in the context of solving explicit examples.

This is done by estimating the leading divergence of each diagram as we
approach the boundary of moduli space.
To do this we need the estimate of the propagator and the three point
interaction of the massive modes (the massless modes
do not propagate in loops).
Let us denote the leading divergence of the propagator by $P$, of the massive
vertex by $V_{MMM}$, and of the vertex with two massless and one massive mode
as $V_{ttM}$. Using the topology of $\phi^3$ graphs at $g$ loops we estimate
the Kodaira--Spencer partition function $F_g$ to behave as
\eqn\fgvmmm{F_g\sim P^{3g-3} (V_{MMM})^{2g-2}  }
We want to express this in terms of $C_{ttt}$, the leading divergence in the
Yukawa coupling for the massless modes {\it written in the canonical
coordinate} $t$.  The $n$--point functions at tree level, are given by
$\partial_t^n C_{ttt}$. Using the tree--level KS perturbation theory, we learn
that the four point function of the massless modes behaves as
$$\partial_t C_{ttt}\sim P\, (V_{ttM})^2,$$
while the six point function goes like
$$\partial_t^3 C_{ttt}\sim P^3\, (V_{ttM})^3\, V_{MMM}.$$
Eliminating $V_{ttM}$ from these two equations and using \fgvmmm\ we learn that
\eqn\estimate{F_g\sim {[\partial_t^3 C_{ttt}]^{2g-2}\over [\partial_t
C_{ttt}]^{3g-3}}.}
Note that the estimate \estimate\ is independent of the definition of the
canonical coordinate $t$ or the gauge for the line bundle $\CL$
as it should be.

\vskip .15in

\subsec{BV formalism and closed string field theory}

In this section we quantize the KS action using the BV formalism
which is particularly well suited to string theory.
The interpretation of the KS theory as string field theory turns out to be very
useful.  In this interpretation the KS
field $A'$ is identified with the string
field.  But in string theory there are `ghost' states, which
mean that we are not restricted to ghost number $(1,1)$.  Translated
to the geometry of Calabi-Yau, this means that we should broaden
the range of $A$ so that $A \in \Omega ^{(0,p)} (\wedge^q
T_M)$;
the ghost counting coincides with the fermion counting and is equal to
$F_L+F_R=(p+q-3)$. The original KS field $A'$ has the ghost number $2$.

The consistent scheme for quantization string field theory is given by
Batalin--Vilkovisky
(BV) formalism \ref\bat{I. A. Batalin and G. A. Vilkovisky,
Phys. Rev {\bf D28}
(1983) 2567}\ref\hen{M. Henneaux,
{\it Lectures on the antifield-BRST formalism for
gauge theories}, Proc. of XXII GIFT Meeting}.
In the Batalin-Vilkovisky formalism one has to
relax the condition for the ghost numbers of string field and
include all possible fields with arbitrary ghost numbers.
The fields $A$ with ghost numbers $q(A) \leq 2$ are called fields,
while the  fields $A^*$ with ghost numbers $q(A) > 2$ are called
antifields.
The space of functionals of fields-antifields is equipped with odd
antibracket
$\lbrace{ ~,~ \rbrace}$.
The BRST symmetry is a canonical transformation in the antibracket. The BRST
variations of the fields are given as follows
$$\delta_{BRST} {\cal A}=\lbrace{ {\cal A},S \rbrace}$$
The original action is replaced by full action which
depends on both fields and antifields. The full action satisfies two
conditions.
When all antifields are
set to zero the full action reduces to the original one. The full action also
satisfies Batalin-Vilkovisky master equation
\eqn\bvq{\lbrace{ S,S \rbrace}=\hbar \Delta S~,}
where $\Delta$ is the natural Laplacian on the space of fields-antifields.
The r.h.s. of \bvq\ is contribution coming from the path integral measure.
At the classical level $(\hbar=0)$ the Batalin-Vilkovisky equation is
nothing else but the condition that full action is gauge invariant.
The gauged fixed action is determined by
an odd functional $\Psi(A)$ and is given
by $S_{\Psi}(A)=S(A,A^*=\delta \Psi / \delta A)$.

In the case of the KS theory the full space of fields is
a subspace  ${\cal H}$ of $\oplus_{p,q} \Omega ^{(0,p)} (\wedge^q T_M)$
satisfying the constraints \sc. The space
$$ \oplus_{p+q \leq 2}
\Omega ^{(0,p)} (\wedge^q T_M)$$
is the space of fields, while
$$\oplus_{p+q > 2}
\Omega ^{(0,p)} (\wedge^q T_M)$$
is the space of antifields.
Note that not all $(p,q)$
are allowed, and the projection of $\cal H$ on $\oplus
\Omega ^{(0,p)} (\wedge^3 T_M)$
is empty. Taking into account that both fields and antifields satisfy
constraints \sc\ we get exactly the same number of fields and antifields.
Fields and antifields are paired with each other
$$  A \in \Omega ^{(0,p)} (\wedge^q T_M)
\longleftrightarrow A^* \in \Omega ^{(0,3-p)}
(\wedge^{(2-q)} T_M)~.$$
and obey opposite statistics.
The odd bracket structure on the space of field-antifields is given by
$$\lbrace A_p ^q(z), A_{\tilde p} ^{\tilde q *} (w) \rbrace =
 \delta_{p+\tilde p,3}  \delta_{q+\tilde q,2}
\Omega^{-1} \partial \delta(z,w)\bar \Omega~,$$
where $\delta (z,w)$ is the delta function on
the Calabi--Yau manifold, defined as
follows
$$\int_M \delta (x,y) \Omega(x)\wedge \bar \Omega(x)=1$$
This structure is promoted to a
canonical antibracket on the space of functionals
and formally may be written as follows
$$\lbrace{ F,L \rbrace}=
\sum \int \left(   {\partial} \left(  {\delta F \over \delta A}
\right)^{\vee}
 {\delta L \over \delta A^{*}} -
{\delta F \over \delta A^{*}}
 {\partial} \left(
 {\delta L \over \delta A}\right)^{\vee} \right)^{\vee} ~.$$
It is quite remarkable that the full KS action is given by the same
expression as the original KS action, but without any restrictions on the
ghost numbers. Indeed, the ghost number conservation requires that either all
fields in the action are elements of $\Omega ^{(0,1)} (T_M)$, or at
least
one field has ghost number greater than $2$ and therefore this field is
an antifield.
When all antifields are put to zero the only contribution to the action comes
from the original field $A \in \Omega ^{(0,1)} (T_M)$.
It is a tedious but straightforward check that the full action
is invariant under the nonlinear gauge transformation.
The proof is based on generalized Tian's Lemma \tn\ for arbitrary
$(p,q)$ forms and the generalized Jacobi identity \jac.

The naive definition of the Laplacian turns out to be the correct one:
$$\Delta =\int \left({\delta \over \delta A^*} \partial \left(
{\delta \over \delta A} \right)^{\vee} \right)^{\vee}.$$
To verify that this definition is indeed covariant one has to take into account
that
$\delta A_p ^q (x)/ \delta A_r ^s (y)=\delta_{p,r} \delta_{q,s} \delta (x,y)
\Omega \wedge \bar \Omega$. Now we can check whether the full Kodaira
Spencer action $S(A,A^*)$ satisfies the master equation \bvq.
The gauge invariance of the full action implies that l.h.s of
\bvq\ is equal to zero. The r.h.s can be easily computed and it is equal
$$\Delta S  \sim \int \partial (\Omega A_0 ^1)\wedge \bar \Omega =0$$
Indeed, $\partial (\Omega A_0 ^1)= \partial( A_0 ^1)'=0 $ due to constraint
\sc. The above discussion implies that quantum corrections are
not needed for maintaining the gauge invariance of the
KS theory.

\vskip .15in

\subsec{Open string field theory}

In the case of the open string, the resulting string field theories
were studied
in detail by \witcs .   There it is shown that the space--time physics of the
$A$--model, defined on
the non--compact Calabi--Yau
$3$--fold $T^*L$ (where $L$ is any {\it real\/} $3$--fold),
is equivalent to the usual Chern--Simons field theory on the real
three--manifold $L$. Instead the $B$--model is classically equivalent to the
following field theory on the original Calabi--Yau manifold $M$
\eqn\openB{S=\half\int\limits_M \Omega\wedge \Tr\left(B\wedge \overline\partial
B+{2\over 3} B\wedge B\wedge B\right),}
where the field $B$ is a one--form on $M$ of type $(0,1)$ taking values in
${\rm End}(E)$ and $\Omega$ is the holomorphic $(3,0)$ form. The classical
solutions of \openB\ are the possible inequivalent holomorphic structures we
can put on the bundle $E$. We thus see the space--time
interpretation of the {\it closed\/} $B$--model string, i.e. the
Kodaira--Spencer theory
is very reminiscent of
\openB; in particular, the classical solutions will correspond to the possible
inequivalent holomorphic structures we can put on the manifold $M$ itself.
To make this analogy even more striking it turns out that
the KS action itself may
be viewed as a CS action where
the gauge group of the open string is replaced by an infinite
dimensional group of $\Omega $--preserving diffeomorphisms of the
$3$--fold.   This point we will now explain.

Let us consider a $6$--real--dimensional symplectic manifold
$M$ which consists of a $3$-dimensional base space $X$ and
a $3$-dimensional internal space $Y$.
This symplectic manifold may be regarded as an ``analytic continuation''
of a Calabi-Yau manifold, where we relate the complex coordinates
$(z, \overline{z})$ of the Calabi-Yau to a pair
of real coordinates $(x,y)$  of $M$
($x \in X, y \in Y$).
The K\"ahler structure on the Calabi-Yau is inherited on $M$
as the symplectic structure, the holomorphic and the anti-holomorphic
$3$-forms on the Calabi-Yau become the volume forms on the base $X$ and
on the fiber $Y$. There is also analog of the $'$-operation on $M$
which is realized by a multiplication by the volume form on the fiber.
Consider the Lie algebra ${\cal L}$ of the volume preserving vector fields
(satisfying condition $d_y A'=0$) along the fiber with
coefficients in $1$-forms on the base. We also assume that the space of ${\cal
L}'$ is orthogonal to $H_1$ on $M$. An invariant Killing form for
this Lie algebra is given as follows
$$Tr A B = \int_Y d^3 y A' {1 \over d_y} B'$$
In this notations it is easy to see that KS action coincides with CS action for
${\cal L}$
$$\lambda^2 S(A,0)={1 \over 2}\int_X d^3 x Tr A \wedge d_x A +
 {1 \over 3}\int_X d^3 x Tr A \wedge A \wedge A$$

\vskip .2in

\subsec{\bf{Kodaira-Spencer theory at one--loop}}
In this section we will discuss the computation of Kodaira-Spencer
theory partition function at one--loop.  In order to do this, and
in view of more general applications, we will first discuss the
{\it holomorphic Ray-Singer Torsion}.

\vskip .15in

\subsec{Holomorphic Ray-Singer torsion}
Consider a K\"ahler manifold $M$ with a holomorphic vector bundle
$V$ on it equipped with a norm and a connection
compatible with it.  Let $\overline \partial_V$
denote the del-bar operator coupled with the vector bundle acting
\foot{
We are abusing the notation of denoting the section of the bundle and the
bundle both by $V$.}

$$\overline \partial_V:\qquad
\wedge^p {\overline T}^* \otimes V \rightarrow \wedge^{p+1}
{\overline T}^* \otimes V$$
where $p$ runs from 0 to ${\rm dim}(M)-1$.
  Let $\Delta_V=\Delta_1 +\Delta_2$ denote the corresponding
Laplacian where $\Delta_1=
\overline\partial_V \overline \partial_V^\dagger$
and $\Delta_2=\overline \partial_V^\dagger \overline \partial_V$.
Let us consider the spectrum of $\Delta_V^{(p)}$ acting on $\wedge^p \bar T^*
\otimes V$.
By Hodge decomposition we can find the
non-zero spectrum of the Laplacian by
finding the spectra of $\Delta_1^{(p)}$ and $\Delta_2^{(p)}$.
Note that the spectra of $\Delta_1^{(p)}$ and $\Delta_2^{(p-1)}$
are the same, as are the spectra of $\Delta_2^{(p)}$ and
$\Delta_1^{(p+1)}$.  Let us denote the spectrum of $\Delta_2^{(p)}$
by $\{ \lambda_{p,p+1} \}$.  In constructing a determinant of
the Laplacian acting on forms of all degree it is natural to consider
an alternating product of spectra raised to the power of $\pm 1$
depending on the parity of the form, deleting the zero modes.
  However this will just give the
net answer 1, because the spectra of Laplacian coming from $\Delta_1^{(p)}$
will cancel with those of $\Delta_2^{(p-1)}$ and those from $\Delta_2^{(p)}$
will cancel with those of $\Delta_1^{(p+1)}$.  To avoid this trivial
cancellation we can consider instead
$$\prod_{p=0}^{n-1} \{ \lambda_{p,p+1} \}^{-(-1)^p}$$
This can also be written, taking into account the Hodge decomposition, as
\eqn\rs{I(V)=\prod ({\rm det'}\Delta_V^{(p)})^{(-1)^p p}}
where $'$ denotes deleting the zero modes.
The appropriately regularized $I(V)$ is known as the holomorphic Ray-Singer
torsion for this vector bundle \ref\RS{D. B. Ray and I. M. Singer,
Ann. Math. 98 (1973) 154}.  The main
theorem in \RS\ asserts that $I(V_1)/I(V_2)$ is independent
of the choice of K\"ahler metric on $M$ though it does depend
on the choice of complex structure on $M$
(the case considered in \RS\ is when there are no zero modes).
Morally speaking
we should think of $I(V)$ as the $\prod ({\rm det'} \overline
\partial_V^{(p)})^{(-1)^p}$.
Note that formally one may write
\eqn\rslog{{\rm log}I(V) =\int_{\epsilon}^{\infty} {ds\over s}
{\rm Tr}'(-1)^p p \ {\rm exp}(-s H )}
where the ${\rm Tr}'$ is over all degree forms in the positive
eigenspace of $H$ where
$H=\Delta_V$.  This integral is
regularized by taking $s$ to run from $\epsilon >0$ to $\infty$.

The main technique to compute the Ray-Singer holomorphic
torsion has been recently developed in connection with
Quillen's holomorphic anomaly \ref\bif{J.M. Bismut and D.S. Freed,
Comm. Math. Phys. 106 (1986) 159; 107 (1986) 103\semi
J. -M. Bismut, H. Gillet and C. Soule,
Comm. Math. Phys. 115 (1988) 49, 79, 301
\semi J.M. Bismut and K. K\"ohler, {\it Higher analytic
torsion forms for direct images and anomaly formulas}, Univ. de
Paris-sud, preprint 91-58.}.
Consider a family of complex structures on $M$ parametrized
by a complex parameter $t$.  Let us assume that there are no
jumps in the zero modes of $\overline \partial_V$.  Choose
a holomorphic basis for the zero modes of $\overline \partial_V$ and
let $d_p={\rm log}({\rm det}g^{(p)})$ denote the determinant of the inner
product in the subspace of $\wedge^p T^* \otimes V$ of the
kernel of $\overline \partial_V$.
Then it turns out that using Quillen
anomaly in this context one can show \bif\
\eqn\qanom{\partial \overline \partial [{\rm log} I(V)]
=\partial \overline \partial \sum_p (-1)^p \ d_p
+2\pi i\int_M {\rm Td}(T) {\rm Ch}(V)\Bigg|_{(1,1)}}
where $T$ is the tangent
bundle of $M$ viewed as a bundle over $M$ times the complex moduli
space, ${\rm Td}$ denotes the Todd class
$${\rm Td}[T]=
\det\left[{R / 2\pi i \over 1-{\rm exp}(-R / 2\pi i )}\right]$$
where $R$ is the curvature form for the tangent bundle,
and ${\rm Ch}(V)=\tr\,{\rm exp}(F/2\pi i)$ denotes the Chern class of the
vector
bundle $V$, viewed as a bundle over $M$ times the complex moduli space
where $F$ is the curvature of $V$.
The symbol $|_{(1,1)}$ in the above formula means that we take the
$(n+1,n+1)$ form of the integrand and integrate over $M$
to be left with a $(1,1)$ form on the complex moduli.
The basic idea behind \qanom\ is that, if we ignore
the zero modes that are present, if we integrate both
sides over 1 dimensional complex moduli, the
left--hand side (l.h.s) gives $2\pi i$ times the total number of zero modes of
$\overline \partial$ (weighted with $\pm$ sign) and r.h.s.
is the family's index for the $\overline \partial_V$ operator,
and thus counts precisely the same as the l.h.s..  The main
non-trivial content of \qanom\ is that it is true even
{\it before} integration over moduli space (this can also
be argued using the integrated version by taking various
interesting limits).  The terms corresponding to the
determinant of the norm of the zero modes in \qanom\
is also familiar from the Quillen anomaly and come about
because we are dealing with the determinant of Laplacian
with the zero modes deleted (see e.g. \ref\AMV{A. Belavin and V. Knizhnik,
Phys. Lett. 168B (1986) 201 \semi
L. Alvarez-Gaume,
G. Moore and C. Vafa, Commun. Math. Phys. 112 (1987) 503}).

\vskip .15in

\subsec{KS theory at one--loop and RS torsion}

Having developed the notion of the RS torsion, we are now ready
to compute the partition function of the
KS theory at one--loop.  In fact we will be more general
as the computation can be carried out in the $B$--model version of
any Calabi-Yau $n$--fold
and not just the $3$--fold.  From the formula for $F_1$ given
by \fone\ it is possible to extract the large volume behaviour, in which
case $F_L$ and $F_R$ as noted in section 2 are given by
$$F_{L,R}={1\over 2}(i(k-k^\dagger)\pm (p-q))$$
inserting $F_L F_R$ in the trace, and using the
$sl(2)$ invariance of K\"ahler manifolds we can
replace $-(k-k^\dagger)^2$ in the trace with $(p+q-n)^2$,
and noting that insertions of $p^2$ or $q^2$ alone in the
trace are independent of the moduli (as they would
be index computations) leads us to the statement that the
insertion of $F_LF_R$ is equivalent (as far as the moduli
dependence is concerned) with insertion of $p\cdot q$.  Now using
the form of $F_1$ and comparison with \rslog\ leads us to
$$F_1={1\over 2}\sum_q (-1)^q q\ {\rm log}\,I(\wedge^q T^*)$$
Now, according to \bcov\ we have a formula for $\partial \overline
\partial$ anomaly of $F_1$.  On the other hand, using the Quillen
anomaly discussed above for $I(V)$, we can compute the anomaly in another
way.  The fact that the two are the same is a
very interesting check on these ideas, and in particular is the
`mirror' version of the conformal theory statement of the anomaly.
There were two terms in the anomaly discussed in \bcov ,
as there are two terms for the anomaly \qanom .  The
first term in each of these two is the same, and simply is the
contribution of the volume of zero modes to the anomaly.   The
more subtle term is the second one
which comes from the contact terms both in string theory
and in the computation of Quillen anomaly. As shown in \bcov\ the
second term there is $\chi (M)\cdot G /24$ where $G$ is the
K\"ahler form for the Zamolodchikov metric on moduli space.
Therefore we wish to prove the following equation
\eqn\rseq{\boxeqn{\ 2\pi i \int_M{\rm Td}(T)\sum_{p=0}^n (-1)^p p\,
{\rm Ch}(\wedge^{n-p} T^*)\Bigg|_{(1,1)-\rm part}
 ={1\over 12}\, \chi(M)\, G\ }}

We start by recalling a few facts \ref\hirz{F. Hirzebruch, {\it
Topological methods in
Algebraic Geometry\/}, Springer-Verlag, 1966 }. First of all,
$${\rm Td}(T)\sum_{p=0}^n (-1)^p {\rm Ch}(\wedge^{p} T^\ast) =
c_n(T)$$
($T^\ast$ is the cotangent bundle).
This is (a special case of)
theorem 10.1.1 in \hirz .
Now we apply the Hirzebruch argument to our case.
Let $\gamma_i$ be the eigenvalues of the curvature form.
Consider the identity (notation as in the proof of th.10.1.1)
\eqn\identity{\sum_{r=0}^n(-1)^r x^r\,
 {\rm Ch}(\wedge^r\xi)=\prod_{i=1}^n(1- x
e^{-\gamma_i}).}
One has
$$\sum_{p=0}^n (-1)^p p\,
 {\rm Ch}(\wedge^p\xi)= {\partial\over\partial x}
\sum_{r=0}^n (-1)^r x^r {\rm Ch}(\wedge^r\xi)\Big|_{x=1}.$$
Using the identity \identity, the r.h.s. becomes
$$n \prod_j(1-e^{-\gamma_j})-\sum_j\prod_{i\not=j}(1-e^{-\gamma_i}).$$
Imitating the proof of the quoted theorem,
we consider ($\xi\mapsto T^*$)
\eqn\previous{{\rm Td}(T)\, \sum_{p=0}^n(-1)^p
 p\, {\rm Ch}(\wedge^r T^\ast)=
n\,  c_n(T)-\sum_j {\gamma_j\over 1
-e^{-\gamma_j}}\prod_{k\not=j}\gamma_k.}
Now,
$${\gamma_j\over 1-e^{-\gamma_j}}=1+\half \gamma_j+{1\over
12}\gamma_j^2+\dots,$$
where $\dots$ means higher degree. Inserting
 this expansion in \previous\
we get
$${\rm Td}(T)\, \sum_{p=0}^n(-1)^p p\, {\rm Ch}(\wedge^p T^\ast)=
{n\over 2}
 c_n(T)-c_{n-1}(T)-{1\over 12} c_n(T) c_1(T)+\dots.$$
We have to take the $(n+1,n+1)$ component of the r.h.s. which is
$$-{1\over 12} c_n(T)
\, c_1(T).$$
Now using the fact that by the discussion in section 2,
$c_1(T)=-c_1(T^*)=-G/2\pi i$, and the fact that $\int c_n(T)=\chi (M)$ we get
\rseq\ which is what we wished to show.

\vskip .15in

\subsec{One--loop topological open string amplitudes}
If we consider the open string version of the $N=2$ twisted model
coupled to gravity, as mentioned before,
it turns out that the space of vacua are related to
a choice of a holomorphic vector bundle $V$ over $M$ \witcs .  In such a case
taking the large volume limit in the $B$-version
of the model would lead us to \rslog .  Thus the one-loop partition
function of open string is exactly the holomorphic Ray-Singer torsion,
$F_1=I(V)$,
and is thus computable again using the Quillen anomaly \qanom .

Note in particular the computation in section 4 at one--loop
amplitude of open string gives the same answer as the first
term in the Quillen anomaly for Ray-Singer torsion.  The contact
terms were not considered in section 4, but since they can
be computed in this field theory setup, they must be the
same as the ones leading to the index integral.

\def\CV{{\cal V}}
\def\CN{{\cal N}}
\def\Pr#1{{\bf P}^#1}
\def\CM{{\cal M}}

\vskip .2in

\subsec{\bf{The geometrical information encoded in} $F_g$ {\bf for the
A--model}}

In this subsection we describe the geometrical information encoded in $F_g$ for
the $A$--model defined on a Calabi--Yau $3$--fold $M$. As discussed in
section 2.2. the $A$--model action reads
$$S=\sum_i t^i\int (\omega_i)_{\alpha\bar\beta}\,\partial X^\alpha \bar\partial
\bar X^{\bar\beta}+
\sum_i \bar t^i \int (\omega_i)_{\alpha\bar\beta}\,\partial \bar
X^{\bar\beta}\bar\partial X^\alpha +{\rm fermions},$$
where the integral forms $\omega_i$ span $H^{1,1}(M)$. As we know from the
discussion in section
3, $F_g$ is not a holomorphic section of $\CL^{2-2g}$, and
therefore $F_g$ depends on a choice of a base point $\bar t^i$ in the moduli
space $H^{1,1}(M,{\bf C})$. The meaning of $F_g$ is particularly transparent if
we choose the base point to be at infinity i.e. to correspond to positive
infinite volume. Then we set $\bar t^i=\bar t\, m^i$, where $\sum_i
m^i\omega_i$ is a positive K\"ahler form $\omega$, and then send $\bar
t\rightarrow +\infty$. Of course, in this process the $t^i$'s are still kept
arbitrary. Since in the weak coupling ($=$ infinite volume)
limit the $A$--model
correlations reduce to classical geometric objects, this is the choice of base
point for which the geometric nature of $F_g$ is more evident.

Indeed, as $\bar t\rightarrow +\infty$ the action becomes
\eqn\genArg{S= \bar t \int \omega_{\alpha\bar\beta}\, \partial \bar
X^{\bar\beta}\bar\partial X^\alpha +\dots.}
Then all finite action configurations satisfy $\bar\partial X^\alpha=0$, i.e.
correspond to holomorphic maps from the Riemann surface $\Sigma_g$ to the
Calai--Yau
space $M$. Thus the $g$--loop amplitudes for the $A$--model with base point at
infinity are exactly given by sums over holomorphic maps $X$ from genus $g$
surfaces to the Calabi--Yau space $M$ of the form
$$F_g= \sum_n^a N^g_{n_1n_2\dots n_h} q_1^{n_1}q_2^{n_2}\cdots q_h^{n_h},$$
where, as in section 2, $q_k=\exp[-t^k]$ and $n_k=\int X^\ast \omega_k$. The
coefficients $N^g_{n_1n_2\dots n_h}$ are related to the `number' of maps in the
given topological class as we will discuss below.
This means that we can use the $A$--model partition functions  to `count' the
number of such maps, or equivalently the number of genus $g$ holomorphic curves
lying on $M$. This counting was done for the special cases $g=0$ and $g=1$ in
Ref. \cand\ and \bcov , respectively.

In general, given a Riemann surface $\Sigma_g$ the existence of holomorphic map
of a given degree into $M$ depends on the complex structure of $M$. Because of
the absence of mixed anomalies (section
3), $F_g$ is independent of the complex
structure of $M$.  Then in order to get the number of
curves from the $A$--model
it is crucial that we integrate over the complex moduli of $\Sigma_g$, i.e.
that the $A$--model is coupled to topological gravity. Then the `number' of
holomorphic maps $\Sigma_g\rightarrow M$ summed over the moduli space of
$\CM_g$ is independent of the complex structure of $M$.

In order to extract from $F_g$ the number of maps of a given type, we need to
know for each kind of map
(including multi--covers and singular ones) how the coefficient
$N^g_{n_1n_2\dots n_h}$ in the $q$--expansion of $F_g$ is related to the actual
number of holomorphic curves. This requires doing an explicit path--integral
around an instanton of the given type. The rest of this section
and appendix A are dedicated to
such path--integral computations. In fact, this section is rather technical.
We will limit ourselves to smooth manifold $M$ and not
deal with spaces such as orbifolds, though many of the techniques
we discuss can be easily adapted to such cases.
The limit $\bar t\rightarrow \infty$ is implicit throughout.

We recall that $F_g$ is given by $F_g=\int_{\CM_g}{\cal Z}_g$, where ${\cal
Z}_g$ is
the following top form over ${\cal M}_g$ (for $g\geq 2$)
\eqn\whatnot{\eqalign{ (3g-3)!\,
(2\pi i)^{3g-3}&\, {\cal Z}_g=\cr
&=\left\langle\left[dm^a
\Big(\int \mu_a \psi_\alpha\partial X^\alpha \Big)\bigwedge
  d\bar m^{\bar b}\Big(\int \bar\mu_{\bar b}\bar
\psi_{\bar \beta} \bar\partial \bar X^{\bar \beta}\Big)
\right]^{3g-3}\right\rangle_g ,\cr}}
and $m^a$ are coordinates on $\CM_g$ associated to the Beltrami differentials
$\mu_a$.

\vglue 14pt
\subsec{\it Contribution to $F_g$ from an isolated genus $g$ curve}
\vglue 10pt

If $F_g$ has to `count' the number of genus $g$ curves lying on the
Calai--Yau manifold
$M$, in particular it should be true that the contribution to $F_g$ from an
{\it isolated} such curve $\CC_g$ is given by
\eqn\isocont{\exp\left[-\sum_i t^i \int_{\CC_g}\omega_i\right],}
with coefficient $1$. Here we check explicitly this property of $F_g$. The
assumption of $\CC_g$ being isolated is rather unrealistic; for $g>1$ the
holomorphic curves typically belong to multi--parameter families. Below we
shall drop this assumption.

Let ${\cal T}_g$ be the Teichm\"uller space of genus $g$ curves. Clearly
counting holomorphic maps $\Sigma_g(m)\rightarrow \CC_g$ for $m\in {\cal M}_g$
is
equivalent to counting holomorphic maps {\it homotopic to the identity} but
with $m\in {\cal T}_g$. We shall take this second viewpoint\foot{It is also
convenient (although not necessary) to modify the metric
$\omega_{\alpha\bar\beta}$ in a tubular neighborhood of $\CC_g$ such that the
induced metric on $\CC_g$ has constant curvature. We are free to do this, since
a deformation of the metric preserving the K\"ahler class is a $D$--term
perturbation which does not affect any topological quantities.}.
We take as base
point in ${\cal T}_g$ the point corresponding to the complex structure of
$\CC_g$ (for some choice of marking); hence for $m^a=0$ we have a holomorphic
map $\Sigma_g(0)\rightarrow \CC_g\subset M$ homotopic to the identity. By the
general argument around eq.\genArg\ the contribution from $\CC_g$ to $F_g$ has
support at
$m^a=0$, so  in the following we take $m^a$ to be very small.

Our action can be rewritten as
\eqn\action{S=\half(t+\bar t)\int d^2 z\, \omega_{\alpha\bar \beta}\,
\partial_\mu X^\alpha
\partial^\mu {\bar X}^{\bar\beta}+\half (t-\bar t) \int X^\ast \omega+\rm
fermions,}
where $\omega$ is the K\"ahler form of $M$. We are interested in the limit
$\bar t\rightarrow +\infty$ at $t$ fixed. The second term in the r.h.s.
of \action\ is independent of the smooth map $X$, as long as its image is in
the homology class of $\CC_g$.
Hence
the minimum of the action in this topological class is obtained by minimizing
the first term i.e. by the corresponding harmonic map. Here we are interested
only in $m^a$ small. In this case the harmonic map has the form
$X(z)^\alpha+\delta X^\alpha$,
 where $X(z)^\alpha$ is the map $\Sigma_g(0)\rightarrow\CC_g$. We can decompose
the variation $\delta X^\alpha$
into a component perpendicular to $\CC_g$ and one along $\CC_g$. The component
perpendicular is an element of $H^0(\Sigma_g(0),T_M)$ and hence vanishes
by the rigidity assumption. Then,
to the first order, our harmonic map can still be seen as map
$X\colon\Sigma_g(m)\rightarrow \CC_g$.
It is a theorem by Shoen and Yau \ref\shoen{R. Shoen and S.T. Yau,
Invent. Math. 16 (1972) 161.} that
there exists a unique harmonic map $\Sigma_g(m)\rightarrow \CC_g$ (homotopic to
the identity).
Neglecting higher orders in $m^a$, the value of the action at the extrema is
then
$$S_{\rm min}=\half(t+\bar t) E(m^a,\bar m^b)+\half (t-\bar t)d,$$
where $E(m^a,\bar m^b)=\int d^2z\, g_{i\bar j}\, \partial_\mu X^i
\partial^\mu {\bar X}^j$ is the Shoen--Yau ``energy"
as a function of the moduli and
$$d=\int X^\ast \omega={1\over t}\sum_i t^i\int_{\CC_g}\omega_i,$$
 is the `degree' of $\CC_g$.

Of course, one has
$$E(m^a,\bar m^b)\geq d,$$
with equality if and
only if the corresponding harmonic map is holomorphic, which
happens only for $m^a=0$.
The function $E(m^a,\bar m^b)$ is the K\"ahler potential for the
`Weil--Petersson' (WP) metric $W_{a\bar b}$ at the base point, i.e.
\ref\tromba{
J.
Tromba, Man. Math. 59 (1987) 249.}
$$\partial_a\bar\partial_b E\Big|_{m=0}= W_{a\bar b}.$$
More precisely,
$W_{a\bar b}$ is the usual Weil--Petersson metric if we have chosen the
metric on $M$ so that the induced metric on $\CC_g$ has constant curvature (see
previous footnote). Otherwise, $W_{a\bar b}$ is some metric on ${\cal T}_g$;
our computations below are valid for any choice of the metric.

Consider the Shoen--Yau solution $X(z,\bar z)$. By definition, this smooth
function is holomorphic with respect
 the
complex structure defined by $\CC_g$. Then, applying the Kodaira--Spencer
machinery to the variation of the complex structure
of the Riemann surface, we see that
\eqn\KStwo{\bar\partial X- \Phi\, \partial X=0,}
where $\Phi$ is the KS vector defining the complex structure of $\CC_g$ in
terms
of that of $\Sigma_g(m)$.

Now, let us consider the derivative $\partial_{m^a}X$ of $X$ at $m=0$. If
$\mu_a$ is the Beltrami differential corresponding to an infinitesimal
variation
of the moduli $\delta m^a$, one has $\partial_{m^a}\Phi=\mu_a$.
Then, taking the derivative of \KStwo,  we get
$$\partial_{m^a}\partial_{\bar z} X\Big|_{m=0}= \mu_a \partial_z X\Big|_{m=0} +
\Phi\Big|_{m=0} \partial_{m^a}\partial X=\mu_a,$$
where we used the fact that at $m=0$ the Shoen--Yau map is the identity, i.e.
$X(z)|_{m=0}=z$, and $\Phi|_{m=0}=0$.
The same argument give $\bar\partial_{\bar m^a}\bar\partial X=0$.

Then\foot{There are two terms in the definition of $E(m,\bar m)$. However,
since their difference is just the topological invariant $d$, for computing the
variation of $E(m,\bar m)$ we can replace their sum by twice the first term.}
$$\partial_a\bar\partial_{\bar b} E\Big|_{m=0}= \int_{\Sigma}
\omega_{\alpha\bar\beta}\, \partial_{m^a}\bar\partial X^\alpha\wedge
\bar\partial_{\bar m^b} \partial\bar X^{\bar\beta}\Big|_{m=0}= (\mu_b,\mu_a),$$
where $(\cdot, \cdot)$ is the Hodge inner product on $\bar K\otimes K^{-1}$
with respect to the metric $\gamma_{z\bar z}$ on $\Sigma_g$
induced by the imbedding in $M$ i.e. $\gamma_{z\bar
z}=\omega_{\alpha\bar\beta}\partial_z X^\alpha\partial_{\bar z}\bar
X^{\partial\beta}$. If we choose this metric to be constant curvature, this
inner product is (by definition) the Weil--Petersson metric on $\CM_g$.

As $\bar t\rightarrow\infty$, the contribution of $F_g$ from the curve $\CC_g$
is concentrated at the point $m=0$ in moduli space. Hence we can assume $m$ to
be small. In this case, one has
\eqn\bosdelta{\eqalign{e^{-S_{\rm bos}}\Big|_{\bar t\rightarrow\infty}&\approx
e^{- t\, d}
\exp\big[ -\half\bar t\, W_{a\bar b}\, m^a \bar m^b\big]\Big|_{\bar t
\rightarrow\infty}\cr
&=\left({2\pi\over \bar t}\right)^{3g-3} \big(\det W_{a\bar b}\big)^{-1}
  e^{-d\, t}\, \delta(m_a)\delta(\bar m_b)\cr
&=\left({2\pi\over \bar t}\right)^{3g-3} e^{-d\, t} \delta_W(m).\cr}  }
where $\delta_W(m)$ is covariant $\delta$--function for the WP metric,
i.e. such that
$\int d\mu_{WP} f(m) \delta_W(m-a)= f(a)$ where $d\mu_{WP}$ is the WP volume
form.

{}From \bosdelta\ we see that only the identity map contributes to the
integral.
In the pre--exponential factor in \whatnot\ we can replace $X^\alpha(z)$ by
this identity map.

Let $i\colon\CC_g\rightarrow M$ be the imbedding map and
$\eta_{A\, \alpha}$ ($A=1,\dots, 3(g-1)$) be a basis\foot{In fact, since the
map $X\colon \Sigma_g(0)\rightarrow\CC_g$ is the identity, $\eta_A$ are just
the ordinary quadratic differentials on $\Sigma_g$.} of
$H^0(\CC_g,K\otimes i^\ast T^\ast)$ orthonormal in the sense that
$$\int_\Sigma \omega^{\alpha \bar \beta}\, \bar\eta_{\bar B \bar \beta}\wedge
\eta_{A\, \alpha}=\delta_{A\bar B}.$$
Let $\mu_a$ be the Beltrami's corresponding to the moduli $m^a$, chosen to be
harmonic with respect to
the metric $\gamma_{z\bar z}$.
Then consider the quantity
$$\omega_{\alpha\bar \beta} \bar\mu_{\bar a}\bar\partial \bar X^{\bar \beta}
dz$$
it belongs to $H^0(\CC_g,K\otimes i^\ast T^\ast)$ and hence has an expansion in
terms of
the $\eta_A$ basis of the form $\bar B_{\bar a}^{\ A}\eta_A$ for some
coefficients $B_{\bar a}^{\ A}$.

Let us expand $\psi(z)$ as\foot{The factor $\sqrt{\bar t}$ arises because
physically we have to
normalize the Fermi zero--modes with respect to the true metric
$(t+\bar t)g_{i\bar j}$, rather than with respect the reference one
$g_{i\bar j}$.}
$$\psi(z)= \sqrt{\bar t}\,\psi^A\eta_A(z).$$
Then
$$\eqalign{dm^a \Big(\int \mu_a \psi_\alpha\, \partial X^\alpha \Big)
&=
\sqrt{\bar t}\, \psi^B dm^a B_a^{\bar A}
\Big(\int \omega^{\bar \beta \alpha}\, \bar \eta_{\bar A\, \bar \beta}\wedge
\eta_{B\, \alpha} \Big)\cr
&= \sqrt{\bar t}\, dm^a B_a^{\bar A}\, \psi^B\, \delta_{\bar A B}
.\cr}$$
Then the expression
$$\left[dm^a\Big(\int \mu_a \psi_\alpha\partial X^\alpha \Big)\bigwedge
  d\bar m^{\bar b}\Big(\int \bar\mu_{\bar b}\bar \psi_{\bar \beta}
\bar\partial \bar X^{\bar \beta}\Big)\right]^{3g-3}\Bigg|_{\rm zero-modes}$$
after the integration over the $\psi$ zero--modes becomes
\eqn\fermiout{(3g-3)!\, |\bar t|^{3g-3} |\det[B]|^2 \prod dm_a d\bar m_{\bar
b}= (3g-3)! |\bar t|^{3g-3} d\mu_{WP}.}
Here the last equality follows since $|\det[B]|^2$ is nothing else than
$\det[W]$, where $W$ is the WP metric.
Indeed,
$$\eqalign{(\mu_a,\mu_b)&=\int \omega^{\alpha\bar \gamma}(\omega_{\alpha\bar
\beta} \bar\mu_{\bar a}
\bar\partial \bar X^{\bar \beta})(g_{\bar \gamma \delta} \mu_{b}
\partial X^\delta)\cr
&= \bar B_{\bar a}^A B_b^{\bar B} \int \omega^{\bar \gamma \alpha} \eta _{A\
\alpha}\wedge \bar\eta_{\bar B\ \bar \gamma}=
(\bar B B^t)_{\bar a b}.\cr}$$

Finally from \whatnot, \bosdelta, and \fermiout, we get
\eqn\finally{{\cal Z}_g\Big|_{{\rm isolated\ curve}}= e^{-d\ t}
\delta_W(m) d\mu_{WP}}
By definition of $\delta_W(m)$, the integral of the r.h.s. in any domain of
${\cal M}_g$ containing our base point $m=0$ is just $\exp[-d\, t]$,
that is the contribution of an isolated genus $g$ curve to ${\cal F}_g$ is
given by eq.\isocont.

\vglue 14pt
\subsec{\it Contribution to $F_g$ from a continuous family of curves}
\vglue 10pt

Typically the holomorphic maps are not isolated but belong to a family. We have
to say how we `count' instantons in this case. In general a direct
path--integral computation is quite hard. However, general principles
\ref\genarguments{E. Witten, Nucl. Phys. 371 (1992) 191\semi E. Witten, in {\it
Essays on Mirror Manifolds}, ed. by S.T. Yau, International Press} lead to
an abstract formula for
$N^g_{n_1,\dots,n_h}$ which is valid in full generality.
In the $A$--model on a
Calai--Yau $3$--fold this formula is as follows.
Assume we have a family of holomorphic maps from genus $g$ surfaces to $M$,
$$f_s\colon\Sigma_g(s)\rightarrow M,\hskip 1.8cm s\in {\cal S},$$
where ${\cal S}$ is the space of parameters for the family. Over ${\cal S}$ we
define the bundle $\CV$ whose fiber at $s$ is the vector space
$$\CV\Big|_s= H^0(\Sigma_g(s), K\otimes f_s^\ast T^\ast_M),$$
and let $r={\rm rank}(\CV)$. Then the contribution of this family to $F_g$
reads
\eqn\astractformula{\exp\left[- t^a\int_\Sigma f^\ast_s\omega_a\right]\,
\int\limits_{\cal S} c_r(\CV),}
that is the coefficient is just the integral over the moduli space $\cal S$ of
the Euler class of the bundle $\CV$.
It is using this abstract formulation that Aspinwall and Morrison
\aspmor\ were able to prove the formula for
contribution from multi--covers in genus zero that we mentioned in section 2.

\vglue 14pt
\subsec{\it Contribution to $F_g$ from constant maps}
\vglue 10pt

We wish to compute the limit of $F_g$ when $\bar t^{\bar j}$ {\it and} $t^i$ go
to infinity. The result of this computation will be needed below to fix part of
the ambiguities arising in the solution of the anomaly equation. In this limit
only the constant maps contribute to $F_g$.

The moduli space of constant maps from a genus $g$ surface to $M$ is given by
$X=\CM_g\otimes M$. There are three $\chi$ zero--modes spanning the fiber of
the vector bundle\foot{$\pi_i$ is the projection into the $i$--th factor space
of $X=\CM_g\otimes M$.} $\pi_2^\ast  T_M$ over $X$, while the $3g$ $\psi$
zero--modes span the fiber of the vector bundle
$$\CV= \pi_1^\ast\CH\otimes\pi_2^\ast  T_M^\ast,$$
where $\CH$ is the Hodge bundle over $\CM_g$ (i.e. the bundle whose fiber at
$m$ is $H^0(\Sigma_g(m),K)$).

Then the general formula \genarguments\ give
\eqn\largettbar{F_g\Big|_{t,\bar t\rightarrow\infty}=\int\limits_X
c_{3g}(\CV).}
It is easy to recover \largettbar\ by a direct path integral computation. In
order to do this, we introduce some notation. Let
$\omega_A$ ($A=1,\dots, g$) be
a basis of holomorphic one--forms on $\Sigma_g(m)$, and the $N^{\bar A B}$ be
the inverse matrix of $N_{A\bar B}={\rm Im}\,\Omega_{AB}=\int
\omega_A\wedge\bar\omega_B$. Then we put
\eqn\whatA{A_{a\, AB}=\int (\mu_a \omega_A)\wedge \omega_B,}
and $\CA_{AB}=dm^a A_{a\ AB}$. From the theory of variations of Hodge structure
(which is essentially the same thing as the $tt^\ast$ equations) we know that
the curvature of the Hodge bundle is $P_{A\bar Ba \bar b}dm^a\wedge d\bar
m^{\bar b}$ where\foot{Notice that if you interpret $A_{a AB}$ as the
$3$--point function on the sphere, this is just the $tt^\ast$ equation for the
curvature.}
$$P_{A\bar B a\bar b}=- A_{a AC} N^{C\bar D} A^\ast_{\bar b \bar D\bar B}.$$
Then the curvature of $\CV$ is given by
\eqn\curva{{\cal R}^{\ (B,\beta)}_{(A,\alpha)}=\Big[\delta_\alpha^{\ \beta}
P_{A \bar C\,  a\bar b}\, N^{\bar CB} dm^a\wedge
d\bar m^{\bar b}+ \delta_A^{\ B} R_{\alpha\bar \gamma \sigma \bar \rho}\,
G^{\bar \gamma \beta}
dx^\sigma\wedge d\bar x^{\bar \rho}\Big].}

As $t,\bar t\rightarrow\infty$ the theory gets
coupled in a weaker and weaker way,
and we can use perturbation theory (that is free fields).
Then the derivatives of the scalars can
be eliminated using the free contraction \ref\hamidi{S. Hamidi and C. Vafa,
Nucl. Phys. B279 (1987) 465}
\eqn\free{\langle\partial X^\alpha(z)\bar\partial
\bar X^{\bar\beta}(\bar w)\rangle_g= G^{\bar \beta \alpha} \omega_A(z)N^{A\bar
B}\,
\overline\omega_{\bar B}(\bar w).}
We denote the $3g$ $\psi$ zero modes,
by $\psi_\alpha^A$ ($\alpha=1,2,3$, $A=1,\dots, g$) with
\eqn\expp{\psi_\alpha(z)=\psi_\alpha^A\, \omega_a(z).}
In addition we have three {\it constant} zero--modes $\chi^\alpha$. The extra
Fermi zero--modes are absorbed by $3$ factors of $(\int {R_{\alpha\bar
\beta}}^{\gamma\bar \delta}\chi^\alpha\bar\chi^{\bar \beta} \,
\psi_\gamma\wedge\bar\psi_{\bar \delta})$
extracted from the exponential of the action. Using \expp\ and the definition
of $N_{A\bar B}$ we get
$$\int {R_{\alpha\bar \beta}}^{\gamma\bar \delta}\chi^\alpha\bar\chi^{\bar
\beta} \,
\psi_\gamma\wedge\bar\psi_{\bar \delta}={R_{\alpha\bar \beta}}^{\gamma\bar
\delta}\chi^\alpha\bar\chi^{\bar \beta}
\, \psi_\gamma^A \bar\psi_{\bar \delta}^{\bar B}\, N_{A\bar B}.$$
Then using \whatA\ and \expp, the contribution to \whatnot\ from the constant
maps is reduced\foot{As always, non--zero modes  cancel by topological
invariance.} to an integral over the zero--modes $\psi^A_\alpha$, $\chi^\alpha$
and $x^\alpha$ of the following quantity
$${(-i)^3\over 3! (3g-3)! \, (2\pi i)^{3g}} \Big[\psi_{\alpha}^{A} \CA_{AB}
 \bar \psi_{\bar \beta}^{\bar C} \CA^\ast_{\bar C \bar D} G^{\bar \beta \alpha}
N^{B\bar D}\Big]^{3g-3}\Big({R_{\alpha\bar \beta}}^{\gamma\bar
\delta}\chi^\alpha\bar\chi^{\bar \beta}
\psi_\gamma^A \bar\psi_{\bar \delta}^{\bar B} N_{A\bar B}\Big)^3 .$$
After integrating away the $\chi$'s, we remain with the integral
over the $\psi$'s and the bosons of
\eqn\thisis{{(-1)^g\over (3g)!\ (2\pi i)^{3g}}\Bigg(\Big[G^{\alpha\bar \beta}
P_{A\bar B \,a\bar b} dm^a\wedge d\bar m^b+ N_{A\bar B}
 {R^{\alpha\bar \beta}}_{\gamma \bar \delta}
dx^\gamma\wedge d\bar x^{\bar\delta}\Big] \psi_\alpha^A\bar \psi_{\bar
\beta}^{\bar B}\Bigg)^{3g}.}
Comparing with \whatnot\ we see that
\eqn\final{\eqalign{\lim_{t,\bar t\rightarrow\infty} F_g &\equiv
\lim_{t,\bar t\rightarrow\infty}\int\limits_{\CM_g}{\cal Z}_g=\cr
&=(-1)^g \int\limits_{\CM_g\otimes M}
\det \left[{{\cal R}\over 2\pi i}\right]=\int\limits_{\CM_g\otimes M}
c_{3g}\big(\CV\big) ,\cr}}
which is eq.\largettbar.

The class $c_{3g}(\CH\otimes T^*_M)$ can be related to the Chern classes of
$T_M$ and $\CH$ using the `splitting principle'.
We use $x_i$ (resp. $y_a$) to denotes the `eigenvalues' of the curvature of the
bundle  $\CH$ (resp. $T^\ast_M$).
We start from the identity
\eqn\identity{\eqalign{\prod_{i=1}^g \prod_{a=1}^3(x_i+y_a)&=
\prod_{a=1}^3 \sum_{r=0}^g y_a^{g-r}\sigma_r(x_i)\cr
&=\sum_{r_1,r_2,r_3=0}^g y_1^{g-r_1} y_2^{g-r_2} y_3^{g-r_3}\sigma_{r_1}(x_i)
\sigma_{r_2}(x_i)\sigma_{r_2}(x_i),\cr}  }
where $\sigma_r$ are the elementary symmetric functions.
For $c_{3g}(\CV)$ we are interested in the terms in \identity\
homogeneous of degree $3$ in the $y$'s. For $g>2$ they are given by
$$\eqalign{\sum_{r_1+r_2+r_3=3} & y_1^{r_1}y_2^{r_2}y_3^{r_3}
\sigma_{g-r_1}(x_i)\sigma_{g-r_2}(x_i)\sigma_{g-r_3}(x_i)\equiv\cr
&\equiv [\sigma_1(y)]^3 \sigma_{g-3}(x)[\sigma_g(x)]^2
+\cr
&\quad+\sigma_1(y) \sigma_2(y)[\sigma_{g-2}(x)\sigma_{g-1}(x)
\sigma_{g}(x)-3\sigma_{g-3}(x)\sigma_g(x)^2]+\cr
&\quad+\sigma_3(y) [\sigma_{g-1}(x)^3-3\sigma_{g-2}(x)
\sigma_{g-1}(x)\sigma_{g}(x)+3\sigma_{g-3}(x)
\sigma_g(x)^2].\cr}$$
In particular, this identity says that the component of $c_{3g}(\CH\otimes
T^*_M)$ which is a $(3,3)$--form on $M$ reads
$$c_3(T^*_M)\, \big[c_{g-1}^3-3 c_{g-2}\, c_{g-1}\, c_g + 3 c_{g-3}\,
c_g^2\big]+\ {\rm terms\ proportional\ to}\ c_1(T^\ast_M),$$
where $c_k\equiv c_k(\CH)$. Since for a Calabi--Yau
manifold $c_1(T_M)=0$, we have
\eqn\againf{\lim_{t,\bar t\rightarrow\infty} F_g=
-\chi(M)\, \,  \int\limits_{{\cal M}_g}\big\{c_{g-1}^3-3
c_{g-2}\wedge c_{g-1}\wedge
c_g+3 c_{g-3} \wedge c_g^2\big\}.}
The integral in the r.h.s. can be simplified. Indeed, as it is well known, the
$c_k$'s are not all independent. Let $\bf H$ be the
de Rham bundle (i.e. the bundle over ${\cal M}_g$ with fiber
$H^1(\Sigma_g,{\bf C})$). Obviously, $\CH$ is a holomorphic subbundle\foot{This
is the same situation we encountered in section 2.1 in the context of special
geometry in section
2.1. The trivial bundle $\bf H$ plays here the same role as
$H^3(M,{\bf C})$ in section 2.} of ${\bf H}$
and we have the exact sequence
$$0\rightarrow \CH \rightarrow {\bf H} \rightarrow \CH^\ast\rightarrow 0.$$
Now, from the $tt^\ast$ geometry discussed in section 2, we know that $\bf H$
comes with
a natural flat connection --- the $\nabla$--connection --- and hence
$$1=c({\bf H})= c(\CH)\, c(\CH^\ast).$$
Then we have the following relation between the $c_k$'s
\eqn\ckrels{\Big(1+\sum_{k=1}^{g} c_k\Big)\Big(1+\sum_{h=1}^g (-1)^h
c_h\Big)=1.  }
In particular, eq.\ckrels\ gives the $c_{2k}$'s as polynomials in the
$c_{2m+1}$. The first few relations are
\eqn\relexamples{c_2=\half (c_1)^2,\qquad
c_4=  c_1 c_3-{1\over 8} (c_1)^4.}
Equating the two sides of
eq.\ckrels\ in degree $4g$ we get
\eqn\twogrel{c_g^2=0,}
while in degree $4(g-1)$ we get
$$2 c_g c_{g-2}- c_{g-1}^2=0.$$
Using these two relations eq.\againf\ reduces to
\eqn\rfinal{\boxeqn{\ \lim_{t,\bar t\rightarrow\infty} F_g=
\half\, \chi(M)\,  \int\limits_{{\cal M}_g} c_{g-1}^3.\ }}

The integrals $\int_{\CM_g} c_{g-1}^3$, can be easily computed if we know the
Chow ring of $\CM_g$. In fact, by definition our $c_k$ are represented in the
Chow ring by the tautological classes $\lambda_k$ (notations as in
ref.\ref\mumford{D. Mumford,
in {Arithmetic and Geometry}, papers dedicated to I.R. Shafarevitch, eds.
M. Artin and J. Tate (Boston, Birk\"auser, 1983).}).

The Chow ring of $\CM_g$ is explicitly known for $g=2$ \mumford\ and for $g=3$
\ref\faber{C. Faber, Ann. Math. 132 (1990) 331.}.
In particular, for $g=2$ th. 10.1 of ref.\mumford\ gives
$$\int_{\CM_2} (c_1)^3= {1\over 2880},$$
while for $g=3$ using eq.\relexamples\ and Table 10 of ref.\faber\ we have
$$\int_{\CM_3} (c_2)^3= {1\over 8}\int_{\CM_3} (c_1)^6= {1\over 8\cdot
90720}={1\over 725760}.$$
Then
\eqn\largeex{\eqalign{F_g\Big|_{t,\bar t\rightarrow\infty} & = {\chi(M)\over
5760} \hskip 2cm {\rm for}\ g=2\cr
&= {\chi(M)\over 1451520} \hskip 1.7cm {\rm for}\ g=3.\cr}  }

\vfill
\eject

\newsec{Solution to the anomaly equation and Feynman rules for $F_g$}

In this section, we develop a systematic method to solve the
holomorphic anomaly equation
\eqn\holanomaly{
   \overline{\partial}_{\bar i} F_g = {1 \over 2}
\overline{C}_{\bar{i}\bar{j}\bar{k}} e^{2K} G^{j \bar j} G^{k \bar k}
\left( D_j D_k F_{g-1} + \sum_{r=1}^{g-1} D_j F_r D_k F_{g-r} \right).
}

\vskip .15in

\subsec{Feyman rules at $g=2, 3$}

As a warm-- up, let us start with the genus--$2$ case.
\eqn\genustwo{
   \overline{\partial}_{\bar i} F_2 = {1 \over 2}
\overline{C}_{\bar{i}\bar{j}\bar{k}} e^{2K} G^{j \bar j} G^{k \bar k}
\left( D_j \partial_k F_{1} + \partial_j F_1 \partial_k F_1 \right). }
Interestingly enough, a key to solving this equation lies in a
genus-$0$ object. Because the Yukawa--coupling $\overline{C}_{\bar{i}\bar{j}
\bar{k}}$ is totally symmetric in its indices and satisfies
$$ D_{\bar i} \overline{C}_{\bar{j}\bar{k}\bar{l}} = D_{\bar j}
\overline{C}_{\bar{i}\bar{k}\bar{l}}, $$
we can always {\it integrate} the Yukawa coupling locally as
\eqn\blobs{ \overline{C}_{\bar{i}\bar{j}\bar{k}} = e^{-2K} D_{\bar{i}}
D_{\bar{j}} \bar{\partial}_{\bar{k}} S. }
where $S$ is a local section of ${\cal L}^{-2}$.  In fact, in all the
examples we will discuss later, it is possible to construct $S$
globally on the moduli space of the topological theories. We will
present such constructions later in this section\foot{In refs.
\ref\cov{L. Castellani, R. D'Auria and S. Ferrara, Class. Quant. Grav. 7
(1990) 1767.}, a
solution to \blobs\ is constructed using particular coordinates on the
moduli space.  However $S$ constructed there does not behave nicely under
the modular transformation, and thus is not globally defined on the
moduli space.  The explicit expressions of $S$ we obtain in various
examples later differ from those obtained in these references.}.  To
simplify the expressions below, we use the following notation.
\eqn\notations{\eqalign{& S_{\bar i} \equiv \bar{\partial}_{\bar i} S \cr
 &S_{\bar i}^j \equiv \bar{\partial}_{\bar i} S^j,~~ {\rm where} ~~
S^j \equiv G^{j \bar j} S_{\bar j} \cr}}
In this notation,
\eqn\furthernotations{
 \overline{C}_{\bar i}^{jk} = \bar{\partial}_{\bar i} S^{jk}}
where
$$ \overline{C}_{\bar i}^{jk} \equiv \overline{C}_{\bar{i}\bar{j}\bar{k}}
e^{2K}
G^{j\bar j}G^{k \bar k}, ~~~ S^{jk} \equiv G^{j \bar j} S_{\bar j}^k
.$$

We now solve the genus-$2$ equation \genustwo\ by
``integration-by-parts''. We first rewrite \genustwo\ using
\furthernotations\ as
$$ \overline{\partial}_{\bar i} \left[ F_2 - {1 \over 2} S^{jk} ( D_j
\partial_k F_{1} +  \partial_j F_1  \partial_k F_1 ) \right]  =-  {1 \over
2} S^{jk} \overline{\partial}_{\bar i} ( D_j \partial_k F_{1} + \partial_j
F_1
\partial_k F_1 )
$$
The r.h.s. can be evaluated using the holomorphic anomaly of
$F_1$ and the special geometry relation for $[
\overline{\partial}_{\bar{i}} , D_j]$ as
$$\eqalign{ & - {1 \over 2} S^{jk}
\overline{\partial}_{\bar i}
( D_j \partial_k F_{1} +  \partial_j F_1 \partial_k
F_1 )= \cr &= -{1 \over 2} \overline{C}_{\bar{i}}^{mn} S^{jk} \left( {1
\over 2} C_{nmjk} + C_{mnj} \partial_k F_1 + C_{jkm} \partial_n F_1
\right) +{\chi
\over 24} S_{\bar{i}}^j \partial_j F_1 \cr} $$
Now we repeat the integration-by-parts.
$$ \eqalign{ \overline{\partial}_{\bar i} & \Bigg[ F_2 - {1 \over 2} S^{jk}
\left( D_j \partial_k F_{1} + \partial_j F_1
\partial_k F_1  \right) + \cr &~~~~~~ +{1 \over 4} S^{mn}
S^{jk} \left( {1 \over 2} C_{nmjk} +2 C_{mnj} \partial_k F_1 \right) -
{\chi \over 24} S^j \partial_j F_1 \Bigg] = \cr &={1 \over 4} S^{mn}
S^{jk} \overline{\partial}_{\bar{i}} \left( {1 \over 2} C_{nmjk} +2 C_{mnj}
\partial_k F_1 \right) -  {\chi \over 24} S^j \overline{\partial}_{\bar i}
\partial_j F_1 . \cr}$$
It turns out that r.h.s. of this equation can also be
written as a total derivative with respect to $\overline{t}^i$. By using
the genus--$1$ anomaly and the special geometry, we find
$$ \eqalign{ &{1 \over 4} S^{mn} S^{jk} \overline{\partial}_{\bar{i}}
\left( {1 \over 2} C_{nmjk} + 2 C_{mnj} \partial_k F_1
\right) -  {\chi
\over 24} S^j \overline{\partial}_{\bar i} \partial_j F_1= \cr
&=\overline{\partial}_{\bar i} \Bigg[ S^{jk}S^{pq}S^{mn} ({1 \over 8}
C_{jkp}C_{mnq} +{1 \over 12} C_{jpm}C_{kqn}) -  \cr
&~~~~~~~~~~-{\chi \over 48} S^j C_{j kl} S^{kl} +{\chi \over
24}({\chi \over 24} -1) S \Bigg] .\cr} $$
Thus the iteration stops here.  We have converted the genus-two
anomaly equation \genustwo\ into the following form.
\eqn\genustwograph{ \eqalign{  &\overline{\partial}_{\bar{i}} F_2 = \cr
&= \overline{\partial}_{\bar{i}}
\Bigg[ {1 \over 2} S^{jk} D_j \partial_k F_1 +
       {1 \over 2} S^{jk} \partial_j F_1 \partial_k F_1 - {1 \over 8}
S^{jk}S^{mn}C_{jkmn} -  \cr &~~~~~~~-{1 \over 2}
S^{jk}C_{jkm}S^{mn} \partial_n F_1 +{\chi \over 24} S^j \partial_j F_1
+ \cr &~~~~~~~ +{1 \over 8} S^{jk} C_{jkp} S^{pq} C_{qmn} S^{mn} + {1
\over 12} S^{jk}S^{pq}S^{mn}C_{jpm}C_{kqn} -\cr &~~~~~~~ - {\chi
\over 48} S^j C_{jkl} S^{kl} +{\chi \over 24}({\chi \over 24}-1) S
\Bigg] .\cr} }

Now one can easily integrate this equation as
\eqn\genustwosolution{ \eqalign{F_2 &=
 {1 \over 2} S^{ij} C^{(1)}_{ij} + {1 \over 2} C^{(1)}_i S^{ij}
C^{(1)}_j - {1 \over 8} S^{jk}S^{mn}C_{jkmn} - \cr &~~~~-{1 \over 2}
S^{ij}C_{ijm}S^{mn} C^{(1)}_n +{\chi \over 24} S^i C^{(1)}_i + \cr
&~~~~ +{1 \over 8} S^{ij} C_{ijp} S^{pq} C_{qmn} S^{mn} + {1 \over 12}
S^{ij}S^{pq}S^{mn}C_{ipm}C_{jqn} -\cr &~~~~ - {\chi \over 48} S^i
C_{ijk} S^{jk} +{\chi \over 24}({\chi \over 24}-1) S + f_2(t) \cr} }
where we used the notation $C_{i_1 \cdots i_n}^{(g)} = D_{i_1} \cdots
D_{i_n} F^{(g)}$. This equation can
be expressed graphically as in \tfig\FigureSixOne.

\ifigure\FigureSixOne{The terms obtained by solving the anomaly
equation for genus 2 have a strong resemblance to Feynman graphs
with correct symmetry factors and with an
appropriate definition of vertices.  This correspondence
can be made precise as discussed in the text.
We identify the solid lines with the massless moduli
modes and the dotted lines with the dilaton field.
These graphs fix $F_2$ up to a holomorphic
function of moduli represented by $f_2(t)$.}{Fig61}{4.3}

Here $f_2(t)$ is some meromorphic object which is
not fixed at this stage.  Since both $F_2$ and $S$ are section of
${\cal L}^{-2}$ and $C_{ijk}$ is a section of ${\cal L}^2 \times {\rm
Sym}^3 T$ on the moduli space, $f_2$ must be a meromorphic section of
${\cal L}^{-2}$.  Although we cannot determine $f_2$ from the
holomorphic anomaly alone, the holomorphicity gives rather stringent
constraints on $f_2$ and, in many cases, almost uniquely determines it.
In the case of the topological sigma--model, we can exploit the
geometric meaning of $F_2$ studied in Section 4 to fix $f_2$. In the
next section, we will demonstrate this procedure in various examples.

This method also works in the case of $g=3$.  After six iterations of
integration-by-parts, we obtain
\eqn\genusthree{\eqalign{
F_3 =& {1 \over 2} S^{ij} C^{(2)}_{ij} + C_i^{(1)} S^{ij} C_j^{(2)} +
({\chi \over 24} + 2) S^{i} C_i^{(2)} + \cr & + 2 F_2 S^i C_i^{(1)} -
{1 \over 2} S^{ij} C_{ijk} S^{kl} C^{(2)}_l - {1 \over 4} S^{ij}
S^{kl} C_{ijkl}^{(1)} - \cr & - {1 \over 2} S^{ij} C^{(1)}_{ijk}
S^{kl} C_l^{(1)} -{1 \over 4} S^{ij}S^{kl} C_{ik}^{(1)} C_{jl}^{(1)} +
\cr & + \cdots ({\rm it~would~take~five~more~pages~to~write~them~all})
\cdots + f_3(t). \cr}}
Here $f_3(t)$ is a meromorphic section of ${\cal L}^{-4}$.  Genus 3
contribution is presented in \tfig\FigureSixTwo.

\ifigure\FigureSixTwo{Some of the Feynman graphs
which emerge in solving the genus 3 anomaly equation.}{Fig62}{2.8}

One may observe that the equations \genustwosolution\ and
\genusthree\
have a strong resemblance
to the Feynman rule (see Fig. 17). Consider a finite
dimensional
quantum system with $(-S^{ij})$ as a propagator connecting the indices
$i$ and $j$, $C_{ijk}$, $C_{ijkl}$,...as classical vertices,
$C^{(1)}_i$, $C^{(1)}_{ij}$,...  as one-loop corrected vertices etc,
and compute two- and three-loop partition functions according to the
Feynman rule. If we multiply an overall factor of $(-1)$ after the
computation, we reproduce all the terms in \genustwosolution\ and
\genusthree\
including all the symmetry factors, except for those containing $S^i$,
$S$ and the holomorphic sections $f_2$ and $f_3$.

The terms involving $S^i$ and $S$ can also be recovered if we
introduce one more degree of freedom $\varphi$ and extend the Feynman
rule as follows. The propagators are given as
\eqn\propagators{ K^{ij} = -S^{ij},~~~K^{i\varphi} =
-S^i,~~~K^{\varphi,\varphi} =-2S } and the vertices are given by
\eqn\vertices{ \eqalign{& \widetilde{C}^{(g)}_{i_1 \cdots i_n, \varphi^{m+1}}
    = (2g-2+n+m) \widetilde{C}^{(g)}_{i_1 \cdots i_n , \varphi^{m}}
\cr &\widetilde{C}^{(g)}_{i_1 \cdots i_n} = C^{(g)}_{i_1 \cdots i_n}
,~~~~~~~~~ \widetilde{C}^{(1)}_{\varphi} = {\chi \over 24} -1 . \cr &
\widetilde{C}^{(0)}_{\varphi^m}=0,~~ \widetilde{C}^{(0)}_{i ,
\varphi^m}=0,~~
\widetilde{C}^{(0)}_{ij , \varphi^m}=0,~~ \widetilde{C}^{(1)}=0 .\cr}
 }
Compute two- and three-loop partition functions using this Feynman rule
and multiply the overall factor of $(-1)$ after the computation. By
adding the meromorphic sections $f_2$ and $f_3$, we recover the
expressions \genustwosolution\ and \genusthree .  The definition
\vertices\ of the vertices reminds us of the puncture equation in the
topological gravity.  In fact, we will now identify the variable
$\varphi$ with the dilaton which is the first topological descendant
of the puncture operator \wittop\ $\sigma_1(P)$.
All the other topological descendants decouple from the correlation
functions simply by the $U(1)$ charge conservation
and thus {\it the only non-vanishing
correlation functions involve those of marginal fields and the dilaton field}.
So far we have only discussed the marginal fields.  To properly
discuss the dilaton field coupling we need to enlarge the field
space from that of pure topological theory. However luckily the
correlation for the dilaton field can quite generally be eliminated
from correlation functions by the recursion relations.  In fact
the first equation in \vertices\ is precisely the general
recursion relation of \wittop\ and so $\varphi$ is indeed the
dilaton field.

\vskip .15in

\subsec{Feynman rules for arbitrary $g$}

The emergence of the Feynman rule is rather mysterious from the way we
discovered it at $g=2$ and $3$. It would be extremely difficult to
prove the Feynman rule for $g \geq 4$ by using the method in the above
since the number of iterations would grow exponentially.  Thus we will
develop another technique which enables us to prove the Feynman rule
directly for all $g$. We will do so by reducing the Feynman rule for
$F_g$ to the Schwinger--Dyson equation of the finite dimensional
system.  In Section 3, we introduced the generating function
$W(\lambda, x ;t,\overline{t})$ for $C^{(g)}_{i_1 \cdots i_n}$ which
satisfies
$$ \eqalign{& {\partial \over \partial \overline{t}^i} \exp(W) =\cr &=
\left[ {\lambda^2 \over 2} \overline{C}_{\bar i}^{jk} {\partial^2 \over
\partial x^j \partial x^k} - G_{\bar{i} j} x^j \big(\lambda {\partial
\over \partial \lambda} + x^k {\partial \over \partial x^k} \big) \right]
\exp(W) .\cr} $$
To prove the Feynman rule, it is more useful to consider a generating
function $\widetilde{W}(\lambda,x,\varphi ,t,\overline{t})$ for the
vertices $\widetilde{C}^{(g)}_{i_1 \cdots i_n ,\varphi^m}$ of the
Feynman rule.
$$ \widetilde{W}(\lambda,x,\varphi;t,\overline{t}) = \sum_{g=0}^{\infty}
\sum_{n,m=0}^{\infty} {1 \over n! m!}\lambda^{g-1}
\widetilde{C}^{(g)}_{i_1 \cdots i_n; \varphi^m} x^{i_1} \cdots x^{i_n}
\varphi^m $$
By the definition of the vertices \vertices , $\widetilde{W}$ is
related to $W$ as
$$ \eqalign{& \widetilde{W}(\lambda,x,\varphi;t,\overline{t}) = \cr &=
\sum_{g=0}^{\infty} \sum_{n}^{\infty} {1 \over n!} \lambda^{2g-2}
C^{(g)}_{i_1 \cdots i_n} x^{i_1} \cdots x^{i_n} \left( {1 \over 1 -
\varphi} \right)^{2g-2+n} +({\chi \over 24} -1) \log \left( {1 \over 1
- \varphi} \right) \cr & = W\left( {\lambda \over 1-\varphi} , {x
\over 1 - \varphi} ; t , \overline{t}\right) - \left( {\chi \over 24} -1
\right) \log \lambda.
\cr} $$
Thus $\widetilde{W}$ satisfies the equation
\eqn\tildedelbar{
{\partial \over \partial \overline{t}^i} \exp(\widetilde{W}) = \left[
{\lambda^2 \over 2} \overline{C}_{\bar i}^{jk} {\partial^2 \over \partial
x^j \partial x^k} - G_{\bar{i} j} x^j {\partial \over \partial
\varphi} \right] \exp(\widetilde{W}) .}

It turns out that there is another function of $x^i$ and $\varphi$
which satisfies almost the same equation as \tildedelbar . It is given
as follows.
\eqn\kinetic{
   Y(\lambda, x,\varphi;t,\overline{t}) = - {1 \over 2 \lambda^2} \left(
\Delta_{ij} x^i x^j + 2 \Delta_{i \varphi} x^i \varphi +
\Delta_{\varphi\varphi} \varphi^2 \right) + {1 \over 2} \log( {\det
\Delta \over \lambda^2} ) }
Here $\Delta$ is an inverse of the propagator $K$ defined by
\propagators , i.e.
\eqn\inverse{
\eqalign{ & S^{ij} \Delta_{jk}  + S^i \Delta_{k\varphi}  =
                      - \delta^i_k \cr & S^{ij} \Delta_{j\varphi} +
S^i \Delta_{\varphi \varphi} = 0 \cr & S^i\Delta_{ij} + 2S \Delta_{j
\varphi} = 0 \cr & S^i \Delta_{i\varphi} + 2S \Delta_{\varphi \varphi}
= -1. \cr}}
Thus $Y$ may be regarded as a kinetic term for the finite dimensional
system of $x^i$ and $\varphi$.
The most important properties of these inverse propagators are
$$\eqalign{& \overline{\partial}_{\bar i} \Delta_{jk} = \overline{C}_{\bar
i}^{mn} \Delta_{mj} \Delta_{nk} + G_{\bar{i} j} \Delta_{k \varphi} +
G_{\bar{i}j} \Delta_{j \varphi} \cr & \overline{\partial}_{\bar i}
\Delta_{j \varphi} = \overline{C}_{\bar i}^{mn} \Delta_{mj} \Delta_{n} +
G_{\bar{i} j} \Delta_{\varphi \varphi} \cr & \overline{\partial}_{\bar i}
\Delta_{\varphi \varphi} = \overline{C}_{\bar i}^{mn} \Delta_{m \varphi}
\Delta_{n \varphi} \cr}$$
which we can derive from \notations\ and \furthernotations .  Just as
the anomaly equations for $C^{(g)}_{i_1 \cdots i_n}$ are encoded in
\tildedelbar , the above equations for $\Delta$'s can be written as a
differential equation for $Y$.
\eqn\kineticequation{
{\partial \over \partial \overline{t}^i} \exp(Y) = \left[- {\lambda^2 \over
2} \overline{C}_{\bar i}^{jk} {\partial^2 \over \partial x^j \partial x^k}
- G_{\bar{i} j} x^j {\partial \over \partial \varphi} \right] \exp (Y)
}

Now we consider the following integral.
\eqn\formalintegral{
    Z = \int dx d \varphi \exp(Y+\widetilde{W}) }
Although this integral itself may be divergent, we can compute its
perturbative expansion with respect to $\lambda$.  The integral $Z$
may be regarded as a partition function of a finite dimensional
quantum system with dynamical variables $x^i$ and $\varphi$, and the
perturbative expansion of $Z$ can be evaluated using the standard
technique of the Feynman rule as
\eqn\expansion{
 \eqalign{ \log Z &= \lambda^2 \left[ F_2 - {1 \over 2} S^{ij} C_{ij}^{(1)}
-{1 \over 2} C_i^{(1)} S^{ij} C^{(1)}_j + \cdots \right] + \cr & +
\lambda^4 \left[ F_3 - {1 \over 2} S^{ij} C_{ij}^{(2)} - C_i^{(1)}
S^{ij} C_j^{(2)} + \cdots \right] + \cr & + \lambda^6 \left[ F_4 - {1
\over 2} S^{ij} C_{ij}^{(3)} - C_{i}^{(1)} S^{ij} C_{j}^{(3)} - {1
\over 2} C_{i}^{(2)} S^{ij} C_j^{(2)} + \cdots \right] + \cr & +
\cdots + \lambda^{2g-2} \left[ F_g -{1 \over 2} S^{ij} C_{ij}^{(g-1)}
- {1 \over 2} \sum_{r=1}^{g-1} C_{i}^{(r)} S^{ij} C_{j}^{(g-r)} +
\cdots \right] + \cdots \cr}}
where $(\cdots)$ in the coefficient of $\lambda^2$ represents the
terms in the r.h.s. of \genustwosolution , $(\cdots)$ in the
coefficient of $\lambda^4$ represents those in \genusthree , and so
on.

Previously we found, by the iterative method, that the coefficients of
$\lambda^2$ and $\lambda^4$ in the perturbative expansion of $Z$ are
holomorphic in $t$. We can now prove the holomorphicity of $Z$ to all
order in the perturbation as the Schwinger--Dyson equation of the
finite dimensional system. By using
\tildedelbar\ and \kineticequation, we obtain
$$ \eqalign{ {\partial \over \partial \overline{t}^i} Z = & \int dx
d\varphi ~e^{Y} \left[ {\lambda^2 \over 2} \overline{C}_{\bar i}^{jk}
{\partial^2 \over \partial x^j \partial x^k} - G_{\bar{i} j} x^j
{\partial \over \partial \varphi} \right] e^{\widetilde{W}} + \cr &+
\int dx d\varphi ~ e^{\widetilde{W}} \left[ -{\lambda^2 \over 2}
\overline{C}_{\bar i}^{jk} {\partial^2 \over \partial x^j \partial x^k} -
G_{\bar{i} j} x^j {\partial \over \partial \varphi} \right] e^{Y} \cr
=& {\lambda^2 \over 2} \overline{C}_{\bar i}^{jk} \int dx d \varphi \left[
{\partial \over \partial x^j} \left( e^{Y} {\partial \over \partial
x^k} (e^{\widetilde{W}}) \right) -{\partial \over \partial x^j} \left(
e^{\widetilde{W}} {\partial \over \partial x^k} (e^{Y}) \right)
\right] - \cr & ~~~- G_{\bar{i} j } \int dx d\varphi ~ {\partial \over
\partial \varphi}\left[ x^j e^{Y + \widetilde{W}} \right] .\cr} $$
The point is that the integrand in the r.h.s. of this
equation is total derivative with respect to $x^i$ and $\varphi$.  In
the perturbative expansion, we are free to perform the
integration-by-part and drop boundary terms since integrals involved
in the perturbation are all Gaussian. Thus we have derived
$$ {\partial \over \partial \overline{t}^i} Z = 0 $$
As is evident from the expansion \expansion , the holomorphicity of
$Z$ means that we can express $F_g$ as a meromorphic section $f_g$ of
${\cal L}^{2-2g}$ minus a sum over the Feynman graphs constructed from
the propagators
\propagators\ and the vertices \vertices .

\vskip .15in

\subsec{Construction of propagators}

So far we have assumed that there is a global section $S$ of ${\cal
L}^{-2}$ which satisfies \blobs . Now we are going to construct such
an object.  The important ingredient is again the special geometry
relation
\eqn\specialrelation{
R_{i \bar{j}l}^{~~~k} = - \overline{\partial}_{\bar j} \Gamma_{il}^k =
G_{i \bar j} \delta^k_l + G_{k \bar j} \delta^k_i - C_{ilm}
\overline{C}_{\bar j}^{km}. }
Since $G_{i \bar j} =
\partial_i \overline{\partial}_{\bar j} K$, this can be rewritten as
$$\overline{\partial}_{\bar i} [ S^{jk}  C_{klm}]
= \overline{\partial}_{\bar i}
\left[ \partial_l K \delta_m^j + \partial_m K \delta_l^j +
\Gamma_{lm}^j \right].$$
This can be easily integrated as
\eqn\onceint{ S^{ij} C_{jkl} =
 \delta_l^i \partial_k K + \delta_k^i \partial_l K + \Gamma_{kl}^i +
f_{kl}^i }
where $f_{kl}^i$ is some meromorphic object which should compensate
for the non-covariance of $\partial_k K$ and $\Gamma_{kl}^i$ in the
r.h.s. side. We can express $f_{kl}^i$ as
$$ f_{kl}^i = \delta_l^i \partial_k \log f + \delta_k^i \partial_l
\log f - \sum_{a=1}^n v_{l,a} \partial_k v^{i,a} + \widetilde{f}_{kl}^i ,$$
where $f$ is a meromorphic section of ${\cal L}$, $\{ v^{i,a}
\}_{a = 1 ,..., n}$ ($n$ is the dimensions of the moduli space)
are meromorphic
tangent vectors which are linearly independent almost everywhere on
the moduli space, $v_{i,a}$ are inverse of $v^{i,a}$ ($\sum_a v_{i,a}v^{j,a}
= \delta_j^i$) and $\widetilde{f}_{kl}^i$ is a meromorphic section of
$T \times {\rm Sym}^2 T^*$.
In general, \onceint\ has ${1 \over 2} n^2(n+1)$ equations
for ${1 \over 2} n(n+1)$ variables $S^{ij}$ and it is over-determined
when $n > 1$. Thus we should
make an appropriate choice of $\widetilde{f}^i_{kl}$ to ensure
that \onceint\ is solvable with respect to $S^{ij}$.

The situation is much simpler in the one-modulus case since there
is only one equation in \onceint\ and there is no constraint on
$\widetilde{f}_{11}^1$.
In order to construct $F_g$
by using the Feynman rule,
\genustwosolution\ and \genusthree\ for example,
we do not need the most general solution to $\overline{\partial} S^{11} =
\overline{C}_{\bar 1}^{11}$ since any holomorphic ambiguity in $S^{11}$ is
absorbed into the holomorphic section $f_g$ which we add to $F_g$ at
the end of the computation. Thus we can, for example, choose
$\widetilde{f}_{11}^1 = 0$.  With this choice, $S^{11}$ becomes
\eqn\onemoduluspropagator{
   S^{11} = {1 \over C_{111}} \left[ 2 \partial \log (e^{K} |f|^2) -
(G_{1 \bar{1}} v)^{-1} \partial (v G_{1 \bar{1}}) \right] }

To find $S^i$, we need to integrate
$$ \overline{\partial}_{\bar i} S^j = G_{i \bar k} S^{jk} .$$
Substituting \onemoduluspropagator\ into this,
we obtain
$$ \eqalign{ \overline{\partial} S^1 = & {1 \over C_{111}} \left[ 2
\partial \log (e^{K} |f|^2 )G_{1 \bar 1} - v^{-1} \partial (v
G_{1 \bar{1}}) \right] \cr = & {1 \over C_{111}} \overline{\partial} \left[
( \partial \log (e^{K} |f|^2) )^2 - v^{-1} \partial (v
\partial K) \right] .\cr} $$
A special solution to this equation can be easily found as
\eqn\onemodulustadpole{ S^1 = {1 \over C_{111}}
 \left[ \left( \partial \log (e^{K} |f|^2) \right)^2 - v^{-1}
\partial \left(v \partial \log (e^K |f|^2) \right) \right] }

Finally we need to find $S$ which satisfies
\eqn\blobequation{ \overline{\partial} S = G_{1 \bar 1} S^{\bar 1}. }
A special solution to this equation is given by
\eqn\onemodulusblob{
  \eqalign{ S = &\left[ S^1  - {1 \over 2}
D_1 S^{11}  - {1 \over 2}
(S^{11})^2 C_{111} \right] \partial \log (e^K |f|^2) + \cr
 &+ {1 \over 2} D_1 S^1
 + {1 \over 2} S^{11} S^1 C_{111}. \cr   }}
Let us check that this indeed satisfies \blobequation .
We first note that the
following special combination of $S^1$ and $S^{11}$ is holomorphic
$$  \eqalign{ & \overline{\partial} \left[ S^1 - {1 \over 2} D_1 S^{11}
 - {1 \over 2} (S^{11})^2
C_{111} \right] = \cr & = G_{1 \bar 1} S^{11} -
{1 \over 2} [ \overline{\partial} , D_1 ] S^{11} - {1 \over 2}
(G^{1 \bar 1})^2 \partial \overline{C}_{\bar{1}\bar{1}\bar{1}}
        -  \overline{C}_{\bar 1}^{11} S^{11} C_{111} \cr
 &=0 ,\cr}
$$
where we used the special geometry relation\foot{
$D_1 S^{11}
 = (\partial - 2\Gamma_1^{11} -2\partial K) S^{11}$.
Therefore
$[ \overline{\partial} , D_1] S^{11} = 2G_{1 \bar 1} S^{11}
-2C_{111}\overline{C}_{\bar 1}^{11}
S^{11}$.} \specialrelation , the definitions of $S^{1}$ and $S^{11}$
and $\partial \overline{C}_{\bar{1}\bar{1}\bar{1}} = 0$.
Now it is straightforward to check the equation \blobequation\ as
$$ \eqalign{
 \overline{\partial} S =  & G_{1 \bar 1} S^1
        - {1 \over 2}G_{1 \bar 1} D_1 S^{11}
     -{1 \over 2} G_{1 \bar 1} (S^{11})^2 C_{111}+ \cr&
 +{1 \over 2} [ \overline{\partial} , D_1 ] S^1 +
    {1 \over 2} D_1 S_{\bar 1}^1 +
    {1 \over 2} \overline{C}_{\bar 1}^{11} S^1 C_{111}
     + {1 \over 2} S^{11} S_{\bar 1}^1 C_{111}
    \cr
 =&   G_{1 \bar 1} S^1 +   {1 \over 2} [ \overline{\partial} , D_1 ] S^1 +
     {1 \over 2} \overline{C}_{\bar 1}^{11} S^1 C_{111}
\cr
=& G_{1 \bar 1} S^1.   \cr}$$
Here we once again used the special geometry relation\foot{
$D_1 S^1 = (\partial - \Gamma_1^{11} -2 \partial K) S^1$.
Therefore
$[\overline{\partial} , D_1] S^1 =
   -  C_{111}\overline{C}_{\bar 1}^{11} S^1$.}.

To summarize, in the one-modulus case, the propagators
$S^{11}$ , $S^1$, and $S$ are given as
$$ \eqalign{
   S^{11} & = {1 \over C_{111}} \partial \log \left[
 2 \partial \log (e^K |f|^2) -
 (G_{1 \bar 1} v)^{-1} \partial ( v G_{1 \bar 1} ) \right] \cr
   S^1 & = {1 \over C_{111}}
 \left[ \left( \partial \log (e^{K} |f|^2) \right)^2 - v^{-1}
\partial \left(v \partial \log (e^K |f|^2) \right) \right] \cr
 S&= \left[ S^1  - {1 \over 2}
D_1 S^{11}  - {1 \over 2}
(S^{11})^2 C_{111} \right] \partial \log (e^K |f|^2) + \cr
 &~~~~+ {1 \over 2} D_1 S^1
 + {1 \over 2} S^{11} S^1 C_{111}. \cr} $$

In the multi-moduli case, \onceint\ gives
\eqn\multipropagator{
  S^{ij} C_{jkl} =
    (\delta_l^i \partial_k + \delta_k^i \partial_l)
    \log (e^{K}|f|^2)
     - \sum_{a=1}^n v_{l,a} G^{i \bar{i}} \partial_k ( v^{m,a}
                           G_{m \bar{i}} )
           +\widetilde{f}_{kl}^i .}
In order to obtain an expression for $S^{ij}$ from this equation,
we need to ``invert'' the Yukawa coupling. Although we do not
know if it is possible to do so in general,  it is certainly possible
for the $A$-model discussed in section 4.
In this model, each chiral field corresponds to a K\"ahler form
in the target space and, in the large volume limit,
the Yukawa coupling $C_{ijk}$ is given
as an intersection of the three K\"ahler forms.
There is a distinguished K\"ahler modulus $t^1$ in this model corresponding
to an overall scaling of the target space metric. In the large
volume limit $t^1 \rightarrow \infty$,
the Yukawa coupling $C_{ij 1}$ then gives the inner product
of the two K\"ahler forms $k_i$ and $k_j$, and it
is non-degenerate as an $n \times n$ matrix, $\det ( C_{ij1}
)_{i,j=1,...,n} \neq 0$.
Since $\det ( C_{ij1} )$ is holomorphic in $t$, this means that
$\det( C_{ij1})$ should be non-zero almost everywhere
on the moduli space. Therefore we can invert $C_{ij1}$
in \multipropagator\ to find an expression for $S^{ij}$, provided
we made an appropriate choice of $\widetilde{f}_{kl}^i$.

As in the one-modulus case, we substitute \multipropagator\
into $\overline{\partial}_{\bar i} S^j = G_{\bar{i}i} S^{ij}$ to obtain
$$\eqalign{
 \overline{\partial}_{\bar i} [S^{j}]C_{jkl} =&
       G_{\bar{i} l} \partial_k \log (e^{K} |f|^2)
        + G_{\bar{i} k} \partial_l \log (e^{K} |f|^2) + \cr
        &~~~     -\sum_{a=1}^n
            v_{l,a} \partial_k( v^{m,a} G_{\bar{i}m})
   +\widetilde{f}^{i}_{kl} G_{\bar{i}i} .\cr}$$
This can be easily integrated as
\eqn\multitadpole{ \eqalign{
   S^i C_{ijk} = &
    \partial_j \log (e^{K} |f|^2)  \partial_k \log (e^{K} |f|^2)
       - \sum_{a=1}^n v_{k,a} \partial_j \left[
            v^{l,a} \partial_l \log (e^{K}|f|^2) \right] + \cr
  &~~~+ \widetilde{f}_{jk}^l \partial_l \log( e^{K} |f|^2)
         + \widetilde{f}_{jk} .\cr}}
Here $\widetilde{f}_{jk}$ is a meromorphic section of
${\rm Sym}^2 T^*$. As in the case of $S^{ij}$ in \multipropagator,
with an appropriate choice of $\widetilde{f}_{jk}$,
we can invert the Yukawa coupling in the above and obtain
an expression for $S^i$.

To complete the Feynman rule, we need $S$ which satisfies
\eqn\multiblobequation{
\overline{\partial}_{\bar i} S = G_{\bar{i} i} S^i .}
A special solution to this equation is given by
\eqn\multiblob{\eqalign{
  S = & {1 \over 2n} \left[
           (n+1) S^i - D_j S^{ij} - S^{ij} S^{kl} C_{jkl} \right]
           \partial_i \log (e^K |f|^2) + \cr
  &~~+{1 \over 2n}
       \left( D_i S^i + S^i S^{jk} C_{ijk} \right) . \cr} }
Let us check that this satisfies \multiblobequation .
As in the case of one-modulus, the following combination of
$S^i$ and $S^{ij}$ is holomorphic due to
the special geometry relation and
$\overline{\partial}_{\bar i} C_{jkl} = 0$.
$$\eqalign{ & \overline{\partial}_{\bar i}
  \left[ (n+1) S^j - D_k S^{jk} - S^{jk} S^{mn} C_{jmn} \right] = \cr
& = (n+1) G_{\bar{i} k} S^{jk} - [ \overline{\partial}_{\bar i} , D_k ]
           S^{jk} - G^{j \bar j} G^{k \bar k} \partial_k
                 \overline{C}_{\bar{i}\bar{j}\bar{k}} - \cr
&~~~~~~- \overline{C}_{\bar{i}}^{jk} S^{mn} C_{kmn} -
         \overline{C}_{\bar{i}}^{mn} S^{jk} C_{kmn} \cr
&=0 .\cr}$$
We can then compute $\overline{\partial}_{\bar i} S$ as
$$ \eqalign{
 \overline{\partial}_{\bar i} S
    =&{n+1 \over 2n} S_{\bar i} +\cr
 &+ {1 \over 2n}
         \left(
  [ \overline{\partial}_{\bar i} , D_j ] S^j
   + D_j S_{\bar i}^j
   + S_{\bar i}^j S^{kl} C_{jkl} + S^j \overline{C}_{\bar i}^{kl} C_{jkl}
      - S_{\bar i}^jS^{kl} C_{jkl}  \right) \cr
=& G_{\bar{i} j} S^j .\cr} $$
Here again, we used the special geometry relation for
$[ \overline{\partial}_{\bar i} , D_j]$.

Thus we have prepared all the ingredients we need for the Feynman rule of
$F_g$. In the next section, we will construct $F_g$ explicitly
in several examples.

\vfill
\eject

\newsec{Examples --- the experimental evidence}
In this section we show how to compute higher loop partition function
$F_g$ (for small $g$) for some examples.  We will elaborate in details
how the perturbation theory developed in previous section works.
The simplest and most trivial example would be a three dimensional
complex torus. In this case there is nothing to compute. All
loop partitions functions are identically equal to zero due
to fermion zero modes. The simplest way to get a non zero answer is to
orbifoldize the model. Below we will consider two examples ---
the ${\bf Z}_3 \otimes {\bf Z}_3$ orbifold model and the quintic ---
in detail,
and comment on some other models also at the end.

\vskip .15in

\subsec{Orbifold}
Let us start with some definitions. ${\bf Z}_3 \otimes {\bf Z}_3$ orbifold is
obtained by
dividing $ T^2 \times T^2 \times T^2$, with each torus having a ${\bf Z}_3$
symmetry,
by the discrete group
generated by
$g={\rm diag}\,(1,\omega,\omega^2)$ and $h={\rm diag}\,(\omega,\omega^2,1)$.
This model has $3$ untwisted K\"ahler moduli corresponding
to the moduli of each of the tori and $81$ corresponding
to the blow up modes.  This orbifold is rigid and has no complex moduli. The
Euler
characteristic $\chi=168$.  We will denote the K\"ahler
moduli of each of the three tori by $\tau_a$ ($a=1,2,3$).
The K\"ahler potential is given as follows
$$e^{-K(\tau_i, \bar \tau_i)}=
 i \prod_{i=1} ^3 (\tau_i - \bar \tau_i).$$
The only non zero component of Yukawa coupling is $C_{123}=1$. Zamolodchikov's
metric is diagonal and is equal to
$$G_{a \bar b}=-{\delta_{ab} \over (\tau_a -\bar \tau_a)^2}$$
We also need the genus one partition function, which is equal to
$$F_1=-\kappa \sum_a log (\tau_a -\bar \tau_a) |\eta^2 (\tau_a)|^2~,$$
where $\kappa=4$ for this orbifold.
In spite of the fact that it is easy to solve the equations for $F_2$ and $F_3$
directly, we first review the ingredients of perturbation technique.
In case of orbifold the equations for different components of propagator
$S^{ab},~S^a$  and $S$ are very simple
\eqn\orbprop{\bar \partial_c S^{ab}=-{ 1 \over (\tau_c - \bar \tau_c)^2} ~~,~~~
\bar \partial_b S^a =-{S^{ab} \over (\tau_b -\bar \tau_b)^2} ~~,~~~
\bar \partial_a S=-{S^a  \over (\tau_a -\bar \tau_a)^2 } ~, }
where $(abc)$ is a permutation of $(123)$ and $S^{ab}=0$ for $a=b$.
Integrating this equations we obtain
\eqn\orbsol{\eqalign{S^{ab}=&
-\left( {1 \over (\tau_c - \bar \tau_c)} + 2{\eta'(\tau_c) \over \eta(\tau_c)}
\right)\cr
S^{a}=&
\left( {1 \over (\tau_b - \bar \tau_b)} + 2{\eta'(\tau_b) \over \eta(\tau_b)}
\right)
\left( {1 \over (\tau_c - \bar \tau_c)} + 2{\eta'(\tau_c) \over \eta(\tau_c)}
\right)\cr
S=& -\prod_a
\left( {1 \over (\tau_a - \bar \tau_a)} + 2{\eta'(\tau_a) \over \eta(\tau_a)}
\right) ~,  \cr}}
where $(abc)$ is a permutation of $(123)$. At every integration step
the holomorphic piece was fixed by modular invariance.
For example, integrating the equation for $S^{ab}$ we obtain
$S^{ab}=-1/(\tau_c -\bar \tau_c)^2 + f(\tau_c)$. The untwisted
moduli space of K\"ahler
structures for this orbifold  is the product of three copies of fundamental
domain in the upper half plane modulo symmetry group.
The condition of modular invariance fixes $f(\tau)=2\eta'(\tau)/\eta(\tau)$.
Similar arguments lead to the answers \orbsol.
It is easy to verify that all diagrams give rise to the same type of
contribution and therefore $F_2$ is proportional to $S$ (this is in fact a
peculiarity of orbifold example).
One can also solve the equation for genus two directly. In this case the
anomaly
equation reads
$$\bar \partial_a F_2=-{1 \over 2 }{1 \over (\tau_a - \bar \tau_a)^2}
\partial_b
F_1 \partial_c F_1$$
Taking into account the explicit form of $F_1$, one can easily integrate the
above equation\foot{If we considered the non-abelian orbifold obtained by the
above one modded out by a further symmetry which permutes the three
tori, we would have ended up with one untwisted modulus
and then $F_g$ would be a modular function of weight $6g-6$
with respect to this modulus.  In this case even at genus 2 we
would have to fix the coefficient of holomorphic contribution
to $F_2$ as there is a modular form of weight 6.}
$$F_2={1 \over 2 \kappa}\prod_a \partial_a F_1={\kappa^2 \over 2} \prod_a
\left( {1 \over (\tau_a - \bar \tau_a)} + 2{\eta'(\tau_a) \over \eta(\tau_a)}
\right)$$

The equation for genus $3$ is given as follows
\eqn\orbgt{\eqalign{\bar \partial_a F_3 = &{1 \over 2}
    {1 \over (\tau_a - \bar \tau_a)^2}
\lbrack  (\partial_b + {2 \over \tau_b - \bar \tau_b})
           (\partial_c + {2 \over \tau_c - \bar \tau_c}) F_2 +\cr
  \partial_b F_1
          &    (\partial_c + {2 \over \tau_c - \bar \tau_c}) F_2 +
    \partial_c F_1
          (\partial_b + {2 \over \tau_b - \bar \tau_b}) F_2
                \rbrack ~\cr}}
As usual $(abc)$ is a permutation of $(123)$.
After substitution the genus-two solution obtained above the  equation  \orbgt\
becomes
$$\eqalign{\bar \partial_a F_3 = &{1 \over 4 \kappa}
    {1 \over (\tau_a - \bar \tau_a)^2} \partial_a F_1
\lbrack  (\partial_b + {2 \over \tau_b - \bar \tau_b}) \partial_bF_1
           (\partial_c + {2 \over \tau_c - \bar \tau_c}) \partial_c F_1 +\cr
  (\partial_b F_1)^2
          &    (\partial_c + {2 \over \tau_c - \bar \tau_c}) \partial_c F_1+
   (\partial_c F_1)^2
              (\partial_b + {2 \over \tau_b - \bar \tau_b}) \partial_b F_1
              \rbrack     \cr}
 $$
The expression in brackets does not depend on $\tau_a$ and therefore we only
need to solve the equation
\eqn\s{\bar \partial_a f = {1 \over 4 \kappa}
    {1 \over (\tau_a - \bar \tau_a)^2} \partial_a F_1}
We must be careful at this point since the solution is not unique. The reason
for this is the existence of modular form of weight four
$\eta''(\tau)/\eta(\tau)-3(\eta'(\tau)/\eta(\tau))^2$.
The general solution for \s\ is given as follows
$$f=\sum_{a=1}^3 x(\partial_a F_1)^2 + ({1 \over 8\kappa}-\kappa x)
             (\partial_a + {2 \over \tau_a -\bar \tau_a}) \partial_a F_1~,$$
where $x$ is an arbitrary parameter.
The condition of permutation symmetry forces the coefficients in front of
the two
terms to be equal to each other. As a result we obtain a fully symmetric
solution for
genus-three partition function
$$\eqalign{F_3=&{1 \over 8\kappa (\kappa +1)}\sum_{a=1}^3 (\partial_a +
                     {2 \over \tau_a -\bar \tau_a}) \partial_a F_1
  (\partial_{a+1} +{2 \over \tau_{a+1} -\bar \tau_{a+1}}) \partial_{a+1} F_1
       (\partial_{a+2} F_1)^2  + \cr
    & + C_0 \prod_{a=1} ^3  \left(
      {\eta''(\tau_a) \over \eta(\tau)} -3
     \left(  {\eta'(\tau_a) \over \eta(\tau)} \right)^2 \right)  ~,   \cr}$$
where $C_0$ is an arbitrary constant and it can not be determined from the
anomaly equation.

This phenomenon persists at every genus whenever there is a modular form of
appropriate weight. Unfortunately  we do not know the asymptotic behavior of
$F_g$ for the orbifold to fix the ambiguity.  An analysis along the line
of section 5 done for the orbifolds would be needed to fix this ambiguity.

\vskip .15in

\subsec{Quintic}

Quintic hypersurface can be described as the vanishing locus of a homogeneous
polynomial of degree $5$ of five variables $W(x_i)=0$
which determines the embedding of complex $3$--fold in ${\bf P}^4$.
This Calabi Yau $3$--fold has $101$ complex moduli,
all these moduli can be thought as the coefficients of the polynomial,
and $1$ K\"ahler moduli, which can be thought as the K\"ahler
class of ${\bf P}^4$.

To construct the mirror manifold one starts with a $1$-parameter subfamily
of quintic hypersurfaces given by
$$W(x_i)\equiv\sum x_i^5-5 \psi x_0 x_1 x_2 x_3 x_4 =0~.$$
All these $M_{\psi}$ hypersurfaces are invariant under the discrete group
${\bf Z}_5 ^3$. Following construction of \ref\grpl{B. R. Plesser and M. R.
Greene, Nucl. Phys. B338 (1990) 15} one may obtain the mirror family
$W_{\psi}$ by dividing out the
discrete symmetries.

The mirror manifold $W$ has only $1$ complex moduli and $101$ Kahler moduli.
One can describe the mirror family $W_{\psi}$ using the same complex parameter
$\psi$ as for $M_{\psi}$
(for construction of mirror map see \cand). The variations of $\psi$ can be
identified with
deformations of complex structure of the mirror $W$.
The multiplication of $\psi$ by a fifth root of unity $\alpha=e^{2 i \pi/5}$
can
be always undone
by appropriate change of variables $x_i$  and therefore
$\psi \rightarrow \alpha \psi$ is a modular transformation. All physical
observables are invariant under $\psi \rightarrow \alpha \psi$.
The modular parameter $\psi$  describes a degenerate Calabi--Yau $3$--fold
only for $\psi=1$  and $\psi=\infty$.
For  $\psi=1$ the corresponding Calabi--Yau is conifold, while
for $\psi=\infty$ the corresponding Calabi--Yau manifold is a singular quintic.
In spite of the fact that $\psi$ and $\alpha \psi$ correspond to the same
complex structure the point $\psi=0$ is a regular point corresponding
to one of Gepner's model.

The purpose of this section is to present the computations of numbers of
holomorphic curves of low genus in the quintic hypersurface.
To be more precise there is no holomorphic isolated curves for genus
bigger then one. The numbers we compute are in fact the Euler characteristics
of the corresponding families as discussed in section 5.10.
We will follow the following logic in this section. We first compute
the elements of the diagram technique for fixed $\psi$ but in the limit
$\bar \psi \rightarrow \infty$. We discuss the holomorphic ambiguity
by requiring regularity $F_g (\psi)$ everywhere except $\psi=1$ and
$\psi=\infty$. Then by making a mirror transform and expanding in instantons
we extract the numbers in question.

The holomorphic three form $\Omega$ is taken in the gauge
$$\Omega=5 \psi {x_4 dx_0 dx_1 dx_2 \over \partial W/ \partial x_3}$$
In the same gauge the Yukawa coupling is equal to
$$C_{\psi \psi \psi}=-\int \Omega \wedge
{\partial^3 \Omega \over \partial \psi^3}=
\left( {2 \pi i \over 5} \right)^3  { 5 \psi^2  \over 1-\psi^5}.$$

Different components of the propagator are expressed in terms of
the K\"ahler
potential, Zamolodchikov's metric and two sections $f \in {\cal L}$ and
$v \in T^*$ (see formulas \onemoduluspropagator, \onemodulustadpole\ and
\onemodulusblob).
The K\"ahler potential always enters into invariant combination
$e^{K}|f|^2$, while metric enters in the invariant combination
$G_{\psi \bar \psi}|v|^2$. As $\psi$ goes to $0$, $e^{K}$ diverges as
$|\psi|^{-2}$, while the
metric remains finite. The condition of regularity at the
origin implies that
$f$ should necessarily have a zero at $\psi=0$, while $v$ remains finite.
The regularity condition at the origin and the absence of any additional
singularities except possibly at $\psi=\infty$ and $\psi=1$ implies the
following ansatz for $f$ and $v$
$$f(\psi)=\psi (1 - \psi^5)^a ~~,~~~v(\psi)=(1-\psi^5)^b~,$$
where $a$ and $b$ are some constants.
The precise choice of these sections is irrelevant since any holomorphic
ambiguity can be reabsorbed into the section $f_2$ which we add to
the final answer $F_2$.
{}From the general formulas
\onemoduluspropagator, \onemodulustadpole\ and
\onemodulusblob\ one can immediately deduce that for small $\psi$
$$S^{\psi \psi}\left( {\partial \over \partial \psi} \right)^2 \sim
\psi^2 \left( {\partial \over \partial \psi} \right)^2~,~~~
S^{\psi} {\partial \over \partial \psi}\sim \psi
{\partial \over \partial \psi}~~{\rm and}~~~~S \sim const~ .$$

The behavior of the
perturbation series near singularity $\psi=1$ follows from the
general arguments presented in the Section 5 (see
equation \estimate ).  In order to apply the formula
\estimate\ we need to find the canonical coordinate near
$\psi =1$ which is an interesting example of how
different a canonical coordinate can be from the special coordinate.
In fact the canonical coordinate at this point is just
$$t\sim -{\rm log}(1-\psi^5).$$
as can be seen from the fact that in this coordinate $\Gamma_{tt}^t$
and all its holomorphic derivatives go
to zero as $\psi\rightarrow 1$.
Taking into account the explicit form of the Yukawa
coupling which in the $\psi$ coordinate behaves as
$$C_{\psi \psi \psi}\sim {1\over (1-\psi^5)}$$
and that
$$C_{ttt}= \Big[{\partial \psi\over \partial t}\Big]^3
C_{\psi \psi \psi}\sim
(1-\psi^5)^2$$
and using formula \estimate\ we find
that
$$F_g \sim {\Big[\partial_t ^3C_{ttt}\Big]^{2g-2}
\over \Big[\partial_t C_{ttt} \Big]^{3g-3}}\sim{a_g\over (1-\psi^5)^{2g-2}}$$
as $\psi \rightarrow 1$.

To discuss the large $\bar \psi$ limit
let us recall that special coordinates of special geometry are nothing else but
canonical coordinates around infinity. One may regard the mirror map
$\psi \rightarrow t$ as a transformation to canonical coordinates. Using
the general properties of canonical coordinates we conclude that
Zamolodchikov's metric  $G_{\psi \bar \psi}$ and K\"ahler potential
$K(\psi, \bar \psi)$
have the following expansion
$$\eqalign{G_{\psi \bar \psi} d \bar \psi= &C {dt \over d \psi} {d \bar \psi
\over \bar \psi ^2} + o(\bar \psi^{-3})\cr
K(\psi, \bar \psi)=&- \log \varpi_0(\psi) +  o(\bar \psi^{-1})\cr}~,$$
where $C$ is some constant and $\varpi_0 (\psi)$ is the solution of
Pickard-Fuchs equation (we are following the notations of \cand ).
The passage to canonical coordinate implies the change of gauge in such
a way that all holomorphic derivatives of $K$ vanish. Namely
$$K \longrightarrow K + \log \varpi_0 + const$$
(see the discussion at the end of section 2).
The choice of $const$ is equivalent to the choice of string coupling
constant, and we will choose it in such a way that Yukawa coupling has
an integral expansion ($const=3 \log(2\pi i/5)$).

In computing the higher genus amplitudes of this example it is
convenient first to take the $\bar t \rightarrow \infty$ while fixing
$t$.  This is useful because in this limit as discussed in section
5.10, there is some information about the behaviour of $F_g$ (as
counting of holomorphic maps of genus $g$ in Calabi--Yau).
We use this correspondence
to fix the holomorphic ambiguity in integrating the anomaly equation.

To consider the $\bar t \rightarrow \infty$ we use the results
discussed at the end of section 2 to
simplify the formulas for
different components of propagator.
Indeed, plugging these expansions into
\onemoduluspropagator\
and \onemodulustadpole\  we obtain the following result
$$\eqalign{S^{\psi \psi}=&\left( {5 \over 2 \pi i} \right)^3
{1-\psi^5 \over 5 \psi^2}\partial_{\psi} \log \left( {dt \over d \psi}v
\left({f \over \varpi_0}\right)^2  \right)\cr
S^{\psi}=&\left( {5 \over 2 \pi i} \right)^3
{1-\psi^5 \over 5 \psi^2}
\left[ \left( \partial_{\psi} \log (f / \varpi_0) \right)^2 +
v^{-1} \partial_{\psi} v  \partial_{\psi} \log (f / \varpi_0)   \right]\cr
}$$
There is not much simplifications in the expression for $S$.
Namely,
$$\eqalign{S&= \left[
 S^{\psi}  - {1 \over 2}
D_{\psi} S^{\psi \psi}  - {1 \over 2}
(S^{\psi \psi})^2 C_{\psi \psi \psi}  \right]
\partial_{\psi} \log (f/ \varpi_0 ) \cr
 &~~~~~~+ {1 \over 2} D_{\psi} S^{\psi}
 + {1 \over 2} S^{\psi \psi} S^{\psi} C_{\psi \psi \psi}~.\cr}$$
In the large volume limit ($\psi \rightarrow \infty$) the propagators
$S^{\psi \psi} \sim \psi^2$, $S^{\psi} \sim \psi$ and $S \sim const$ and
therefore all $F_g$ go to $const$.

The genus zero and one have already been discussed in
\cand \bcov\ respectively.  So we consider the genus 2 for which the
techniques developed in this paper is crucial.
The genus two partition function is given by equation \genustwosolution\
$$F_2=\left( {1 \over 2} S^{\psi \psi}  C^1 _{\psi \psi}+
{1 \over 2}C^1 _{\psi} S^{\psi \psi} C^1 _{\psi} -
{1 \over 8} S^{\psi \psi} S^{\psi \psi} C_{\psi \psi \psi \psi}+...\right)
+f(\psi)~,$$
where $f(\psi)$ is holomorphic ambiguity. The most general form of
holomorphic ambiguity consistent with the asymptotic behavior of $F_g$
is given as follows
\eqn\genambig{f_2(\psi)=A+ {B \over (1 - \psi^5)} + {C \over (1 - \psi^5)^2}~.}

Now we are almost done. We just need to transform $F_2$ to canonical
coordinate $t$ and canonical section for the bundle.
Note that $F_2$ is a section of a line bundle
${\cal L}^{-2}$. Taking into account the change in the gauge in going
to canonical coordinates we obtain $F_2$
\eqn\expone{F_2 (\psi) \longrightarrow \left( \left({2 \pi i \over 5}
\right)^3 \varpi_0(\psi(t)) \right)^2  F_2(\psi(t)) }
Note that the ambiguities in the choice of the sections $f$ and $v$
given by two coefficients $a$ and $b$ should
simply shift $F_2$ by a holomorphic function and thus should be possible
to absorb in $A,B$ and $C$.  That this should be possible leads
to a strong check both for the Feynman graph techniques discussed
in section 6 in solving the $F_g$, as well as for the computer
code we wrote.  So we set $a=b=0$ and we are thus left
to fix the three unknown coefficients $A,~B$ and $C$.

To do this we need to know the structure of instanton expansion.
First of all there are no genuine genus two curves of degree $1,~2$ and $3$.
The contribution of degree $1,~2$ and $3$ comes entirely
from the bubbling of the
sphere or a torus (in case of degree 3).
We take into account these bubblings and demand
that the denominators of coefficients of such terms can at most
be  $1/5760$ consistent with some characteristic
class computation on moduli space of genus 2.  Now it is
natural to expect that after subtraction of these contributions
the rest of the expansion be with integral coefficients
as would follow from `counting' holomorphic curves.  Of course
there is no guarantee that this is correct to impose, because
indeed there are continuous families of holomorphic curves
and we are computing the appropriate Euler character as
discussed in section 5, and these could be fractional
if the corresponding moduli space has orbifold points.  Anyhow
to proceed we assume that at least in the case of the quintic
these coefficients are integral and we end up uniquely fixing
all the coefficients.  We obtain $A=-71375/288,~B=-10375/288,~
C= 625/48$, and get
\eqn\strexp{F_2(q)=-{5 \over 144} + {1 \over 240} \sum_n ^{\infty} {d_n q^n
\over (1-q^n)^2} +
\sum_r D_r q^r ~,}
where $d_n$ counts the number of holomorphic rational curves of degree $n$,
$D_n$ the number of holomorphic curves of genus $2$.
We found that there is no toroidal bubbling, which can also be
argued on physical grounds\foot{
  If there
were toroidal bubbling then we would end up with a moduli
space which has as a factor a torus.  However, since
we have to bring down factors of curvature from the action
to absorb the fermion zero modes, and since the curvature
vanishes for the torus, we just get zero.}.
In the above search we {\it did not\/} impose by hand the large volume
behavior $t,\overline t \rightarrow \infty$ computed in section 5.
Indeed it was shown there that the leading term should be $\chi(M)/5760$
which in our case is $-5/144$ in agreement with what we found, thus
lending further support to the assumptions we made in fixing
the coefficients of bubbling.  Moreover the number $1/240$ is also
very natural as it is minus the Euler character of
moduli space of genus 2 curves.  It would be very interesting
to understand this.  Also the structure of the multi-bubbling
is very simple, though different from what has been encountered
in genus 0 \cand\ and 1 \bcov .  It would also be important
to derive this structure.  At any rate
the results of this computation for $D_n$ are summarized in
table 1.
\vfill
\eject

\vskip .5in

\centerline{\vbox{\offinterlineskip
\hrule
\halign{&\vrule#&
   \strut\quad\hfil#\quad\cr
height2pt\cr
&Degree\hfil&&$g=0$&&$g=1$&\cr
height2pt&\omit&\cr
\noalign{\hrule}
height2pt&\omit&\cr
&n=0&&5&&50/12&\cr
&n=1&&2875&&0&\cr
&n=2&&609250&&0&\cr
&n=3&&317206375&&609250&\cr
&n=4&&242467530000&&3721431625&\cr
&n=5&&229305888887625&&12129909700200&\cr
&n=6&&248249742118022000&&31147299732677250&\cr
&n=7&&295091050570845659250&&71578406022880761750&\cr
&n=8&&375632160937476603550000&&154990541752957846986500&\cr
&n=9&&503840510416985243645106250&&324064464310279585656399500&\cr
&...&&...&&...&\cr
& large n && $a_0n^{-3}(\log n)^{-2} e^{2 \pi n \alpha}$
&& $a_1n^{-1} e^{2 \pi n\alpha}$ &\cr
height2pt&\omit&\cr
\noalign{\hrule}
height2pt&\omit&\cr
\noalign{\hrule}
&Degree\hfil&&$g=2$&&$g$&\cr
height2pt&\omit&\cr
\noalign{\hrule}
height2pt&\omit&\cr
&n=0&&-5/144&&-$100\cdot [c_{g-1}^3]$&\cr
&n=1&&0&& &\cr
&n=2&&0&& &\cr
&n=3&&0&& &\cr
&n=4&&534750&& &\cr
&n=5&&75478987900&& &\cr
&n=6&&871708139638250&& &\cr
&n=7&&5185462556617269625&& &\cr
&n=8&&90067364252423675345000&& &\cr
&n=9&&325859687147358266010240500&& &\cr
&...&&...&&...&\cr
& large n && $a_2 n (\log n)^2 e^{2 \pi n\alpha}$ &&
$a_g n^{2g-3} (\log n)^{2g-2} e^{2 \pi n \alpha}$ &\cr
height2pt&\omit&\cr}\hrule}}

\centerline{Table 1. \#\ curves of genus $g$ on quintic hypersurface}

\vfill
\eject

As we have just seen the knowledge of the instanton expansion allows
us to fix holomorphic ambiguity. Holomorphic
ambiguity  at genus $g$ can be written as follows
$$f_g(\psi)=\sum_{g=0} ^{2g-2} {A_g \over (1-\psi^5)^{2g-2} }$$
In general there are $2g-1$ unknown parameters. To fix this ambiguity
uniquely one need to know the precise structure of the instanton
expansion. What is lacking in particular is how the lower genera
contribute to genus $g$ (bubbling).  Even if this is fixed,
to completely fix the $A_g$ we need to know the first few
coefficients for the number of holomorphic curves of genus $g$
to fix all the rest.

The asymptotic behavior of $D_n(g)$ (i.e. the coefficient of
asymptotic expansion for large $n$ and fixed $g$)  is determined
by the structure of singularity around $\psi =1$.
As was argued the asymptotic behavior of $F_g$ as $\psi
\rightarrow 1$ is given by
$$F_g (t) \rightarrow {A_{2g-2} \over (1-\psi^5)^{2g-2}}
   \left( \left( {2 \pi i \over 5}\right)^{3} \varpi_0 (\psi) \right)^{2g-2}$$
The last factor $((2 \pi i/5)^3 \varpi_0 )^{2g-2}$
is nothing else but the gauge transformation.
In the limit $\psi \rightarrow 1$ this factor tends to a constant and
therefore it does not affect the asymptotic behavior.
On the other hand the structure of singularity around $\psi=1$
is dictated by asymptotic behavior of $D_n(g)$ coefficients.
Assuming the reasonable ansatz
$D_n(g)\sim n^{\rho}  (\log n)^{\sigma} e^{2 \pi n t(1)}$
we immediately get\foot{The contribution
from lower genera is subleading for $\psi\rightarrow 1$.}
$$\eqalign{F_g (\psi) &\sim
     \int dn\, n^{\rho}  (\log n)^{\sigma} e^{-2 \pi n(t(\psi) -t(1))}\sim  \cr
     & \sim  \left(  {1 \over \psi-1} \right)^{\rho+1}
      \left[   \log(\psi-1)     \right]^{\sigma-\rho-1}\cr}$$
Comparing the last two formulas we obtain $\rho=2g-3$ and $\sigma=2g-2$.
Thus the asymptotic behavior of $D_n(g)$ is given as follows
\eqn\scal{D_n (g)=a_g  n^{2g-3}  (\log n)^{2g-2} e^{2 \pi n \alpha}~,}
where $a_g$ and $\alpha=t(1)$ are constants which are not universal
in the sense that they depend on the manifold under consideration.
Morally speaking the degree of the map $n$ coincides with the notion of the
area of the embedding measured in some units.
In this interpretation $D_n$ is nothing else but a fixed area partition
function. Asymptotic dependence \scal\
of $D_n(g)$ on the degree of the map is the same as the area  dependence for
$c=1$ model coupled to gravity.
It even reproduces  correctly the logarithmic scaling
violation\ref\lsv{V. Kazakov, A. Migdal, Nucl. Phys. B311 (1989) 171 \semi
J. Polchinski, Nucl. Phys. B346 (1990) 253}, which is specific for
the $c=1$ model.
This fact is not very surprising; ${\hat c}=3$  $N=2$ topological models are
closely related  to $c=1$ model coupled to gravity.  In fact
it has been shown that a particular $\hat c=3$ twisted $N=2$ theory
is equivalent to $c=1$ model coupled to gravity \ref\mv{S. Mukhi and C. Vafa,
{\it Two-Dimensional Black Hole as a Topological Coset Model of C=1
String Theory} HUTP-93-A002, hep-th/9301083}.
To see the logarithmic scaling violation,
consider $F_g$  as a function of cosmological constant $\Delta$
which can be identified with $2 \pi (t-\alpha)$.
For large areas ($n$) one can replace the summation by integral
$$F_g (\Delta) \sim \int dn\, n^{2g-3}  (\log n)^{2g-2}
   e^{- n \Delta } \sim
     \left( { \Delta \over \log\,  \Delta } \right)^{2-2g}~,$$
The real scaling behavior is
determined not by $\Delta$
but by $\mu=\Delta/\log\, \Delta$ exactly like in $c=1$ model.
The $\Delta$ dependence of $F_g$ coincides with $t$-dependence (up to
irrelevant shift and rescaling).
Then the logarithmic scaling violation is entirely
due to the structure of canonical map around $\psi=1$.
Indeed, $(\psi-1) \sim (t-\alpha)/\log (\psi-1) \sim (t-\alpha)/\log (t-\alpha)
\sim \mu$
around $\psi=1$ (where $t$ here is the canonical coordinate defined for
$\overline t\rightarrow \infty$).

In cases where there are more than one K\"ahler moduli, fix a
direction in the K\"ahler cone of $H^{1,1}(M,Z)$.
Let us denote this direction by $(n_1,...,n_r)$ where
$n_i$ are integers.   For large $n$ the asymptotic behavior for $D_{n\cdot
(n_1,...,n_r)}(g)$ for fixed $(n_1,...,n_r)$
is thus expected to be given by
the expression \scal . The exact values $a_g$ and $\alpha$ clearly
depend on $M$ and the direction chosen in $H^{1,1}$, while the
powers $2g-3$ and $2g-2$ are expected to be
universal.  It would be important
to check this conjecture in full generality.

\vskip .15in

\subsec{Other examples of Calabi-Yau models}

Here we briefly describe the results of genus two
calculations for some other Calabi-Yau models.  Let us first
consider some hypersurfaces in projective spaces.
These Calabi--Yau
spaces are described as the vanishing locus of a quasihomogeneous
polynomial which describes (up to deformation) the embedding of Calabi--Yau
$3$--fold in a
weighted
projective space
\eqn\models{\eqalign{
  k=5~~:\qquad\quad
W_0&=z_0^5+z_1^5+z_2^5+z_3^5+z_4^5=0 \cr
  k=6~~:\qquad\quad
W_0&=2z_0^3+z_1^6+z_2^6+z_3^6+z_4^6=0 \cr
  k=8~~:\qquad\quad
W_0&=4z_0^2+z_1^8+z_2^8+z_3^8+z_4^8=0 \cr
  k=10:\qquad\quad
W_0&=5z_0^2+2z_1^5+z_2^{10}+z_3^{10}+z_4^{10}=0 \cr}}
These models were earlier investigated in connection with $g=0$
holomorphic maps in
\ref\moris{D.R. Morrison, in {\it Essays on Mirror Manifolds}, ed. by
S.T. Yau, International Press, 1992.}
\ref\klemm{A. Klemm and S. Theisen,
{\it Considerations of One-Modulus Calabi-Yau Compactifications:
Pickard-Fuchs Equations, K\"ahler Potentials and Mirror maps }, KA-THEP-03/92,
TUM-TH-143-92}. The
higher genus computations for these
models are parallel to the quintic case.
The most general holomorphic ambiguity consistent with asymptotic behavior
is given by \genambig (with 5 replaced by
$k$). Again to fix the ambiguity we must know
some additional data (the large volume behavior of genus
two partition function fixes only $A$). In case of the quintic we knew that
there
are no genuine genus two curves of degree $1,~2$ and $3$.
Now there is no such information available. It is known that there are families
of genus two curves of degree $1$ and $2$ for cases $k=6$ and $k=8$.
There are no a reasons to believe that their contribution to genus two
partition function is zero. We denote their contributions by $N$ and
$M$ respectively.

As in the quintic case the genus two partition function has the
structure
$$F_2(q)={\chi(M) \over 5760} + {1 \over 240} \sum_n ^{\infty} {d_n q^n
\over (1-q^n)^2} +
\sum_r D_r q^r ~.$$
The coefficient $1/240$ in front of the spherical bubbling is universal and
independent of the model. We found that after subtraction the genus zero
contribution (bubbling) ${\tilde F}_2 (q)=\chi(M)/5760 + \sum_r D_r q^r$
has almost the integral expansion, except for $k=6$ model.
The results of calculations are summarized in the following $q$-expansions
\eqn\othermodsix{\eqalign{&~~~ {\tilde F}_2 ^{k=6}(q)=-{{17}\over {480}} + N\,q
+ M\,{q^2} + \cr
    & +( 14735432142 + 18504\,M -
       97465842\,N  ) \,{q^3} + \cr
   & +\left( {{512439449683401}\over 2} + 239228316\,M -
       1652255019168\,N \right) \,{q^4} + \cr
    &+(3199366969602589296
      + 2654549098512\,M \, - \cr
      &~~~~~~~~~~~~~~~~~~~~~~~~~~- 20399446637531235\,N  \,) q^5 + \cr
   & +(34720817411136316872780
   + 27042685856051310\,M \, - \cr
      &~~~~~~~~~~~~~~~~~~~~~~~~~~- 219919127006205233856\,N \, )q^6
+ \cdots \cr}}
for $k=6$ model,
\eqn\othermodsev{\eqalign{&~~~{\tilde F}_2 ^{k=8}(q)= -{{37}\over {720}} + N\,q
+ M q^2  + \cr
    & + \left( 2297430758208 + 102816\,M -
       2982239872\,N \right) \,{q^3} + \cr
  & + \left( 222468094578584808 + 7410413536\,M -
       282015713196032\,N \right) \,{q^4} + \cr
  &  +( 15516453237414083197120 \,+
     459069253511168\,M \,- \cr
      &~~~~~~~~~~~~~~~~~~~~~~~~~~~- 19447231842568395440\,N)  \,{q^5} +\cr
  & +( 941762378252908894389530784 \,
      - 26129248919673002880\,M \, - \cr
    &~~~~~~~~~~~~~~~~~~~~~~~~~~-  1171714563944600408125440 \,N )\,{q^6}
+\cdots  \cr}}
for $k=8$ and
\eqn\othermodten{\eqalign{&~~~{\tilde F}_2 ^{k=10} (q)= -{1\over {20}} + N\,q +
M\,{q^2} + \cr
   & +(  2869664890712800 + 1271200\,M -
    447052624000\,N  ) \, {q^3} + \cr
   & +(  3508008133715103890200  +
    1143497004000\,M \, - \cr
   &~~~~~~~~~~~~~~~~~~~~~- 529021878501120000\,N  )\, {q^4} + \cr
   & +(3098620653232515436678572256\, +
   887703919048960000\,M\,- \cr
   &~~~~~~~~~~~~~~~~~~~- 457872639654043275150000\,N)\,{q^5} + \cr
   &+( 2385179845759540102344438070862400\, +
   634572439637621668400000\,M\,- \cr
   &~~~~~~~~~~~~~~~- 346846888907287393959739633664\,N)\,{q^6} + \cdots
\cr}}
for $k=10$.
In fact we checked that all coefficients are integer up to
$q^{10}$, except
for the coefficient $q^4$ in $k=6$ model (provided that $N$ and $M$ integers).
This in particular suggests that there must be continuous families
of holomorphic maps in this case where they have at least
${\bf Z}_2$ orbifold points, and
they contribute a $1/2$ to the coefficient of $q^4$.
It would be interesting to verify this.

Another example which is amusing is the ${\bf Z}_3$ orbifold which is
obtained by modding out $T^2\times T^2 \times T^2$ by a diagonal
$(\omega ,\omega ,\omega )$.  In this case explicit computation
of $F_1$ shows that it is zero.   Now the anomaly formula
for $F_2$ implies that $F_2$ is purely holomorphic, and indeed
this is exactly zero as can be seen by a direct computation
of the orbifold model at all $g$.  This in particular means
that even though there was room for $F_g$ to be non-zero
consistent with the anomaly equation, as there
are appropriate holomorphic functions, nevertheless it vanishes.

There are other interesting models that one may wish to consider.
A particularly interesting class is where there are {\it no} marginal
operators in the twisted theory.  This can happen, for example,
in the context of $B$--models for Calabi--Yau which are rigid.  In such
cases the $F_g$ is simply a number (up to multiplication by the
string coupling constant $\lambda^{2g-2}$), and summing over all $g$ will lead
to a function $F(\lambda )$.  This may be an easier case to study.  In
particular since there are no marginal directions, there are no anomalies
either.  A simple realization of this type of model is again given
by the ${\bf Z}_3 \times {\bf Z}_3$ orbifold model discussed in this section,
but
with the $B$--twist instead of the $A$--twist.  The ${\bf Z}_3$ orbifold in the
$B$--twist is also rigid but in this case one can show again
by explicit computation that $F_g=0$.

Note that in the $A$--model twisting and for smooth manifolds, we can compute
$F_g$ for
all $g$ up to exponentially small corrections, in the limit of large
volume, as $F_g\rightarrow {1\over 2}\chi(M) [c_{g-1}^3]$, in terms
of some cohomology computation on the moduli of Riemann surfaces.
In particular if $\chi(M)\not= 0$ (and barring an accidental zero of
$[c_{g-1}^3]$) we see that $F_g\not=0$.   It would be interesting
in this connection to study Calabi--Yau
manifolds with $\chi =0$, as this argument
also shows that $F_g=0$ up to exponentially small terms in the large
volume limit.

\vfill
\eject

\newsec{Physical implications of topological
amplitudes}
One of the main motivations to study $N=2$ SCFT's comes from
the fact that they serve as building blocks for string vacua.
In this connection particular objects which have natural interpretations
for the $N=2$ SCFT's turn out to also have some interesting
phenomenological implications in string models.  One such object
is the Yukawa coupling.  If one considers heterotic strings
compactified on a Calabi--Yau $3$--fold,
with gauge connection identified with the spin connection
of Calabi--Yau, then the
chiral primary fields of charge 1 give rise to massless generations
and the chiral ring
coefficients $C_{ijk}$ give the Yukawa couplings between the different
generations.  Given the fact that Yukawa couplings are simply the three
point function of topological gravity, it is natural to expect
that all the other computations of twisted $N=2$ theories coupled
to gravity also have similar physical significance for an appropriate string
theory.  In particular we would like to discuss the significance
of $F_g$ in connection with standard string theories.
Before we discuss this let us note where we could
look for such contributions in the effective field theories
arising from string theory.

Let us note that the massless fields $t^i(x^\mu)$ in general end up as
lowest component of chiral superfields from the spacetime point of view.
In general in supersymmetric theories we can have F-terms, i.e.,
superpotential terms, which
involve only chiral superfields, i.e. are holomorphic functions in $t^i$.
Now morally we expect $F_g$ to be a holomorphic function of $t^i$
(ignoring the holomorphic anomaly) and so we expect that $F_g$ is a
contribution to a superpotential.  This observation,
 together with the fact that
$F_g$ is a section of a particular bundle essentially fixes what
term we get in the effective Lagrangian. However instead of guessing
we will show this more directly below.  So before we proceed further,
the dictionary we expect is
$${\rm Topological\ Computations}~\leftrightarrow~
F-{\rm terms\ in\ field\ theories}.$$
We will discuss both the case of closed and open strings.
Afterwards we consider the
computation of threshold corrections for heterotic strings at one-loop
and its relation to $F_1$ and Ray--Singer torsion.  The case
of the closed string has also been recently discussed
in detail by \ref\nar{I. Antoniadis, E. Gava, K. S. Narain and T. R. Taylor,
{\it Topological Amplitudes in String Theory}, hep-th/9307158}.

\vskip .15in

\subsec{Type II string Interpretation}

We start by asking which string theory $F_g$ should be
related with? Given the fact that it is
left--right symmetric, and it is related
to the twisting of a supersymmetric sigma--model for closed
string theory, one is naturally led to consider type II strings
compactified from 10 to 4 on a ${\hat c}=3$ internal theory.  We thus
are searching for low energy effective field theory terms that
$F_g$ is computing.  Compactifying type II on $\hat c=3$ theory
gives rise to a low energy field theory in four dimensions
with $N=2$ supergravity.  The chiral fields $t^i$
are scalar fields for this supergravity (for
aspects of $N=2$ supergravities that one obtains by compactifying
on Calabi-Yau manifolds see \ref\low{N. Seiberg, Nucl. Phys. B303
(1988) 206\semi
S. Cecotti, S. Ferrara and L. Girardello, J. Mod. Phys. A4 (1989) 2475.s}).
The $N=2$ supergravity multiplet in particular contains a Maxwell
field which is called gravi--photon.  We will denote the field strength
for this field by $T$.
This field arises from the
Ramond--Ramond sector of type II string and the vertex operator
for this field, in the limit of vanishing momentum
$k\rightarrow 0$ is proportional to
$$V_T^{\pm \pm}=k_{\pm \pm}  { S^\pm}{\overline S}^\pm
 \sigma \overline
\sigma e^{(-\phi /2)}$$
where $\phi$ is part of the the bosonized $\beta, \gamma$
field \ref\fms{D. Friedan, E. Martinec and S. Shenker,
Phys. Lett. 160B (1985) 55}\ and
where $S$ ($\overline S$) denote the left--moving
(right-moving) 4d spinor vertex operators and $\sigma$
($\overline \sigma$) denotes the
unique vertex operator for the left--moving (right--moving)
charge $3/2$ (3/2) Ramond vacuum state
for the internal $N=2$ theory (with $\hat c=3$).
Indeed this vertex operator is the same as the FMS \fms\
spin operator (taking into account the fact that
the internal theory is a general $\hat c=3$ rather than
flat space).  Note that
$\sigma ,\ S$ and ${\rm exp}(-\phi/2) $ (together
with their right--moving counterparts)
generate the
spectral flow from the $NS$ sector to the $R$ sector.  We will use
this vertex operator to go between the twisted theory and
the untwisted theory.

There are a number of differences between the twisted theory
and the ordinary type II strings.  First of all there are more
fields in the ordinary theory.  In addition to an
{\it untwisted} $N=2$
SCFT with $\hat c=3$, in type II strings we have the
fermionic diffeomorphism ghosts $(b,c)$ of spin $(2,-1)$, the bosonic
super--diffeomorphism
ghosts $(\beta ,\gamma )$ of spin $(3/2,-1/2)$, and the
space--time fields, which we take to be two complex bosons ${X^i}$
of spin $0$ and
two complex fermions ${\psi ^i}$ and their conjugates $\chi_{\overline i}$
of spin $1/2$ with $i=1,2$.  Of course the same content of fields is
needed for the right-moving part which we denote by barred fields.
If we could {\it twist} the 1/2 integral spin fields by half a
unit, then their spins would be the same as the integral
spin fields but with opposite statistics, so they would
tend to cancel out of the partition function.
  In addition we
would need to twist the internal $N=2$ theory
which is also the same as shifting the 1/2-integral fermion
spins of the internal theory.  Both of these
can be accomplished by insertion of
$(2g-2)$ vertex operators for gravi-photon $V_T^{++}$
(modulo some subtleties mentioned below).
The way to see this is that the spin content of fields can be
changed by addition to the action of

\eqn\boso{{1\over 2}\int R\varphi}
where $\varphi$ denotes the bosonized version of the fields.  We can
choose the curvature $R$ to have delta--function like support at
$2g-2$ points.
But each such point is equivalent to the insertion of $V_T^{++}$ as mentioned
before.  However, to write it in a conformally meaningful way,
given that $V_T^{++}$ is dimension $(1,1)$ we have
to integrate it over the surface (which is equivalent to choosing
the delta-function support for $R$ by averaging over all points);
we thus have found the dictionary that
\eqn\twisting{\big\langle \Big[ \int V_T^{++} \Big]^{2g-2}\cdots
\big\rangle_{\rm
untwisted}=\langle \cdots \rangle_{\rm twisted}}
This means that the determinant of non-zero modes of the extra
fields which were not in the original twisted internal
$N=2$ theory cancel out,  leaving us with the twisted
internal theory.  However, we have to pay particular attention
to the zero modes of the extra fields
we have introduced:  There are zero modes for $b,\beta $ and the
$\psi ,\chi$ system that have to be absorbed in order for
the partition function not to vanish.  Let us first deal with
the ghost zero modes.

The $b$ zero modes give rise to the measure over
moduli space.  In fact if $\mu_i$ denote the basis
for Beltrami-differentials, we have to insert in the
superstring measure a factor of
$$\Big| b(\mu_1) \dots b(\mu_{3g-3})\Big|^2$$
to absorb the $b$ zero modes.  For the $\beta$ zero modes
we usually have to insert $2g-2$ factors of $\delta(\beta)\cdot G$
where $G$ is the $N=2$ supersymmetry current for the {\it full\/}
theory.
But that is true for the partition function with no operators inserted.
In our case inserting $2g-2$ vertex operators $V_T^{++}$ which are
in the $-1/2$ picture means that we need to insert $3g-3$ factors
of $\delta (\beta)$.  Moreover the fact that $\beta ,\gamma$
is effectively twisted means that we can choose the same basis
for the Beltrami differentials to fold with them.
Moreover by charge conservation for the internal twisted
theory only the $G^-$ component of the internal
theory gives non-vanishing amplitude, so we end up with
$$\Big|\delta \big(\beta(\mu_1)\big)\cdots
\delta\big(\beta (\mu_{3g-3} )\big)\Big|^2
|G^-(\mu_1)\cdots G^-(\mu_{3g-3})|^2$$
With this choice of Beltrami-differential, the zero modes
of $b$ and $\delta(\beta )$ give opposite contribution
and thus $b,c$ and $\beta ,\gamma$ completely drop out
of the picture, having left us with the twisted $N=2$ theory
with $3g-3$ insertions of $G^-$ which is precisely the prescription
we had for computing $F_g$ of the twisted string coupled to gravity.
However we still have to get rid of the space--time fermion zero modes.
There are $g$ of $\psi^i$ (which has spin 1) and one of $\chi_{\overline i}$
(which has spin zero)
zero mode for each $i$ (and similarly for the right movers).
To absorb the $\chi$ zero mode and one $\psi$ zero mode we can
insert the operator
\eqn\ins{
\epsilon_{ij}\epsilon_{i'j'}\epsilon_{\overline i
\overline j}\epsilon_{\overline i' \overline j'}
\int \psi^i\chi_{\overline i}{\overline \psi}^{i'}
{\overline \chi}_{\overline i'}
\int \psi^j\chi_{\overline j}{\overline \psi}^{j'}
{\overline \chi}_{\overline j'}}
Note that up to factors of momentum, this operator
is precisely the insertion of two graviton vertex operators.
We are left to absorb $g-1$ extra zero modes of $\psi^i$.
Taking into account that after twisting $\psi^i$ has spin
1, one is tempted to introduce $g-1$ operators of the
form $\int \psi^i {\overline \psi}^j$ but unfortunately
this does not have a well-defined meaning as a vertex
operator for the untwisted theory.  Instead, motivated
by a suggestion from authors of \nar , we can
make the insertion of $g-1$ of $\psi^1\psi^2{\overline \psi}^1{\overline
\psi}^2$ operators at
$g-1$ of the points where we have taken
the delta--function curvature singularity. This choice
will have the property of absorbing the unwanted $\psi$
zero modes, without getting an operator which does not make
sense in the untwisted theory. This is because choosing
this position for the $g-1$  curvature singularities will convert
$g-1$ of $V_T^{++}$ to $V_T^{--}$ which is the
vertex for gravi--photon field with opposite
self--duality property.  In this way we can absorb all the zero modes
and end up with $F_g$.  We thus see, putting all this together, that
\eqn\fundfi{F_g=\big\langle \Big[ \int V_T^{++}\Big]^{g-1} \Big[ \int V_T^{--}
\Big ]^{g-1}
\epsilon_{ij}\epsilon_{i'j'}\epsilon_{\overline i
\overline j}\epsilon_{\overline i' \overline j'}
\int \psi^i\chi_{\overline i}{\overline \psi}^{i'}
{\overline \chi}_{\overline i'}
\int \psi^j\chi_{\overline j}{\overline \psi}^{j'}
{\overline \chi}_{\overline j'}\big\rangle}
Putting the momentum factors this means that $F_g$ is the coefficient
in the low energy effective action for a term of the form $R^2(T^2)^{g-1}$.
This completes the derivation of relation between topological
partition function and field theory.  However we should note that
in the above derivation we were somewhat careless in some points:
We assumed that we can twist fields simply by adding ${1\over 2}\int R \varphi$
term to the action, but as is well known this is true up to boundary terms.
The boundary terms are in fact responsible for picking which point
on the Jacobian of the twisted field we end up (i.e. the
choice of the flat bundle) --- we have to make sure that
we end up with the trivial flat bundle tensored with the appropriate
power of the canonical bundle.  Secondly, a point which is
related to this, is the fact that we have to sum over spin structures
in the untwisted theory.  Somehow this is already taken into
account in the twisting, because
viewing the twisting as choosing a background
gauge field set equal to half the gauge connection is ambiguous up to
a choice of a ${\bf Z}_2$ bundle, which is just the choice of spin structure.
This ambiguity should translate to a sum over spin structure to get
a correspondence between the twisted and untwisted theory.  To make
sure that these points do not affect our argument one will have
to go to more detail and check the explicit factors arising
in the twisting.  Fortunately this has been considered very
carefully in \nar\ using bosonization techniques which confirms the
above heuristic arguments.

As argued at the beginning of this section we should
expect a term in the superpotential which give rise to the
effective action of the form $R^2 (T^2)^{g-1}$.
In fact one can find an $F$-term which gives rise to such a term:
\eqn\sugra{[F_g ({\cal W}^2)^{g}]_F}
where ${\cal W}^2$ is the square of the Weyl superfield (${\cal W}^2$ is a
composite chiral superfield of weight $2$),
and $[\cdots]_F$ is the $F$--density for conformal $N=2$
 supergravity. Notice
that this coupling makes sense since, $F_g$
is a section\foot{In more
traditional terms, $F_g$ is represented in superspace by a homogeneous function
of the vector fields $X_I$ of weight $2-2g$. The $F_g$ we use throughout the
paper is obtained from this homogeneous function by choosing a gauge for the
line bundle $\CL$.}  of $\CL^{2-2g}$, which --- in the language of conformal
$N=2$ tensor calculus \dewit\
 --- means that it is a chiral field of weight $2-2g$, so
that the combination $F_g ({\cal W}^2)^g$ has weight $2$ and hence defines an
invariant $F$--term \dewit.

However, eq.\sugra\ makes sense only if $F_g$ is a chiral superfield, which
happens only if $F_g$ is a holomorphic function of the chiral fields $t^i$. But
as discussed in section 3, $F_g$ is not holomorphic because of anomalies. Then
\sugra\ cannot be the correct form of the supergravity coupling corresponding
to the amplitudes we discussed above.  However we have to recall how
one deals with a field theory which has flat directions, as is the case here.
In such cases there are {\it inequivalent} vacua determined by what
the expectation value of the massless fields are.  Suppose we have
chosen such an expectation value, which we denote by $(t_0, \overline t_0)$.
Then we can expand $F_g$ {\it holomorphically} about this base point.
What this means is that we consider (in canonical coordinates)
$$F_g(x+t_0,{\overline t_0}) =\sum_{i} {1\over n!}
x^{i_1}...x^{i_n}D_{i_1}...D_{i_n}
F_g(t_0,\overline t_0)=$$
$$=
\sum_{i} {1\over n!} x^{i_1}...x^{i_n}\partial_{i_1}...\partial_{i_n}
F_g(t_0,\overline t_0).$$
Thus $F_g$ is now a {\it holomorphic} function of
superfields $x^i$, and we are thinking
of $(t_0,\overline t_0)$ as a base point for expansion
of $F_g$ and not as a superfield.  This
view of the effective Lagrangian we are presenting is motivated
from the fact that in the construction of solutions to the anomaly
equation, discussed in section 5, a function $ W$ was introduced
which was a {\it holomorphic} function of $x^i$.  So in particular
we end up with the superpotential for the $N=2$ supergravity,
including all loop contributions:
\eqn\sugar{\sum_g [F_g(x+t_0,\overline t_0){\cal W}^{2}(\lambda
\, {\cal W}^2)^{g-1}]_F
=[{\cal W}^2\, { W}({\lambda  {\cal W}},x;t_0,\overline t_0 )]_F}
where here $\lambda^{-1} $ is a section of ${\cal L}^{-1}$ and plays
the role of compensating field in the supergravity theory \dewit\
(one--loop contribution can also be included here by
addition of a term proportional to ${\rm log}\, \lambda {\cal W}$).

\vskip .15in

\subsec{Open superstring interpretation}
As discussed in the previous section, we can also consider
the twisted $N=2$ theory for the open strings.  It is also
natural to ask what is the interpretation of
the $F^g_h$ in the low energy effective theory of some
superstring theory.  The natural superstring theory to
look for in this context is the 10-dimensional open superstrings
compactified on an internal $N=2$ SCFT with ${\hat c}=3$.
This theory gives rise to a 4-dimensional low energy theory
of $N=1$ supersymmetric Yang-Mills coupled to supergravity.
Actually as is well known to get a consistent theory we need to
consider unoriented strings.  Also if we wish to avoid
anomalies we need to take the gauge group $O(32)$
which brings us to one of the most interesting superstring
theories.  Our considerations in the following will also apply to
the more general gauge group of $O(N)$.

Unoriented strings will have worldsheets which include
both orientable and non-orientable surfaces.
Let us concentrate on the contribution from orientable surfaces which we
have discussed for the twisted $N=2$ theories.  To simplify further
let us first consider the case with no handles $g=0$ with $h$ boundaries.
We will use the same idea as in the closed string case, in other
words add the extra fields which are present in the superstring
compared to the $N=2$ twisted topological model, and then put
appropriate insertions to twist the $\half$-integral spin
fields to obtain integral fields which cancel among each other
except for zero modes, which have to be checked separately.  The
field analogous to the graviphoton in the open string case is the
gaugino field, which we denote by the vertex operator $V_\Psi^{\pm}$,
at zero momentum.  This operator is the spectral flow operator
in the internal $N=2$ SCFT, combined with the operator which
twists the spins of $\beta, \gamma$ ghosts and space-time
fermionic fields $\psi,\chi$.  In particular this operator is inserted
where we choose curvature singularities of appropriate strength.

Let  us consider the open string worldsheet shown in \tfig\FigureEightOne.

\ifigure\FigureEightOne{The worldsheet for open strings with
$g=0$ and $h=5$ boundaries, two of which are the two boundaries
of the cylinder and three of them ($S_1,S_2,S_3$) are slits
on the cylinder.  The twisted theory corresponds to putting
gaugino vertex operators $V_\Psi$ on the end points of the slits
and gauge field vertex operators $V_F$ on the boundary of the
cylinder.}{Fig81}{1.7}

This is a cylinder with two boundaries and with $h-2$ slits cut on it.
We also mean this geometrically, i.e., that the metric be the flat
metric on the cylinder.  However, note that this introduces curvature
singularities at the two end points of each of the $h-2$
slits.   The reason for this is that the zero curvature
on the boundary corresponds to $\pi$ radians, but here at the two
end points we get $2\pi $ radians of worldsheet.  So we insert
$V_{\Psi}^+$ operators at each of the two end points of the
$(h-2)$ slits.  This takes care of the twisting of the internal theory;
the ghost zero modes also cancel leaving us with the measure for
the twisted $N=2$ theory coupled to gravity.  So we only
need to consider the space-time fermion zero modes.  There are
$h-1$ zero modes for each of the two $\psi^i$ and 1 zero mode
for each of the $\chi_{\overline i}$.  The $\chi_{\overline i}$ zero modes can
be absorbed by
adding the operator
$$\epsilon^{\overline i \overline l}\epsilon_{jk}
\oint \chi_{\overline i} \psi ^j \oint \chi_{\overline l}\psi^k$$
Each of these is the vertex operator of a gauge field $V_F$
(up to momentum factors) at zero momentum.  Again as in the
closed string case we need to absorb the remaining $h-2$
zero modes for each of the $\psi^i$.
Again, this can be done
in a conformally meaningful way only by including them
at one of the two end points of each of the $h-2$ slits converting
$h-2$ of the $V_{\Psi}^+$ operators to $V_{\Psi }^-$
operators.  This will thus conclude absorbing zero modes, and so
we end up with
$$F^0_h=\langle [\oint_{S_i} V_\Psi^+ \oint_{S_i}V_\Psi^-]^{h-2}
\oint_S V_F \oint_S V_F \rangle_{untwisted}$$
where $S_i$ denote the interior slits and the $S$ denotes one
of the two boundaries of the cylinder.
Thus we see, taking into account the structure of the insertions
at the boundaries in taking the trace, that this gives
rise to a term in the effective lagrangian of the form
$$F^0_h\cdot
{\rm Tr}\, F^2 [\Tr\, {\Psi ^2}]^{h-2}.$$
 The non--orientable worldsheets
do not contribute to this amplitude because the absorption of
fermion zero
modes does not have the right structure.

As discussed in the introduction we expect that the topological
theory is computing the coefficient of a
a superpotential term.  Indeed there is a superpotential term
which give rise to the above interaction and that is given by
\eqn\magic{\int d^2\theta F^0_h(W_\alpha  W^\alpha )^{h-1}.}
As discussed in the previous section, the partition function $F^0_h$
is now going to be a section of ${\cal L}^{2-h}$.  This is consistent
with the fact that $W^2$ is a section of $\CL$
(which is also related to the fact that closed
string coupling is the square of the open string coupling constant).
The discussion we had regarding the non--holomorphicity
of $F_g$ in the closed string case applies word for word in the
present situation and we will thus not repeat it.

The appearance of \magic\ as a topological amplitude,
which is in principle exactly computable possibly
using anomaly techniques discussed for open strings,
is very interesting.  This is because
such an interaction has a strong bearing on the question of
gaugino condensates, which has been proposed \ref\gaug{J. P. Derendinger,
L. E. Ibanez and H. P. Nilles, Phys. Lett. 155B (1985) 65 \semi
M. Dine, R. Rohm, N. Seiberg and E. Witten, Phys. Lett. 156B (1985) 55}\ as a
mechanism to break supersymmetry
in the context of superstrings!  This would be very interesting
to pursue in detail.  Also the heterotic version of this
would have to be investigated \ref\NNar{This is currently being
pursued by  Narain et. al.}.

\vskip .15in

\subsec{Threshold corrections for heterotic strings}

In the context of heterotic strings the one--loop contribution
to threshold correction for gauge coupling is related to the
topological amplitude we have been discussing.  In fact it has
been shown in
\ref\thres{V. Kaplunovsky, Nucl. Phys. B307 (1988) 36   \semi
L.J. Dixon, V.S. Kaplunovsky and J. Louis,
Nucl. Phys.
B355 (1991) 649\semi  S.Ferrara, C.Kounnas, D.L\"ust and F.Zwirner, Nucl. Phys.
B365 (1991) 431\semi I. Antoniadis, E. Gava and K.S. Narain,
Nucl. Phys. B 383 (1992) 93 and Phys.
Lett. B283 (1992) 209\semi
 J.--P. Derendinger, S. Ferrara, C. Kounnas and F. Zwirner, Nucl. Phys. B372
(1992) 145.
}\
that the one--loop corrected gauge coupling constant which depends
on the moduli of the internal theory can be written as
\eqn\thco{{16 \pi^2\over g^2_a(\mu )}=k_a  {16 \pi^2\over g^2_{GUT}}+
b_a\cdot {\rm log} {M^2_{GUT}\over \mu ^2} +\Delta_a}
where $a$ denotes the gauge group in question,
$k_a$ is the level of the group, $b_a$ denotes
the contribution of massless modes to the threshold, and
$\Delta_a$ which includes contribution of internal stringy
states is given by
\eqn\dedel{\Delta_a =\int {d^2\tau \over \tau_2}
{\rm Tr}' (-1)^{F_L}F_L Q^2_aq^{H_L }{\overline q}^{
H_R}}
where the trace is
in the R-R sector and is over the massive modes of the internal theory,
including the right--moving gauge group contribution and
the four dimensional modes, $Q_a$ denotes a gauge group generator
for the group $a$, and the integral is over fundamental domain of
moduli of tori.   The $b_a$ in the above formula reflects
the fact that the zero modes lead to a divergence in the above
formula which can be removed by defining a running scale $\mu$,
and so
$$b_a =Tr(-1)^{F_L}F_LQ^2_a\Bigg|_{massless \ modes}$$
It was shown \thres\ that $\Delta_a$ satisfies an anomaly equation
in terms of its dependence on moduli of the internal theory.  Moreover
it was shown that in the case of identifying gauge connection with the
spin connection of the Calabi-Yau, which breaks $E_8\times E_8$ heterotic
string to $E_6\times E_8$, $\Delta({E_6})-\Delta ({E_8})$ satisfies
the same anomaly equation as $12 \cdot F_1$, where $F_1$ is the
genus one topological partition function defined in section 2.  It would
be interesting to show this fact directly and moreover show
that they also have the same holomorphic piece, i.e. not only
$\partial \overline \partial [\Delta({E_6})-\Delta ({E_8})]=12
\partial \overline \partial F_1$, but that
$\Delta({E_6})-\Delta ({E_8})=12 F_1$.

In order to argue this, it is worthwhile deriving the more general
formula for the behavior of $\Delta_a$ even if the internal theory
is not $(2,2)$, i.e. when the gauge connection is not identified
with the spin connection of the Calabi--Yau but belongs
to some bundle $V$.  The bundle $V$ needs to be stable and
${1\over 2}c_2(V)={1\over 2}c_2(M)$ for a consistent heterotic string vacuum
\ref\wit{E. Witten, Nucl. Phys. B268 (1986) 79}.  To be able to relate
$\Delta_a$ to what we have computed and in particular
to the Ray--Singer torsion, in this generality,
we need to take a particular limit, namely the limit of large
volume of the Calabi--Yau.  We will compute
the dependence of $\Delta_a$ on the complex moduli of
Calabi--Yau in this limit.  Actually taking the large
volume limit in the case  $V=T(M)$ is not a restriction
as is well known that the complex structure dependence
and K\"ahler structure dependence of
$\Delta_a$ decouple in this case.  So for this case our remarks
are quite general.  We suspect our answer are also independent of this
limit in the more general case but we do not have a rigorous argument.

 Let us consider the internal theory to be a Calabi--Yau
manifold with a vector bundle $V$ on it.
Let $H$ denote the holonomy of this bundle.  This means
that the first $E_8$ gets broken down to
$$E_8\rightarrow G\times H$$
where $G$ is the maximal remaining group for which $G\times H$
can be imbedded in $E_8$.  We will for simplicity
of notation take $G$ to be
a simple Lie group, otherwise we can do what we are about to do
for each simple factor of $G$.
 Thus the unbroken gauge group
in 4 dimensions is $G\times E_8$.   Now the adjoint representation
of $E_8$ breaks under this decomposition to
$$(248)\rightarrow \sum_\alpha (R_\alpha ,r_\alpha )$$
where $R_\alpha$ ($r_\alpha$) denotes the $G$ ($H$) representation.

Now consider the limit of infinite
volume on the Calabi--Yau with arbitrary complex structure.  In this
limit the computation of $\Delta_a$ is easy
to do, because by adapting the argument used in
the derivation of the Kodaira--Spencer theory to the present case,
the interior part of the moduli
space make no contribution to the answer, and only degenerate
Riemann surfaces contribute.   In the case of moduli of torus, this
means that only $\tau_2 \rightarrow \infty$ contributes, in which case
(and after integrating over $\tau_1$) only the massless modes
of the right--moving sector contributes and the internal theory simply becomes
the same computation as the Ray--Singer torsion discussed in section 5.7.
We thus see that
\eqn\trers{\Delta (G) =\sum_{\alpha} {\cal T} (R_\alpha) I(V_{r_\alpha}) }
where $I({V_{r_\alpha}})$ denotes the Ray-Singer torsion
for the vector bundle $V$ with representation $r_\alpha$
and ${\cal T}(R_\alpha)$ denotes the index of representation
$R_\alpha $ (coming from $Q^2_a$).
Also note that similarly for the unbroken $E_8$ we have
\eqn\trerrs{\Delta (E_8) ={\cal T} (E_8) I_0 }
where $I_0$ denotes the Ray-Singer torsion with the trivial
bundle.  As discussed in \thres\ only the difference between
the threshold corrections is meaningful, and so physically
we should only consider $\Delta (G)
-\Delta (E_8)$.  This is our general result.  Now we specialize
to the case where $V=T(M)$, in which case $G=E_6$.  We have
the decomposition
\eqn\decoe{248\rightarrow (78,1)\oplus (27,3)\oplus ({\overline {27}}
,\overline 3)\oplus (1,8)}
Also we note that since the spin connection is identified with the
gauge connection we have
\eqn\sping{I(V_0)=I_0\qquad I(V_3)= I(T^*)\qquad
I(V_{\overline 3})=I(T^*\wedge T^*)}
Using the values ${\cal T}(E_8)=30,{\cal T}(E_6)=12,{\cal T}(27)=3$, and making
use of \sping,  \decoe, \trers\ and \trerrs\ we find
$$\Delta(E_6)-\Delta (E_8)=(12-30)I_0 +3\big( I(T^*)+I(T^*\wedge T^*)
\big)=$$
$$=6 \big(-3I_0+{1\over 2}I(T^*)+{1\over 2}I(T^*\wedge T^*)\big)=$$
$$=6\big({-3\over 2}I_0 +{1\over 2}I(T^*)+
{1\over 2}I(T^*\wedge T^*)-{3\over 2}I(T^*
\wedge T^* \wedge T^*)\big)=$$
\eqn\fintre{=
{12\over 2} \sum_p (-1)^p\left(p-{3\over 2}\right)I(\wedge^p T^*)
=12 F_1}
where we used the fact that $I(T^*
\wedge T^* \wedge T^*)=I_0$.
This is what we wished to show.  Even though
we derived this in the context of complex structure dependence
of the threshold corrections, by mirror transform, it may also be viewed
as the K\"ahler structure dependence.  If we view it in this way
we can then use the result of \bcov\ to estimate the dependence of $F_1$
for large volume of Calabi-Yau.  It was shown there that (taking
into account the factor of 2 difference in the definition of $F_1$)
$$F_1\buildrel k\gg 1\over \longrightarrow {1\over 24}\int_M k\wedge
c_2$$
where $k$ denotes the K\"ahler class of the Calabi-Yau
manifold.  Note that (as discussed in \bcov ) $\int_M k\wedge c_2 >0 $.
So we have
$$\Delta(E_6)-\Delta (E_8)={1\over 2}\int_M k\wedge c_2 >0$$
Now we can use \thco\ to see that the effect of changing $k$ beyond
the Planck scale is the same as getting an
$M_{GUT}^{effective}$ according to
$$M_{GUT}^{effective}=M_{GUT}\cdot{\rm exp}({\Delta \over 2 b})=M_{GUT}
\cdot {\rm exp}[{\int_M k\wedge c_2\over 2b}]$$
where in this case $b=54+3(h_{1,1}+h_{1,2})$.  We thus see that
the effective grand unification scale for this relatively
general class of string compactifications is extremely sensitive
to the size of the internal manifold and moreover when
we increase the size of the internal manifold above Planck scale
it tends to increase exponentially fast!

\vfill
\eject

\newsec{Open problems}

In this section we discuss open problems and directions for future
research.  Let us first summarize some of the main results of this  paper.
We have considered $N=2$ twisted topological strings. The
partition function of these theories
at a given genus $g$ is formally a holomorphic modular
function of weight $2g-2$ on moduli space of the conformal theory. However
we find that there is an anomaly and that the partition function is not
necessarily holomorphic.  What goes wrong with the formal argument
of holomorphicity is the assumption that total derivative
terms vanish upon integration over the moduli space of Riemann
surfaces.  Using the geometry of moduli space of
Riemann surfaces and the structure of the $N=2$ twisted
theories one can compute the boundary contributions
and they turn out to be expressible as (products of) lower
genus correlation functions.
This can be summarized as a
second order linear differential equation,  the {\it master anomaly
equation}, for the full partition function of the theory summed over all
genera.
 This recursion relation
for the antiholomorphic dependence of the partition function
can be solved by introducing Feynman rules which can be expressed
as an integral over an auxiliary
space which includes the dilaton and the marginal fields
as propagating degrees of freedom and whose
vertices are the correlation functions of the lower genus and the
propagators are made of a {\it canonical}\foot{Which
is different from the ones previously used in the literature.}\ prepotential,
for the anti-topological theory, and its
derivatives.   This fixes the genus $g$ partition function
up to a holomorphic modular form, which are typically finite
in number and thus reduces the computation of the partition
function to fixing the coefficients of these functions.
In concrete examples using mirror symmetry these coefficients
can also be fixed, at least for low genus.

We discussed the realization of $N=2$ SCFT's in terms of
sigma--models on Calabi--Yau manifolds.  There are two
different ways to twist such theories, the $A$--twist (the K\"ahler
twist) or the $B$--twist (the complex twist).  In the case
of $A$--twist a particular limit of the topological string theory
computes the number of holomorphic maps (or an appropriate Euler
character on the moduli space of holomorphic maps) from the
Riemann surfaces to the Calabi--Yau.  In the case of the $B$--twist
the target space theory of the topological theory
may be described as an ordinary field theory, actually
a topological field theory, which quantizes the complex structures on
Calabi--Yau,  which we called the Kodaira--Spencer theory of gravity.

We also found an interpretation of the computations of the
topological partition function $F_g$ as the genus--$g$
correction to four dimensional low energy lagrangian generating
 superpotential terms that arise upon compactification of
superstrings on internal theory with $\hat c=3$ from
10 to 4 dimensions.
In the open string case this term will be relevant for the gaugino
condensates.
 Also we related $F_1$, the genus
one partition function, to the threshold corrections for
gauge group couplings for heterotic strings in the `standard'
compactification scenario (identifying the gauge connection
with the spin connection of the Calabi--Yau).  This shows
a surprisingly universal {\it exponential}
 dependence of the effective GUT
scale with respect to the volume of the Calabi--Yau manifold.

This was  a basic summary of some of the main results.  Let
us now discuss some directions for future research.  One of the
most significant aspects of the master anomaly equation is that
it captures the anomaly to all orders in string perturbation theory
and is thus a way even to proceeds towards non--perturbative
formulation of it.  It would be interesting to compare
how the non-perturbative aspects of the topological strings
discussed here compare with those of some other string theories
discussed in \ref\shenke{S. Shenker, Proceedings of Careges Workshop
on Random Surfaces, Quantum Gravity and Strings, 1990.}.
   A first
step in this direction is to find exact solutions to the anomaly equation
to all orders. In this paper we saw how we can do it order by order
in perturbation theory (using the Feynman graph technique discussed in the
text) but we did not manage to find a simple closed form for any example which
would be valid to all orders.
The simplest example to consider
in this connection is the toroidal example discussed in section
7.  One might well imagine that a theta function like solution
may exist to the master anomaly equation, though we were
not able to find one.  The attempt is complicated by the fact
that not only one wishes to find a solution to the master
equation, but a solution which satisfies the correct
boundary conditions (dictated by the genus--$1$ answer).
Of course finding a solution to the master anomaly equation, even if
it satisfies the correct boundary condition is no guarantee
to be the correct amplitudes given by the string amplitudes, just
because the anomaly equation only captures the anti-holomorphic
dependence of the partition function on the moduli.  One still
has the freedom to correct it order by order by addition of
holomorphic terms.  Indeed changing the holomorphic dependence
at a given genus will affect even the non-holomorphic dependence
for any higher genus computation.  Finding a nice
way to fix the holomorphic dependence, even though it
just means fixing a finite number of coefficients at each
order, is a major challenge.  The most logical way to proceed,
in the case of the B-model
is to study the Kodaira--Spencer perturbation theory which naturally
will also give the holomorphic part as well as
the anomalous part of the amplitudes.  Otherwise we have rather
limited resources to fix the holomorphic part of the amplitudes.
Mirror symmetry helps, as
it did in the examples considered in section 7, in fixing
some of the low genus answers by relating it to counting holomorphic
maps of genus $g$ to a target space.
  But even in these examples the attempt was complicated by the
fact that for a given genus $g$ the lower genus holomorphic maps
may contribute as a kind of degenerate contribution to the
genus $g$ amplitude (which in the case of genus 0 contribution
of degree one is called the `bubbling').   A deeper understanding
of these general bubbling phenomena would be greatly helpful
in fixing the holomorphic ambiguity of the solution
to the anomaly equation.  The situation in understanding
these contributions can significantly improve through
collaboration between algebraic geometers and physicists.  Some
discussions of the bubbling phenomena appears in appendix A.

One of the most mysterious aspects which emerged in the
course of solving the anomaly equation (see section 6) was the
appearance of Feynman rules involving
propagation of massless modes and the dilaton.
This was rather unexpected and
needs to be understood better.  In a sense it seems
to suggest that effectively we can add the massless modes
as dynamical fields to the string field theory
despite the fact that we had to delete
them in order to write the string field action in
section 5. In this interpretation putting
back the massless fields in the theory is effectively
a way to restore background independence and so would
suggest that including the massless modes would
simply lead to answers which are independent of $t,\overline t$
thus explaining the Feynman graph rules we found for computing $F_g$.
  In fact the propagator we have for the
marginal fields, which is formally identified with $b_0\bar b_0/L_0$
and is ill-defined $0/0$ is effectively `regularized' by the propagator
$S^{ij}$ introduced in section 6.  In fact one can `formally' derive
the defining property of $S^{ij}$ from this definition using
a $tt^*$-type argument.  It would be interesting to develop
this further as well as see how the propagators involving
the dilaton field will appear.  At any rate demystification
of the Feynman rules that we found is a very important
hint in progress in a better understanding of these theories.

Perhaps the most important aspect of the present work is the
discovery of a new topological gravity theory in six dimensions, the
Kodaira--Spencer theory.  It is topological in the sense that it is
independent of the metric of the Calabi--Yau manifold, though it depends on
the complex structure chosen.  This topological theory is the target space
description of a topological worldsheet theory on a Calabi--Yau. The fact
that there is a string theory description of this theory makes us believe
that the ultra-violet
divergencies of the KS theory are not a real obstacle to its
existence and strings can be viewed as effectively giving a `nice
regularization' of the theory (deforming it from the manifold space to the
loop space).  Nevertheless it should be interesting to regularize the KS
theory using the more standard regularization techniques of field
theories.  In particular it should be possible to derive the holomorphic
anomaly master equation directly in this field theory set up for all loops.
The one--loop version of the anomaly
was checked explicitly to agree with field theory one using the zeta function
regularization techniques (which were used in \bif ).

  In more than one way the KS theory in 3 complex dimensions mirrors its
cousin the Chern--Simon theory in 3 real dimensions.  It is a closed string
version of Chern--Simon theory. Thus just as one has interesting
topological invariants in the Chern--Simon theory, giving link invariants on
three manifolds, one also expects the same here in the context of
invariants associated to Calabi--Yau $3$--folds (or more abstractly
classification question of variation of Hodge structures which arise in
superconformal theories).  This aspect is worth more thought.  Also the
open string version of strings on Calabi--Yau is a mirror to ordinary
Chern--Simon theory. So in this setup the coupling of this mirror theory to
KS theory is interesting to study.  In particular the holomorphic
anomalies in the open string sector discussed in this paper
 should be the mirror transformed versions of (a certain limit of)
Chern-Simon theory's anomalous dependence on the metric of the 3 manifold
which has been studied recently \ref\axsin{S. Axelrod and I. Singer,
{\it Chern-Simons Perturbation Theory }, MIT preprints, hep-th/9110056,
hep-th/9304087}\ to all loops.
It would be interesting to work out the detail of the anomaly equation
for the open string case which, except for the one--loop case which
we computed in detail, we just briefly discussed
in this paper.
This is more urgent in view of the fact that gaugino condensates which
are believed to be a mechanism to break supersymmetry in string theories
will be strongly affected by such terms.  This
aspect of the present work, which may have
potential relevance in questions of phenomenology, i.e., the fact
that topological partition functions may also be viewed as
particular computations in certain string models
compactified on the corresponding topological theory we find rather
significant. Not only topological theories can be used to compute
some amplitudes in ordinary strings, but the amplitudes
that they compute are the most interesting ones to compute, i.e.
the superpotential terms.
This opens the door to exact computations in string theories
using topological techniques.  Amplitudes which
are computable, at least in the context of open superstrings,
 will be of interest also
in connection with gaugino condensates which has been proposed as a mechanism
to break supersymmetry.  In fact it would be quite satisfactory
that deep facts such as supersymmetry breaking be linked
to very natural topological computations.
It would be nice to extend these computations to the heterotic
case in view of the potential phenomenological implications.
The fact that heterotic strings morally should behave
like the open strings suggests that even in this case
the topologically formulated heterotic string should compute
similar superpotential terms, as would be interesting in questions
of gaugino condensates.  At any rate it would
be very important to determine the consequences of such terms
in the supersymmetry breaking scenarios in string theory.

The crucial link needed to establish
topological theories with conventional superstring computations
 was the observation that basically the twisting of
an ordinary superstring is equivalent to insertion of
an appropriate number of FMS spin operators which twists
the field measure.  Even though topological amplitudes correspond
to very special amplitudes in string theories it is natural
to ask whether one can formulate {\it arbitrary} amplitudes
in superstring theories using the twisted topological models
by inclusion of non--topological operators (including
conjugate FMS spin operators to untwist the measure).
If such a formulation can be done it would be a step forward
in that one would not have to deal with issues of summing
over spin structures or the question of splitness of supermoduli
space, both of which are naturally absent in the topological
theory because the spin of all the fields are integral.
In fact results of \ref\Berk{N. Berkovitz, Nucl. Phys.
B395 (1993) 77}\ suggest that
this should be possible.

Another aspect of the present work was the fact that in all the
examples studied, the large area behavior of the genus $g$ partition
function of the topological theory on a Calabi--Yau $3$--fold is
in the same universality class as the $c=1$ theory coupled to
gravity (i.e. has the same exponents).  This result shows
that the identification of the $c=1$ theory coupled to gravity
with a particular supersymmetric coset representation
of black hole with $\hat c=3$ discovered
in \mv\
which is a non-unitary $N=2$ twisted model, is actually only the
tip of the iceberg.  Indeed what we have found seems to strongly
suggest that the universality class of $c=1$ strings is the
same as that of topologically twisted $\hat c=3$ theories.
It would be interesting to study this connection further.  In particular
for each Calabi--Yau manifold $M$ the large `worldsheet'
area $A\gg 1$ behaviour should
go like
$$a_g(M)\, A^{2g-3}{\rm log}^{2-2g} (A)\, {\rm exp}[b(M) A].$$
It would be interesting to compute $a_g(M)$ for all
$g$ and for all Calabi--Yau manifolds ($b(M)$ can be computed from the genus
zero result if one knows the mirror manifold).  For a fixed
$g$ how does the number $a_g(M)$ depend on $M$? Also
for a fixed $M$ how do the numbers $a_g(M)$ depend on $g$?  Do
they satisfy recursion relations of the type encountered in
topological theory coupled to gravity?

There are even more connections with $c=1$.  Indeed as
pointed out in \ref\witzwi{E. Witten, Nucl. Phys. B373 (1992) 187\semi
E. Witten and B. Zweibach,
Nucl. Phys. B377 (1992) 55}\ the target
space physics of $c=1$ strings has the symmetry of volume preserving
diffeomorphism.  As discussed in section 5
this is precisely the gauge symmetry of the
Kodaira--Spencer theory which is the target space physics of the
critical topological strings.    This relation is also worth further
investigation and is suggestive of the universal relation between
$c=1$ strings and $\hat c=3$ topologically twisted theories.

The special status $\hat c=3$ topological string
enjoys among more general topologically twisted theories,
is very much analogous to the special status $c=1$ strings enjoys
among all the theories with $c\leq 1$ coupled to gravity.
It is natural to ask if what we have been discussing in connection
with unitary $\hat c=3$ twisted theories has any bearing on the
more general classes of possibly non--unitary theories (as
is the case with the theory discussed in \mv ) or twisted
theories with $\hat c <3$ (as is the case for the minimal
$N=2$ twisted theories which is related to the $(1,p)$ theories
coupled to gravity \Li, \Vers ).  The central
question is whether the anomaly equation should exist in these
cases.  In fact morally it should be true but to make it  precise
a few technical obstacles should be overcome:  In the context
of non--unitary theories one has to argue that the cohomology
elements are the only ones that contribute for long tubes
(this is no longer guaranteed in the non-unitary case).
In the context of $\hat c<3$
models one has to recall that in order to get non--zero amplitudes
one will have to perturb the corresponding conformal theory in two
directions:  The massive direction, as well as turning on the
gravitational (or topological) descendants.  Turning on relevant
perturbations which makes the theory massive raises the question of
whether we can still integrate over conformally inequivalent classes
of metric.  Even if this can be done, we will have to know the analog
of Zamolodchikov metric for these massive theories.  One would
imagine that the analog of $tt^*$ equations which is
also known for the massive \cv\
case should be relevant
(in fact the results of \ref\ising{S. Cecotti and  C. Vafa, {\it
Ising Model and
N=2 Supersymmetric Theories}, HUTP-92-A044, hep-th/9209085}\
suggest that the one--loop partition function should be
related to the tau-function).  However it is not
completely straightforward because as we discussed in section 2
the Zamolodchikov connection and $tt^*$ connection differ
by a term involving the connection on the line bundle $\cal L$.
Unfortunately in the massive case the line bundle $\cal L$ is not
a holomorphic sub-bundle of the vacuum bundle and so this prevents
one from constructing canonical connections on it\foot{In the context
of integrable massive perturbations $\cal L$ is typically a sub-bundle
because of discrete symmetries and
one can define the corresponding line bundle connection.
It would be interesting to study this particular class further.}.
This will have to be better understood.   Another direction of
perturbation is turning on the gravitational descendants (which
are in particular needed for a non--vanishing amplitude at
higher genus for the twisted minimal models coupled to gravity).
The correlations involving topological descendants can typically
be viewed as boundary contributions to the amplitudes \Vers .
Thus one would expect an interesting mixture with the anomaly
discussed in this paper.  In this connection the Landau-Ginzburg
formulation of the descendants may be particularly useful \ref\loeg{A.
Losev,{\it Descendants constructed from matter fields in topological
Landau-Ginzburg theories coupled to topological gravity}, hepth/9211089
\semi
 A. Losev and I. Polyubin, {\it
On Connection between Topological Landau-Ginsburg
Gravity and Integrable Systems}, hep-th/9305079\semi
T. Eguchi, H. Kanno, Y. Yamada and S. -K. Yang, {\it
Topological Strings, Flat
Coordinates and Gravitational Descendants}, hep-th/9302048}.

Typically string theories have infinitely many particles.
However there are some cases known where string theory
has only a finite number of particles.  Precisely in these cases
the string theory
seems also to be related to topological theories
both in the sense of world sheet {\it and} in the sense of target theory.
Let us summarize some of the known examples and speculate
on the relation between them.

Let us summarize some of the most important known topological
field theories:  Apart from the 6-dimensional one that
we discovered in this paper,
and its open string analog \witcs , there
are two important topological theories in 4-dimensions,
topological gravity and topological Yang-Mills theory
(Donaldson theory)\ref\wit4d{E. Witten, Comm. Math. Phys. 117 (1988) 117 \semi
E. Witten, Phys. Lett. 206B (1988) 601 },
in 3-dimensions one has the Chern--Simon theory \ref\witcsor{E. Witten,
Comm. Math. Phys. 121 (1989) 351}\ and in 2-dimensions one has topological
sigma--models
and topological gravity theories, and topological Yang-Mills theories
\wittop.
Amazingly enough {\it almost all} of these theories seem to be
describing the target space physics of some string theory:  The
6-dimensional KS theory is the target space physics of critical
topological strings as we have discussed in this paper.
The 4-dimensional topological theories seem also to be
related to target space of $N=2$ strings \ov\ in that the
relevant target space geometry in both case involves
self-dual geometries\foot{This connection needs to be
clarified further.}.  The 3-dimensional CS theory is
equivalent to open string topological theory \witcs\ .
Finally the 2d topological YM theories, which are equivalent
to ordinary 2d YM theories may also be viewed as a
string theory using the results of \ref\gr{D. Gross, Nucl. Phys.
B400 (1993) 161\semi
D. Gross and W. Taylor, Nucl. Phys. B400 (1993) 181,
Nucl. Phys. B403 (1993) 395}\
which could also be viewed even as a topological
string theory (a deformed topological sigma--model
coupled to gravity \bcov\ \ref\di{R. Dijkgraaf, K. Intriligator and
 R. Rudd unpublished}).
So it seems that many of these topological field theories
are string field theories of string theories which themselves
are topological (i.e. are coupling of 2d topological theories
to topological gravity).  The completion of this picture
suggests that the $N=2$ strings
should have a reformulation
as a topological theory in the worldsheet sense, and should also be
able to obtain
topological sigma--models and topological gravity theories
in 2 dimensions as effective target space theories for
a topological world sheet theory.
Having such a unified picture also raises the question
of what are the relations between various topological
theories, and also their relation to integrable theories.
In a sense the six--dimensional topological theories
should play a key role in connecting them.  In particular self-dual
geometries arise naturally from considerations of holomorphic
vector bundles in six dimensions, through twistor transform.
One could speculate whether this formulation can be used to connect
it to the topological theory describing the open strings
on Calabi--Yau \witcs, which has as a solution
an arbitrary holomorphic vector bundle in six dimensions.
Similarly one may expect that the KS theory which characterizes
the complex structures of a six dimensional space be related
to the topological gravity theories in 4d, which characterize
self-dual geometries.  Clearly a lot more work
remains to be done.  We hope to have taken
one small step which may be helpful in the final emergence
of a unified picture.

\vskip .1in

This research was supported in part by NSF grants
PHY-87-14654 and PHY-89-57162 and a
Packard fellowship. H.O. is also supported
in part by grant-in-aid for scientific research on
priority area ``Infinite Analysis'' from the
ministry of science, education and culture of Japan.

\vskip .1in

We would like to thank S. Axelrod, S. Cordes,
C. Faber, S. Hosono,
 K. Intriligator, S. Katz, K.S. Narain,
I. Singer, T. Taylor,
G. Tian, E. Witten and S.T. Yau for valuable discussions.
H.O. also thanks the Lyman Laboratory of Physics, Harvard
University where most of this work was carried out, and
the Institute for Theoretical Physics
at Santa Barbara where a part of this work was done.
C.V. also thanks the Rutgers physics department for its
hospitality and for providing stimulating environment
during the final stage of this work.
Also we wish to thank K. House for her assistance in the preparation
of figures appearing in this paper.

\vfill
\eject

\appendix{A}{The Bubbled Torus}

In this appendix we shall rederive from a physical viewpoint the
result by Katz (see appendix of \bcov)
on the bubbling of spheres in genus $1$. Some preliminary computation
for $g>1$ is also presented.

As before, $\psi$ denotes a fermi field which is a section\foot{Strictly
speaking, this description is adequate for {\it non--degenerate}
instantons only.} of $K\otimes f^\ast T_M^\ast$, whereas $\chi$ is
a section of $f^\ast T_M$.

The basic degenerate instanton in genus one is given by
(here $D\subset M$ is a rational curve rigid in $M$ and $C$ is the
world--sheet torus)
$$\Gamma_{pq}\subset C\times D\subset C\times M,$$
given by
\eqn\dege{(C\times \{q\}) \cup (\{p\}\times D).}

As always, the $\chi$ zero--modes are in one--to--one correspondence with
the `collective coordinates' describing the given family of instantons.
In the present case we have just two of them, corresponding to the freedom
of choosing the two points $p$ and $q$ in $C$ and $D$, respectively.
We are also interested in finding the zero--modes of $\psi$ in this
configuration.
 From the viewpoint of an `observer' in a `generic' point of the
torus\foot{I.e. not at the point $p$ where the `bubble' is attached.},
the situation looks as follows.
The torus $C$ gets mapped into the point $q\subset D\subset M$. Then ---
as for any constant map --- the pullback of $T_M$ to $C$ is trivial,
and ${\rm dim}\, H^1(C,T_M)=3$. However, there are not really three
obstructions, since the deformation of $q$ in the direction tangent
to $D$ is `not obstructed'. So we remain with the two zero--modes
which are orthogonal to $D$ at $q$, in agreement with the index theorem
which predicts an equal number of $\chi$ and $\psi$ zero--modes. Our
observer on $C$ understands this as follows.
For him the instanton $\Gamma_{pq}$ arises
by the following limiting process:
One constructs an {\it approximate} solution mapping the world--sheet to the
(rational) curve $D$ by taking an usual instanton on the plane, of
scale\foot{By conformal invariance, there are instanton of any arbitrary
small scale.} $a$ much smaller than the periods of the torus $C$ and
`gluing' it at the point $p$. Then letting $a\rightarrow 0$ we get
a true solution which --- to our observer --- looks like a `delta--function'
instanton centered at $p$.
To be specific, let us identify $D$ with
${\bf P}^1$ in such a way that the point
$q$ is taken as the origin. Then our approximate instanton reads (for $w\sim
p$)
$$ f(w)= {a\over w-p}.$$
The pullback K\"ahler form for $w\sim p$ reads
(say, taking $D$ to be a line, and the metric to be the one induced
 by Fubini--Study)\foot{The normalization of
the $\delta$'s is such that the integral of the r.h.s. is just $1$.}
$${a^2 dz\wedge  d\bar z\over (2\pi i) (|z|^2+a^2)^2}\Bigg|_{a\rightarrow 0}
= {1\over i}\delta(z)\, \delta(\bar z) dz\wedge d\bar z ,$$
where $z\equiv w-p$. Of course this is just the statement that we have
a $\delta$--function instanton. Clearly if $D$ is a degree $k$ rational
curve this generalizes to
\eqn\observer{\Bigg({\rm the\ pulled\ back\ Kahler\ class\atop
as\ seen\ by\ our\ observer\ on\ {\it C}}\Bigg)= -i k\, \delta(z)\,
\delta(\bar z) dz\wedge d\bar z.}
 From the viewpoint of this observer, as $a\rightarrow 0$ the instanton
disappears, leaving a local operator inserted at the point $p$. This operator
implements a boundary condition at $p$ for the $\psi$ zero modes; it is
this condition that gets rid of the tangent component of $\psi$ leaving
just the two `normal' components.

To see the nature of the above boundary condition we have to discuss the
situation from the viewpoint of a second observer on $D$. From the point
of view of this observer, the limit $a\rightarrow 0$ is accompanied by
a compensating conformal rescaling by $a^{-1}$, so that to him the instanton
looks to have a finite size in the limit. However, at $a=0$ he happens to
be in a different $2d$ `universe' with respect to the other guy (i.e. on $D$).
 For this observer the pulled back K\"ahler form is
\eqn\secondguy{\Bigg({\rm the\ pulled\ back\ Kahler\ class\atop
as\ seen\ by\ our\ observer\ on\ {\it D}}\Bigg)= i^\ast \omega_M,}
where $\omega_M$ is the K\"ahler form for $M$
and $i\colon~D~\rightarrow~M$
is the embedding. Putting together the two `universes' one gets
\eqn\kahler{{\rm the\ pulled\ back\ Kahler\ class}=-i k\, \delta(z)\,
\delta(\bar z) dz\wedge d\bar z+i^\ast \omega_M,}
which should be compared with Katz's result, i.e. $E_1+h$
(see appendix of \bcov).

 From the viewpoint of the second observer, there are no zero--modes for
$\psi$, since on the sphere
$$H^1(T_M)\simeq H^0(K\otimes t^\ast T^\ast_M)=0.$$
In order for a zero--mode to be regarded as vanishing by the observer
on $D$, it should have a vanishing invariant norm as $z\rightarrow p$.
Let $\psi_=$ be the tangent would--be zero--mode. Near $p$ its norm reads
$$\|\psi_=\|^2=\left(1+{a^2\over |z|^2}\right)^2 |\psi_=|^2$$
which is divergent as $z\rightarrow 0$, unless $\psi_=(0)=0$. But a
holomorphic function\foot{Recall that from the viewpoint of the first
observer $T_X$ is trivial and so is $K$.} vanishing at one point vanishes
everywhere. Instead for the `normal' zero modes\foot{We use that
$$T_X|_D={\cal O}(2)\oplus {\cal O}(-1)\oplus{\cal O}(-1).$$}
$$\|\psi_\perp\|^2= {|\psi_\perp |^2\over (1+ a^2/|z|^2)},$$
which vanishes at $z=0$ {\it for any} $\psi_\perp$. This shows that we have
just two $\psi$ zero modes (as required by the index theorem). The structure
of these zero--modes is as predicted by Katz.

The moduli space of the above configuration is ${\cal M}_{1,2}
\times {\bf P}^1$
(here we identify $D\simeq {\bf P}^1$).
As we vary the point in the moduli
space, the zero modes $\psi_{ia}$ ($a=1,2$)
will also vary, giving a bundle $\CB$ over the above moduli space.
By construction
$$\CB= \pi_1^\ast\CH\otimes \pi_2^\ast\CN,$$
where $\CH$ is the Hodge (line) bundle  over ${\cal M}_{1,1}$ whose
fiber is spanned the holomorphic one--forms for the corresponding
elliptic curve, and $\CN$ is the normal bundle to $D$ in $X$.
The curvature of $\CB$ has the structure $1\otimes P+ \tilde R\otimes 1$,
where $P$ is the Hodge bundle  curvature as computed in \S.1.1, and
$\tilde R$ is the curvature of the normal bundle ${\cal N}=T_M/T_D$.
As it is well known
\eqn\second{\tilde R= R\Big|_{\cal N}-\theta\wedge\theta^\dagger}
where $\theta$ is the 2nd--fundamental form of
$$0\rightarrow T_D\rightarrow T_M\rightarrow {\cal N}\rightarrow 0.$$
Thus, in general is not true that the curvature of the normal bundle
is the restriction of the curvature of the tangent bundle to the
normal directions. However, is $D$ a rigid sphere on some Calabi--Yau
$3$--fold $M$,
$$T_M\simeq {\cal O}(2)\oplus {\cal O}(-1)\oplus {\cal O}(-1),$$
and hence the bundle splits and the extra term in \second\ should be
an exact form. Then the curvature is\foot{We do not write indices for
$N$ and $P$ since in the present case the indices can take only one value.}
\eqn\errebi{\Big({\cal R}_\CB\Big)^i_{\ j}= \delta^i_{\ j} N^{-1}P
dy\wedge d\bar y+R^i_{\ j k\bar l}dx^k\wedge d\bar x^{\bar l}+\dots,}
where $\dots$ means exact terms and $x$ are the coordinates of the point
$q$ in $M$.

Let us summarize: there are $2$ zero--modes for
$\chi$ corresponding to the deformations of the
 point $p$ and $q$, and two zero modes for $\psi_i$.
Then consider the quantity
$$-\int\limits_{{\cal M}_{1,1}}\left\langle \Big(dy \int\limits_C
\mu \psi_i\partial X^i\Big)\bigwedge
\Big(d\bar y \int\limits_C \bar\mu \bar\psi_{\bar j}\bar\partial
\bar X^{\bar j}\Big)
\big(k_{m\bar l}\chi^m\bar\chi^{\bar l}\big) \int\limits_C{R_{k\bar
h}}^{n\bar p}\chi^k\bar\chi^{\bar h}\psi_n\bar\psi_{\bar p}\right\rangle$$
i.e. the $g=1$ one--point function. We wish to compute the contribution
of the (single) degenerate instanton to the above quantity.
The subtle identity is \kahler. It means that we have the replacement
\eqn\strange{k_{i\bar j}(0)\chi^i\bar\chi^{\bar j}\mapsto {\rm deg}(D)\
\delta(z)\delta(\bar z)\chi^z\bar\chi^{\bar z}+ \tilde k
\chi^x\bar\chi^{\bar x},}
where $\chi^z$ (resp. $\chi^x$) is the zero mode associated
to the variation of the coordinate $z$ (resp. $x$)
of the point $p$ (resp. $q$) on $C$ (resp. $D$),
and $-i\tilde k dx\wedge d\bar x\equiv i^\ast \omega_M$.

Integrating away the $\chi$'s and the $\psi$'s (using
the same formulae as in \S.1.2) we reduce to an integration
over the boson zero--modes, of the expression
$${{\rm deg}(D)\over (2\pi i)^2} \int\limits_{{\cal M}_{1,2}\otimes D}
\delta(z)\delta(\bar z)dz\wedge d\bar z \det[{\cal R}_\CB],$$
where ${\cal R}_\CB$ is given in \errebi. This can be rewritten as
$${\rm deg}(D)\ \int\limits_{{\cal M}_{1,1}\otimes D}
c_2(\CH\otimes{\cal N}).$$
A simple computation gives (recall ${\cal N}={\cal O}(-1)\oplus {\cal O}(-1)$)
$$c_2(\CH\otimes {\cal N})= 2\, c_1(\CH)\, c_1({\cal O}_D(-1)).$$
Finally from
$$\eqalign{&2\, {\rm deg}(D)\ \int\limits_{{\cal M}_{1,1}}
c_1(\CH) \int\limits_D c_1({\cal O}(-1))= \cr
&=2\, {\rm deg}(D)\, {\rm deg}({\cal O}(-1))\, \chi({\cal M}_{1,1}) =
{2\over 12}{\rm deg}(D).\cr}$$
\vglue 14pt
{\it Preliminary Considerations for Genus $g>1$}
\vglue 10pt

If $M$ is simply--connected it is also algebraic. Then let $\omega_0$ be the
K\"ahler form induced by the imbedding of $M$ inside
$\Pr N$. By degree of a curve $\CC$ lying on $M$ we mean $\int_\CC\omega_0$.
Then a curve of degree one is a line in $\Pr N$ and hence it is necessarily
rational.
Therefore for all $g$ the $O(e^{-t})$ contribution to $F_g$
should arise from maps of the form
$$\Sigma_g\buildrel f \over \rightarrow D\buildrel i \over \hookrightarrow M,$$
where $D$ is a degree $1$ rational curve on $M$, $i$ the inclusion and $f$ some
degree $1$ holomorphic map. However,
for $g>0$ there is no such a thing as a degree $1$ meromorphic function. Thus
at first sight, it may seem that for $g>0$ the
$O(e^{-t})$ term in $F_g$ should vanish. However, it is not so as was shown
explicitly in ref.\bcov\ for $g=1$. The point is that although there is no
smooth
instanton in this topological class, we can construct an approximate solution
mapping $\Sigma_g$ to $D$
by taking an usual instanton on the plane for the $\Pr 1$ sigma--model, of
scale\foot{By conformal invariance, there are instanton of any arbitrary
small scale.} $a$ much smaller than the periods of the curve $\Sigma_g$ and
`gluing' it at the point $p\in \Sigma_g$. The approximation gets better and
better as $a\rightarrow 0$.
In the limit we get
a solution which looks like a `delta--function'
instanton centered at $p$. However, there is a better viewpoint. By conformal
invariance, while we let $a\rightarrow 0$ we can do a compensating scale
transformation in a neighborhood of $p$ such that the instanton remains of a
finite scale in the limit. In this picture, as $a\rightarrow 0$ a sphere will
`bubble off' the world--sheet. In terms of the graph $\Gamma$ of the map
$\Sigma_g\rightarrow D$, the resulting degenerate instanton will be ($q$ is a
point in $D$)
$$\Gamma_{pq}\subset \Sigma_g\times D\subset \Sigma_g\times M,$$
given by
\eqn\dege{(\Sigma_g\times \{q\}) \cup (\{p\}\times D).}
That such singular instantons like $\Gamma_{pq}$ should be taken into account,
follows from Gromov's theory of symplectic invariants; as $\bar t\rightarrow
\infty$ the functional measure gets concentrated on the critical points only if
the integration space (the `space of all maps') is compactified. Otherwise the
instanton may `escape to infinity'. Now, these configurations \dege\ belong to
the Gromov compactification of the `space of all maps'.

Comparing with the genus one case, it appears
that the following computation should be relevant for the
higher genus bubbling
\eqn\bubbleout{\int_{\CM_{g,1}\otimes D}
C_{3g-1}(\CB).}
where
$$\CB=\pi^\ast_1\CH\otimes \pi_2^\ast {\cal N}^\ast\oplus
\pi^\ast_1\tilde\CH \otimes \pi_2^\ast T_D^\ast.$$
Here $\CN$ is the normal bundle to $D$ in $M$ and $\CH$ is the Hodge vector
bundle as before.
The fiber of $\CH$ is $H^0(\Sigma_g,K)$. Instead $\tilde\CH$ is the bundle with
fiber $\Gamma({\cal O}(-p)\otimes K)$ --- that is the holomorphic one--forms
vanishing at $p\in \Sigma_g$. Obviously we have the following
exact sequence
\eqn\sequence{0\rightarrow \tilde \CH\rightarrow\CH\rightarrow L\rightarrow 0,}
where $L$ is the line bundle over ${\cal M}_{g,1}$ whose fiber is $T^\ast_p$.

{}From the definition of $\CB$ one has
$$c(\CB)= c(\CH\otimes {\cal N}^\ast)\, c(\tilde\CH\otimes T_D^\ast)$$
and then
$$c_{3g-1}(\CB)= c_{2g}(\CH\otimes {\cal N}^\ast)\, c_{g-1}(\tilde\CH\otimes
T_D^\ast)$$

Since $D$ is one--dimensional
$$\eqalign{&c_{2g}(\CH\otimes \CN)= c_g(\CH)^2-2 c_1(\CN)\, c_g(\CH)\,
c_{g-1}(\CH)=2 c_1(\CN)\, c_g(\CH)\, C_{g-1}(\CH)\cr
&c_{g-1}(\tilde\CH\otimes T_D)=c_{g-1}(\tilde\CH)-c_{g-2}(\tilde \CH)\,
c_1(T_D).\cr}$$
where, in the first line we used \twogrel.
Now we see that
$$c_1(\CN)=2\, c_1({\cal O}_D(-1)),\qquad c_1(T_D)=c_1({\cal O}_D(2)).$$
On the other hand, from \sequence\
$$c(\CH)=c(\tilde\CH)\, c(L)=c(\tilde\CH)\, \big(1+c_1(L)\big),$$
or equivalently,
$$c(\tilde H)=c(\CH)\, \Big(1+\sum_{k=1}^g (-1)^k c_1(L)^k\Big),$$
which in particular gives
$$c_{g-1}(\tilde\CH)=\sum_{k=0}^g (-1)^k c_{g-k-1}(\CH)\, c_1(L)^k.$$
Then
$$\eqalign{\int_D c_{3g-1}(\CB)&=\int_D \left[ c_{2g}(\CH\otimes \CN^\ast)\,
c_{g-1}(\tilde\CH\otimes
T_D^\ast)\right]=\cr
&=4\, c_g(\CH)\, c_{g-1}(\CH)
\sum_{k=0}^g (-1)^k c_1(L)^k\, c_{g-k-1}(\CH).\cr}$$
The term with $k=0$  vanishes, since the integrand is the pull--back of a
$(3g-2)$--form on ${\cal M}_{g,0}$. Also the term with $k=g$ vanishes for
trivial
reasons. Comparing with \mumford\ this can be rewritten in terms
of Mumford classes as
$$4\sum_{k=1}^{g-1}(-1)^k \int_{\CM_g} \lambda_g\wedge
\lambda_{g-1}\wedge\lambda_{g_k-1} \wedge \kappa_{k-1}.$$
It remains to understand the precise relation between this
Chern class computation and the actual bubbling coefficient.

\vfill
\eject

\appendix{B}{Further analysis on the master anomaly equation}

In section 3, we found that the generating function
\eqn\appengenerating{ W(\lambda,x;t,\overline{t}) =
 \sum_{g=0}^\infty \sum_{n=0}^\infty {1 \over n!} \lambda^{2g-2}
                    C_{i_1 \cdots i_n}^{(g)} x^{i_1} \cdots x^{i_n}
           + \left( {\chi \over 24} -1 \right) \log \lambda }
is characterized by the two equations,
\eqn\appendelbar{\eqalign{ &
  {\partial \over \partial \overline{t}^i} \exp(W) = \cr
 & = \left[ {\lambda^2 \over 2} \overline{C}_{\bar{i}\bar{j}\bar{k}}
           e^{2K} G^{j\bar j} G^{k \bar k}
    {\partial^2 \over \partial x^j \partial x^k}
     - G_{\bar{i} j} x^j
  \left( \lambda {\partial \over \partial \lambda}
         + x^k {\partial \over \partial x^k}\right) \right] \exp(W) \cr}}
and
\eqn\appendel{\eqalign{ &
 \left[ {\partial \over \partial t^i} + \Gamma_{ij}^k x^j
     {\partial \over \partial x^k} + \partial_i K
    \left( {\chi \over 24} -1-\lambda {\partial \over \partial \lambda}
\right) \right] \exp(W) = \cr
& = \left( {\partial \over \partial x^i}
   - \partial_i F_1 - {1 \over 2 \lambda^2}
         C_{ijk} x^j x^k \right) \exp(W) .\cr}}
The first equation \appendel\ summarizes
the holomorphic anomaly equations for $C_{i_1 \cdots i_n}^{(g)}$, and
the second equation \appendel\ implies that
$C_{i_1 \cdots i_n}^{(g)}$ in $W(\lambda ,x;t, \overline{t})$
are given by derivatives of the partition function $F_g$.
In section 6, we developed a method to solve the holomorphic
anomaly equation order by order in $g$. In this appendix,
we analyze the two equations \appendelbar\ and \appendel\
directly to all order in $g$.
We hope that the method presented here would be useful
to understand non--pertubative aspects of the
string theory.

Let us first solve the anomaly equation \appendelbar\
without imposing \appendel. This turned out to be possible
by the Borel transformation in the string coupling constant $\lambda$
and by the Fourier transformation in $x^i$.
\eqn\borel{
  \exp( W(\lambda, x; t, \overline{t}))
  = \int dp\, dq \exp(  - \lambda^{-1} q + i \lambda^{-1} x^ip_i
                       + \Gamma(q,p; t , \overline{t}) ). }
The anomaly equation \appendelbar\ for $W$ is transformed
into the following first-order linear differential equation
\eqn\transmaster{
\left( {\partial \over \partial \overline{t}^i}+i
  G_{\bar{i} j} q {\partial \over \partial p_j} \right) \Gamma
  =- {1 \over 2} \overline{C}_{\bar{i}\bar{j}\bar{k}}
              e^{2K} G^{j\bar j} G^{k \bar k} p_j p_k }
A special solution to this equation is easily found as
$$ \Gamma_0(q,p;t,\overline{t}) =
   - {1 \over 2} S^{ij}
            p_i p_j + i S^i p_i q
               +  S q^2 .$$
This satisfies \transmaster\ by
the definitions of $S^{ij}$, $S^i$
and $S$
$$ \eqalign{ & S^{ij} = \overline{C}_{\bar{i}\bar{j}} e^{2K} G^{j \bar j}
                G^{k \bar k}~,~~
                S^i = \overline{C}_{\bar i} e^{2K} G^{i \bar i} ~,~~
                 S= \overline{C} e^{2K} \cr
            & \overline{C}_{\bar{i}\bar{j}\bar{k}} =
                D_{\bar i} \overline{C}_{\bar{j}\bar{k}}~,~~
             \overline{C}_{\bar{j}\bar{k}}= D_{\bar j}
\overline{C}_{\bar k}~,~~
               \overline{C}_{\bar k} = D_{\bar k} \overline{C} .\cr} $$
Since \transmaster\ is first-order
and linear, its general solution can be expressed as
\eqn\appengeneral{
\exp(\Gamma(q,p; t, \overline{t} )) = \vartheta( q,
 p -iq \partial \log( e^{K} |f|^2) ;
                               t) \exp( \Gamma_0(q,p;t, \overline{t})) }
where $f(t)$ is a meromorphic section of ${\cal L}$,
and $\vartheta$ does not depend on $\overline{t}$ except through
$e^{K}$ in the second argument.

The general solution \appengeneral , after the Borel transformation,
does not necessarily have the form \appengenerating . So we need
to impose the second equation \appendel . After the Borel
transformation \borel, \appendelbar\ becomes
$$ \eqalign{
  & \left[ {\partial \over \partial t^i}
            - \Gamma_{ij}^k p_k {\partial \over \partial p_j} - \Gamma_{ij}^j
             - \partial_i K \left(q {\partial \over \partial q}
                                + p_j {\partial \over \partial p_j}
            + n+2 - {\chi \over 24} \right) \right]
         \exp( \Gamma) = \cr
  & = \left( i p_i {\partial \over \partial q}
          - \partial_i F_1 +{1 \over 2} C_{ijk}
         {\partial^2 \over \partial p_j \partial p_k} \right)
         \exp(\Gamma) .\cr} $$
where $n$ is the dimensions of the moduli space of the $N=2$ theory.
Substituting \appengeneral\ in the above, we obtain a differential
equation for $\vartheta$ as
\eqn\heatequation{
  \eqalign{
 \Big( & {\partial \over \partial t^i}
     +f_{ij}^k(t)
       \tilde{p}_k {\partial \over \partial \tilde{p}_j} -
     i f_{ij}(t) \tilde{q} {\partial \over \partial \tilde{p}_j}
           -i \tilde{p}_i
             {\partial \over \partial \tilde{q}} \Big)
            \vartheta = \cr
&=\Big( {1 \over 2 } \tilde{C}_{ijk}(t)
          {\partial^2 \over \partial \tilde{p}^j \partial \tilde{p}^k}
       - {1 \over 2} e_i^{jk}(t)
          \tilde{p}_j \tilde{p}_k
       +i e_i^j(t) \tilde{p}_j \tilde{q}
     + e_i (t) \tilde{q}^2 + h_i(t)
  \Big) \vartheta ,\cr }}
where
$$ \tilde{p}_i = f^{-1} \left( p_i - i q \partial_i
\log(e^{K}|f|^2) \right),
{}~~~\tilde{q} = f^{-1} q,~~\tilde{C}_{ijk} = f^{-2} C_{ijk} . $$
Due to the special geometry relation, the coefficients
$$
\eqalign{ f_{ij}^k(t) =&
    C_{ijl} S^{lk} - \Gamma_{ij}^k
           - \delta_i^k \partial_j \log (e^K |f|^2) -
            \delta_j^k \partial_k \log (e^K |f|^2) \cr
    f_{ij}(t) =&
     C_{ijk} \left[ S^k - S^{kl} \partial_l \log(e^K |f|^2) \right]
     + D_i \partial_j \log ( e^K |f|^2) + \cr
      &+ \partial_i \log ( e^K |f|^2) \partial_j \log ( e^K |f|^2) \cr}
$$
are holomorphic in $t$, and so are
$$
\eqalign{
 e_i^{jk}(t)  =f^{-2}
    \Big[ & D_i S^{jk} + C_{imn} S^{mj} S^{nk} -
            \delta_i^j S^k - \delta_i^k S^j \Big] \cr
 e_i^j(t) =f^{-2}  \Big[ &
     D_i S^j + C_{imn} S^m S^{nj}- 2 S \delta_i^j + \cr
  & + (2 S^j \delta_i^k - D_i S^{jk}
        - C_{imn} S^{mj} S^{nk}) \partial_k \log (e^K|f|^2) \Big] \cr
e_i(t) = f^{-2} \Big[ &
 D_i S - {1 \over 2} C_{ijk} S^j S^k - \cr
& -( D_i S^j + C_{imn} S^m S^{nj} - 2 S \delta_i^j)
    \partial_j \log(e^K |f|^2) + \cr
  & + ( {3 \over 2} D_i S^{jk} +
           {1 \over 2} C_{imn} S^{mj} S^{nk}
         - S^j \delta_i^k) \times \cr
& ~~~~\times \partial_j \log(e^K |f|^2)
        \partial_k \log(e^K |f|^2) \Big]. \cr}
$$
Due to the genus-$1$ anomaly equation, $h_i(t)$ given by
$$ h_i(t) = \partial_i F_1 + {1 \over 2} C_{ijk} S^{jk}
              -\partial_i K \left( n+2 - {\chi \over 24}
 \right) - \Gamma_{ij}^j $$
is also holomorphic.

For each $t^i$,
the equation \heatequation\ is of the form of the
Schr\"odinger equation for a particle moving in an $(n+1)$-dimensional
space of $\tilde{p}^i$ and $\tilde{q}$ in a $t$-dependent harmonic
oscillator potential and a $t$-dependent constant magnetic field.
Since it is a first order differential equation in $t$,
we can solve it uniquely once we know $\vartheta$ at particular
value of $t$. The situation is similar to the case of the Wess-Zumino-Witten
(WZW) model on a Riemann surface
where the partition function satisfies the heat equation
with the moduli of the surface $\Sigma$ being time-like variables and
the the moduli of the holomorphic vector
bundle on $\Sigma$ being space-like variables.
It is known that the WZW model
is related to the three-dimensional Chern--Simons (CS) theory, and
the heat equation in the WZW model is identified as
the physical state condition for a wave--function in the CS theory.
The similarity between the WZW model and the Kodaira--Spencer
theory suggests that the Schr\"odinger type
equation \heatequation\ for the (Borel--transformed) generating
function $\vartheta$ may also be derived as a physical state
condition of some higher-dimensional system. It would be very
interesting to identify such a system. This would also explain
the origin of the finite dimensional quantum system discussed
in \witt .

\listrefs

\end